\documentclass[11pt]{article} 
\usepackage[utf8]{inputenc}
\usepackage{multirow}
\usepackage{amsfonts}
\usepackage{amsmath}
\usepackage{amssymb}
\usepackage{amsthm}
\usepackage{marvosym}
\usepackage{fontawesome5}
\usepackage{esint}
\usepackage{bbold}
\usepackage{bm}
\usepackage{booktabs, array}

\newcolumntype{M}{>{$}c<{$}}  
\usepackage[export]{adjustbox}
\usepackage[english]{babel}
\usepackage{authblk}
\usepackage{tikz}
\usetikzlibrary{shapes.misc,decorations.markings,arrows,decorations.pathmorphing,backgrounds,matrix,patterns,positioning,decorations.pathreplacing,calc}
\usetikzlibrary{arrows.meta,calc,patterns.meta,positioning}
\usepackage{xcolor}
\definecolor{axisgray}{HTML}{28323C}
\definecolor{highblue}{HTML}{BFE3F2}
\definecolor{highbluetext}{HTML}{24718F}
\definecolor{lowgreen}{HTML}{CFE8D2}
\definecolor{weakviolet}{HTML}{BEA2E3}
\definecolor{weakviolettext}{HTML}{7445A3}
\definecolor{criticalred}{HTML}{C43D3D}
\definecolor{unstablered}{HTML}{F4D5D2}
\definecolor{excludedgray}{HTML}{D9DDE1}
\definecolor{dgreen}{rgb}{0,0.5,0}
\definecolor{darkblue}{rgb}{0,0,0.6}
\definecolor{lblue}{rgb}{0.5,0.5,1.0}
\definecolor{purple}{rgb}{0.4,.2,0.7}
\definecolor{ggray}{rgb}{0.85,.85,0.85}
\definecolor{dpink}{rgb}{1,0.5,0.5}

\numberwithin{equation}{section}

\makeatletter
\usepackage{comment}
\usepackage{float}
\usepackage{caption}
\usepackage[margin = 2.5cm]{geometry}
\usepackage{graphicx}
\usepackage[hyperfootnotes = true, colorlinks = true, linkcolor = blue, citecolor = purple]{hyperref}
\usepackage{subcaption}
\usepackage{blkarray}
\def\tr{\mathrm{tr}}

\newcommand{\bes}{\begin{equation} \begin{split} }	
	\newcommand{\ees}{\end{split} \end{equation} }
	
	\renewcommand{\i}{\mathrm{i}}

	\renewcommand{\O}{\mathcal{O}}

	\newcommand{\evs}[1]{\langle #1 \rangle} %

	\newcommand{\inv}{^{-1}}

	\usepackage{physics}

\usepackage{amsfonts}
\usepackage{amsmath}
\usepackage{amssymb}
\usepackage{braket}
\usepackage{cite}
\usepackage{enumitem}
\usepackage{float}
\usepackage{graphicx}
\usepackage{mathrsfs}
\usepackage{setspace}
\usepackage{physics}
\usepackage{subcaption}
\usepackage{titlesec}
\usepackage{comment}
\usepackage{makecell}
\usepackage{marvosym}
\usepackage{wasysym}
\usepackage{simplewick}

\graphicspath{{./figures/}}

\usepackage[margin = 2.5cm]{geometry}
\usepackage{color}
\definecolor{darkblue}{rgb}{0,0,0.6}
\definecolor{purple}{rgb}{0.4,.2,0.7}
\definecolor{darkgreen}{rgb}{0,0.5,0}

\newcommand{\M}{\mathcal{M}}

\renewcommand{\i}{\mathrm{i}}
\renewcommand{\d}{\mathrm{d}}

\begin{document}

\thispagestyle{empty}
\begin{center}
    ~\vspace{5mm}
    
    {\LARGE \bf 
     Non-singlets and the phase diagram\\ of Matrix Quantum Mechanics
     }

   \vspace{0.5in}

Igor R. Klebanov,$^{a,b}$
Henry W. Lin,$^{a}$
Pavel Meshcheriakov,$^{a}$
and Jessica Yeh$^{c}$

\vspace{0.2in}

{\it
    $^{a}$Joseph Henry Laboratories and Leinweber Forum for Theoretical Physics,\\
    Princeton University, Princeton, NJ 08544, USA
    }

\vspace{0.1in}

{\it
    $^{b}$ Princeton Center for Theoretical Science,\\
    Princeton University, Princeton, NJ 08544, USA
    }
    
    \vspace{0.1in}

{\it
    $^{c}$Leinweber Institute for Theoretical Physics,\\
    Stanford University, Stanford, CA 94305, USA
    }
    \vspace{0.5in}

\end{center}

\vspace{0.5in}

\begin{abstract}
The quantum mechanics of one $N\times N$ Hermitian matrix in the infinite $N$ limit is well known to define a two-dimensional string theory when the quartic coupling constant $g$ is tuned to a special negative value $g_*$. We study the thermodynamics of this Matrix Quantum Mechanics (MQM), treating the SU$(N)$ symmetry as a global symmetry. Crucially, the non-singlets, i.e. states in non-trivial representations of SU$(N)$, implement the effects of vortices on the discretized worldsheets. This random lattice model has the low-temperature $c=1$ and high-temperature $c=0$ phases, with the Berezinskii-Kosterlitz-Thouless (BKT) transition separating them. We map out the phase diagram in the coupling--temperature plane, which is separated by the critical stability curve $T_c(g)$ extending from $g_*< g <0$. We study this curve using the non-perturbative matrix bootstrap and low- and high-temperature perturbation theory, which explain the thermal instability. At low temperatures, thermally excited non-singlets backreact on the Fermi sea of matrix eigenvalues, causing it to spill over. At high temperatures, the theory is described by an effective matrix integral obtained by dimensional reduction on the thermal circle; the eigenvalue distribution of the zero mode becomes unstable at the critical coupling. We conjecture that the BKT transition occurs at a special point on the critical curve and estimate its position. Finally, we switch to the regular square lattice and study the MQM-inspired Villain model, where the link factor is the one-dimensional propagator on a circle. Using Monte Carlo simulations, we map out the BKT transition in this model. 
\end{abstract}

\vspace{1in}

\pagebreak

\setcounter{tocdepth}{3}

\tableofcontents
\section{Introduction and summary}
In theories with $N\times N$ matrix degrees of freedom, the 't Hooft large $N$ limit restricts the sum over Feynman diagrams to the planar ones \cite{tHooft:1973alw}. Such diagrams may be thought of as discretized string worldsheets, so this provides a path towards connecting matrix models with the Polyakov path integral for non-critical string theory \cite{Polyakov:1981rd}. 
In this paper, we will be concerned with 
the quantum mechanics of a single $N\times N$ Hermitian matrix, which was introduced in the seminal paper by Br\'ezin, Itzykson, Parisi and Zuber (BIPZ) \cite{Brezin:1977sv}. This model, often referred to as Matrix Quantum Mechanics (MQM), has the
Hamiltonian\footnote{Following BIPZ, we focus on the quartic potential, although our methods extend to other metastable potentials with similar critical behavior. Here and throughout the paper $\tr$ denotes the normalized trace, $\tr\, \mathcal{O}=\frac{1}{N}\Tr\,\mathcal{O}=\frac{1}{N}\sum_i\mathcal{O}_{ii}$. Thus the effective Planck constant of the eigenvalue problem is $\hbar_{\mathrm{eff}}=1/N$ and $V$ is a polynomial whose coefficients are independent of $N$ in the ’t Hooft limit. In these notations $[P_{ij},X_{kl}]=-\frac{i}{N}\delta_{il}\delta_{jk}$. In Section \ref{sec:small-g-F}, we temporarily use canonically normalized matrix variables.}
\begin{equation}\label{MQM-Hamiltonian-def}
    H=N^2\, \tr\left(\frac{1}{2}P^2+V(X)\right)\ ,\qquad V(X)=\frac{1}{2}X^2+gX^4\ .
\end{equation}
Its SU$(N)$ invariant (singlet) sector is solvable because the $N$ eigenvalues of $X$ act as non-interacting fermions \cite{Brezin:1977sv}. In particular, the ground state is a singlet and is obtained by filling the $N$ lowest energy levels. The Euclidean path integral for MQM at zero temperature corresponds to the sum over Feynman graphs with each link connecting vertices at times $\tau_i$ and $\tau_j$ contributing the factor
$\frac{1}{2}e^{-|\tau_i - \tau_j|}$, which is the one-dimensional massive Green function. Although a standard discretization of the Polyakov string action \cite{Polyakov:1981rd} would instead contain a Gaussian link factor \cite{Klebanov:1991qa}, there is strong evidence that the exponential link factor found in MQM produces a theory in the same universality class.\footnote{For the theory on a fixed lattice with link factor $\frac{1}{2}e^{-|\tau_i - \tau_j|}$, it has been proven that a massless Gaussian free field emerges in the thermodynamic limit \cite{Armstrong:2023arxiv}.} 

MQM may be gauged by introducing the non-dynamical gauge field $A$, which restricts the dynamics to the singlet sector. 
We will instead be studying the original ungauged model \eqref{MQM-Hamiltonian-def}, which includes the richer physics of the non-singlet states. These states, whose study was initiated by Marchesini and Onofri (MO) \cite{Marchesini:1979yq}, are crucial for understanding the thermodynamic properties of the ungauged MQM in the large $N$ limit. We will revisit the conjectures \cite{Gross:1990ub, Gross:1990md} concerning the role of non-singlets for the Berezinskii-Kosterlitz-Thouless (BKT)  \cite{Berezinskii:1970pzv, Kosterlitz:1973xp} phase transition in MQM 
and obtain new insights into it.  

In the 't Hooft limit (in our conventions, $N\to \infty$ holding $g$ fixed), MQM \eqref{MQM-Hamiltonian-def} has a stable ground state for $g>g_*$,\footnote{For $g<0$, this refers to the perturbatively defined large-$N$ branch. The finite-$N$ decay is nonperturbative in $1/N$ and will not play a role in our analysis.}
\begin{equation}\label{g-crit}
  g_*=-\frac{\sqrt 2}{6\pi}\approx-0.075026\ . 
\end{equation}
As $g$ approaches this critical value from above, appropriate correlation functions are dominated by planar diagrams whose typical size diverges. 
In this limit, the distance $\mu$ of the Fermi energy to the top of the potential approaches zero\footnote{Furthermore,
there is an interesting double scaling limit where $N \to \infty, \mu \to 0$ holding $N\mu$  fixed \cite{Brezin:1989ss,Gross:1990ay,Ginsparg:1990as, Gross:1990ub}. It corresponds to string theory in two dimensions with string coupling $\sim \frac{1}{N\mu}$ (for a review, see \cite{Klebanov:1991qa}). In the strict 't Hooft limit we have a tree-level or classical string theory.}.  
The spectrum of adjoint states obtained from the MO equation \cite{Marchesini:1979yq} in the limit $\mu\rightarrow 0$ was recently found to have a nice correspondence with Regge trajectories of the oscillating folded strings \cite{Klebanov:2026vmk}.

\begin{figure}
    \centering
    \input{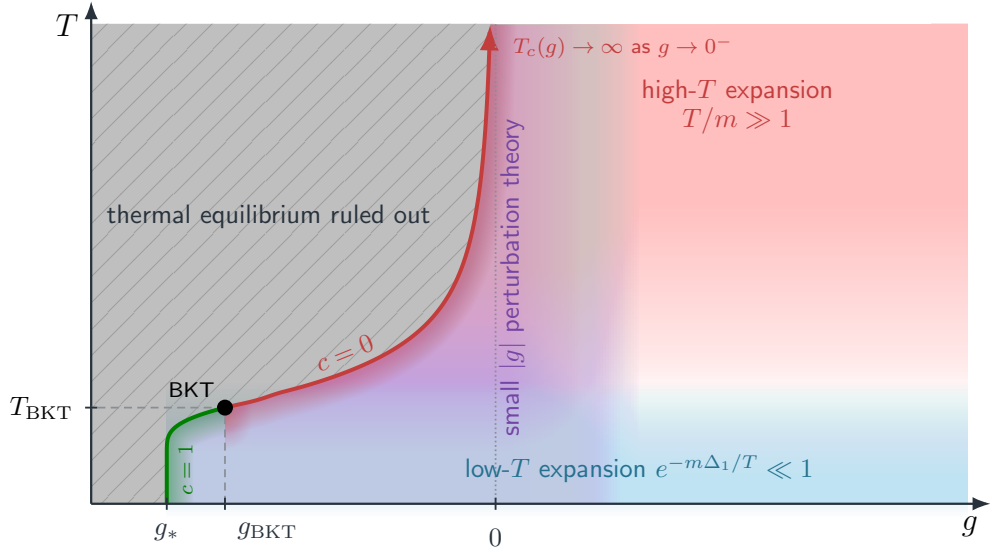}

    \begin{tikzpicture}[
      x=1cm,
      y=1cm,
      >={Latex[length=2.4mm,width=1.7mm]},
      axis/.style={draw=axisgray,line width=0.85pt,-Latex},
      region label/.style={
        align=center,
        font=\sffamily\small,
        text=axisgray,
        inner sep=1pt
      },
      note/.style={
        font=\sffamily\scriptsize,
        text=axisgray!80,
        align=center
      }
    ]

    \coordinate (gc) at (-4.35,0);
    \coordinate (origin) at (0,0);
    \coordinate (bkt) at (-3.580003,1.272141);

    \begin{scope}
      \clip
        plot[smooth] coordinates {\TcCurveAllCoordinates}
        -- (6.25,6.35)
        -- (6.25,0)
        -- (-4.35,0)
        -- cycle;

      \foreach \i in {0,...,59} {
        \pgfmathsetmacro{\bandhalfwidth}{1.90 - 0.037*\i}
        \fill[weakviolet,opacity=0.044]
          (-\bandhalfwidth-6,0)
          rectangle
          (\bandhalfwidth,6.35);
      }
    \end{scope}

    \begin{scope}
      \clip
        plot[smooth] coordinates {\TcCurveAllCoordinates}
        -- (6.25,6.35)
        -- (6.25,0)
        -- (-4.35,0)
        -- cycle;

      \foreach \i in {0,...,59} {
        \pgfmathsetmacro{\highbandbase}{3.35 + 0.050*\i}
        \fill[pink,opacity=0.025]
          (-4.2,\highbandbase-2)
          rectangle
          (6.25,6.35);
      }
    \end{scope}

    \foreach \i in {0,...,59} {
      \pgfmathsetmacro{\lowbandheight}{1.6 - 0.030*\i}
      \fill[highblue,opacity=0.04]
        (-4.35,0)
        rectangle
        (6.25,\lowbandheight);
    }

    \path[fill=gray,opacity=0.50]
      plot[smooth] coordinates {\TcCurveAllCoordinates}
      -- (0,6.35)
      -- (-5.35,6.35)
      -- (-5.35,0)
      -- cycle;

    \path[
      pattern={Lines[
        angle=45,
        distance=7pt,
        line width=0.35pt
      ]},
      pattern color=black!38
    ]
      plot[smooth] coordinates {\TcCurveAllCoordinates}
      -- (0,6.35)
      -- (-5.35,6.35)
      -- (-5.35,0)
      -- cycle;

    \begin{scope}

      \clip (-5.35,0) rectangle (6.25,6.35);

      \clip (-4.40,0) rectangle (-3.580003,6.35);

      \clip
        plot[smooth] coordinates {\TcCurveAllCoordinates}
        -- (6.25,6.35)
        -- (6.25,0)
        -- (-4.35,0)
        -- cycle;

      \clip
        plot[smooth] coordinates {\TcCurveGreenCoordinates}
        -- (-3.580003,0)
        -- (-4.35,0)
        -- cycle;

      \foreach \i in {1,...,24} {
        \pgfmathsetmacro{\greenfadewidth}{21 - 0.8*\i}
        \draw[
          dgreen,
          opacity=0.018,
          line width=\greenfadewidth pt,
          line cap=butt,
          line join=round
        ]
          plot[smooth] coordinates {\TcCurveGreenCoordinates};
      }
    \end{scope}

    \begin{scope}

      \clip (-5.35,0) rectangle (6.25,6.35);

      \clip (-3.580003,0) rectangle (6.25,6.35);

      \clip
        plot[smooth] coordinates {\TcCurveAllCoordinates}
        -- (6.25,6.35)
        -- (6.25,0)
        -- (-4.35,0)
        -- cycle;

      \clip
        plot[smooth] coordinates {\TcCurveRedCoordinates}
        -- (0,6.35)
        -- (6.25,6.35)
        -- (6.25,0)
        -- (-3.580003,0)
        -- cycle;

      \foreach \i in {1,...,24} {
        \pgfmathsetmacro{\redfadewidth}{21 - 0.8*\i}

        \draw[
          criticalred,
          opacity=0.018,
          line width=\redfadewidth pt,
          line cap=round,
          line join=round
        ]
          plot[smooth] coordinates {\TcCurveRedCoordinates};

        \draw[
          criticalred,
          opacity=0.013,
          line width=\redfadewidth pt,
          line cap=round
        ]
          (-3.580003,1.08) -- (bkt);
      }
    \end{scope}

    \draw[dgreen,line width=1.5pt]
      plot[smooth] coordinates {\TcCurveGreenCoordinates};

    \draw[criticalred,line width=1.5pt,-Latex]
      plot[smooth] coordinates {\TcCurveRedCoordinates};

    \draw[
      axisgray!55,
      densely dotted,
      line width=0.65pt
    ]
      (0,0) -- (0,6.35);

    \draw[axis]
      (-5.35,0) -- (6.55,0)
      node[below left=1pt,font=\large] {$g$};

    \draw[axis]
      (-5.35,0) -- (-5.35,6.65)
      node[below left=1pt,font=\large] {$T$};

    \draw[axisgray,line width=0.7pt]
      (-4.35,0.08) -- (-4.35,-0.08)
      node[below=1pt,font=\small] {$g_*$};

    \draw[axisgray,line width=0.7pt]
      (0,0.08) -- (0,-0.08)
      node[below=3pt,font=\small] {$0$};

    \node[region label] at (-3.0,3.8)
      {thermal equilibrium ruled out};

    \node[
      region label,
      text=criticalred
    ] at (3.20,5.25)
      {high-$T$ expansion\\[-1pt]
       $T/m\gg1$};

    \node[
      region label,
      text=weakviolettext,
      rotate=90
    ] at (0.22,3)
      {small $|g|$ perturbation theory};

    \node[
      region label,
      text=highbluetext
    ] at (1.90,0.47)
      {low-$T$ expansion
       $e^{-m\Delta_1/T}\ll1$};

    \node[
      font=\sffamily\scriptsize,
      align=center,
      text=dgreen,
      rotate=90
    ] at (-4.10,0.45)
      {$c=1$};

    \node[
      font=\sffamily\small,
      text=criticalred,
      inner sep=1pt,
      rotate=22
    ] at (-2.0,2.0)
      {$c=0$};

    \draw[
      axisgray!55,
      densely dashed,
      line width=0.65pt
    ]
      (bkt |- origin) -- (bkt);

    \draw[axisgray,line width=0.7pt]
      (bkt |- origin)
      ++(0,0.08) -- ++(0,-0.16)
      node[below right=1pt,font=\small]
      {$g_{\rm BKT}$};

    \coordinate (Taxis) at (-5.35,0);
    \coordinate (bktT) at (Taxis |- bkt);
    
    \draw[
      axisgray!55,
      densely dashed,
      line width=0.65pt
    ]
      (bkt) -- (bktT);
    
    \draw[axisgray,line width=0.7pt]
      (bktT)
      ++(-0.08,0) -- ++(0.16,0);
    
    \node[left=3pt,font=\small] at (bktT)
      {$T_{\rm BKT}$};

    \fill[black] (bkt) circle (3pt);

    \node[
      note,
      text=black,
      anchor=south east
    ] at (bkt)
      {BKT};

    \node[
      note,
      text=criticalred,
      anchor=west
    ] at (0.10,6.04)
      {$T_c(g)\to\infty$ as $g\to0^{-}$};

    \end{tikzpicture}

    \caption{Schematic phase diagram of the matrix quantum mechanics with potential $V=\tfrac{m^2}{2} X^2 + g X^4$ in the infinite $N$ 't Hooft limit. Here $g_c(T=0)=g_*=-\frac{\sqrt{2}}{6\pi}$ (in units $m=1$). The planar MQM is singular on the critical curve $T_c(g)$ and only makes sense below it. We list various analytic techniques and their regimes of validity. The bootstrap method can be used anywhere in the   phase diagram where thermal equilibrium exists.}
    \label{fig:phase-diagram}
\end{figure}

Our paper is devoted mainly to studying MQM at finite temperature $T=1/\beta$, which is described by
the path integral with compact Euclidean time, $\tau \sim \tau+\beta$. In the 't Hooft large $N$ limit, 
it is described by the sum over planar Feynman diagrams with the Green function
\begin{equation}\label{MQMvillain}
    G_\beta(\tau) = \frac{1}{2}\sum_{n\in\mathbb Z} e^{-|\tau+n\beta|}=\frac{\cosh\left(\frac{\beta}{2}-\tau\right)}{2\sinh(\beta/2)}\ , \qquad 0\leq\tau<\beta\ .
\end{equation}
As in the Villain model \cite{Villain:1974ir}, this involves a sum over an integer.
From the point of view of summing the planar diagrams, the critical behavior is achieved as $g$ is tuned to the critical value $g_c < 0$, which is a function of the parameter $T$. At $g= g_c(T)$, the planar theory is singular, so it makes sense only for $g> g_c(T)$.
For $T= 0$, the sum over planar diagrams with the Green function 
$G_\infty(\tau)=\frac{1}{2}e^{-|\tau|}$
can be evaluated using the ground state energy of the free fermions \cite{Brezin:1977sv}, leading to the exact result
$ g_c(T = 0)= g_*$. We will find that the critical coupling $g_c(T)$ increases monotonically as $T$ increases, approaching zero in the $T\rightarrow \infty$ limit, where it behaves as $-\frac{1}{48 T}.$ The problem of finding the critical curve $g_c(T)$, which is shown schematically in Figure \ref{fig:phase-diagram}, does not appear to admit an exact solution. Nevertheless, we will make significant progress towards determining it. To this end, we combine analytical approaches and novel matrix bootstrap methods, obtaining good agreement between the two.  

The study of MQM at finite temperature was initiated in \cite{Gross:1990ub}. There  it was found that, in the double scaling limit, the thermal free energy of the singlet sector exhibits T-duality, which
maps $R\rightarrow\alpha'/R$ where $R=\frac{\beta}{2\pi}$.\footnote{The extra symmetry is related to forbidding the BKT vortices on the discretized worldsheets \cite{Gross:1990ub}. This restriction can be implemented also on regular lattices and is now known as Modified Villain Model \cite{Sulejmanpasic:2019ytl, Gorantla:2021svj }. }
Therefore, in the double-scaled gauged MQM, there is no thermal phase transition. 
In contrast to this, in the original ungauged model \eqref{MQM-Hamiltonian-def}, there is a thermal phase transition at a critical temperature of the order $N^0$, which can be interpreted as a BKT transition on a random lattice \cite{Gross:1990ub,Gross:1990md}. It takes place at temperature $T_{\rm BKT}$ and is related to the liberation of vortices present on the discretized worldsheet. Remarkably, the physics of vortices is encoded in the non-singlet states of MQM 
\cite{Gross:1990ub,Gross:1990md, Boulatov:1991fp, Boulatov:1991xz, Kazakov:2000pm}. For $T<T_{\rm BKT}$ the double scaled MQM is expected to describe a compact $c=1$ CFT coupled to 2D quantum gravity, while for $T>T_{\rm BKT}$ the CFT becomes gapped, so that MQM describes pure 2D quantum gravity, i.e. the $c=0$ universality class. Based on continuum world sheet considerations, the conventional Gaussian BKT temperature is $T^{(0)}_{\rm BKT}=\frac{1}{4\pi\sqrt{\alpha'}}$ \cite{Gross:1990ub, Gross:1990md} ($\alpha'=\frac{1}{2}$ in our convention \eqref{MQM-Hamiltonian-def}). We argue that the transition temperature $T_{\rm BKT}$ found in MQM need not coincide exactly with this value, since in lattice models there is typically a renormalization of the radius due to the effects of vortices. Indeed, RG flow of the radius is known to occur in the $O(2)$ model on a regular lattice \cite{Kosterlitz:1974nba, Amit:1979ab}, and  the actual BKT transition temperature is found to be lower than the continuum value $T_{\rm BKT}^{(0)}$.
This was observed in explicit numerical simulations \cite{Janke:1993va, Otsuka:2023far, Okabe:2025rps, Mochizuki_2025}. On a random lattice, a similar shift of critical temperature may occur as well \cite{Schmidhuber:1993qe}. 
In this paper, we will use a combination of methods to obtain a more detailed understanding of the BKT transition in MQM. 

Another large $N$ matrix model that exhibits the BKT transition, while being simpler to analyze, is the infinite matrix chain \cite{Parisi:1990jy, Gross:1990ub}. While on the lattice generated by the planar Feynman diagrams, this model has a discretized one-dimensional target space with lattice spacing $\epsilon$, on the dual lattice it has compact target space. As shown in \cite{Gross:1990ub, Klebanov:1991qa}, for $\epsilon$ smaller than the special value $\epsilon_{\rm BKT}$, this large $N$ matrix model is in the $c=1$ universality class, while for $\epsilon>\epsilon_{\rm BKT}$ it is in the $c=0$ universality class, as expected from the BKT transition. Deeper insights into this transition in the matrix chain model were later obtained by Matytsin and Zaugg \cite{Matytsin:1996tx, Matytsin:1997uj}. More recently, the seven-vertex model on dynamical lattices was studied in \cite{Kostov:2025awi, Junca:2026uqv}, providing a complementary matrix-model realization of sine-Liouville gravity with massive, dilute, and dense critical regimes. These results offer another useful perspective on the role of winding perturbations in random-lattice models.

Recently, interest in large $N$ matrix models has been revitalized due in large part to the successes of the large $N$ matrix bootstrap \cite{Lin:2020mme,Han:2020bkb,Kazakov:2021lel, Cho:2025vws,Lin:2025srf}. In particular, the bootstrap methods have been applied to the ungauged MQM with both positive and negative $g$ \cite{Cho:2024kxn, Adams:2025nww}. The work \cite{Cho:2024kxn} demonstrated using bootstrap that the planar quartic model has a high-temperature instability for negative $g$ and constrained the critical stability curve $T_c (g)$, $g_* <g<0$. $T_c(g)$ is the inverse function of $g_c(T)$, which we discussed earlier. The planar theory is singular on the critical curve and makes sense only for $T<T_c(g)$.  
In this paper, we obtain a deeper understanding of the critical curve using bootstrap in conjunction with various analytic approximation schemes. This combined approach leads us to the thermal phase diagram of the ungauged planar MQM shown schematically in Figure \ref{fig:phase-diagram}. 

Our analytic methods include perturbation theory in powers of $g$, keeping the full temperature dependence at each order; a high-temperature expansion; and a low-temperature expansion in the Boltzmann weights of non-singlet states. The latter two approaches determine the critical curve near $g=0$ and $g=g_*$, respectively. In particular, for small negative $g$, we obtain $T_c(g)=\frac{1}{48|g|}+\mathcal O(|g|)$. This high-temperature expansion is parametrically reliable because it leads, via dimensional reduction on the thermal circle, to the solvable matrix integral with negative quartic coupling \cite{Brezin:1977sv} (the potential also contains multi-trace terms which can be treated using standard large $N$ methods). Including higher orders allows us to describe the critical curve progressively farther from $g=0$. The radius of convergence of this expansion remains unknown, but appears to be not far from $g_*$.

In Section \ref{sec:backreaction}, we study the opposite end of the critical curve. We develop an approximation valid at small $\delta=1-g/g_*$, where the theory is in the $c=1$ universality class. In this regime, $T_c(g)\sim \frac{\Delta_{1}}{|\log\delta|}$, where $\Delta_{1}$ is the first eigenvalue of the MO equation (the gap between the adjoint and singlet ground states), evaluated at criticality;\footnote{Its numerical value is $\Delta_1(g=g_*)\approx  0.7415737$ \cite{Cho:2025vws, Klebanov:2026vmk}.} see \eqref{Tc-lowT-final} for more details.\footnote{ We note that $T_c''(g)$ is very negative for $g\rightarrow g_*^+$, but very positive for $g\rightarrow 0^-$. If the critical curve is smooth, then there must be an inflection point.} In Figure \ref{fig:bkt-scenarios}, we illustrate three scenarios regarding the position of BKT transition on the curve $T_c(g)$. The logarithmic behavior of $T_c(g)$ shows that the critical curve moves away from $g_*$ for any nonzero temperature, ruling out the vertical segment in the phase diagram depicted in scenario I of Figure \ref{fig:bkt-scenarios}. 

Importantly, there is a special point $\left(g_{\rm BKT},T_{\rm BKT}\right)$ on the critical curve, which separates the $c=1$ and $c=0$ phases (this is analogous to what happens in the matrix chain model \cite{Matytsin:1996tx,Matytsin:1997uj}). While its precise location remains unknown, we use the Gaussian BKT temperature $T_{\rm BKT}^{(0)}$ as a reference. Combining it with our low-temperature critical curve gives $g_{\rm BKT}^{(0)}\equiv g_c\left(T_{\rm BKT}^{(0)}\right)\approx0.991g_*\approx-0.0744$. As discussed in Section \ref{sec:BKT-point}, the renormalization of the compactification radius suggests that the physical transition occurs slightly below the Gaussian value, $T_{\rm BKT}<T_{\rm BKT}^{(0)}$, provided the weak-winding approximation applies. This disfavors scenario II, which corresponds to a zero-temperature BKT transition. Taken together, the low- and high-temperature critical behaviors favor scenario III in Figure \ref{fig:bkt-scenarios}, with a finite-temperature BKT point separating the two universality classes. Within the same weak-winding approximation, we expect this point to lie close to the Gaussian estimate, although its precise location remains to be determined microscopically.

We summarize the key findings of our work and the organization of the paper below: 
\vspace{-0.5cm}
\begin{enumerate}
    \item In the next two sections, we collect some old and new facts about the energy spectrum in low-lying representations, including the singlet, adjoint, and biadjoint sectors (a casual reader may wish to skip these sections on a first read). \S \ref{sec:singlet-sector} reviews the free-fermion description of the singlet sector of MQM and the planar limit of the quartic model. \S \ref{sec:Non-singlets} develops the non-singlet spectrum, focusing on the adjoint and the four traceless irreducible biadjoint representations of SU$(N)$. While finite-$N$ corrections to adjoint gaps involve only even powers of $1/N$ \eqref{adj-gaps}, multi-adjoint gaps may also contain odd powers, depending on the representation (see \eqref{biadjoint-finiteN-expansion} and \eqref{mixed-biadjoint-parity-pattern}). This somewhat surprising fact turns out to be important for the low-temperature expansion of thermodynamic observables at leading order in the large-$N$ limit.

    Complementary Appendices \ref{app:singlet-WKB}--\ref{app:representation-characters} develop the microscopic particle-hole, oscillator, and representation-theoretic descriptions underlying these results, making explicit how the low-lying singlet and non-singlet states, their $1/N$ splittings, and their degeneracies arise.

    \item \S \ref{sec:small-g} develops weak-coupling perturbation theory for the thermal free energy, keeping the full temperature dependence at each order in $g$. These results provide independent cross-checks of the low- and high-temperature expansions developed in the following sections, while the expansion in thermal winding number establishes a direct connection with the non-singlet spectrum.
    
    \item We develop the low-temperature expansion in detail in \S \ref{sec:lowTexpansion}. 
    The low-temperature expansion involves systematically incorporating the contributions of the adjoints, biadjoints, and higher irreps to the thermal free energy. In particular, the leading correction to thermodynamic quantities is controlled by $e^{-\Delta_1 \beta}$, where $\Delta_1$ is the lowest eigenvalue of the MO equation. It might seem counterintuitive that $\Delta_n \sim \mathcal O(1)$ energies should appear in the planar free energy, but note that $\mathcal O(N^2)$ degeneracy of the adjoint spectrum compensates for this. Continuing this line of reasoning, the small $\sim 1/N$ splittings in the energies of biadjoint irreps also contribute to the planar free energy, because these representations have even larger degeneracies.    
    This refines and corrects the low-temperature treatment of \cite{Cho:2024kxn, Adams:2025nww}.
    
    Complementary Appendix \ref{app:non-interacting-ungauged-states} studies a toy model in which the $1/N$ splittings are neglected. This approximation can be analyzed explicitly and provides a useful benchmark for understanding the organization of the non-singlet contribution to planar thermodynamics and the role of interactions between adjoint excitations.

    We introduce a backreaction mechanism for the low-temperature thermodynamics of ungauged MQM: thermally excited non-singlet states deform the singlet Fermi sea through a temperature-dependent effective potential. Solving the resulting backreacted saddle determines the leading low-temperature shift of the critical curve $T_c(g)$ near $g_*$. Appendix \ref{app:backreaction} gives an equivalent linear-response derivation of this backreaction.
    
    \item \S \ref{sec:highTexpansion} derives the high-temperature effective theory by integrating out the nonzero Matsubara modes and develops the high-temperature/weak-coupling expansion of $T_c(g)$. Along this high-temperature branch, the instability at $T_c(g)$ corresponds to criticality in the zero-dimensional matrix integral, a $c=0$ phase.

    Appendix \ref{app:highT} gives the details of the high-temperature matching, including the corrections generated by the nonzero Matsubara modes and an explicit comparison with the finite-temperature small-$|g|$ perturbative expansion. The corresponding results for simple thermal observables in the weak-coupling, low-temperature, and high-temperature regimes are collected in Appendix \ref{app:simple-observables}.

    \item \S \ref{sec:BKT} discusses the BKT transition between the low-temperature $c=1$ and high-temperature $c=0$ branches. We estimate a reference point $\left(g_{\rm BKT}^{(0)},T_{\rm BKT}^{(0)}\right)$ on the critical curve, compare it with bootstrap results in \S \ref{sec:bootstrap}, and discuss the possible shift of the physical BKT transition. We also revisit the correspondence between non-singlet states and worldsheet vortices, including the interpretation of the adjoint spectrum and the relation between microscopic MQM and continuum winding couplings. Appendix \ref{app:gravity-description} reviews the continuum gravity description and the BKT RG equations.

    \item \S \ref{sec:bootstrap} formulates canonical and microcanonical matrix bootstrap, providing non-perturbative numerical bounds to validate our analytic approximations. In particular, the two bootstrap methods give estimates of $T_c(g)$ consistent with the analytic low- and high-temperature approximations. Further details on the multi-adjoint positivity constraints and their implementation at finite bootstrap level are collected in Appendix \ref{app:bootstrap-details}.
    \item \S \ref{app:MQMinspired-Villain} is devoted to a Monte Carlo analysis of the lattice model with the link factor taken to be the MQM propagator \eqref{MQMvillain}. For such a model on the regular square lattice, which we call the ``MQM-inspired Villain model," we find good numerical evidence for the existence of a massless compact-boson regime separated from a gapped phase by the BKT transition at a special value of $\beta$. These results are qualitatively similar to those found with more conventional link factors \cite{Janke:1993va,Saito:1981, Otsuka:2023far, Okabe:2025rps, Mochizuki_2025 }.  
\item Finally, \S \ref{sec:discussion} discusses possible generalizations of our results and directions for future work.  Although our work is mainly focused on solving MQM, some of our results are expected to be universal and, therefore, should have contact with the worldsheet approach to the 2D string theory.
\end{enumerate}

\begin{figure}
    \centering
\resizebox{\columnwidth}{!}{
\begin{tikzpicture}[
    x=0.92cm,
    y=0.92cm,
    axis/.style={draw=black!72, line width=0.85pt, -stealth},
    critical/.style={draw=red!68!black, line width=1.45pt},
    string/.style={draw=green!48!black, line width=1.45pt},
    phase label/.style={font=\sffamily\small, text=black!72, inner sep=1pt},
    panel label/.style={font=\bfseries\large, text=black!72}
  ]

    \foreach \panelx/\roman in {0/I,6.6/II,13.2/III} {
    \begin{scope}[shift={(\panelx,0)}]
        \draw[axis] (-1.20,0) -- (4.65,0) node[above left=1pt,font=\large] {$g$};
        \draw[axis] (-1.20,0) -- (-1.20,4.65) node[above left=1pt,font=\large] {$T_c$};
        \draw[black!72,line width=0.65pt] (0,0.08) -- (0,-0.08) node[below=3pt,font=\small] {$g_*$};
        \draw[black!72,line width=0.65pt] (4.00,0.08) -- (4.00,-0.08) node[below=3pt,font=\small] {$g=0$};
        \node[panel label,anchor=south east] at (4.45,4.40) {\roman};
    \end{scope}
  }

    \begin{scope}[shift={(0,0)}]
        \draw[string] (0,0) -- (0,1.35);
        \fill[black] (0,1.35) circle (2.25pt);
        \draw[critical] (0.06,1.4) .. controls (0.38,1.78) and (1.05,1.92) .. (1.76,2.22) .. controls (2.62,2.57) and (3.34,3.18) .. (3.77,4.16);
        \node[phase label,text=green!42!black, rotate=90] at (-0.30,0.77) {$c=1$};
        \node[phase label,text=black,anchor=south east] at (-0.08,1.36) {BKT};
        \node[phase label,text=red!68!black,rotate=26] at (1.90,2.55) {$c=0$};
    \end{scope}

    \begin{scope}[shift={(13.2,0)}]
    
        \draw[string]
        plot[smooth] coordinates {
            (0.0000,0.0000)
            (0.0000,0.1600)
            (0.0000,0.3200)
            (0.0000,0.4800)
            (0.0000,0.6400)
            (0.0001,0.7200)
            (0.0014,0.8000)
            (0.0094,0.8800)
            (0.0340,0.9600)
            (0.0830,1.0400)
            (0.1640,1.1200)
            (0.2830,1.1800)
            (0.4500,1.2300)
            (0.7200,1.2800)
        };
    
      \draw[
        black!55,
        densely dashed,
        line width=0.65pt
      ]
        (0.72,0) -- (0.72,1.28);
    
      \draw[black!72,line width=0.65pt]
        (0.72,0.08) -- (0.72,-0.08)
        node[below right=2pt,font=\scriptsize]
        {$g_{\rm{BKT}}$};
    
      \fill[black]
        (0.72,1.28) circle (2.25pt);
    
      \draw[critical]
        plot[smooth] coordinates {
          (0.790,1.290)
          (0.880,1.305)
          (0.990,1.325)
          (1.120,1.350)
          (1.270,1.380)
          (1.440,1.420)
          (1.620,1.470)
          (1.810,1.530)
          (2.000,1.600)
          (2.190,1.690)
          (2.370,1.800)
          (2.550,1.930)
          (2.720,2.090)
          (2.880,2.290)
          (3.020,2.530)
          (3.150,2.820)
          (3.270,3.170)
          (3.380,3.580)
          (3.480,4.100)
        };
    
      \node[
        phase label,
        text=green!42!black,rotate=90
      ] at (-0.25,0.5)
        {$c=1$};
    
      \node[
        phase label,
        text=black,
        anchor=south east
      ] at (0.64,1.30)
        {BKT};
    
      \node[
        phase label,
        text=red!68!black, rotate=22
      ] at (2,2)
        {$c=0$};
    
    \end{scope}

    \begin{scope}[shift={(6.6,0)}]
    
    \draw[critical]
    plot[smooth] coordinates {
        (0.0000,0.0000)
        (0.0000,0.1600)
        (0.0000,0.3200)
        (0.0000,0.4800)
        (0.0000,0.6400)
        (0.0001,0.7200)
        (0.0014,0.8000)
        (0.0094,0.8800)
        (0.0340,0.9600)
        (0.0830,1.0400)
        (0.1640,1.1200)
        (0.2830,1.1800)
        (0.4500,1.2300)
        (0.7200,1.2800)
    };
    
    \draw[critical]
    plot[smooth] coordinates {
        (0.720,1.280)
        (0.790,1.290)
        (0.880,1.305)
        (0.990,1.325)
        (1.120,1.350)
        (1.270,1.380)
        (1.440,1.420)
        (1.620,1.470)
        (1.810,1.530)
        (2.000,1.600)
        (2.190,1.690)
        (2.370,1.800)
        (2.550,1.930)
        (2.720,2.090)
        (2.880,2.290)
        (3.020,2.530)
        (3.150,2.820)
        (3.270,3.170)
        (3.380,3.580)
        (3.480,4.100)
    };
    
      \fill[black] (0,0) circle (2.25pt);
    
      \node[
        phase label,
        text=green!42!black,
        anchor=south west
      ] at (0.08,0.15)
        {$c=1$};
        
    \node[
        phase label,
        text=black,
        anchor=south east
      ] at (-0.05,0.15)
        {BKT};
        
      \node[
        phase label,
        text=red!68!black,
        rotate=22
      ] at (2.00,2.00)
        {$c=0$};
    
    \end{scope}
        
\end{tikzpicture}
}
\caption{Three possible near-critical behaviors of the thermal critical curve. In scenario I, there is a vertical segment at $g=g_*$, corresponding to $c=1$ criticality, which ends at the BKT point. In scenario II, the critical curve $T_c(g)$ remains analytic down to $g_*$. Since the high-temperature branch corresponds to $c=0$ gravity, the entire finite-temperature critical curve would belong to the $c=0$ universality class, while the $c=1$ theory is reached only at zero temperature. This scenario may therefore be interpreted as a zero-temperature BKT transition. In scenario III, the critical curve rounds smoothly to the right near $g_*$, with the $c=1$ and $c=0$ branches meeting at a finite-temperature BKT point (see \cite{Matytsin:1996tx,Matytsin:1997uj} for an analogous scenario in the matrix-chain model).}
\label{fig:bkt-scenarios}
\end{figure}
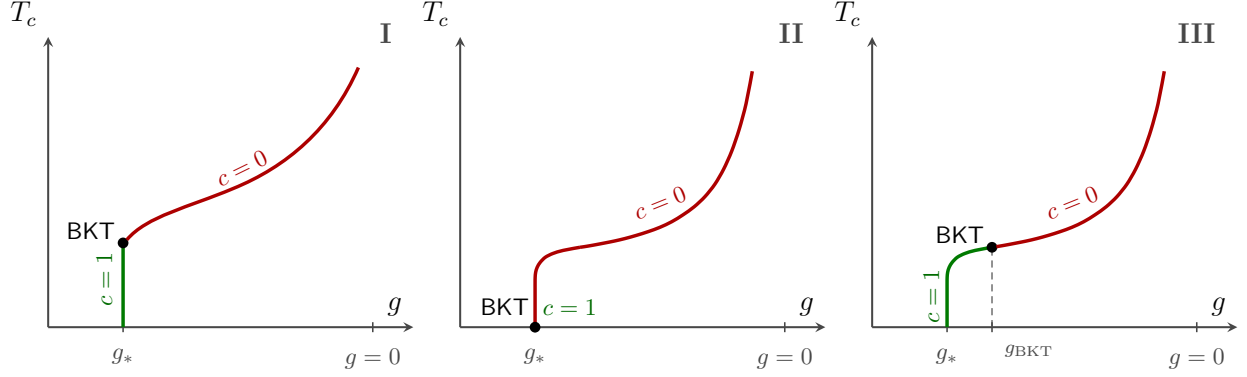
\begin{figure}[h!]
    \centering
\includegraphics[width=0.8\linewidth]{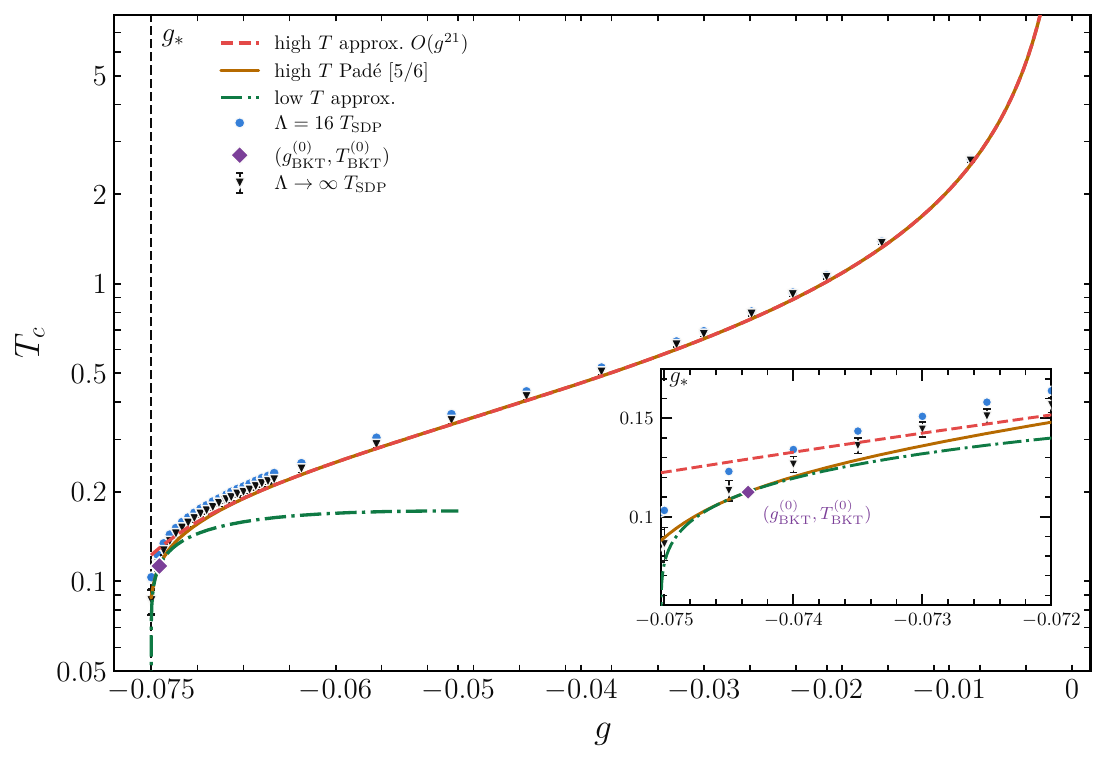}
    \caption{Critical temperature in the quartic theory for $g_*\leq g<0$. The high-temperature approximation shows the perturbative prediction of the critical temperature up to $\mathcal{O}\left(g^{21}\right)$ \eqref{highT-Tc-long-series}. The blue dots are the finite-level thermal-bootstrap upper bounds at $\Lambda=16$, while the black dots are estimates obtained by extrapolating to $\Lambda\to\infty$. The error bars indicate the spread between different extrapolation fits. The solid brown curve is a Pad\'e continuation of the high-temperature series, which lies closer to the low-temperature approximation \eqref{Tc-lowT-final}, shown in dashed green, than the truncated series does.}
    \label{fig:tc-critical-line}
\end{figure}
\section{The singlet sector and free fermions}\label{sec:singlet-sector}
The simplest sector of MQM is the singlet sector, consisting of states invariant under the conjugation symmetry $X\to UXU^\dagger$. Equivalently, it is the physical Hilbert space of the gauged theory. After diagonalizing $X$ and absorbing the Vandermonde determinant into the eigenvalue wave function, the latter becomes antisymmetric, and the singlet problem reduces exactly to $N$ non-interacting fermions in the one-particle potential $V(x)$ \cite{Brezin:1977sv}.

In the normalization of \eqref{MQM-Hamiltonian-def}, the singlet Hamiltonian is
\begin{equation}\label{singlet-fermion-H}
    H_{\rm{sing}}=N\sum_{i=1}^{N}h_N(x_i)\ ,\qquad h_N=-\frac{1}{2N^2}\frac{\d^2}{\d x^2}+V(x)\ .
\end{equation}
The many-body singlet spectrum is obtained by filling one-particle levels. If
\begin{equation}
    h_N\psi_n(x)=\epsilon_n\psi_n(x)\ , \qquad n=0,1,2,\ldots\ ,
\end{equation}
the ground state is obtained by filling the first $N$ of them
\begin{equation}\label{singlet-ground-state-sum}
    E_{\mathrm{gs}}=N\sum_{n=0}^{N-1}\epsilon_n\ .
\end{equation}
At large $N$, the effective Planck constant is $\hbar_{\mathrm{eff}}=1/N$, and the ground state becomes a semiclassical Fermi droplet. Denoting the planar Fermi energy by
$\mu_F=\lim_{N\to\infty}\epsilon_{N-1}$, the occupied phase-space region is
\begin{equation}\label{phase-space}
    \frac{p^2}{2}+V(x)\leq \mu_F\ .
\end{equation}
We restrict to regular one-cut backgrounds, for which the projection of this region onto the $x$ axis is a single interval $[x_1,x_2]$, with
\begin{equation}
    V(x_1)=V(x_2)=\mu_F\ .
\end{equation}
Generally, we denote by $x_{1,2}(e)$ the turning points at one-particle energy $e$, defined by $V(x_{1,2}(e))=e$.

It is useful to introduce the semiclassical action, equivalently the phase-space volume below energy $e$,
\begin{equation}\label{singlet-action}
    \mathcal I(e)\equiv\int\frac{\d x\d p}{2\pi}\,\Theta\left(e-\frac{p^2}{2}-V(x)\right)=\frac{1}{\pi}\int_{x_1(e)}^{x_2(e)}\limits \d x\,\sqrt{2\bigl(e-V(x)\bigr)}\ ,
\end{equation}
where $\Theta$ is the Heaviside step function.
At leading order, $N\mathcal I(e)$ counts the one-particle levels below $e$, so the Fermi energy is fixed by $\mathcal I(\mu_F)=1$. Equivalently,
\begin{equation}\label{density-MQM}
    1=\int_{x_1}^{x_2}\limits\d x\,\rho(x)\ ,\qquad
    \rho(x)=\frac{1}{\pi}\sqrt{2\bigl(\mu_F-V(x)\bigr)}\ .
\end{equation}
Thus the planar eigenvalue density is the projection of the filled Fermi droplet onto the $x$ axis. This differs from the ordinary Hermitian matrix integral, where the density is determined by a static Coulomb-gas saddle.

The ground-state energy admits the topological expansion
\begin{equation}\label{MQM-gs}
    E_{\mathrm{gs}}=\sum_{h=0}^{\infty}N^{2-2h}\mathcal E_h\ ,
\end{equation}
with the leading in large-$N$ limit planar term
\begin{equation}\label{MQM-planar-gs-0}
    \mathcal E_0=\int_0^1\limits\d\mathcal I\,e(\mathcal I)=\mu_F-\int_{x_1}^{x_2}\limits\frac{\d x}{3\pi}\left[2\bigl(\mu_F-V(x)\bigr)\right]^{3/2}\ .
\end{equation}

\begin{figure}[t]
    \hspace{2cm}
   \begin{tikzpicture}[
        xscale=0.3, yscale=0.3,
        potential/.style={thick, black},
        boundary/.style={dashed, thin, gray!70!black},
        declare function={
            V(\x) = -6 + 0.5*\x*\x - 0.006*\x*\x*\x*\x;
            tp(\E) = sqrt((0.5 - sqrt(0.25 - 0.024*(\E + 6)))/0.012);
        }
    ]

    \pgfmathsetmacro{\muF}{0.0}
    \pgfmathsetmacro{\et}{1.30}
    \pgfmathsetmacro{\vmax}{4.4166667}
    \pgfmathsetmacro{\xmax}{6.4549722}
    \pgfmathsetmacro{\xmu}{tp(\muF)}
    \pgfmathsetmacro{\xt}{tp(\et)}

    \begin{scope}[shift={(0,0)}]

        \fill[gray!18]
            (-\xmu,\muF) -- (\xmu,\muF)
            plot[domain=\xmu:-\xmu, smooth, samples=120] (\x,{V(\x)})
            -- cycle;

        \draw[potential]
            plot[domain=-7.8:7.8, smooth, samples=160] (\x,{V(\x)});

        \draw[boundary] (-\xmu,\muF) -- (\xmu,\muF);
        \node[anchor=west] at (\xmu+0.25,\muF) {$\mu_F$};

        \node[gray!70!black] at (0,-1.8) {\small Fermi sea};

    \end{scope}
    \hspace{2cm} 
    \begin{scope}[shift={(14.8,0)}]

        \fill[gray!18]
            (-\xmax,\vmax) -- (\xmax,\vmax)
            plot[domain=\xmax:-\xmax, smooth, samples=160] (\x,{V(\x)})
            -- cycle;

        \fill[white]
            (-\xt,\et) -- (\xt,\et)
            plot[domain=\xt:-\xt, smooth, samples=120] (\x,{V(\x)})
            -- cycle;

        \draw[potential]
            plot[domain=-7.8:7.8, smooth, samples=160] (\x,{V(\x)});

        \draw[boundary] (-\xmax,\vmax) -- (\xmax,\vmax);
        \draw[boundary] (-\xt,\et) -- (\xt,\et);

        \node[anchor=west] at (\xmax+0.25,\vmax) {$V_{\rm max}$};
        \node[anchor=west] at (\xt+0.25,\et) {$e_t$};

        \node[gray!70!black] at (0,3.05) {\small filled upper shell};

    \end{scope}
    \end{tikzpicture}
    \caption{{\it Left}: the ground state where all energies are occupied up to $\mu_F$, the surface of the Fermi sea. More generally, there are high energy singlet states which include ``bubbles'' of unfilled energies in the Fermi sea. {\it Right}: The highest-energy filling $\mathcal E_{\rm top}$ occupies the upper shell between $e_t$ and $V_{\max}$. Above this energy, the Fermi sea would overflow the potential.
    }
    \label{fig:quartic-ground-top}
\end{figure}
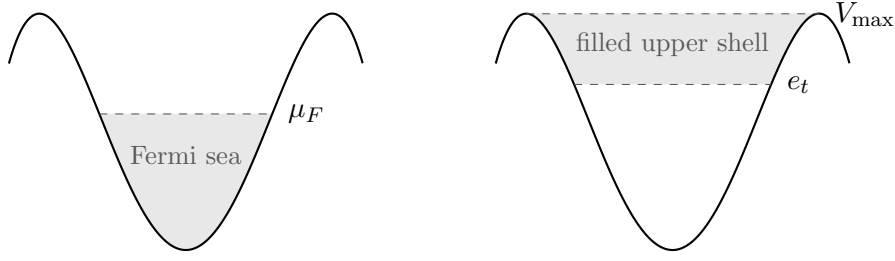

For a potential that is unbounded below, let $\mathcal I_{\rm max}\equiv \mathcal I(V_{\max})$ and $\mathcal I(e_t)=\mathcal{I}_{\rm max}-1$. The ground state occupies the lowest unit interval in action, while the highest-energy filling confined within the potential well occupies the uppermost one, $\mathcal{I}_{\rm max}-1<\mathcal I<\mathcal{I}_{\rm max}$, as illustrated in Figure \ref{fig:quartic-ground-top}. Its planar energy is therefore
\begin{equation}\label{Etop-general}
    \mathcal E_{\rm top}=\int_{\mathcal I_{\rm max}-1}^{\mathcal I_{\rm max}}\limits\d\mathcal I\,e(\mathcal I)\ .
\end{equation}
Thus the metastable planar singlet band extends from $\mathcal E_0$ to $\mathcal E_{\rm top}$. At the critical coupling $\mathcal{I}_{\rm max}=1$, and the interval collapses, $\mathcal E_{\rm top}=\mathcal E_0$.

The free-fermion description also fixes the low-lying singlet spectrum. For a regular Fermi surface, finite-energy excitations probe only the local spacing of one-particle levels near $\mu_F$. At planar order they form an equally spaced tower,
\begin{equation}\label{singlet-planar-spectrum}
    \Delta_s^{(0)}=s\,\omega_F\ , \qquad s=1,2,\ldots\ ,
\end{equation}
where $\omega_F=\omega(\mu_F)$ is the classical frequency at the Fermi surface,
\begin{equation}\label{frequency-def}
    \frac{\pi}{\omega(e)}=\int\limits_{x_1(e)}^{x_2(e)}\frac{\d x}{\sqrt{2\bigl(e-V(x)\bigr)}}\ .
\end{equation}

Microscopically, the states in \eqref{singlet-planar-spectrum} are particle-hole excitations of the filled Fermi sea. Several distinct particle-hole configurations may have the same planar energy $s\omega_F$; their explicit construction and the first finite-$N$ splitting of this degeneracy are reviewed in Appendix \ref{app:singlet-WKB}. The same planar spectrum follows from the linearized collective-field description of the Fermi surface \cite{Jevicki:1979mb,Das:1990kaa}.

The regular expansion is not uniform near the top of the potential well. As the Fermi surface $\mu_F$ approaches a local maximum $V_{\max}$ of the one-particle potential, the classical period diverges logarithmically,
\begin{equation}\label{frequency-critical}
    \mu\equiv V_{\rm max}-\mu_F \to0^+ \quad\Rightarrow\quad \omega_F \approx \frac{2\pi}{n_{\rm end}\sqrt{\alpha'}\log(\Lambda/\mu)}\ , \qquad \alpha'=\frac{1}{|V''_{\rm max}|}\ , \quad n_{\rm end}\in\{1,2\}\ .
\end{equation}
Here $n_{\rm end}$ counts the critical endpoints of the classically allowed interval and $\Lambda$ is a non-universal scale determined by the global shape of the potential. Thus every fixed-$s$ singlet excitation becomes logarithmically light near criticality. The detailed weak-coupling and near-critical behavior for the quartic potential is collected in Appendix \ref{app:singlet-WKB}.

We now turn from the spectrum to thermodynamics. Since the physical one-particle energies are $N\epsilon_n$, a fixed temperature $T=\mathcal{O}\left(N^0\right)$ smears the occupation numbers only within
\begin{equation}
    \delta\epsilon\sim\frac{T}{N}
\end{equation}
of the Fermi surface.  For a regular one-cut background,
\begin{equation}
    \frac{\d n}{\d\epsilon}\simeq\frac{N}{\omega(\epsilon)}\qquad\Rightarrow\qquad\delta n\sim\frac{T}{\omega_F}=\mathcal{O}\left(N^0\right)\ .
\end{equation}
Only an $N$-independent number of particle-hole levels is therefore thermally populated, and the thermal correction starts at order $N^0$, consistently with the genus expansion,
\begin{equation}
    F_{\rm{sing}}(\beta)\equiv-\frac{1}{\beta}\log Z_{\mathrm{sing}}(\beta)=N^2\mathcal{E}_0+\mathcal{F}_{1,\rm{sing}}(\beta)+\mathcal{O}\left(N^{-2}\right)\ .
\end{equation}
In particular, the planar singlet free energy is independent of temperature.

Near criticality, the logarithmic softening \eqref{frequency-critical} enhances the number of thermally active levels $\delta n\sim\frac{T}{\omega_F}\sim\frac{n_{\rm end}\sqrt{\alpha'}\,T}{2\pi}\log\frac{\Lambda}{\mu}$, but this remains $\mathcal O\left(N^0\right)$ at fixed $\mu>0$. The fixed-coupling large-$N$ expansion becomes non-uniform only when the distance between the Fermi energy and the potential maximum becomes comparable to the thermal window, $\mu\lesssim T/N$. This is a correlated critical scaling regime rather than a finite-temperature planar transition. Hence, at fixed $g>g_*$ and $T=\mathcal O\left(N^0\right)$, the gauged singlet sector has no finite critical temperature \cite{Gross:1990ub}; a genuine thermal transition of the ungauged model must instead involve the angular non-singlet sectors, whose representation degeneracies can contribute at order $N^2$.

The discussion above applies to a generic one-cut potential with a non-multicritical endpoint. We now specialize to the quartic model \eqref{MQM-Hamiltonian-def}
\begin{equation}
    V(x)=\frac{1}{2}x^2+gx^4\ , \qquad g_*\leq g<0\ .
\end{equation}

For this even potential the one-cut solution is symmetric, $x_2=-x_1=a$. On this branch, the Fermi curve and planar density are
\begin{equation}
    y^2(x)=2\bigl(\mu_F-V(x)\bigr)=-2g(a^2-x^2)(b^2-x^2)\ , \qquad \rho(x)=\frac{y(x)}{\pi}\ ,\qquad-a\leq x\leq a\ .
\end{equation}
Introducing the elliptic modulus
\begin{equation}\label{elliptic-modulus}
    k=\frac{a}{b}\ , \qquad 0<k<1\ ,
\end{equation}
the two positive roots of the Fermi curve and the Fermi energy are then
\begin{equation}
    a^2=-\frac{k^2}{2g(1+k^2)}\ , \qquad b^2=-\frac{1}{2g(1+k^2)}\ ,\qquad \mu_F=-\frac{k^2}{4g(1+k^2)^2}\ .
\end{equation}
The normalization condition \eqref{density-MQM} fixes the coupling and Fermi energy as
\begin{equation}
\begin{aligned}\label{g-muF-as-modulus-k}
    g(k)
    &=
    -\frac{k^2}{4(1+k^2)^{3/2}}\,{}_2F_1\left(-\frac{1}{2},\frac{1}{2};2;k^2\right)=-\frac{(1+k^2)\mathrm{E}(k)-(1-k^2)\mathrm{K}(k)}{3\pi(1+k^2)^{3/2}}\ ,\\
    \mu_F(k)
    &=
    \frac{1}{\sqrt{1+k^2}\,{}_2F_1\left(-\frac{1}{2},\frac{1}{2};2;k^2\right)}=\frac{3\pi k^2}{4\sqrt{1+k^2}\left[(1+k^2)\mathrm{E}(k)-(1-k^2)\mathrm{K}(k)\right]}\ ,
\end{aligned}
\end{equation}
where $\mathrm K$ and $\mathrm E$ are complete elliptic
integrals of the first and second kind, respectively.

The Fermi-surface frequency \eqref{frequency-def} is
\begin{equation}\label{frequency-analyt}
    \omega_F(k)=\frac{\pi}{2\sqrt{1+k^2}\,\mathrm K(k)}\ ,
\end{equation}
while the planar ground-state energy \eqref{MQM-planar-gs-0} is determined by
\begin{equation}\label{MQM-planar-gs-quartic}
    \mathcal{E}_0(k)=\mu_F(k)-\frac{k^4}{32g(k)^2(1+k^2)^{7/2}}\, {}_2F_1\left(-\frac{3}{2},\frac{1}{2};3;k^2\right)\ .
\end{equation}
The weak-coupling limit $g\to0$ corresponds to $k\to0$, whereas the critical endpoint of the one-cut branch $g\to g_*=-\frac{\sqrt{2}}{6\pi}$ is reached at $k\to1$. Useful expansions of physical quantities of the quartic theory in both limits are presented in Appendix \ref{app:singlet-WKB}. Below, we use the shorthand $\omega(g)\equiv\omega_F(k(g))$ and $\mathcal E_0(g)\equiv\mathcal E_0(k(g))$.

For the quartic potential, the two maxima are located at $x=\pm a_{\max}$, with $a_{\max}^2=-1/(4g)$ and $V_{\max}=-1/(16g)$. Its phase-space volume is
\begin{equation}\label{quartic-Isep}
    \mathcal I_{\rm max}(g)=\frac{1}{\pi}\int_{-a_{\rm max}}^{a_{\rm max}}\limits\d x\,\sqrt{2\bigl(V_{\rm max}-V(x)\bigr)}=\frac{g_*}{g}=-\frac{\sqrt2}{6\pi g}\ .
\end{equation}

The upper edge of the planar singlet band can be expressed directly through the ground-state function \eqref{MQM-planar-gs-0},
\begin{equation}\label{Etop-quartic}
    \mathcal E_{\rm top}(g)=\left(\frac{g_*}{g}\right)^2\mathcal E_0(g_*)-\left(\frac{g_*-g}{g}\right)^2
    \mathcal E_0(g_*-g)\ ,
\end{equation}
where $\mathcal E_0(g_*)=57\pi/(280\sqrt2)$. Thus $\mathcal I_{\rm max}\to1$ and $\mathcal E_{\rm top}\to\mathcal E_0$ as $g\to g_*$.
\section{Non-singlet sectors}\label{sec:Non-singlets}
We now turn to the ungauged MQM, whose Hilbert space contains both singlet and non-singlet states. To clarify the distinction between the gauged and ungauged theories, consider the Euclidean action with a non-dynamical U$(N)$ gauge field $A_\tau$,
\begin{equation}
    S_E[X,A_\tau]=N^2\int_0^\beta\limits\d\tau\,\tr\left[\frac12(D_\tau X)^2+V(X)\right]\ ,\qquad D_\tau X=\dot X-i[A_\tau,X]\ .
\end{equation}
The ungauged theory is obtained by setting $A_\tau=0$, without imposing the Gauss-law constraint, whereas the gauged theory is defined by integrating over $A_\tau$ modulo gauge transformations:
\begin{equation}
    Z_{\rm ung}(\beta)=\int_{X(\beta)=X(0)}\mathcal DX\,e^{-S_E[X,0]}\ ,\qquad
    Z_{\rm gauged}(\beta)=\frac{1}{\operatorname{Vol}\mathcal G}\int\mathcal DA_\tau\,\mathcal DX\,e^{-S_E[X,A_\tau]}\ .
\end{equation}
Here $X(\tau)$ is an $N\times N$ Hermitian matrix, periodic in Euclidean time. In the ungauged theory, the conjugation symmetry
\begin{equation}
    X(\tau)\longmapsto UX(\tau)U^\dagger\ ,\qquad U\in U(N)\ ,
\end{equation}
is global. The overall phase in U$(N)$ acts trivially on $X$, so the faithful global symmetry is
\begin{equation}
    PU(N)=U(N)/U(1)\cong SU(N)/\mathbb Z_N\ .
\end{equation}
We will nevertheless label the sectors by irreducible representations of U$(N)$, and denote by $\mathcal R_0$ the set of representations occurring in the MQM Hilbert space. Necessarily, every $R\in\mathcal R_0$ is trivial under the central U$(1)$ and hence descends to a representation of PU$(N)$; in SU$(N)$ language, this condition corresponds to vanishing $N$-ality. We will characterize the allowed representations more precisely below.

In one dimension, the gauge field has no propagating degrees of freedom. On the thermal circle, however, its holonomy
\begin{equation}
    \Omega=\mathcal P\exp\left(i\int_0^\beta\limits\d\tau\,A_\tau\right)
\end{equation}
cannot generally be gauged away by a periodic gauge transformation. Instead, eliminating $A_\tau$ is equivalent to imposing twisted boundary conditions on the matrix field. This motivates introducing the twisted partition function
\begin{equation}
    Z(\beta,\Omega)=\int_{X(\beta)=\Omega^\dagger X(0)\Omega}\mathcal DX\,e^{-S_E[X,0]}\ .
\end{equation}
The ungauged partition function corresponds to trivial holonomy, while the gauged theory is obtained by averaging over all holonomies,
\begin{equation}
    Z_{\rm ung}(\beta)=Z(\beta,\mathbb 1)\ ,\qquad Z_{\rm gauged}(\beta)=\int_{U(N)}\d\Omega\,Z(\beta,\Omega)\ ,
\end{equation}
where $\d\Omega$ is the normalized Haar measure.

The twisted partition function admits an expansion in irreducible characters,
\begin{equation}
    Z(\beta,\Omega)=\Tr_{\mathcal H}\left(e^{-\beta H}\widehat\Omega\right)=\sum_{R\in\mathcal R_0}\chi_R(\Omega)\Tr_{\mathcal H_R}e^{-\beta H_R}\ ,
\end{equation}
where $\widehat\Omega$ implements the conjugation symmetry on the Hilbert space, $\chi_R(\Omega)$ is the character of $R$, and $\mathcal H_R$ is the space of independent states in representation $R$, with the global multiplet degeneracy factored out.\footnote{For completeness, a non-singlet sector also admits a gauged-theory interpretation in terms of an external color source. A Wilson line $W_{\bar R}=\Tr_{\bar R}\mathcal P\exp\left(i\int_0^\beta\d\tau\,A_\tau\right)$ in the conjugate representation $\bar R$ wrapping the Euclidean-time circle carries the external charge, while the Gauss-law constraint requires the MQM degrees of freedom to transform in $R$, so that the combined state is gauge invariant \cite{Minahan:1993mv,Maldacena:2005hi,Maldacena:2018vsr}. Accordingly, $Z_{\rm gauged}[W_{\bar R}]=\int\d\Omega\,\chi_{\bar R}(\Omega)Z(\beta,\Omega)=\Tr_{\mathcal H_R}e^{-\beta H_R}$. More generally, the winding-deformed construction of Kazakov, Kostov, and Kutasov 
\cite{Kazakov:2000pm} uses the same holonomy but introduces independent winding couplings through the weight $\exp[\sum_{n\neq0}\lambda_n\Tr\Omega^n]$ under the holonomy integral. In contrast, the ungauged MQM considered here has $Z_{\rm ung}=Z(\beta,\mathbb 1)$, with all non-singlet representations contributing with their fixed multiplicities $d_R$, without independently adjustable winding couplings.} Using $\chi_R(\mathbb 1)=d_R\equiv\dim R$ and the orthogonality of characters, we obtain
\begin{equation}\label{Z-decomposition}
    Z_{\rm ung}(\beta)=\sum_{R\in\mathcal R_0}d_R\Tr_{\mathcal H_R}e^{-\beta H_R}\ ,\qquad Z_{\rm gauged}(\beta)=\Tr_{\mathcal H_{\mathbb 1}}e^{-\beta H_{\mathbb 1}}\ .
\end{equation}
Thus, gauging projects onto the singlet sector, whereas the ungauged theory retains all non-singlet sectors with their representation-theoretic multiplicities. The character decomposition of the twisted partition function is developed further for the free matrix oscillator in Appendix \ref{app:representation-characters}. In the following we will suppress the subscript ``ung".

It is also useful to introduce an observable that directly measures the non-singlet content of a state. The conjugation symmetry $X\to UXU^\dagger$ has conserved charges $J^A$, associated with the traceless Hermitian generators $T^A$ of $\mathfrak{su}(N)$. We normalize them as
\begin{equation}\label{charges-Casimir}
    \Tr(T^AT^B)=\frac{1}{2}\delta^{AB}\ ,\qquad J^A=iN\,\Tr\left(T^A[X,P]\right)\ ,\qquad A=1,\ldots,N^2-1\ .
\end{equation}
The corresponding quadratic Casimir operator is
\begin{equation}\label{C2-commutator}
    C_2\equiv\sum_AJ^AJ^A=-\frac{N^2}{2}\left(\Tr\left([X,P]^2\right)+N\right)\ .
\end{equation}
The last term is the operator-ordering contribution, following from $\Tr[X,P]=iN$. Since $C_2$ commutes with the Hamiltonian and with the $U(N)$ action, it takes the fixed value $C_2(R)$ on the sector transforming in $R$. With the normalization above, $C_2(\mathrm{Adj})=N$. Thus $C_2$ provides a natural observable for resolving the singlet and non-singlet content of the spectrum and of thermal states. Some useful formulas for the quadratic Casimir are collected in Appendix \ref{app:casimir}.

To determine the allowed representations and the Hamiltonian in each sector, we diagonalize
\begin{equation}\label{diagonalization}
    X=U\Lambda U^\dagger\ , \qquad \Lambda=\mathrm{diag}(x_1,\ldots,x_N)\ .
\end{equation}
The eigenvalues $x_i$ are the radial variables, while the dependence on $U$ carries the representation data. After quotienting the overall $U(1)$, this parametrization retains the residual redundancy
\begin{equation}
    U(1)^{N-1}\rtimes S_N\ .
\end{equation}
Diagonal rotations leave $\Lambda$ unchanged and therefore require the angular wave function to be neutral under all Cartan generators. Thus $\mathcal R_0$ consists of representations with a non-vanishing zero-weight subspace \cite{Gross:1990md,Boulatov:1991xz}. The Weyl group $S_N$ acts by simultaneously permuting the eigenvalues and the corresponding color indices; it is a redundancy of the eigenvalue parametrization rather than an additional global symmetry. 

After absorbing the Vandermonde determinant into the radial wave function, the Hamiltonian in a fixed sector $R$ takes the spin-Calogero form
\begin{equation}\label{spin-Calogero-Hamiltonian}
   H_R=N\sum_{i=1}^{N}\left[-\frac{1}{2N^2}\frac{\partial^2}{\partial x_i^2}+V(x_i)\right]+\frac{1}{2N}\sum_{i\neq j}\frac{T^R_{ij}T^R_{ji}}{(x_i-x_j)^2} \ .
\end{equation}
Here $T^R_{ij}$ denotes the action in the representation $R$ of the matrix-unit generator $E_{ij}$, restricted to the zero-weight subspace. The first term describes the radial eigenvalue motion, while the second couples the eigenvalues to the internal non-singlet degrees of freedom through an inverse-square interaction. It vanishes in the singlet sector, recovering the free-fermion Hamiltonian of the previous section. 

This structure has a close analogue in two-dimensional Yang-Mills theory. On a cylinder, the latter can be reformulated as a one-dimensional unitary matrix model whose eigenvalues behave as fermions on a circle \cite{Minahan:1993np,Caselle:1993mq}. Calogero-Moser and Sutherland-type systems can also be obtained by Hamiltonian reduction of two-dimensional Yang-Mills theory with appropriate sources \cite{Gorsky:1993pe}; in particular, a timelike Wilson line in an arbitrary representation $R$ leads to a generalized Sutherland system with internal degrees of freedom \cite{Minahan:1993mv}. At the level of the representation-dependent angular interaction, equation \eqref{spin-Calogero-Hamiltonian} is the corresponding rational Hermitian-matrix counterpart: the eigenangles of the unitary model are replaced by the real eigenvalues $x_i$, while the trigonometric Sutherland interaction is replaced by the inverse-square Calogero interaction \cite{Polychronakos:1999sx}. 

The representation-theoretic structure of non-singlet sectors was  discussed in \cite{Boulatov:1991xz} and developed further in \cite{Gaiotto:2005gd,Hatsuda:2006xr,Bourgine:2007ua}, where the angular degrees of freedom were treated systematically and the spin-Calogero form of the eigenvalue Hamiltonian played a central role. Related spin-Calogero realizations of non-singlet MQM, based on gauged matrix quantum mechanics coupled to additional fundamental degrees of freedom, were studied in \cite{Betzios:2017yms}.

The allowed sectors are conveniently organized in the box--antibox notation shown in Figure \ref{fig:young-diagrams_box_antibox}. A box represents a fundamental index and an antibox an anti-fundamental one. In $U(N)$ notation, neutrality under the central $U(1)$ requires an equal number of boxes and antiboxes.\footnote{The corresponding statement for irreps of SU$(N)$ is that their $N$-ality vanishes: $n_{\mathrm{box}}-n_{\mathrm{antibox}}=0$ mod $N$.} Thus, for example, the singlet and adjoint representations are allowed, while the fundamental one is not. 

For fixed level $L$ and sufficiently large $N$, the completely traceless irreducible sectors are labeled by a pair of Young diagrams \cite{Koike:1989}
\begin{equation}\label{general-irrep}
    R=\left(Y,\overline{Y}\right)\ , \qquad |Y|=|\overline{Y}|=L\ ,\qquad R\in\mathcal R_0\ .
\end{equation}
The diagrams $Y$ and $\overline{Y}$ specify the permutation symmetries of the upper and lower indices, respectively. Contractions between an upper and a lower index produce representations belonging to lower levels and are therefore removed from the level-$L$ sector. At finite $N$, a given pair is admissible provided $\ell(Y)+\ell(\overline Y)\leq N$, where $\ell(Y)$ is the number of rows; in particular, all pairs at level $L$ are present once $N\geq2L$. 

The first nontrivial example is the adjoint,
\begin{equation}
    R_{\mathrm{adj}}=(1,1)\ ,
\end{equation}
while at $L=2$ there are four completely traceless sectors
\cite{Boulatov:1991xz, Hatsuda:2006xr, Bourgine:2007ua},
\begin{equation}\label{biadj-irreps}
\begin{aligned}
    R_{++}=(2,2)\ ,\qquad R_{+-}=\left(2,1^2\right)\ ,\qquad R_{-+}=\left(1^2,2\right)\ ,\qquad R_{--}=\left(1^2,1^2\right)\ .
\end{aligned}
\end{equation}
Here the first sign refers to the symmetry of the upper indices, while the second sign refers to the symmetry of the lower indices. Trace components, obtained by contracting an upper and a lower index, reduce to the adjoint or singlet sectors and are not counted as new biadjoint representations. 

More generally, at fixed $L$ and sufficiently large $N$, the level-$L$ sectors are labeled by pairs of partitions $Y\vdash L$ and $\overline Y\vdash L$, giving $p(L)^2$ irreducible representations, where $p(L)$ is the number of partitions of $L$. This classification is purely representation-theoretic: the Young diagrams determine the color symmetry, dimensions, and permutation structure of the states, while their excitation energies are dynamical. We begin with the adjoint sector, which contains the basic non-singlet excitations, and then study their combination and splitting in the biadjoint sectors. Further details on the relevant representations, dimensions, and characters are collected in Appendix \ref{app:representation-characters}.

Their physical interpretation will be discussed alongside the spectral analysis below. More broadly, analogous questions arise in holographic matrix quantum mechanics, where the comparison between gauged and ungauged theories probes how angular degrees of freedom affect the spectrum, thermodynamics, and the emergence of a gravitational description \cite{Maldacena:2018vsr}.

\tikzset{
    ydbox/.style={
        draw=black,
        fill=white
    },
    ydantibox/.style={
        draw=black,
        fill=gray!18
    }
}

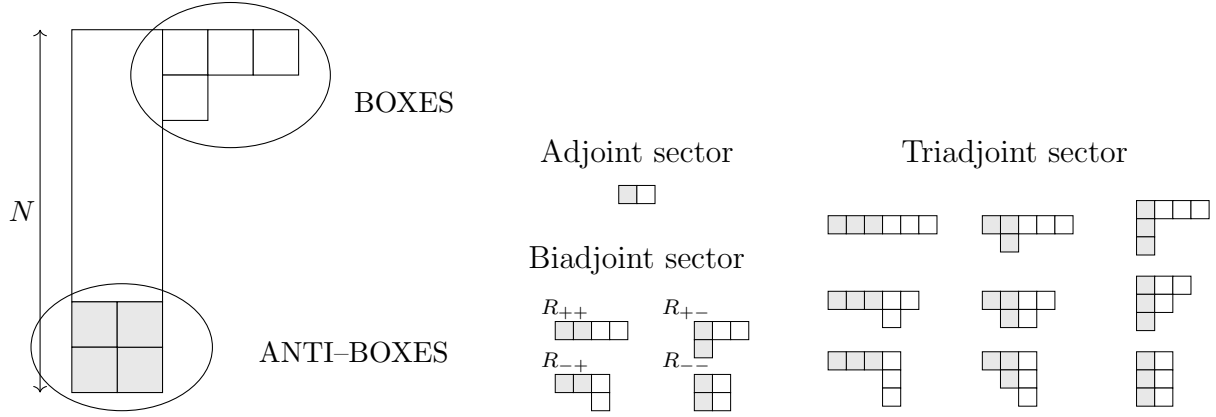
\begin{figure}[t]
    \centering

    \begin{minipage}[t]{0.34\textwidth}
        \centering

        \begin{tikzpicture}[scale=0.6, line width=0.4pt]

            \draw (0,2) rectangle (2,8);

            \draw[<->] (-0.7,0) -- (-0.7,8);
            \node at (-1.1,4) {$N$};

            \draw[ydbox] (2,8) rectangle (3,7);
            \draw[ydbox] (2,7) rectangle (3,6);
            \draw[ydbox] (3,8) rectangle (4,7);
            \draw[ydbox] (4,8) rectangle (5,7);

            \draw (3.5,7.0) ellipse (2.2 and 1.6);
            \node[right] at (6.0,6.4) {BOXES};

            \draw[ydantibox] (1,0) rectangle (2,1);
            \draw[ydantibox] (0,1) rectangle (1,2);
            \draw[ydantibox] (1,1) rectangle (2,2);
            \draw[ydantibox] (0,0) rectangle (1,1);

            \draw (1.1,1.0) ellipse (2.0 and 1.4);
            \node[right] at (3.9,0.9) {ANTI--BOXES};

        \end{tikzpicture}
    \end{minipage}
    \hfill
    \begin{minipage}[t]{0.62\textwidth}
        \centering

        \begin{tikzpicture}[
            x=1cm,
            y=1cm,
            line width=0.38pt,
            font=\small
        ]

        \newcommand{\OneBox}[2]{%
            \begin{scope}[shift={({#1},{#2})}]
                \draw[ydbox]
                    (0,0) rectangle (0.24,0.24);
            \end{scope}
        }

        \newcommand{\TwoSBox}[2]{%
            \begin{scope}[shift={({#1},{#2})}]
                \draw[ydbox]
                    (0,0) rectangle (0.24,0.24);
                \draw[ydbox]
                    (0.24,0) rectangle (0.48,0.24);
            \end{scope}
        }

        \newcommand{\TwoABox}[2]{%
            \begin{scope}[shift={({#1},{#2})}]
                \draw[ydbox]
                    (0,0) rectangle (0.24,0.24);
                \draw[ydbox]
                    (0,-0.24) rectangle (0.24,0);
            \end{scope}
        }

        \newcommand{\ThreeSBox}[2]{%
            \begin{scope}[shift={({#1},{#2})}]
                \draw[ydbox]
                    (0,0) rectangle (0.24,0.24);
                \draw[ydbox]
                    (0.24,0) rectangle (0.48,0.24);
                \draw[ydbox]
                    (0.48,0) rectangle (0.72,0.24);
            \end{scope}
        }

        \newcommand{\ThreeMBox}[2]{%
            \begin{scope}[shift={({#1},{#2})}]
                \draw[ydbox]
                    (0,0) rectangle (0.24,0.24);
                \draw[ydbox]
                    (0.24,0) rectangle (0.48,0.24);
                \draw[ydbox]
                    (0,-0.24) rectangle (0.24,0);
            \end{scope}
        }

        \newcommand{\ThreeABox}[2]{%
            \begin{scope}[shift={({#1},{#2})}]
                \draw[ydbox]
                    (0,0) rectangle (0.24,0.24);
                \draw[ydbox]
                    (0,-0.24) rectangle (0.24,0);
                \draw[ydbox]
                    (0,-0.48) rectangle (0.24,-0.24);
            \end{scope}
        }

        \newcommand{\OneAnti}[2]{%
            \begin{scope}[shift={({#1},{#2})}]
                \draw[ydantibox]
                    (-0.24,0) rectangle (0,0.24);
            \end{scope}
        }

        \newcommand{\TwoSAnti}[2]{%
            \begin{scope}[shift={({#1},{#2})}]
                \draw[ydantibox]
                    (-0.24,0) rectangle (0,0.24);
                \draw[ydantibox]
                    (-0.48,0) rectangle (-0.24,0.24);
            \end{scope}
        }

        \newcommand{\TwoAAnti}[2]{%
            \begin{scope}[shift={({#1},{#2})}]
                \draw[ydantibox]
                    (-0.24,0) rectangle (0,0.24);
                \draw[ydantibox]
                    (-0.24,-0.24) rectangle (0,0);
            \end{scope}
        }

        \newcommand{\ThreeSAnti}[2]{%
            \begin{scope}[shift={({#1},{#2})}]
                \draw[ydantibox]
                    (-0.24,0) rectangle (0,0.24);
                \draw[ydantibox]
                    (-0.48,0) rectangle (-0.24,0.24);
                \draw[ydantibox]
                    (-0.72,0) rectangle (-0.48,0.24);
            \end{scope}
        }

        \newcommand{\ThreeMAnti}[2]{%
            \begin{scope}[shift={({#1},{#2})}]
                \draw[ydantibox]
                    (-0.24,0) rectangle (0,0.24);
                \draw[ydantibox]
                    (-0.48,0) rectangle (-0.24,0.24);
                \draw[ydantibox]
                    (-0.24,-0.24) rectangle (0,0);
            \end{scope}
        }

        \newcommand{\ThreeAAnti}[2]{%
            \begin{scope}[shift={({#1},{#2})}]
                \draw[ydantibox]
                    (-0.24,0) rectangle (0,0.24);
                \draw[ydantibox]
                    (-0.24,-0.24) rectangle (0,0);
                \draw[ydantibox]
                    (-0.24,-0.48) rectangle (0,-0.24);
            \end{scope}
        }

        \node at (1.95,6.95) {\large Adjoint sector};

        \OneAnti{1.95}{6.30}
        \OneBox{1.95}{6.30}

        \node at (1.95,5.55) {\large Biadjoint sector};

        \node[anchor=north west, font=\scriptsize]
            at (0.55,5.17) {$R_{++}$};

        \node[anchor=north west, font=\scriptsize]
            at (2.15,5.17) {$R_{+-}$};

        \node[anchor=north west, font=\scriptsize]
            at (0.55,4.47) {$R_{-+}$};

        \node[anchor=north west, font=\scriptsize]
            at (2.15,4.47) {$R_{--}$};

        \TwoSAnti{1.35}{4.50}
        \TwoSBox{1.35}{4.50}

        \TwoAAnti{2.95}{4.50}
        \TwoSBox{2.95}{4.50}

        \TwoSAnti{1.35}{3.80}
        \TwoABox{1.35}{3.80}

        \TwoAAnti{2.95}{3.80}
        \TwoABox{2.95}{3.80}

        \node at (6.95,6.95) {\large Triadjoint sector};

        \ThreeSAnti{5.20}{5.90}
        \ThreeSBox{5.20}{5.90}

        \ThreeMAnti{7.00}{5.90}
        \ThreeSBox{7.00}{5.90}

        \ThreeAAnti{8.80}{6.10}
        \ThreeSBox{8.80}{6.10}

        \ThreeSAnti{5.20}{4.90}
        \ThreeMBox{5.20}{4.90}

        \ThreeMAnti{7.00}{4.90}
        \ThreeMBox{7.00}{4.90}

        \ThreeAAnti{8.80}{5.10}
        \ThreeMBox{8.80}{5.10}

        \ThreeSAnti{5.20}{4.10}
        \ThreeABox{5.20}{4.10}

        \ThreeMAnti{7.00}{4.10}
        \ThreeABox{7.00}{4.10}

        \ThreeAAnti{8.80}{4.10}
        \ThreeABox{8.80}{4.10}

        \end{tikzpicture}
    \end{minipage}

    \caption{
    Zero-weight representations in box--antibox notation. {\it Left}: a generic Young-diagram representation containing equal numbers of boxes and antiboxes, corresponding to equal numbers of fundamental and anti-fundamental indices. {\it Right}: the lowest non-singlet sectors: the adjoint sector, the four irreps in the biadjoint sector, and the nine irreps in the triadjoint sector. White boxes denote fundamental indices, while gray anti-boxes denote anti-fundamental indices.
    }

    \label{fig:young-diagrams_box_antibox}

\end{figure}
\subsection{The adjoint representation}\label{subsec:adjoint}
We first analyze the adjoint representation. In the large-$N$ limit this sector admits a particularly simple description: the adjoint excitation behaves as a single impurity propagating on top of the singlet eigenvalue background \cite{Marchesini:1979yq}. 

The corresponding planar adjoint gaps $\Delta_n^{(0)}$\footnote{The superscript $(0)$ denotes the planar limit, and for simplicity, we do not use this for $\Phi_n$.} are determined as eigenvalues of the Marchesini-Onofri (MO) equation \cite{Marchesini:1979yq}
\begin{equation}\label{MO-eq}
    \Delta_n^{(0)}\,\Phi_n(x)=\fint\limits_{x_1}^{x_2} \mathrm{d}y\,\rho(y)\,\frac{\Phi_n(x)-\Phi_n(y)}{(x-y)^2}\ ,\qquad n=1,2,\ldots \ .
\end{equation}
Here $\rho(x)$ is the planar eigenvalue density \eqref{density-MQM}, $\Phi_n(x)$ is the wave function of the adjoint impurity  orthonormal with respect to $\rho$, and the principal value prescription is understood. The eigenvalues of \eqref{MO-eq} describe the genuinely adjoint excitations and do not include admixtures with singlet states.

A more precise way to state the result of \cite{Marchesini:1979yq} is that, for singlet states whose excitation does not modify the leading planar fermion density, the adjoint sector forms universal towers over the singlet spectrum.  In particular, for the finite-energy singlet excitations
\begin{equation}
    E_{s,n}^{\mathrm{adjoint}}=E_s^{\mathrm{singlet}}+\Delta_n^{(0)}\ .
\end{equation}
At planar order the gaps $\Delta_n^{(0)}$ are independent of the particular singlet state $s$, since all such states have the same large-$N$ fermion density. In the oscillator language of Appendix \ref{app:operator-formalism}, this is the distinction between an open-chain adjoint excitation and the same open chain dressed by closed singlet chains. This universality need not hold for macroscopically excited singlet backgrounds which modify the planar density; in that case the adjoint gaps should instead be regarded as background-dependent quantities $\Delta_n[\rho]$.

Away from criticality, highly excited adjoint levels are well described by the semiclassical WKB formula
\begin{equation}\label{MO-WKB}
    \Delta_n^{(0)}=(n+1)\omega(g)+\eta(g)+\mathcal {O}\left(\frac{1}{n}\right) \ .
\end{equation}
Here $\omega(g)$ is the singlet energy gap \eqref{frequency-analyt}, while $\eta(g)$ is the period-average of the effective potential
\cite{Marchesini:1979yq,Klebanov:2026vmk}. This approximation is already rather accurate for the low-lying levels at generic coupling. However, as one approaches the critical point $g\to g_*$, the onset of the WKB regime is pushed to increasingly large $n$ \cite{Klebanov:2026vmk}. For fixed $n$, comparison with the perturbative solution \eqref{adjoint-lowest-planar-gaps} shows that the small-$g$ expansion of the WKB expression agrees with the exact planar answer only up to a finite order; the first disagreement appears at order $g^{n+1}$. This mismatch is naturally attributed to subleading $1/n$ corrections to the WKB formula, for which we do not presently have a closed analytic expression.

The critical behavior of the adjoint sector, and more generally of higher non-singlet sectors, is more subtle. Early analyses found that the lowest non-singlet energies separate logarithmically from the singlet spectrum in the double-scaling limit \cite{Gross:1990md,Boulatov:1991xz}. A physical interpretation was proposed in terms of long folded strings extending along the Liouville direction, obtained as a long-string limit of open strings on FZZT branes \cite{Maldacena:2005hi}. This correspondence was developed further through the study of adjoint and multi-adjoint scattering amplitudes \cite{Gaiotto:2005gd,Fidkowski:2005ck,Kostov:2006dy,Bourgine:2007ua,Karczmarek:2008sc,Balthazar:2018qdv,Cho:2024kxn}. Following \cite{Kazakov:2000pm,Kazakov:2001pj}, more recent works \cite{Betzios:2022pji,Ahmadain:2022gfw} have proposed connections between large non-singlet sectors, winding condensates, and possible microscopic descriptions of two-dimensional black-hole states (but see our comment in Section \ref{sec:discussion}). In this form, however, the picture left the fate of the low-lying MO states unresolved. 

Recent work \cite{Cho:2025vws} provided the numerical evidence that this conclusion was incomplete. The lowest adjoint gaps were extracted using the newest bootstrap methods, and were also compared with a direct discretization of the MO equation \eqref{MO-eq}. These results indicated that the low-lying adjoint spectrum remains well-defined near criticality, but by themselves did not provide the analytic mechanism behind this behavior.

The puzzle was resolved in \cite{Klebanov:2026vmk}. By reexamining the adjoint sector in the critical region $g\rightarrow g_*$, both analytically and numerically, it was shown that the adjoint spectrum develops a universal Regge regime $\Delta^2_n\sim n/\alpha'$ rather than simply disappearing from the double-scaled theory. For the quartic model \eqref{MQM-Hamiltonian-def}, in the conventions of \cite{Klebanov:2026vmk}, the critical gaps obey, with exponentially small corrections in $n$ translated to our notation for the critical coupling $g_*$,
\begin{equation}\label{regge-phi4}
    \Delta_n^{\mathrm{Regge}}\approx\sqrt{2n+\frac{2}{3}}-\frac{2\sqrt 2}{\pi} \ , \qquad n\lesssim n_{\mathrm{max}}\approx\frac{1}{4\pi^2}\log^2\frac{4}{e^2|g_*|\mu}-\frac{1}{3} \ .
\end{equation}
In the dual two-dimensional string description, these Regge states are interpreted as oscillating short folded open strings whose endpoints are fixed at the UV wall, while their folds move in the Liouville direction. Slightly away from criticality, sufficiently highly excited states ($\Delta_n=\Delta_{n_{\rm max}}+\mathcal{O}(1)$) cross over to a long-string regime, where the MO wavefunctions probe the endpoints of the finite time-of-flight interval. In the dual picture, these states are interpreted as oscillating long folded strings extending toward the Liouville wall. Unlike the continuum scattering states studied in \cite{Maldacena:2005hi}, they belong to the discrete spectrum of finite-potential MQM. The Euclidean interpretation of adjoints \cite{Gross:1990md,Boulatov:1991fp,Boulatov:1991xz} as vortex-antivortex bound states will be discussed in Section \ref{sec:vortice}.

The finite-gap Regge tower is directly relevant to the thermodynamics of the ungauged model. The thermal trace involves a sum over all irreducible representations $R\in\mathcal R_0$ \eqref{Z-decomposition}, whose dimensions grow as powers of $N$, while the number of allowed sectors proliferates with the multi-adjoint level. Even when the non-singlet energies are approximated as additive and their interactions are neglected, this representation sum can already provide an $\mathcal O\left(N^2\right)$ contribution to the free energy, as discussed in Appendix \ref{app:non-interacting-ungauged-states}. A controlled description of the planar thermodynamics therefore requires not only the adjoint spectrum, but also the representation-dependent splittings of multi-adjoint states, whose first thermodynamic consequences are analyzed in Section \ref{sec:lowTexpansion}.

For the explicit spectral data, we solve the Marchesini-Onofri equation \eqref{MO-eq} perturbatively in the quartic coupling. 
A systematic development of this perturbative construction will be presented in a separate paper, and here we quote the results.
The same method will later be used to obtain the weak-coupling expansion of the biadjoint matrix elements $h_{kl;mn}$ \eqref{h-def}.

For the lowest large $N$ adjoint eigenvalues we find
\begin{equation}\label{adjoint-lowest-planar-gaps}
\resizebox{\textwidth}{!}{$
\begin{aligned}
    \Delta_1^{(0)}&=1+2g-\frac{27}{4}g^2+\frac{337}{8}g^3-\frac{10557}{32}g^4+\frac{373023}{128}g^5-\frac{28413729}{1024}g^6+\frac{284791065g^7}{1024}+\mathcal{O}\left(g^8\right) \ ,
    \\
    \Delta_2^{(0)}&=2+5g-\frac{75}{4}g^2+\frac{1003}{8}g^3-\frac{66183}{64}g^4+\frac{1219179}{128}g^5-\frac{24048975}{256}g^6+\frac{1988141061g^7}{2048}+\mathcal{O}\left(g^8\right) \ ,
    \\
    \Delta_3^{(0)}&=3+8g-\frac{63}{2}g^2+\frac{875}{4}g^3-\frac{119079}{64}g^4+\frac{1125573}{64}g^5-\frac{90821961}{512}g^6+\frac{1914537129g^7}{1024}+\mathcal{O}\left(g^8\right) \ .
\end{aligned}
$}
\end{equation}
For general fixed $n>3$, the first terms take the form
\begin{equation}\label{adjoint-levels-perturb}
    \Delta_n^{(0)}=n+(3n-1)g-\frac{3}{4}(17n-9)g^2+\frac{125}{4}(3n-2)g^3-\frac{105}{64}(509n-393)g^4+\mathcal {O}\left(g^5\right) \ .
\end{equation}
Several of the leading coefficients in these expansions can be obtained independently. The oscillator perturbation theory of Appendix \ref{app:operator-formalism} reproduces the first planar terms, while the lowest gap $\Delta_1^{(0)}$ is also recovered from the small-$g$ winding expansion in Section \ref{seq:small-g-winding-exp}.

At finite $N$, higher adjoint levels can mix with singlet-dressed states, making their analysis more involved. Nevertheless, for nondegenerate adjoint levels, the genus expansion of the thermal free energy implies that their gaps contain only even powers of $1/N$, as we explain in Section \ref{sec:low-T-planar-thermodynamics}:
\begin{equation}\label{adj-gaps}
    \Delta_n(g,N)=\Delta_n^{(0)}(g)+\frac{\Delta_n^{(2)}(g)}{N^2}+\frac{\Delta_n^{(4)}(g)}{N^4}+\ldots\ .
\end{equation}

The lowest adjoint mode is particularly simple, since it occurs with multiplicity one at oscillator level one. Ordinary non-degenerate perturbation theory, described in Appendix \ref{app:operator-formalism}, gives\footnote{Note that if one were to impose, for some reason, the additional constraint $\Tr X=0$, the planar MO gaps would remain unchanged, while finite-$N$ corrections would in general differ. Throughout this work no such constraint is imposed, and all finite-$N$ results refer to the unconstrained Hermitian matrix model with the trace mode retained.}
\begin{equation}\label{genus-corrections-Delta1}
\begin{aligned}
    \Delta_1^{(2)}(g)=-\frac{21}{4}g^2+\frac{789}{8}g^3-\frac{49617}{32}g^4+\mathcal{O}\left(g^5\right) \ ,\qquad \Delta_1^{(4)}(g)=-\frac{30885}{64}g^4+\mathcal {O}\left(g^5\right) \ .
\end{aligned}
\end{equation}
The corrections in \eqref{genus-corrections-Delta1} will also enter the biadjoint energies discussed below and contribute to the low-temperature thermodynamics.

Beyond weak coupling, we use planar gaps obtained by direct numerical solution of the MO equation \eqref{MO-eq}, following \cite{Casartelli:1979iz,Klebanov:2026vmk}. The values used in our analysis are collected in Table \ref{tab:delta-h-num}.

\begin{table}[t]
    \centering
    \begin{tabular}{ccccccc}
        \toprule
        & \multicolumn{2}{c}{$\Delta_1$}
        & \multicolumn{2}{c}{$\Delta_2$}
        & \multicolumn{2}{c}{$h_{11;11}$} \\
        $g$
        & num. & PT
        & num. & PT
        & num. & PT \\
        \midrule
        $-1/100$ & 0.9792793 & 0.9792793 & 1.9479882 & 1.9479882 & $-0.01065694$ & $-0.01065694$ \\

        $-1/40$ & 0.9449566 & 0.9449572 & 1.8607943 & 1.8607964 & $-0.02981709$ & $-0.02981556$ \\

        $-1/20$ & 0.8739858 & 0.8742359 & 1.6748646 & 1.6757872 & $-0.07818779$ & $-0.07751761$ \\

        $-0.06$ & 0.8365327 & 0.8379862 & 1.5720536 & 1.5775100 & $-0.1116500$ & $-0.1075407$ \\

        $0.90g_*$ & 0.8006120 & 0.8058480 & 1.4681130 & 1.4883070 & $-0.1536537$ & $-0.1376235$ \\

        $0.95g_*$ & 0.7773826 & 0.7877946 & 1.3960479 & 1.4373794 & $-0.1904583$ & $-0.1559102$ \\

        $0.991g_*$ & 0.7509534 & 0.7717845 & 1.3034751 & 1.3917534 & $-0.2543690$ & $-0.1729189$ \\

        $g_*$ & 0.7415737 & 0.7681102 & 1.2612196 & 1.3812234 & $-0.2955074$ & $-0.1769233$ \\
        \bottomrule
    \end{tabular}
    \caption{
        The two lowest planar adjoint excitation gaps $\Delta_1$ and $\Delta_2$, together with the diagonal two-adjoint matrix element $h_{11;11}$, for the quartic MQM. The numerical values are obtained using the Chebyshev discretization of the Marchesini-Onofri equation described in \cite{Casartelli:1979iz,Klebanov:2026vmk}. We use $M=4000,6000,8000,10000$ and extrapolate the finite-$M$ results to $M\to\infty$ using polynomial fits in $1/M$. The perturbative values (PT) follow from the weak-coupling expansions in \eqref{adjoint-lowest-planar-gaps} and \eqref{h-perturbative-1}. Their agreement with numerics is excellent at small $|g|$ and deteriorates gradually as $g\to g_*=-\sqrt{2}/(6\pi)$.}
    \label{tab:delta-h-num}
\end{table}
\subsection{Biadjoint representations}\label{subsec:biadjoint}
The biadjoint level consists of the four traceless irreducible representations $R_{\sigma\tau}$ introduced in \eqref{biadj-irreps} \cite{Boulatov:1991xz,Gaiotto:2005gd,Hatsuda:2006xr,Bourgine:2007ua}. At large $N$, a biadjoint state may be viewed as a two-impurity state built from adjoint MO modes. Let us denote the MO operator by
\begin{equation}
    (\mathcal K f)(x)=\fint\limits_{x_1}^{x_2}\d y\,\rho(y)\, \frac{f(x)-f(y)}{(x-y)^2} \ ,
\end{equation}
so that the single-adjoint eigenfunctions obey \eqref{MO-eq}
\begin{equation}
    \mathcal K\Phi_n=\Delta^{(0)}_n\Phi_n \ .
\end{equation}
At leading order in the large-$N$ expansion the two impurities are independent. The planar two-impurity Hamiltonian is
\begin{equation}
    H_{\mathrm{biadj}}^{(0)}=\mathcal K_1+\mathcal K_2 \ ,
\end{equation}
and
\begin{equation}
    H_{\mathrm{biadj}}^{(0)}\Phi_m(x)\Phi_n(y)=\left(\Delta^{(0)}_m+\Delta^{(0)}_n\right)\Phi_m(x)\Phi_n(y) \ .
\end{equation}
As we will argue below, the first genuine interaction between the two marked impurities is a connected effect and is suppressed by one power of $N$. Therefore the effective two-impurity Hamiltonian has the form
\begin{equation}\label{biadj-effective-H}
    H_{\mathrm{biadj}}=\mathcal K_1+\mathcal K_2+\frac{1}{N}\mathcal V_{\sigma\tau}+\mathcal O\left(N^{-2}\right) \ .
\end{equation}
The two-body interaction between adjoint impurities arises from the fact that $\mathcal K$ is not a derivative. Indeed, for two functions $f$ and $g$ one finds
\begin{equation}
    \mathcal K(fg)(x)=f(x)(\mathcal K g)(x)+g(x)(\mathcal K f)(x)-\fint\limits_{x_1}^{x_2}\mathrm d y\,\rho(y)\,\frac{\bigl(f(x)-f(y)\bigr)\bigl(g(x)-g(y)\bigr)}{(x-y)^2} \ .
\end{equation}
The last term is the universal two-body kernel. It is important that this is not a multiplicative potential acting on the product $f(x)g(y)$. Rather, it is a bilinear difference operator acting on the two adjoint wave functions.

Following \cite{Balthazar:2018qdv,Cho:2024kxn}, we project this bilinear kernel onto the bra state $\Phi_k(x)\Phi_l(y)$ and define the four-index tensor $h_{kl;mn} = \bra{k,l} \mathcal{V} \ket{m,n}$
\begin{equation}\label{h-def}
    h_{kl;mn}=-\int\limits_{x_1}^{x_2}\d x\int\limits_{x_1}^{x_2}\d y\,\rho(x)\rho(y)\,\Phi_k(x)\Phi_l(y)\frac{\bigl(\Phi_m(x)-\Phi_m(y)\bigr)\bigl(\Phi_n(x)-\Phi_n(y)\bigr)}{(x-y)^2} \ .
\end{equation}
This makes $h_{kl;mn}$ a universal four-index object determined entirely by the planar density \eqref{density-MQM} and the adjoint MO eigenfunctions and independent of the particular color projection at the biadjoint level. Its manifest symmetries are
\begin{equation}\label{h-symmetry}
    h_{kl;mn}=h_{lk;mn}=h_{kl;nm} \ .
\end{equation}

For an even potential, $\rho(x)=\rho(-x)$, the MO eigenfunctions can be chosen with definite parity,
\begin{equation}
    \Phi_n(-x)=p_n\Phi_n(x)\ , \qquad p_n=(-1)^n \ .
\end{equation}
Hence $h_{kl;mn}$ vanishes unless $p_kp_lp_mp_n=+1$ or equivalently $k+l+m+n$ is even.
Although $h_{kl;mn}$ is not generically symmetric under interchange of the bra and ket pairs $kl\leftrightarrow mn$, self-adjointness of the MO operator implies
\begin{equation}
    h_{kl;mn}-h_{mn;kl}=\left(\Delta_k^{(0)}+\Delta_l^{(0)}-\Delta_m^{(0)}-\Delta_n^{(0)}\right)C_{klmn}\ ,\qquad C_{klmn}=\int\d x\,\rho(x)\Phi_k\Phi_l\Phi_m\Phi_n\ .
\end{equation}
Thus the kernel is symmetric on every degenerate planar two-adjoint subspace. For an even potential, both sides vanish identically when $k+l+m+n$ is odd by parity.

The color dependence enters only through the symmetry projection on the two upper and two lower indices. If we denote these symmetries by
\begin{equation}
    \sigma=\pm 1\ , \qquad \tau=\pm 1 ,
\end{equation}
then, in an ordered two-slot basis, the leading interaction in the sector $R_{\sigma\tau}$ is
\begin{equation}\label{biadj-interaction}
    \left(\mathcal{V}_{\sigma\tau}\right)_{kl;mn}=\frac{\sigma+\tau}{2}\,h_{kl;mn} \ .
\end{equation}
Thus, the symmetric-symmetric and antisymmetric-antisymmetric sectors receive opposite order-$1/N$ splittings, while the mixed sectors $R_{+-}$ and $R_{-+}$ have no splitting at this order.

For $i=j$, the mixed sectors are absent by permutation symmetry, so only $R_{++}$ and $R_{--}$ occur. For two identical adjoint modes, the leading biadjoint energies therefore take the form
\begin{equation}\label{biadjoint-two-identical-adj}
\begin{aligned}
    \Delta_{i,i;++}(g,N)&=2\Delta_i(g,N)+\frac{h_{ii;ii}}{N}+\mathcal {O}\left(N^{-2}\right) \ ,
    \\
    \Delta_{i,i;--}(g,N)&=2\Delta_i(g,N)-\frac{h_{ii;ii}}{N}+\mathcal {O}\left(N^{-2}\right) \ .
\end{aligned}
\end{equation}
For $i\ne j$, the ordered states $\ket{ij}$ and $\ket{ji}$ form a degenerate pair with energy $\Delta_i+\Delta_j$. Because $h_{kl;mn}$ is symmetric in each pair of indices \eqref{h-symmetry}, the interaction matrix in this subspace is proportional to the matrix with all entries equal to $h_{ij;ij}$, whose nonzero eigenvalue is twice the single matrix element. Thus,
\begin{equation}\label{biadjoint-two-different-adj}
\begin{aligned}
    &\Delta_{i,j;++}(g,N)=\Delta_i(g,N)+\Delta_j(g,N)+\frac{2h_{ij;ij}}{N}+\mathcal {O}\left(N^{-2}\right) \ ,
    \\
    &\Delta_{i,j;--}(g,N)=\Delta_i(g,N)+\Delta_j(g,N)-\frac{2h_{ij;ij}}{N}+\mathcal {O}\left(N^{-2}\right) \ ,
    \\
    &\Delta_{i,j;\pm\mp}(g,N)=\Delta_i(g,N)+\Delta_j(g,N)+\mathcal{O}\left(N^{-2}\right) \ .
\end{aligned}
\end{equation}
Here $\Delta_i(g,N)$ and $\Delta_j(g,N)$ denote the full energies of the isolated adjoint excitations, including their full large-$N$ corrections. We have also assumed that no additional accidental degeneracy $\Delta_i+\Delta_j=\Delta_k+\Delta_l$, with $\{i,j\}\neq\{k,l\}$, is present. At such a resonance, one must instead diagonalize $h_{kl;mn}$ in the enlarged degenerate subspace. The occurrence of such resonances at general coupling is left for future study\footnote{If the critical Regge formula \eqref{regge-phi4} is naively treated as exact, it produces an infinite family of accidental pair-sum degeneracies. For example, whenever $3n_a+1=s_a^2$ and $s_1+s_2=s_3+s_4$, one has $\Delta_{n_1}^{\rm Regge}+\Delta_{n_2}^{\rm Regge} =\Delta_{n_3}^{\rm Regge}+\Delta_{n_4}^{\rm Regge}$. Since the Regge expression is only approximate, albeit exponentially accurate at sufficiently large $n$, such coincidences need not survive in the exact MO spectrum. Determining which of them persist, therefore, requires a dedicated numerical analysis near $g_*$.}.

For practical purposes, $h_{kl;mn}$ can be obtained either numerically, by inserting the MO eigenfunctions into \eqref{h-def}, or analytically at weak coupling using the same perturbative solution of the MO equation that leads to \eqref{adjoint-lowest-planar-gaps}. We do not reproduce this construction here. A systematic treatment will be presented separately. For $h_{11;11}$, the first few coefficients are also reproduced independently by the finite-$N$ oscillator perturbation theory of Appendix \ref{app:operator-formalism}. For several low-lying modes, we find
\begin{equation}\label{h-perturbative-1}
\resizebox{\textwidth}{!}{$
\begin{aligned}
    h_{11;11}(g)&=g-6g^2+\frac{411}{8}g^3-\frac{1995}{4}g^4+\frac{662355}{128}g^5-\frac{14329461}{256}g^6+\frac{638143329}{1024}g^7+\mathcal{O}\left(g^8\right)\ ,\\
    h_{22;22}(g)&=4g-\frac{57}{2}g^2+\frac{1095}{4}g^3-\frac{23205}{8}g^4+\frac{2068617}{64}g^5-\frac{95111091}{256}g^6+\frac{4466764383}{1024}g^7+\mathcal{O}\left(g^8\right)\ ,
    \\
    h_{33;33}(g)&=7g-54g^2+\frac{2211g^3}{4}-6174g^4+\frac{4605825}{64}g^5-\frac{110129877}{128}g^6+\frac{10712093217}{1024}g^7+\mathcal{O}\left(g^8\right)\ .
\end{aligned}
$}
\end{equation}
and
\begin{equation}\label{h-perturbative-2}
\resizebox{\textwidth}{!}{$
\begin{aligned}
    h_{12;12}(g)&=2g-\frac{51}{4}g^2+\frac{1827}{16}g^3-\frac{73563}{64}g^4+\frac{3147915}{256}g^5-\frac{139892781}{1024}g^6+\frac{6379945173}{4096}g^7+\mathcal{O}\left(g^8\right)\ ,\\
    h_{13;13}(g)&=2g-\frac{27}{2}g^2+126g^3-\frac{41931}{32}g^4+\frac{921783}{64}g^5-\frac{83855337}{512}g^6+\frac{3903276591}{2048}g^7+\mathcal{O}\left(g^8\right)\ ,\\
    h_{23;23}(g)&=5g-\frac{147g^2}{4}+\frac{5811g^3}{16}-\frac{126249g^4}{32}+\frac{11503047g^5}{256}-\frac{539243937g^6}{1024}+\frac{12884893095g^7}{2048}+\mathcal{O}\left(g^8\right)\ .
\end{aligned}
$}
\end{equation}

More generally, for $n>3$, and for $k>n$ (here we can use \eqref{h-symmetry}), the leading small-$g$ behavior is
\begin{equation}
\begin{aligned}
    h_{nn;nn}(g)&=(3n-2)g-\frac{3}{2}(17n-15)g^2+\frac{3}{4}(375n-388)g^3-\frac{3}{16}(17815n-20523)g^4+\mathcal{O}\left(g^5\right)\ , \\
    h_{nk;nk}(g)&=(3n-1)g-\frac{3}{2}(17n-9)g^2+\frac{375}{4}(3n-2)g^3-\frac{105}{16}(509n-393)g^4+\mathcal{O}\left(g^5\right)\ .
\end{aligned}
\end{equation}
These perturbative series provide useful input for the finite-$N$ splitting of the biadjoint levels \eqref{biadjoint-two-identical-adj}-\eqref{biadjoint-two-different-adj} and give an independent check of the numerical evaluation of the two-body kernel. 

In the limit in which the second adjoint mode is taken to be highly excited $k\gg1$, it acts as a WKB probe of the planar density. The diagonal biadjoint matrix element then reduces to the Hellmann-Feynman variation of the adjoint MO gap,
\begin{equation}
    \lim_{k\to\infty} h_{nk;nk}(g)=g\,\partial_g \Delta^{(0)}_n(g) \ .
\end{equation}
Moreover, at finite but sufficiently large $k>n$, the perturbative expansion of $h_{nk;nk}$ agrees with that of $g\,\partial_g \Delta^{(0)}_n$ through order $g^{k-1}$, or equivalently up to corrections of order $\mathcal{O}\left(g^k\right)$.

The discussion above determines the complete order-$1/N$ interaction between two adjoint impurities directly from the planar MO problem. For arbitrary planar modes $i,j$, the corresponding splitting is fixed by the matrix element $h_{ij;ij}$: the $R_{++}$ and $R_{--}$ sectors receive opposite shifts, while the mixed sectors have no order-$1/N$ contact term. 

Beyond order $1/N$, the planar MO kernel alone no longer determines the spectrum. The separation used above has a simple physical interpretation: the adjoint excitations may be regarded as elementary ``atoms", while biadjoint and higher multi-adjoint states form ``molecules". The energy of a molecule consists of the full energies of its isolated adjoint constituents $\Delta_n(g,N)$ together with their residual, representation-dependent interaction. Making this decomposition explicit for a biadjoint state, we write
\begin{equation}\label{biadjoint-finiteN-expansion}
    \Delta_{i,j;\sigma\tau}(g,N)=\Delta_i(g,N)+\Delta_j(g,N)+\sum_{r=1}^{\infty}\frac{\Delta_{i,j;\sigma\tau}^{(r)}(g)}{N^r}\ ,\qquad \sigma,\tau=\pm\ .
\end{equation}
Thus the coefficients $\Delta_{i,j;\sigma\tau}^{(r)}$ contain only the ``molecular" interaction which remains after the two isolated adjoint energies have been subtracted. The leading interaction is determined by the planar matrix element $h_{ij;ij}$ \eqref{h-def}, whereas higher orders require genuinely nonplanar information.

For the lowest mode $\Delta_1$ this decomposition is unambiguous, since the adjoint occurs with multiplicity one at oscillator level one. For higher modes, several copies of the adjoint representation occur and mix at finite $N$, so the corresponding isolated adjoint energies must first be identified by diagonalizing this multiplicity space. We do not solve this general mixing problem here. Its operator realization, together with the explicit finite-$N$ analysis of the lowest adjoint and biadjoint states, is described in Appendix \ref{app:operator-formalism}.

The lowest two-adjoint states are particularly simple. The representations $R_{++}$ and $R_{--}$ first occur at oscillator level two with multiplicity one, so these two branches can be followed unambiguously at finite $N$. Denoting their gaps by $\Delta_{1,1;++}\equiv\Delta_+$ and $\Delta_{1,1;--}\equiv\Delta_-$, the perturbative expansion takes the alternating form
\begin{equation}\label{lowest-biadjoint-allorders}
    \Delta_\pm(g,N)=2\Delta_1(g,N)\pm\frac{A(g)}{N}+\frac{B(g)}{N^2}\pm\frac{C(g)}{N^3}+\frac{D(g)}{N^4}+\ldots\ .
\end{equation}
Here $A(g),B(g),C(g),D(g),\ldots$ are the successive interaction coefficients of the lowest two-adjoint state. The leading one is fixed by the planar two-body kernel, $A(g)\equiv h_{11;11}(g)$ \eqref{h-perturbative-1}, while the higher coefficients are obtained using the finite-$N$ oscillator perturbation method described in Appendix \ref{app:operator-formalism}:
\begin{equation}\label{lowest-biadjoint-interactions}
    \begin{aligned}
        B(g)&=-\frac{21}{4}g^2+\frac{345}{4}g^3-\frac{19449}{16}g^4+\mathcal O\left(g^5\right)\ ,
        \\
        C(g)&=\frac{185}{2}g^3-\frac{10383}{4}g^4+\mathcal O\left(g^5\right)\ ,
        \\
        D(g)&=-\frac{93471}{64}g^4+\mathcal O\left(g^5\right)\ .
    \end{aligned}
\end{equation}
The same construction can be continued systematically to higher orders in $g$ and $1/N$.

The alternating structure in \eqref{lowest-biadjoint-allorders} is not restricted to the lowest biadjoint state. As we show in Section \ref{sec:low-T-planar-thermodynamics}, it follows from the even-power large-$N$ expansion of the thermal free energy, together with the exact representation dimensions. For isolated two-adjoint levels, the residual interactions in the $R_{++}$ and $R_{--}$ sectors satisfy
\begin{equation}\label{biadjoint-parity-pattern}
    \Delta_{i,j;++}^{(2r)}=\Delta_{i,j;--}^{(2r)}\ ,\qquad\Delta_{i,j;++}^{(2r-1)}=-\Delta_{i,j;--}^{(2r-1)}\ ,\qquad r=1,2,\ldots\ .
\end{equation}

The mixed representations $R_{+-}$ and $R_{-+}$ are conjugate and have identical spectra. Applying the same thermodynamic argument to the complete two-adjoint multiplet gives
\begin{equation}\label{mixed-biadjoint-parity-pattern}
    \Delta_{i,j;+-}^{(2r)}=\Delta_{i,j;-+}^{(2r)}\ ,\qquad\Delta_{i,j;+-}^{(2r-1)}=\Delta_{i,j;-+}^{(2r-1)}=0\ ,\qquad i\neq j\ , \qquad r=1,2,\ldots\ .
\end{equation}

A useful analogy is with $k$-strings in confining gauge theories \cite{KorthalsAltes:2005ph,Greensite:2011gg}. At large $N$, a string carrying two units of fundamental flux can be viewed as two non-interacting strings, while at finite $N$ their symmetric and antisymmetric color combinations can have different energies. A similar representation-dependent interaction appears in the biadjoint sector of MQM. The main difference is that each adjoint impurity carries both fundamental and antifundamental indices, allowing independent symmetry projections on the two sets of indices and giving rise to the four representations $R_{\sigma\tau}$.
\section{Thermodynamics at small \texorpdfstring{$g$}{}}\label{sec:small-g}
In this section we develop weak-coupling perturbation theory for the thermodynamics of the quartic matrix model around the exactly solvable harmonic oscillator. At each order in $g$, the dependence on the inverse temperature $\beta$ is kept exact. This gives a useful analytic description of the thermal observables over a broad range of temperatures and will serve as a reference for the approximations developed in the two following sections.
\subsection{Matrix Harmonic Oscillator (\texorpdfstring{$g=0$}{})}\label{sec:MHO-thermo}
We begin with the matrix harmonic oscillator $V(x)=x^2/2$.  Since a Hermitian $N\times N$ matrix contains $N^2$ independent real components, the exact partition function is
\begin{equation}\label{partition-func-MHO}
    Z_0(\beta)=\frac{e^{-\beta N^2/2}}{(1-q)^{N^2}}\ ,\qquad q=e^{-\beta}\ ,
\end{equation}
and hence
\begin{equation}\label{free-energy-oscillator}
    \log Z_0=-\frac{\beta N^2}{2}-N^2\log(1-q)\ .
\end{equation}

To resolve the thermal trace into representation sectors, we factor out the singlet contribution,
\begin{equation}\label{character-expand}
    Z_0=Z_{\mathrm{sing}}\left(1+\sum_{\substack{R\in\mathcal R_0,\,R\ne\mathbb1}}d_R\,P^{(N)}_R(q)\right)\ ,\qquad Z_{\mathrm{sing}}=e^{-\beta N^2/2}\prod_{m=1}^{N}\frac{1}{1-q^m}\ .
\end{equation}
Here $P^{(N)}_R(q)$ denotes the reduced thermal character in the sector $R$, and $d_R$ its dimension. See Appendix \ref{app:representation-characters} for details of the calculation of these objects.

The first contribution comes from the adjoint representation,
\begin{equation}\label{adj-dim-and-character}
    d_{\mathrm{adj}}=N^2-1\ , \qquad P^{(N)}_{\mathrm{adj}}(q)=\frac{q-q^N}{1-q}=q+q^2+\mathcal O\left(q^3\right)\ .
\end{equation}

At the next representation level, there are four representations \eqref{biadj-irreps}, all of which are present for $N\geq4$ and have dimensions of order $N^4$ at large $N$. The two self-conjugate representations $R_{++}$ and $R_{--}$ (called $B_2$ and $C_2$ in \cite{Boulatov:1991xz, Hatsuda:2006xr, Bourgine:2007ua}) have dimensions
\begin{equation}\label{biadj-self-conj-dims}
    d_{++}=\frac{1}{4}N^2(N+3)(N-1)\ ,\qquad d_{--}=\frac{1}{4}N^2(N-3)(N+1)\ ,
\end{equation}
and the corresponding finite-$N$ reduced thermal characters are
\begin{equation}\label{biadj-characters-1}
    P^{(N)}_{++}(q)=\frac{q^2(1-q^{N-1})(1-q^N)}{(1-q)(1-q^2)}\ ,\qquad P^{(N)}_{--}(q)=\frac{q^2(1-q^{N-3})(1-q^N)}{(1-q)(1-q^2)}\ .
\end{equation}
Both begin at order $q^2$. The remaining conjugate pair, $R_{+-}$ and $R_{-+}$ (called $A_2$ and $\bar A_2$ in 
\cite{Boulatov:1991xz, Hatsuda:2006xr, Bourgine:2007ua}),
has
\begin{equation}\label{biadj-characters-2}
    d_{+-}=d_{-+}=\frac{1}{4}(N^2-1)(N^2-4)\ ,\qquad P^{(N)}_{+-}(q)=P^{(N)}_{-+}(q)=\frac{q^3(1-q^{N-2})(1-q^{N-1})}{(1-q)(1-q^2)}\ .
\end{equation}
Their characters begin at order $q^3$; therefore, they do not contribute to the calculation below. Combining all the characters and dimensions, and keeping terms through order $q^2$ in \eqref{character-expand}, gives
\begin{equation}
    \log Z_0=-\frac{\beta N^2}{2}+N^2 q+\frac{N^2}{2}q^2+\mathcal O\left(q^3\right)\ ,
\end{equation}
in agreement with the expansion of the exact result \eqref{partition-func-MHO}. 

We can proceed to the next level and include the $9$ representations whose dimensions grow as $N^6$ at large $N$, which, together with reduced thermal characters, are collected in Appendix \ref{app:representation-characters}. Only the three self-conjugate representations, together with $R_{+-},R_{-+}$, contribute to the next order in $q$ reproducing the expected term $N^2 q^3/3$ in the expansion of $\log Z_0$. These calculations illustrate an important feature of the sum over representations. The individual $p$-adjoint dimensions scale as $N^{2p}$, but their contribution to $\log Z$ combines with products of lower-level terms so that the final answer remains of order $N^2$. The subleading powers of $N$ in the exact representation dimensions are therefore essential and cannot be discarded before performing the full sum and taking its logarithm.
\subsection{Free energy}\label{sec:small-g-F}
We now turn on the quartic interaction. For the perturbative calculation it is convenient to pass from the normalized-trace variables of \eqref{MQM-Hamiltonian-def} to the canonically normalized matrices used in Appendix \ref{app:operator-formalism},
\begin{equation*}
    Q=\sqrt N\,X\ ,\qquad \Pi=\sqrt N\,P\ ,
\end{equation*}
which obey
\begin{equation*}
    [\Pi_{ij},Q_{kl}]=-i\,\delta_{il}\delta_{jk}\ .
\end{equation*}
In these variables the Hamiltonian becomes
\begin{equation*}
    H=\frac{1}{2}\Tr\left(\Pi^2+Q^2\right)+\frac{g}{N}\Tr Q^4\ .
\end{equation*}
To keep the notation light, throughout this subsection we relabel $Q\to X$ and $\Pi\to P$. Thus the matrices appearing below are canonically normalized; we return to the normalized-trace variables of \eqref{MQM-Hamiltonian-def} in the following sections.

In the interaction picture, the partition function can be written as
\begin{equation}
    Z=\Tr e^{-\beta H}=Z_0\cdot\left\langle\mathcal T_\tau\exp\left[-\frac{g}{N}\int_0^\beta\limits\d\tau\, \Tr X(\tau)^4\right]\right\rangle_0,
\end{equation}
where $\mathcal T_\tau$ denotes ordering in Euclidean time and
$\braket{\ldots}_0$ is the thermal expectation value in the free
matrix oscillator.

Taking the logarithm selects connected vacuum diagrams. The corresponding
cumulant expansion is
\begin{equation}\label{small-g-cumulant}
    \log Z={}\log Z_0-\frac{g}{N}\int_0^\beta\limits \d\tau\,\left\langle\Tr X(\tau)^4\right\rangle_0+\frac{g^2}{2N^2}\int_0^\beta\limits\int_0^\beta\limits \d\tau_1\d\tau_2\,\left\langle\mathcal T_\tau\Tr X(\tau_1)^4\Tr X(\tau_2)^4\right\rangle_{0,c}+\mathcal{O}\left(\frac{g^3}{N^3}\right).
\end{equation}

\begin{figure}[h!]
    \centering
    \resizebox{\textwidth}{!}{
    \begin{tikzpicture}[thick]

        \begin{scope}[shift={(0,0)}]
            \draw (0,0) circle (1);
            \draw (2,0) circle (1);
            \filldraw (1,0) circle (2pt);

            \node at (1,-1.5) {(a) figure-eight, $\mathcal{O}(g)$};
        \end{scope}

        \begin{scope}[shift={(6,0)}]
            \draw (-1,0) circle (1);
            \draw (1,0) circle (1);
            \draw (3,0) circle (1);
            \filldraw (0,0) circle (2pt);
            \filldraw (2,0) circle (2pt);

            \node at (1,-1.5) {(b) double snail, $\mathcal{O}\left(g^2\right)$};
        \end{scope}

        \begin{scope}[shift={(11.5,0)}]
            \filldraw (0,0) circle (2pt);
            \filldraw (2,0) circle (2pt);

            \draw (1,0) circle (1);
            
            \draw (0,0) to[bend left=35] (2,0);
            \draw (0,0) to[bend right=35] (2,0);

            \node at (1,-1.5) {(c) melon, $\mathcal{O}\left(g^2\right)$};
        \end{scope}

    \end{tikzpicture}
    }
    \caption{
        Connected vacuum topologies contributing to the free energy through order $g^2$. The figure-eight diagram contributes at order $g$, while the double-snail and melon diagrams contribute at order $g^2$.
    }
    \label{fig:vacuum-diagrams-g2}
\end{figure}
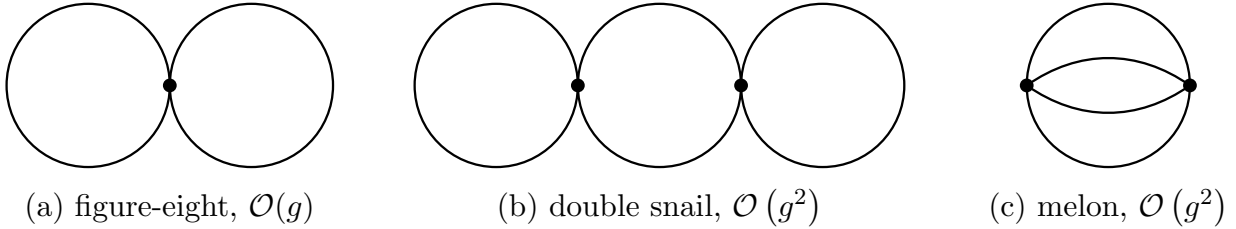

At first order there is a single connected vacuum topology, the figure-eight diagram.  At second order the connected Wick contractions organize into two topologies, the double-snail and melon diagrams. These three contributions are displayed together in Figure \ref{fig:vacuum-diagrams-g2}. The diagrams through order $g^2$ will be evaluated at finite $N$ by applying Wick's theorem with the exact thermal propagator.  At order $g^3$ the number of ribbon contractions increases considerably; we therefore do not display the individual three-vertex diagrams and quote only their complete planar sum.

For the unconstrained Hermitian-matrix oscillator, the matrix is not traceless, and the thermal propagator is
\begin{equation}\label{thermal-matrix-propagator}
    \left\langle\mathcal T_\tau X_{ij}(\tau)X_{kl}(0)\right\rangle_0=\delta_{il}\delta_{jk}\,G_\beta(\tau)\ , \qquad G_\beta(\tau+\beta)=G_\beta(\tau)\ ,
\end{equation}
where \eqref{MQMvillain}
\begin{equation*}
    G_\beta(\tau)=\frac{\cosh\left(\frac{\beta}{2}-\tau\right)}{2\sinh(\beta/2)}=\frac{e^{-\tau}+q\,e^{\tau}}{2(1-q)}\ , \qquad 0\leq\tau<\beta\ .
\end{equation*}

A basic family of thermal integrals, already appearing at order $g^2$ and recurring in higher-order diagrams, is
\begin{equation}\label{Ik-def}
    I_k(\beta)\equiv\int_0^\beta\limits\d\tau\,\big(G_\beta(\tau)\big)^k=\frac{1}{2^k(1-q)^k}\sum_{r=0}^{k}\binom{k}{r}q^r\cdot
    \begin{cases}
        \displaystyle\frac{q^{\,k-2r}-1}{2r-k}\ ,
        & 2r\neq k\ ,
        \\
        \beta\ ,
        & 2r=k\ .
    \end{cases}
\end{equation}
The second line in \eqref{Ik-def} is present only for even $k$ and comes from the time-independent term in the binomial expansion of $G_\beta(\tau)^k$. In particular, the integrals needed at order $g^2$ are
\begin{equation}\label{I2-I4}
    I_2(\beta)=\frac{1-q^2+2\beta q}{4(1-q)^2}\ , \qquad I_4(\beta)=\frac{1+8q-8q^3-q^4+12\beta q^2}{32(1-q)^4}\ .
\end{equation}

The figure-eight diagram gives the first-order contribution
\begin{equation}\label{logZ-g1-1/N}
    \frac{1}{N^2}(\log Z)^{(1)}=-\beta g
    \left(2+\frac{1}{N^2}\right)\big(G_\beta(0)\big)^2\ , \qquad G_\beta(0)=\frac{1+q}{2(1-q)}\ .
\end{equation}
At second order, the double-snail topology produces the term proportional to $G_\beta(0)^2I_2(\beta)$, while the melon topology produces the term proportional to $I_4(\beta)$. Their sum is
\begin{equation}\label{logZ-g2-1/N}
    \frac{1}{N^2}(\log Z)^{(2)}=g^2\beta\left[\left(16+\frac{20}{N^2}\right)\big(G_\beta(0)\big)^2I_2(\beta)+\left(2+\frac{10}{N^2}\right)I_4(\beta)\right]\ .
\end{equation}
Equations \eqref{logZ-g1-1/N} and \eqref{logZ-g2-1/N} give the complete finite-$N$ correction through order $g^2$ for the unrestricted Hermitian matrix model. At higher orders, the same integrals $I_k(\beta)$ reappear as substructures, together with genuinely multivariable Euclidean-time integrals associated with diagrams containing three or more interaction vertices.

It is convenient to introduce the planar thermal free energy
\begin{equation}\label{Free-energy-def}
    \mathcal F(\beta,g)\equiv-\frac{1}{\beta N^2}\log Z(\beta,g)\bigg|_{N=\infty}.
\end{equation}
The free contribution appearing in the planar thermal free energy is \eqref{free-energy-oscillator}
\begin{equation}\label{free-planar-thermal-free-energy}
    \mathcal F_0(\beta)=-\frac{1}{\beta N^2}\log Z_0=\frac{1}{2}+\frac{1}{\beta}\log(1-q)=\frac{1}{\beta}\log\left(2\sinh\frac{\beta}{2}\right)\ , \qquad q=e^{-\beta}.
\end{equation}

Evaluating the connected vacuum diagrams gives the planar thermal free
energy, through third order in $g$, in the form
\begin{equation}\label{small-g-planar-free-energy}
    \mathcal F(\beta,g)=\mathcal F_0(\beta)+\sum_{r=1}^{3}g^r\sum_{s=0}^{r-1}\beta^s\,\mathcal F_{r,s}(q)+\mathcal O\left(g^4\right)\ ,
\end{equation}
where at first order,
\begin{equation}\label{small-g-F-functions-1}
    \mathcal F_{1,0}(q)=\frac{(1+q)^2}{2(1-q)^2}\ .
\end{equation}
At second order,
\begin{equation}\label{small-g-F-functions-2}
    \mathcal F_{2,0}(q)=-\frac{(1+q)(17q^2+40q+17)}{16(1-q)^3}\ , \qquad \mathcal F_{2,1}(q)=-\frac{q(8q^2+19q+8)}{4(1-q)^4}\ .
\end{equation}
At third order,
\begin{equation}\label{small-g-F-functions-3}
    \begin{aligned}
        \mathcal F_{3,0}(q)&=\frac{(1+q)^2(75q^2+224q+75)}{16(1-q)^4}\ ,
        \\
        \mathcal F_{3,1}(q)&=\frac{3q(1+q)(18q^2+49q+18)}{4(1-q)^5}\ ,
        \\
        \mathcal F_{3,2}(q)&=\frac{q(8q^4+99q^3+188q^2+99q+8)}{6(1-q)^6}\ .
    \end{aligned}
\end{equation}

The explicit coefficients exhibit a remarkably simple organization. At order $g^r$, the explicit dependence on $\beta$, apart from that contained in $q=e^{-\beta}$, is polynomial of degree $r-1$, and the coefficient of $\beta^s$ takes the form
\begin{equation}\label{small-g-F-general-pattern}
    \mathcal F_{r,s}(q)=\frac{f_{r,s}(q)}{(1-q)^{r+s+1}}\ ,\qquad q^{r+s+1}f_{r,s}(q^{-1})=f_{r,s}(q)\ , \qquad 0\leq s\leq r-1\ .
\end{equation}
Thus $f_{r,s}(q)$ is reciprocal with respect to the formal degree $r+s+1$. For $s=0$ this coincides with its ordinary polynomial degree, while for $s\geq1$ its degree is $r+s$. Equivalently,
\begin{equation}\label{small-g-F-reciprocity}
    \mathcal F_{r,s}(q^{-1})= (-1)^{r+s+1}\mathcal F_{r,s}(q)\ .
\end{equation}
The coefficients displayed above suggest that this reciprocal structure persists to all orders in the weak-coupling expansion, and it will be interesting to prove this explicitly. Equation \eqref{small-g-planar-free-energy} retains the full temperature dependence at each order in $g$ and does not involve a high- or low-temperature expansion.

As a nontrivial check, in the zero-temperature limit $q\to0$ one finds\footnote{After restoring the first non-planar correction, the ground-state at weak coupling is equal to  
\begin{equation}\label{ground-state-quartic}
    \frac{E_{\rm gs}(g,N)}{N^2}=\frac{1}{2}+\frac{g}{4}\left(2+\frac{1}{N^2}\right)-g^2\left(\frac{17}{16}+\frac{25}{16N^2}\right)+\frac{75}{16}g^3+\mathcal{O}\left(\frac{g^3}{N^2},g^4\right)\ .
\end{equation}}
\begin{equation}\label{E0-small-g-from-thermal}
    \mathcal{E}_0(g)=\lim_{\beta\to\infty}\mathcal F(\beta,g)=\frac{1}{2}+\frac{g}{2}-\frac{17}{16}g^2+\frac{75}{16}g^3+\mathcal{O}\left(g^4\right).
\end{equation}
This precisely reproduces the small-$g$ expansion \eqref{planar-ground-state-small-g} of the known analytic planar expression \eqref{MQM-planar-gs-quartic}.  The agreement provides an independent check of the planar ribbon combinatorics and of the finite-temperature Euclidean-time integrals.
\subsection{Expansion in thermal winding number}\label{seq:small-g-winding-exp}
The same result admits a useful organization in thermal winding number.
After subtracting the interacting ground-state contribution, we write
\begin{equation}
    \frac{1}{N^2}\left(\log Z(\beta,g;N)+\beta E_{\mathrm{gs}}(g,N)\right)=\sum_{n=1}^{\infty}\ell_n(\beta,g;N)\,q^n\ ,
\end{equation}
with
\begin{equation}
    \ell_n(\beta,g;N)=\ell_n^{(0)}(\beta,g)+\frac{1}{N^2}\ell_n^{(2)}(\beta,g)+\mathcal{O}\left(N^{-4}\right)\ .
\end{equation}
In the planar limit, $E_{\mathrm{gs}}(g,N)/N^2\to \mathcal{E}_0(g)$ and $\ell_n(\beta,g;N)\to\ell_n^{(0)}(\beta,g)$.

The planar coefficient, obtained from
\eqref{small-g-planar-free-energy} through $g^3$, is
\begin{equation}\label{elln-planar-g3}
\resizebox{\textwidth}{!}{$
\begin{aligned}
    \ell_n^{(0)}(\beta,g)
    ={}&
    \frac{1}{n}-2n\beta g+g^2\left[\frac{\beta}{8}\left(37n^2+17\right)+\frac{\beta^2n}{24}\left(35n^2+13\right)\right]-
    \\
    &-
    g^3\left[\frac{\beta n}{24}\left(374n^2+637\right)+\frac{\beta^2n^2}{16}\left(85n^2+131\right)+\frac{\beta^3n}{360}\left(201n^4+305n^2-26\right)\right]+\mathcal{O}\left(g^4\right)\ .
\end{aligned}
$}
\end{equation}
The finite-$N$ calculation of \eqref{logZ-g1-1/N} and \eqref{logZ-g2-1/N} determines the genus-one contribution through $\mathcal{O}\left(g^2\right)$,
\begin{equation}\label{elln-nonplanar-g2}
    \ell_n^{(2)}(\beta,g)={}-n\beta g+g^2\left[\frac{5\beta}{8}\left(13n^2+5\right)+\frac{5\beta^2n}{24}\left(11n^2+1\right)\right]+\mathcal{O}\left(g^3\right)\ .
\end{equation}

The expansion in $q$ also provides a useful spectral check. At the first oscillator level,
\begin{equation}
    \mathbf N\otimes\overline{\mathbf N}=\mathbb{1}\oplus\mathrm{Adj}\ ,
\end{equation}
so that the coefficient of the first winding is
\begin{equation}
    N^2\ell_1=(N^2-1)e^{-\beta(\Delta_1-1)}+e^{-\beta(\omega_N-1)}\ ,
\end{equation}
where $\omega_N$ is the lowest singlet gap \eqref{singlet-spectrum}, whose planar limit is $\omega(g)$ \eqref{frequency-analyt}.

At planar order the singlet contribution is suppressed, giving
\begin{equation}\label{Delta1-from-ell1}
    \Delta_1^{(0)}=1-\frac{1}{\beta}\log\ell_1^{(0)}\ .
\end{equation}
Substituting the $n=1$ specialization of \eqref{elln-planar-g3} reproduces the planar adjoint gap through $\mathcal O(g^3)$ quoted in \eqref{adjoint-lowest-planar-gaps}. The same relation provides a nonplanar check: using the genus-one coefficient \eqref{elln-nonplanar-g2}, together with the planar singlet gap, precisely reproduces the $\mathcal{O}(g^2)$ term in $\Delta_1^{(2)}$ quoted in \eqref{genus-corrections-Delta1}. Since the genus-one free energy is known here only through $\mathcal{O}(g^2)$, higher coefficients in $\Delta_1^{(2)}$ cannot be tested by this method with the present data.

Starting at the second winding, the spectral decomposition becomes less direct. Higher adjoint modes coexist with singlet-dressed lower adjoints in the same representation and can mix at finite $N$; in particular, this already affects the continuation of the second adjoint mode $\Delta_2$. The thermal coefficients also contain contributions from multi-adjoint representations. Extracting individual finite-$N$ energies therefore requires additional spectral information and resolution of the corresponding mixing problem, as discussed in Sections \ref{subsec:adjoint}--\ref{subsec:biadjoint} and illustrated in Appendix \ref{app:operator-formalism}.
\subsection{Thermal observables}
Once the thermal free energy is known, several observables follow directly by differentiation. Returning to the normalized-trace variables of \eqref{MQM-Hamiltonian-def} and restoring the oscillator mass in $V(X)=m^2X^2/2+gX^4$, we have
\begin{equation}\label{thermal-observables-from-F}
    \begin{aligned}
        \mathcal E(\beta,m,g)\equiv\frac{\braket{H}_\beta}{N^2}=\partial_\beta\left[\beta\mathcal F(\beta,m,g)\right]\ , \quad \braket{\tr X^2}_\beta=\frac{1}{m}\,\partial_m\mathcal F(\beta,m,g)\ ,\quad \braket{\tr X^4}_\beta=\partial_g\mathcal F(\beta,m,g)\ .
    \end{aligned}
\end{equation}
They also obey the thermal virial relation
\begin{equation}\label{thermal-virial-relation}
    \mathcal E=m^2\braket{\tr X^2}_\beta+3g\braket{\tr X^4}_\beta\ .
\end{equation}

The explicit weak-coupling expansions of $\mathcal E$, $\braket{\tr X^2}_\beta$, and $\braket{\tr X^4}_\beta$ are collected in Appendices \ref{app:thermal-energy} and \ref{app:thermal-moments}. Their zero-temperature limits reproduce the corresponding ground-state results, while their low-temperature behavior will be recovered directly from the non-singlet spectrum in the next section. At high temperature, the free energy can be obtained by dimensional reduction, including the contribution of nonzero Matsubara modes. We compute this explicitly in Appendix \ref{app:free-energy-highT}. 

The quadratic Casimir, which measures the non-singlet content of thermal states, can likewise be studied perturbatively. Its weak-coupling expansion and its low- and high-temperature limits are presented in Appendix \ref{app:casimir}.
\section{Low-temperature expansion}\label{sec:lowTexpansion}
Having determined the structure of the lowest non-singlet states in Section \ref{sec:Non-singlets}, we now use this spectral information to construct the low-temperature expansion of thermodynamic observables. At sufficiently low temperature, only a small number of the lowest non-singlet levels contribute appreciably, while the large-$N$ expansion of their energies determines the corresponding corrections to the planar free energy. In particular, the low-temperature expansion is sensitive not only to the planar excitation gaps, but also to their subleading $1/N$ corrections.
\subsection{Leading spectral contributions}\label{sec:low-T-planar-thermodynamics}
We begin by retaining the two lowest adjoint gaps, $\Delta_1$ and $\Delta_2$, together with the lowest levels in the two self-conjugate biadjoint representations, $\Delta_{1,1;++}$ and $\Delta_{1,1;--}$. For brevity, we denote the latter by $\Delta_+\equiv\Delta_{1,1;++}$ and $\Delta_-\equiv\Delta_{1,1;--}$ throughout this section\footnote{In the final planar thermodynamic expressions, $\Delta_n$ denotes the planar gap $\Delta_n^{(0)}$, unless the finite-$N$ dependence is displayed explicitly.}. Building on the results of Section \ref{sec:Non-singlets}, their large-$N$ expansions take the form
\begin{equation}\label{Deltaexpand}
    \begin{aligned}
        \Delta_{\pm}(g,N)=2\Delta_1(g,N)+\frac{A_\pm(g)}{N}+\frac{B_\pm(g)}{N^2}+\frac{C_\pm(g)}{N^3}+\mathcal{O}\left(\frac{1}{N^4}\right)\ .
    \end{aligned}
\end{equation}

The structure of these expansions is constrained by the genus expansion of the thermal free energy. For a nondegenerate adjoint level, the dimension $d_{\rm adj}=N^2-1$ implies that odd powers of $1/N$ in its energy would generate odd powers of $N$ in the free energy. This explains the even-power structure of \eqref{adj-gaps}. In the biadjoint sectors, odd powers of $1/N$ are allowed, but their contributions must cancel between different representations.

Consider the contribution associated with the planar energy $2\Delta_1$ in $\log Z$. After factoring out $e^{-2\beta\Delta_1}$, its coefficient is
\begin{equation}
    \mathcal W_{11}=d_{++}e^{-\beta(\Delta_+-2\Delta_1)}+d_{--}e^{-\beta(\Delta_--2\Delta_1)}-\frac{1}{2}d_{\mathrm{adj}}^2\ .
\end{equation}
The last term comes from expanding the logarithm of the partition function. Using the exact dimensions \eqref{biadj-self-conj-dims}, general order-$1/N$ shifts $A_\pm/N$ would produce a term $-\beta N^3(A_++A_-)/4$. Its cancellation requires $A_+=-A_-\equiv A$, in agreement with the color projections of the two-impurity interaction \eqref{biadj-interaction}, which also determine $A=h_{11;11}$.

The absence of odd powers of $N$ imposes further constraints. Substituting \eqref{Deltaexpand}, we find that the coefficient of $N$ is
\begin{equation}
    [\mathcal W_{11}]_N=\frac{1}{4}\left[\beta^2 A(B_+-B_-)-\beta\left(2(B_+-B_-)+C_++C_-\right)\right]\ .
\end{equation}
Since this must vanish for arbitrary $\beta$, the coefficients of $\beta$ and $\beta^2$ vanish separately. For generic $A\neq 0$, we therefore obtain $B_+=B_-$ and $C_+=-C_-$. The argument extends to higher orders: the identity $d_{++}(-N)=d_{--}(N)$, together with the distinct order-$1/N$ splittings, requires the two energy branches to be exchanged under the formal transformation $N\to -N$. For $i\neq j$, the same reasoning applies to the complete four-sector multiplet, with the mixed branches invariant under this transformation. This proves the parity relations \eqref{biadjoint-parity-pattern}--\eqref{mixed-biadjoint-parity-pattern}, provided there are no accidental degeneracies between different planar two-adjoint levels.

We henceforth write $B_+=B_-\equiv B$. As discussed in Section \ref{sec:singlet-sector}, thermal singlet excitations contribute only at order $N^0$ and do not enter the planar free energy. Combining the lowest adjoint and biadjoint contributions in \eqref{Z-decomposition}, we obtain
\begin{equation}\label{lowT}
    \begin{aligned}
        N^{-2}\log Z+\beta\mathcal E_0
        ={}&e^{-\beta\Delta_1}+e^{-\beta\Delta_2}+\left[-\frac{1}{2}-\beta A-\frac{\beta}{2}B+\frac{\beta^2A^2}{4}\right]e^{-2\beta\Delta_1}+
        \\
        &+\mathcal O\left(e^{-\beta\Delta_3},\beta^2e^{-\beta(\Delta_1+\Delta_2)},\beta^4e^{-3\beta\Delta_1}\right)\ .
    \end{aligned}
\end{equation}
The $1/N$ and $1/N^2$ corrections to the biadjoint energies thus contribute to the planar free energy at order $e^{-2\beta\Delta_1}$, despite vanishing for individual states as $N\to\infty$. Here and below, the gaps in the final planar expressions are understood as their large-$N$ limits.

Equation \eqref{lowT} also shows that the low-temperature treatment of \cite{Cho:2024kxn,Adams:2025nww} is incomplete already at order $e^{-2\beta\Delta_1}$. Their result contains the interaction term $-\beta A e^{-2\beta\Delta_1}$, but not the full coefficient in \eqref{lowT}. In particular, the non-interacting two-adjoint contribution is $-\frac{1}{2}e^{-2\beta\Delta_1}$, rather than $+\frac{1}{2}e^{-2\beta\Delta_1}$ obtained by treating adjoints as independent matrix oscillators. This discrepancy is already visible at $g=0$: the exact matrix-oscillator partition function \eqref{partition-func-MHO} gives $q+\frac{1}{2}q^2+\mathcal O(q^3)$ for the planar logarithm after subtracting the ground-state contribution, whereas the independent-adjoint counting gives $q+\frac32 q^2+\mathcal O(q^3)$, with $q=e^{-\beta}$. Appendix \ref{app:non-interacting-ungauged-states} explains the origin of this mismatch using a toy model that preserves the exact $U(N)$ representation-theoretic counting while neglecting the residual interactions between adjoint constituents of multi-adjoint states.

The hierarchy among the contributions at order $e^{-2\beta\Delta_1}$ depends strongly on the coupling. Near $g_*$, $A$ approaches a finite nonzero value (see Table \ref{tab:delta-h-num}), while the relevant low-temperature regime has $\beta\gg1$. The contribution $\beta^2A^2/4$ is therefore parametrically enhanced relative to both $-\beta A$ and the constant $-1/2$. Its comparison with the $B$ term depends on their unknown behavior near $g_*$. If $B$ remains finite there, the $\beta^2A^2/4$ contribution necessarily dominates for $\beta\gg1$. Near the free point, the hierarchy is different: $A\sim g$ and $B\sim g^2$, so the relative size of the corrections is controlled by the combination $\beta g$.

The corresponding low-temperature expansion of the average energy follows from \eqref{lowT} and is given by
\begin{equation}\label{lowT-energy}
    \begin{aligned}
        \mathcal E(\beta)
        ={}&\mathcal E_0+\Delta_1 e^{-\beta\Delta_1}+\Delta_2 e^{-\beta\Delta_2}+
        \\
        &+\Bigg[A+\frac{B}{2}-\frac{\beta A^2}{2}+\Delta_1\left(\frac{\beta^2A^2}{2}-2\beta A-\beta B-1\right)\Bigg]e^{-2\beta\Delta_1}+
        \\
        &+\mathcal{O}\left(e^{-\beta\Delta_3},\beta^2e^{-\beta(\Delta_1+\Delta_2)},\beta^4e^{-3\beta\Delta_1}\right)\ .
    \end{aligned}
\end{equation}

Extending the expansion beyond \eqref{lowT} requires including several additional contributions. These include the next adjoint excitation $\Delta_3$, the lowest states in the remaining two biadjoint representations, $R_{+-}$ and $R_{-+}$, as well as the lowest states in the three self-conjugate triadjoint representations, whose planar energies approach $3\Delta_1$. The lowest levels in $R_{+-}$ and $R_{-+}$ have the structure given in \eqref{biadjoint-two-different-adj}, with no order-$1/N$ interaction, in agreement with the parity relations derived above.

At higher orders in the low-temperature expansion, additional non-singlet states contribute, requiring knowledge of their energies to increasingly high orders in $1/N$. The even-power genus expansion imposes relations among these corrections in different representations, but the resulting constraints are unlikely to be sufficient to determine all the independent coefficients without further dynamical input.

It is therefore useful to contrast the full theory with a simplified model in which all subleading $1/N$ corrections to the non-singlet energies are discarded. Despite this simplification, the resulting low-temperature expansion can be resummed exactly. In Appendix \ref{app:non-interacting-ungauged-states}, we derive an analytic expression for the corresponding free energy and discuss its thermodynamic properties. The result differs from the full theory once the residual interactions of multi-adjoint states are included. It also clarifies the mismatch with the free part of the long-string effective action proposed in \cite{Cho:2024kxn}: treating the adjoint excitations as independent matrix oscillators overcounts multi-adjoint states and fails, already at $g=0$, to reproduce the exact matrix-oscillator partition function.
\subsection{Low-temperature expansion of one-point functions}\label{sec:low-T-moments}
We next consider single-trace one-point functions, whose low-temperature expansion can be obtained directly from the non-singlet spectrum by introducing sources for the corresponding operators. We consider a general even potential by writing
\begin{equation}
    H_{\mathbf J}=N^2\tr\left(\frac{1}{2}P^2+\sum_{k=1}^{\infty}J_{2k}X^{2k}\right)\ , \qquad\mathbf J=(J_2,J_4,J_6,\ldots)\ .
\end{equation}
The source derivatives then give
\begin{equation}
    \left\langle\mathrm{tr}\,X^{2k}\right\rangle_{\beta}=\frac{1}{N^2}\frac{\partial F}{\partial J_{2k}}\ .
\end{equation}
At planar order, the leading thermal corrections arise from the lowest non-singlet excitations, while thermal excitations within the singlet sector are suppressed by $1/N^2$. The spectral expansion takes the form
\begin{equation}
    Z(\beta)=e^{-\beta N^2\mathcal{E}_0}\left[1+(N^2-1)\sum_n e^{-\beta\Delta_n}+\ldots\right]\ ,
\end{equation}
and hence
\begin{equation}
    \mathcal{F}(\beta)=\mathcal{E}_0-\frac{1}{\beta N^2}\log\left[1+(N^2-1)\sum_n e^{-\beta\Delta_n}+\ldots\right]\ .
\end{equation}
Differentiating with respect to the source and using $(N^2-1)/N^2=1+\mathcal{O}\left(N^{-2}\right)$ in the planar limit gives
\begin{equation}
    \left\langle\mathrm{tr}\,X^{2k}\right\rangle_{\beta}=\left\langle \mathrm{tr}\,X^{2k}\right\rangle_{T=0}+\sum_n\frac{\partial\Delta_n}{\partial J_{2k}}e^{-\beta\Delta_n}+\ldots\ .
\end{equation}
Thus, the leading thermal correction to a one-point function is controlled not only by the non-singlet gaps themselves, but also by their response to the source coupled to the corresponding operator.

For the quartic potential,
\begin{equation}
    V(x)=\frac{1}{2}m^2x^2+gx^4\ , \qquad J_2=\frac{m^2}{2}\ , \qquad J_4=g\ ,
\end{equation}
the dependence on the oscillator mass can be removed by a rescaling. The adjoint gaps satisfy
\begin{equation}
    \Delta_n(m,g)=m\,\Delta_n(\kappa)\ , \qquad \kappa=\frac{g}{m^3}\ ,
\end{equation}
where $\Delta_n(\kappa)$ on the right-hand side denotes the adjoint sector of the theory with $m=1$. It follows that
\begin{equation}
    \begin{aligned}
        \frac{\partial\Delta_n}{\partial J_2}
        &=
        2\frac{\partial\Delta_n}{\partial m^2}=\frac{1}{m}\left[\Delta_n(\kappa)-3\kappa\Delta_n'(\kappa)\right]\ ,
        \\
        \frac{\partial\Delta_n}{\partial J_4}
        &=
        \frac{1}{m^2}\Delta_n'(\kappa)\ .
    \end{aligned}
\end{equation}

Setting $m=1$ and retaining only the contribution of the lowest non-singlet gap, we obtain the leading low-temperature corrections
\begin{equation}\label{trX2-low-T}
    \boxed{\left\langle \mathrm{tr}\,X^2\right\rangle_{\beta}=\left\langle\mathrm{tr}\,X^2\right\rangle_{\mathrm{T}=0}+\left[\Delta_1(g)-3g\Delta_1'(g)\right]e^{-\beta\Delta_1(g)}+\ldots}
\end{equation}
and
\begin{equation}\label{trX4-low-T}
    \boxed{\left\langle \mathrm{tr}\,X^4\right\rangle_{\beta}=\left\langle \mathrm{tr}\,X^4\right\rangle_{\mathrm{T}=0}+\Delta_1'(g)e^{-\beta\Delta_1(g)}+\ldots}\ .
\end{equation}
For the average energy, however, the dependence on $\Delta_1'(g)$ cancels between the quadratic and quartic contributions. The leading low-temperature correction therefore probes only the gap $\Delta_1(g)$ itself, and the average energy cannot be used to extract $\Delta_1'(g)$ at this order. These expressions extend directly to additional adjoint levels. Incorporating higher non-singlet representations requires including the corresponding multi-adjoint sectors and their representation-dependent interactions, which is a more difficult exercise. The zero-temperature moments appearing in these expressions are given analytically in Appendix \ref{app:thermal-moments}.
\subsection{Backreaction of non-singlets}\label{sec:backreaction}
The low-temperature expansion above treats the non-singlet spectrum on the zero-temperature singlet background. More generally, thermally excited non-singlets backreact on the Fermi sea because their energies depend on the eigenvalue density. While the dynamical coupling of adjoint impurities to collective-field fluctuations was studied in \cite{Karczmarek:2008sc}, here we consider the equilibrium deformation of the Fermi sea within the collective-field description \cite{Jevicki:1979mb,Das:1990kaa}. Keeping the two lightest non-singlet excitations, which belong to the adjoint sector, we write
\begin{equation}\label{backreaction-free-energy}
    \resizebox{\textwidth}{!}{$
    \begin{aligned}
        \mathcal F[\rho]=\mathcal E_0[\rho]-T e^{-\Delta_1[\rho]/T}-T e^{-\Delta_2[\rho]/T}+\mathcal O\left(e^{-2\Delta_1/T}\right)\ ,\qquad\mathcal E_0[\rho]=\int\d x\left[\frac{\pi^2}{6}\rho^3(x)+V(x)\rho(x)\right]\ ,
    \end{aligned}
    $}
\end{equation}
with $\int\d x\,\rho(x)=1$. The functional $\mathcal E_0[\rho]$ is the collective-field representation of the planar singlet ground-state energy \eqref{MQM-planar-gs-0}. Its saddle reproduces the equilibrium Fermi-sea density \eqref{density-MQM}. We retain the two lowest adjoint modes because they provide the dominant low-temperature corrections near criticality.

Including further adjoint levels is straightforward, whereas incorporating higher non-singlet representations requires accounting for their representation-dependent interactions. In particular, the subleading coefficient $B$ in \eqref{Deltaexpand} cannot be determined from the planar density and MO equation alone. Thus, the simple functional \eqref{backreaction-free-energy} does not capture the full higher-order backreaction, which may require additional collective degrees of freedom or a more general effective functional of $\rho$.

Varying \eqref{backreaction-free-energy} with respect to the density gives
\begin{equation}\label{backreaction-saddle}
    \frac{\pi^2}{2}\rho^2(x)+V(x)+y_1Q_1(x)+y_2Q_2(x)+\mathcal{O}\left(y_1^2\right)=\mu_F\ , \quad y_n\equiv e^{-\Delta_n[\rho]/T}\ , \quad Q_n(x)\equiv\frac{\delta\Delta_n}{\delta\rho(x)}\ .
\end{equation}
Here $\mu_F(T)$ is the Lagrange multiplier enforcing $\int\d x\,\rho(x)=1$ and reduces to the zero-temperature Fermi energy as $T\to0$.

For the normalized MO eigenfunction $\Phi_n$, the Hellmann-Feynman theorem gives\footnote{Because $\int\d x\,\delta\rho(x)=0$, the functional derivative $Q_n(x)=\delta\Delta_n/\delta\rho(x)$ is defined modulo an additive constant, which can be absorbed into $\mu_F(T)$.}
\begin{equation}\label{backreaction-kernel}
    Q_n(x)=\int\d z\,\rho(z)\frac{\left(\Phi_n(x)-\Phi_n(z)\right)^2}{(x-z)^2}-\Delta_n\Phi_n^2(x)\ , \qquad \int\d x\,\rho(x)\Phi_n^2(x)=1\ .
\end{equation}

Thus the non-singlets act, to this order, as a temperature-dependent deformation of the potential,
\begin{equation}\label{Veff}
    V(x)\longmapsto V_{\mathrm{eff}}(x,T)=V(x)+y_1Q_1(x)+y_2Q_2(x)+\mathcal{O}\left(y^2_1\right)\ .
\end{equation}
An equivalent derivation of $V_{\rm eff}$, based on the linear response of the zero-temperature singlet density to a general deformation of the one-particle potential, is given in Appendix \ref{app:backreaction}. This provides an independent check of the collective-field argument and also relates the backreacted density directly to the low-temperature one-point functions of Section \ref{sec:low-T-moments}.

The free matrix oscillator provides an exact check of this effective-potential description at the level of the equilibrium eigenvalue distribution. At $g=0$, for any finite $N$, the thermal eigenvalue distribution coincides with the ground-state distribution of $N$ non-interacting fermions in the temperature-dependent harmonic potential
\begin{equation}
    V_{\rm eff}(x,T)\Big|_{g=0}=\frac{x^2}{2}\tanh^2\frac{\beta}{2}\ .
\end{equation}
Its low-temperature expansion, $V_{\rm eff}(x,T)=x^2/2-2e^{-\beta}x^2+\mathcal O(e^{-2\beta})$, agrees with \eqref{Veff}, since $\Delta_1(0)=1$ and \eqref{backreaction-kernel} gives $Q_1(x)=2-2x^2$, with the constant term absorbed into $\mu_F(T)$.

We now expand directly around the critical quartic background, using the parametrization $g=g_*(1-\delta),\,\delta\ll1.$ To linear order in $\delta$ and the fugacities,
\begin{equation}
    V_{\mathrm{eff}}(x)=V_*(x)+\delta\,|g_*|x^4+y_{1,*}Q_{1,*}(x)+y_{2,*}Q_{2,*}(x)+\mathcal{O}\left(\delta\, y_{1,*},y_{1,*}^2\right)\ .
\end{equation}
On the critical line, $\delta$ takes the value $\delta_c(T)\equiv1-\frac{g_c(T)}{g_*},$ and the Fermi level reaches the maximum of $V_{\mathrm{eff}}$. Since $V_*'(a_*)=0$, the corresponding variation of the critical density is
\begin{equation}
    \delta\rho_c(x)=\frac{\delta_c(T)\,|g_*|\left(a_*^4-x^4\right)+\displaystyle\sum_{n=1}^{2}y_{n,*}\big(Q_{n,*}(a_*)-Q_{n,*}(x)\big)}{\pi\sqrt{2\left(V_*(a_*)-V_*(x)\right)}}\ .
\end{equation}
At the accuracy considered here, the non-singlet gaps and the corresponding kernels $Q_n(x)$ entering $V_{\rm eff}$ are evaluated on the undeformed zero-temperature critical background. Requiring the deformed density to remain normalized gives
\begin{equation}\label{backreaction-critical-line}
    \delta_c(T)=\mathcal B_1e^{-\Delta_{1,*}/T}+\mathcal B_2e^{-\Delta_{2,*}/T}+\mathcal O\left(e^{-2\Delta_{1,*}/T}\right)\ ,\qquad \mathcal B_n\equiv-\frac{I_n}{|g_*|I_g}\ ,
\end{equation}
where $\Delta_{n,*}\equiv\Delta_n(g=g_*)$ and
\begin{equation}
    I_n=\int_{-a_*}^{a_*}\limits\d x\,\frac{Q_{n,*}(a_*)-Q_{n,*}(x)}{\sqrt{2\left[V_*(a_*)-V_*(x)\right]}}\ , \qquad I_g=\int_{-a_*}^{a_*}\limits\d x\, \frac{a_*^4-x^4}{\sqrt{2\left[V_*(a_*)-V_*(x)\right]}}=\frac{1}{3\sqrt{2}\,|g_*|^2}\ .
\end{equation}

Numerically solving the MO equation on the critical background \cite{Casartelli:1979iz,Klebanov:2026vmk} and evaluating these integrals numerically gives
\begin{equation}
    I_1 \approx -18.19355\ , \qquad I_2 \approx -194.810 \qquad \Rightarrow \qquad \mathcal B_1 \approx 5.79119\ , \qquad \mathcal B_2 \approx 62.0099\ .
\end{equation}

For a fixed coupling $g$, the low-temperature instability is determined by $\delta=\delta_c(T_c)$. Inverting \eqref{backreaction-critical-line} therefore gives\footnote{For the cubic potential $V(X)=X^2/2+gX^3$, with $g_*^{\mathrm{cubic}}=-1/\sqrt{15\pi}$, the analogous low-temperature critical curve is given by \eqref{Tc-lowT-final}, with $\mathcal{B}_1^{\mathrm{cubic}}\approx 8.20161$ and $\mathcal{B}_2^{\mathrm{cubic}}\approx 210.589$, where $\delta_{\mathrm{cubic}}=1-g/g_*^{\mathrm{cubic}}$ and $\Delta_{n,*}^{\mathrm{cubic}}$ denote the corresponding critical adjoint gaps. The values of these quantities are given in \cite{Klebanov:2026vmk}.}
\begin{equation}\label{Tc-lowT-final}
    T_c(g)\simeq\frac{\Delta_{1,*}}{\displaystyle\log\frac{\mathcal B_1}{\delta}+\frac{\mathcal B_2}{\mathcal B_1}\left(\frac{\delta}{\mathcal B_1}\right)^{\Delta_{2,*}/\Delta_{1,*}-1}}\ ,\qquad \delta=1-\frac{g}{g_*}\ .
\end{equation}
Since $\Delta_{2,*}<2\Delta_{1,*}$, the second adjoint mode gives the leading correction to the one-level logarithmic result, while nonlinear multi-adjoint effects begin at higher order in $\delta$. We will use this expression in Section \ref{sec:BKT} to estimate $g_{\rm BKT}$.

The physical interpretation of \eqref{backreaction-critical-line} is important. The thermal non-singlets deform the potential seen by the singlet fermions, and the instability occurs when the Fermi surface reaches the top of the \emph{effective} potential $V_{\mathrm{eff}}$. Beyond this point, the Fermi droplet can spill over the barrier. This differs from the simpler picture of \cite{Cho:2024kxn} in which thermal fermions spill over the undeformed zero-temperature potential: near criticality the non-singlet backreaction is itself essential for determining where the instability occurs.

Along a fixed-temperature slice, the difference $\delta-\delta_c(T)$ provides a local coordinate transverse to the instability line. We introduce the thermal cosmological constant as the separation between the Fermi level and the maximum of the backreacted effective potential,
\begin{equation}\label{thermal-mu}
    \mu_T\equiv V_{\rm eff}(a_T)-\mu_F\ ,\qquad V_{\rm eff}'(a_T)=0\ .
\end{equation}
Thus both $\delta-\delta_c(T)$ and $\mu_T$ vanish on the instability line. Near the quadratic maximum of the effective potential, normalization of the Fermi sea gives
\begin{equation}\label{thermal-mu-scaling}
    \delta-\delta_c(T)=\left[\frac{1}{\sqrt{2}\pi}+\mathcal O\left(e^{-\Delta_{1,*}/T}\right)\right]\mu_T\left(\log\frac{\Lambda_T}{\mu_T}+1\right)+\ldots\ ,
\end{equation}
where the coefficient receives low-temperature corrections from the backreacted critical background, and $\Lambda_T$ is a non-universal scale with $\Lambda_T\to12\sqrt{2}\pi$ as $T\to0$. Consequently, after subtracting terms analytic in $\delta-\delta_c(T)$, the singular part of the planar free energy has the characteristic $c=1$ form
\begin{equation}\label{lowT-c1-singularity}
    \mathcal F_{\rm singular}(g,T)=\left[\frac{1}{2\sqrt{2}\pi}+\mathcal O\left(e^{-\Delta_{1,*}/T}\right)\right]\mu_T^2\log\frac{\Lambda_T}{\mu_T}+\ldots\sim\frac{(g-g_c(T))^2}{\log(1/(g-g_c(T)))}\ .
\end{equation}
At $T=0$ this reduces to the exact quartic result of Appendix \ref{app:singlet-WKB}. The finite critical adjoint gaps found in \cite{Klebanov:2026vmk} therefore allow thermally excited non-singlets to shift the critical line, even though their occupation is exponentially suppressed at low temperature. The critical singularity nevertheless retains its characteristic $c=1$ form.

\section{High-temperature expansion}\label{sec:highTexpansion}
At high temperature, the thermal path integral is naturally organized in Matsubara modes along the Euclidean time circle. The nonzero-frequency modes become perturbative and can be integrated out, while the zero mode remains nonperturbative and is described by an effective zero-dimensional matrix model. This dimensional reduction provides a systematic description of the weak-coupling end of the critical curve, where $T_c\to\infty$ as $g\to0^-$. In the strict high-temperature limit, the reduced theory is the quartic one-matrix model of BIPZ \cite{Brezin:1977sv}. 

Its one-cut saddle continuously connected to the Gaussian solution terminates at a finite negative value of the quartic coupling, providing the leading approximation to the instability line of MQM. At finite but small $\beta$, integrating out the nonzero Matsubara modes renormalizes the couplings of this matrix model and generates multi-trace interactions. Below we determine these corrections to the order required for the first three terms in the weak-coupling expansion of $T_c(g)$; the details of the hard-mode matching are collected in Appendix \ref{app:highT}.

Throughout this section and Appendix \ref{app:highT} we set $m=1$ and define
\begin{equation}\label{high-T-notations}
    \lambda\equiv\frac{g}{\beta}\ .
\end{equation}
The expansion is organized in powers of $\beta\ll1$ at fixed $\lambda$. Along the critical curve, $\lambda$ approaches the finite BIPZ critical value $\lambda_c\to-1/48$ as $\beta\to0$. Consequently, solving for $\lambda_c(\beta)$ and using $g=\beta\lambda_c(\beta)$ converts the high-temperature expansion into the weak-coupling expansion of $T_c(g)$ near $g=0^-$. To restore dimensions in intermediate formulas, one makes the replacements $\beta\to\beta m$ and $g\to g/m^3$, so that $\lambda\to g/(\beta m^4)=gT/m^4$.
\subsection{Dimensional reduction}
The Euclidean action is
\begin{equation}\label{MQM-action-quartic}
    S[X]=N^2\int_0^\beta\limits\mathrm{d}\tau\,\tr\left[\frac{1}{2}\dot X^2+\frac{1}{2}X^2+gX^4\right]\ .
\end{equation}
Introducing $t=\tau/\beta\in[0,1]$, we separate the static mode from the nonzero Matsubara modes \cite{Cho:2024kxn,Kawahara:2007ib},
\begin{equation}\label{highT-zero-mode-split}
    X(\tau)=\frac{1}{\sqrt{\beta}}\left[X_0+Y(t)\right]\ ,\qquad\int_0^1\limits\mathrm{d}t\,Y(t)=0\ ,
\end{equation}
where
\begin{equation}
    Y(t)=\sum_{n\neq0}Y_ne^{-2\pi int}\ ,\qquad Y_{-n}=Y_n^\dagger\ .
\end{equation}
The static and Gaussian hard-mode actions are
\begin{equation}
    S_0=N^2\tr\left(\frac{1}{2}X_0^2+\lambda X_0^4\right)\ , \qquad S_{\mathrm h}=\frac{N^2}{2}\sum_{n\neq0}\left[1+\left(\frac{2\pi n}{\beta}\right)^2\right]\tr(Y_nY_{-n})\ ,
\end{equation}
with propagator
\begin{equation}\label{highT-hard-propagator}
    \left\langle(Y_n)_{ij}(Y_p)_{kl}\right\rangle_0=\frac{\beta^2}{N}\frac{\delta_{n+p,0}\delta_{il}\delta_{jk}}{(2\pi n)^2+\beta^2}\ .
\end{equation}
Every hard propagator, therefore, carries two powers of $\beta$, which directly organizes the dimensional-reduction expansion.

At fixed $X_0$, integrating out the hard modes defines
\begin{equation}\label{highT-effective-action-definition}
    e^{-S_{\mathrm{eff}}[X_0]}=e^{-S_0[X_0]}\left\langle e^{-S_{\mathrm{int}}[X_0,Y]}\right\rangle_0\ .
\end{equation}
At this stage the Matsubara dependence is kept exact, so the reduced-theory calculation can be compared directly with the fixed-temperature expansion of Section \ref{sec:small-g}. Agreement through the first two orders in $g$ is verified explicitly in Appendix \ref{app:highT}. The high-temperature series is obtained subsequently by expanding in small $\beta$.

We restrict to the symmetric branch continuously connected to the Gaussian theory, for which all odd moments of $X_0$ vanish.\footnote{The condition $\tr X_0=0$ characterizes the symmetric saddle and is not imposed on the matrix integration. To verify stability against a uniform displacement, we write $X_0=\bar x\,\mathbb 1+X'$, with $\tr X'=0$. The curvature in this direction remains positive at the critical point, $m_{\bar x,c}^2=2/3+\mathcal O(\beta^2)>0$. Thus the instability considered below is associated with the BIPZ endpoint rather than the trace mode.} Defining $c_n\equiv\tr X_0^n$ on this branch, the reduced action relevant for the instability takes the form
\begin{equation}\label{reduced-soft-action}
    \frac{S_{\mathrm{red}}}{N^2}=\frac{r}{2}c_2+u\,c_4+v\,c_2^2+\mathcal{O}\left(\beta^6\right)\ ,
\end{equation}
At order $\beta^6$, in addition to further corrections to the displayed couplings, new zero-mode interactions are generated, including sextic single- and multi-trace structures such as $c_6$, $c_4c_2$, and $c_2^3$. These terms are not needed for the accuracy considered here. To the accuracy required for determining $T_c(g)$ through order $|g|^3$, the matched couplings are
\begin{equation}\label{highT-matched-couplings}
    r=1+\lambda\left(\frac{2}{3}\beta^2-\frac{\beta^4}{90}\right)+\mathcal O\left(\beta^6\right)\ , \qquad u=\lambda-\frac{\lambda^2\beta^4}{90}+\mathcal O\left(\beta^6\right)\ , \qquad v=-\frac{\lambda^2\beta^4}{60}+\mathcal O\left(\beta^6\right)\ .
\end{equation}

The double-trace interaction is generated by a planar hard-mode contraction with two color boundaries. Although it is multi-trace, it is not suppressed in the large-$N$ saddle. Its variation can be incorporated into the self-consistent quadratic coupling \cite{Das:1989fq,Sugino:1994zr,Gubser:1994yb},
\begin{equation}\label{double-trace-mass}
    r_{\mathrm{eff}}=r+4v\,c_2^*\ ,
\end{equation}
where $c_2^*\equiv\int\d y\,y^2\varrho(y)$ is the planar saddle value of $c_2=\tr X_0^2$.
\subsection{Static saddle and critical curve}\label{sec:high-T-MM}
After imposing \eqref{double-trace-mass}, the eigenvalue problem is reduced to the ordinary quartic matrix model
\begin{equation}\label{quartic-MM-functional}
    \mathcal S_{\mathrm{MM}}[\varrho;r_{\mathrm{eff}},u]=\frac{r_{\mathrm{eff}}}{2}\int\limits\mathrm{d}y\,y^2\varrho(y)+u\int\limits\mathrm{d}y\,y^4\varrho(y)-\int\limits\mathrm{d}y\,\mathrm{d}z\,\varrho(y)\varrho(z)\log|y-z|\ ,
\end{equation}
with $\int\mathrm{d}y\,\varrho(y)=1$. We use $\varrho$ here to distinguish the zero-dimensional matrix-model density from the Fermi-sea density $\rho$ of Section \ref{sec:singlet-sector}.

Writing the symmetric one-cut support as $[-2a,2a]$, the BIPZ solution \cite{Brezin:1977sv} is
\begin{equation}\label{highT-density}
    \varrho(y)=\frac{1}{\pi}\left(\frac{r_{\mathrm{eff}}}{2}+4ua^2+2uy^2\right)\sqrt{4a^2-y^2}\ ,\qquad |y|\leq2a\ ,
\end{equation}
where
\begin{equation}\label{highT-endpoint}
    r_{\mathrm{eff}}a^2+12ua^4=1\ , \qquad a^2=\frac{\sqrt{r_{\mathrm{eff}}^2+48u}-r_{\mathrm{eff}}}{24u}\ .
\end{equation}
The symmetric one-cut branch terminates when the prefactor of the square root in \eqref{highT-density} first vanishes at the endpoints. This gives
\begin{equation}\label{highT-criticality-condition}
    u_c=-\frac{r_{\mathrm{eff},c}^2}{48}\ , \qquad a_c^2=\frac{2}{r_{\mathrm{eff},c}}\ , \qquad c_{2,c}^*=\frac{4}{3r_{\mathrm{eff},c}}\ .
\end{equation}

Using \eqref{highT-matched-couplings} together with the critical zero-mode moment $c_{2,c}^*$ \eqref{highT-criticality-condition}, the self-consistency condition gives, to the required order
\begin{equation}\label{r-eff-final}
    r_{\mathrm{eff}}=1+\lambda\left(\frac{2}{3}\beta^2-\frac{\beta^4}{90}\right)-\frac{4}{45}\lambda^2\beta^4+\mathcal O\left(\beta^6\right)\ .
\end{equation}
The last term is entirely due to the generated double trace.

It is most transparent to solve the criticality condition first for the reduced quartic coupling,
\begin{equation}\label{highT-critical-lambda}
    \boxed{\lambda_c(\beta)=-\frac{1}{48}+\frac{\beta^2}{1728}-\frac{29\beta^4}{1244160}+\mathcal O\left(\beta^6\right)}\ .
\end{equation}
The leading value $\lambda_c=-1/48$ is precisely the critical coupling of the quartic BIPZ model. Since $\lambda_c=gT_c$ in the $m=1$ variables, restoring the oscillator mass gives\footnote{For the cubic potential $V(X)=m^2X^2/2+gX^3$, the analogous high-temperature expansion is
\begin{equation*}
    T_c(g)=\frac{\sqrt{3}m^6}{324g^2}-\frac{9g^2}{2m^4}-\frac{5832\sqrt{3}g^6}{5m^{14}}+\mathcal O\left(\frac{g^{10}}{m^{24}}\right)\ .
\end{equation*}}
\begin{equation}\label{Tc-high-T}
    \boxed{T_c=\frac{m^4}{48|g|}-\frac{4|g|}{3m^2}-\frac{704|g|^3}{15m^8}+\mathcal O\left(\frac{|g|^5}{m^{14}}\right)}\ .
\end{equation}
The first term is the strict dimensional-reduction result, while the subsequent terms encode the effect of the nonzero Matsubara modes. In particular, the generated double trace first changes the coefficient at order $|g|^3$; omitting it would replace $-704/15$ by $-192/5$. Using a symbolic-algebra implementation of the high-temperature expansion, we extend the series through order $|g|^{21}$; the result is quoted in Appendix \ref{app:highT}.

Although \eqref{Tc-high-T} is obtained as an expansion around $g=0$, we expect its all-orders completion, generated by the effective zero-mode matrix model, to describe the entire $c=0$ segment of the critical curve up to the BKT point $g_{\rm BKT}$. In this sense, retaining more terms should progressively improve the quantitative approximation as one moves away from $g=0$ toward $g_{\rm BKT}$. Whether the same perturbative series admits a continuation beyond the BKT point that also describes the $c=1$ portion of the critical curve toward $g_*$ is presently unclear. Since $g_{\rm BKT}$ and $g_*$ are numerically very close, the available coefficients do not allow us to distinguish whether the corresponding analytic continuation is limited at $g_{\rm BKT}$, at $g_*$, or at another nearby singularity.
\subsection{Critical thermodynamics}
To extract thermodynamic quantities near the endpoint, we also need the planar free energy of the reduced quartic matrix model. For the one-cut solution \eqref{highT-density}, it is \cite{Brezin:1977sv}
\begin{equation}\label{quartic-MM-free-energy}
    \mathcal F_{\mathrm{MM}}(r,u)=\frac{3}{4}+\frac{1}{2}\log r+\frac{1}{24}\left(ra^2-1\right)\left(9-ra^2\right)-\frac{1}{2}\log\left(ra^2\right)\ ,
\end{equation}
where $ra^2+12ua^4=1$.

Recall that the physical planar free-energy density is $\mathcal{F}=-\log Z/(\beta N^2)$. Combining the hard-mode contribution with the reduced matrix model gives
\begin{equation}\label{highT-full-free-energy}
    \beta\mathcal{F}=\log\beta+\mathcal F_{\mathrm h}(\beta,\lambda)+\Big[\mathcal F_{\mathrm{MM}}(r_{\mathrm{eff}},u)-\mathcal F_{\mathrm{MM}}(1,0)\Big]-v\left(c_2^*\right)^2+\mathcal O\left(\beta^6\right)\ .
\end{equation}
The hard-mode contribution $\mathcal F_{\mathrm h}$, including its perturbative evaluation and consistency checks, is derived in Appendix \ref{app:highT}. The subtraction uses $\mathcal F_{\mathrm{MM}}(1,0)=3/4$ as the Gaussian reference point of the reduced matrix integral. At $g=0$,
\begin{equation*}
    \beta\mathcal F(\beta,0)=\log\beta+\log\frac{\sinh(\beta/2)}{\beta/2}=\log\left(2\sinh\frac{\beta}{2}\right)\ ,
\end{equation*}
reproducing the exact matrix harmonic oscillator \eqref{free-planar-thermal-free-energy}.

The limiting energy density is obtained by differentiating at fixed physical coupling $g$ \eqref{thermal-observables-from-F}, and then substituting the critical temperature $T_c$. Restoring the mass, the high-temperature expansion gives
\begin{equation}\label{E_c-high-T}
    \boxed{\mathcal E_c=\frac{13}{12}T_c+\frac{2}{27}\frac{m^2}{T_c}-\frac{17}{9720}\frac{m^4}{T_c^3}+\mathcal O\left(\frac{m^6}{T_c^5}\right)=\frac{13m^4}{576|g|}+\frac{19|g|}{9m^2}-\frac{752|g|^3}{45m^8}+\mathcal O\left(\frac{|g|^5}{m^{14}}\right)}\ .
\end{equation}

The approach to this endpoint exhibits the characteristic BIPZ singularity. Writing $\delta\beta=\beta-\beta_c\to0^+$ on the physical side,
\begin{equation}\label{highT-critical-free-energy}
    \beta\mathcal{F}=\left(\beta\mathcal{F}\right)_c+\mathcal E_c\,\delta\beta-\left[\frac{17}{24}T_c^2+\mathcal O\left(m^2\right)\right]\delta\beta^2+\left[\frac{4}{15}T_c^{5/2}+\mathcal O\left(m^2T_c^{1/2}\right)\right]\delta\beta^{5/2}+\mathcal O\left(\delta\beta^3\right)\ .
\end{equation}
The regular terms depend on the full MQM thermodynamics, while the nonanalytic contribution is fixed by the critical zero-mode saddle.  Isolating the singular part,
\begin{equation}\label{highT-c0-singularity}
    \mathcal F_{\rm singular}(g,T)=
    \left[\frac{4}{15}T_c^{7/2}+\mathcal O\left(m^2T_c^{3/2}\right)
    \right]\left(\beta-\beta_c\right)^{5/2}+\ldots\sim\left(g-g_c(T)\right)^{5/2}\ ,
\end{equation}
where in the last step we used the critical
curve \eqref{Tc-high-T}. This is the characteristic planar $c=0$ (BIPZ) singularity \cite{Brezin:1977sv}, but now with the temperature explicitly included as a parameter.

Although sufficiently strong double-trace interactions can change the universality class of matrix models, leading to branched-polymer criticality \cite{Das:1989fq,Gubser:1994yb,Klebanov:1994pv}, this does not occur in our high-temperature regime. Indeed, using $r_{\rm eff}=r+4vc_2^*$, the susceptibility at fixed $u,v$ is
\begin{equation}
    \frac{\partial c_2^*}{\partial r}=-\frac{a^4}{1+4va^4}\ ,\qquad 1+4va_c^4=1-\frac{\beta^4}{8640}+\mathcal O\left(\beta^6\right)>0\ .
\end{equation}
Thus no additional multitrace instability appears, and the critical behavior remains that of $c=0$ gravity. We expect this universality class to persist up to the BKT point.
\section{BKT transition in thermal MQM}\label{sec:BKT}
We now return to the physical meaning of the special point on the critical curve separating the low-temperature $c=1$ regime from the high-temperature $c=0$ regime. The two limits analyzed above provide complementary descriptions of the same critical curve. Near $g=g_*$, the low-temperature expansion is governed by the compact $c=1$ string, whereas for small negative $g$, which corresponds to high temperature, dimensional reduction leads to $c=0$ critical behavior found in a matrix integral. The change of universality class between these two portions of the critical curve at some special point is naturally associated with the BKT transition expected from the continuum worldsheet description.
\subsection{The BKT point}\label{sec:BKT-point}
In the continuum worldsheet description of compactified MQM, the Euclidean time coordinate becomes a compact scalar $X$\footnote{Here and in Appendix \ref{app:gravity-description}, $X$ denotes the compact worldsheet scalar, not the matrix variable of MQM.} with periodicity $X\sim X+\beta$ and radius $R=\beta/(2\pi)$. The compactness of $X$ allows worldsheet vortices, around which the field winds an integer number of times around the target-space circle. At low temperatures these vortices are irrelevant, but they become increasingly important as the temperature increases. In the Gaussian $c=1$ theory, the lowest winding perturbation becomes marginal at $R=2\sqrt{\alpha'}$. This gives the conventional estimate of the BKT temperature \cite{Gross:1990ub,Gross:1990md},
\begin{equation}\label{BKT-T0}
    T_{\rm BKT}^{(0)}\equiv\frac{1}{4\pi\sqrt{\alpha'}}=\frac{1}{2\sqrt{2}\pi}\simeq0.11254\ ,\qquad \alpha'=1/2\ .
\end{equation}

We denote by $T_{\rm BKT}$ the physical transition temperature of the ungauged MQM, corresponding to the point $\left(g_{\rm BKT},T_{\rm BKT}\right)$ on the critical curve, where $g_{\rm BKT}\equiv g_c(T_{\rm BKT})$. The two temperatures need not coincide. At nonzero winding coupling, the kinetic term of the compact matter field $X$ is renormalized \cite{Schmidhuber:1993qe}. We denote its renormalization factor by $z_X$, which can equivalently be absorbed into the effective compactification radius $R_{\rm eff}=R\sqrt{z_X}$. Near the Gaussian BKT point, the critical RG trajectory separating the $c=1$ and $c=0$ phases gives (see Appendix \ref{app:gravity-description}),
\begin{equation}\label{BKT-shift}
    T_{\rm BKT}=\frac{\sqrt{z_X^{\rm UV}}}{1+\frac{\pi}{2}\zeta_{\rm UV}+o(\zeta_{\rm UV})}T_{\rm BKT}^{(0)}\ ,
\end{equation}
where $\zeta_{\rm UV}$ and $z_X^{\rm UV}$ are the unit-winding coupling and kinetic-term renormalization factor at the continuum matching scale.

The kinetic operator is winding-neutral, so its renormalization requires at least two winding insertions with opposite charges.  The leading winding-induced correction to the Gaussian kinetic term is therefore of order $\zeta^2$. Assuming a weak-winding expansion around the canonically normalized $c=1$ theory, with $0<\zeta_{\rm UV}\ll1$ and $z_X^{\rm UV}=1+\mathcal O(\zeta_{\rm UV}^2)$, we obtain
\begin{equation}\label{TMQM_BKT<T_BKT}
    T_{\rm BKT}=T_{\rm BKT}^{(0)}\left[1-\frac{\pi}{2}\zeta_{\rm UV}+o(\zeta_{\rm UV})\right]<T_{\rm BKT}^{(0)}\ .
\end{equation}
It is therefore tempting to expect that the BKT transition occurs slightly below $T_{\rm BKT}^{(0)}$. The small Boltzmann weights of the lowest non-singlet states near this temperature support a weak-winding picture, although the relation between these weights and the continuum winding coupling remains unknown.

At each temperature $T$, the critical coupling $g_c(T)$ determines an ensemble of arbitrarily large planar ribbon graphs, representing discretized random worldsheets. Coarse-graining these surfaces defines a worldsheet RG flow, with different points on the critical curve corresponding to different trajectories. On the low-temperature segment, the winding coupling flows to zero and the infrared matter theory is a compact Gaussian $c=1$ CFT with $h_{X,1}^{\rm IR}>1$. At $T=T_{\rm BKT}$, the flow approaches the marginal Gaussian fixed point with $h_{X,1}^{\rm IR}=1$. In both cases $\zeta_{\rm IR}=0$, although at the BKT point the approach is logarithmically slow. On the high-temperature segment, the winding coupling flows to strong coupling and the compact matter is expected to become massive, leaving $c=0$ gravity.

We therefore use $T_{\rm BKT}^{(0)}$ as a reference temperature. Evaluating the low-temperature critical line \eqref{backreaction-critical-line} at this temperature gives
\begin{equation}
    g_{\rm BKT}^{(0)}\equiv g_c\left(T_{\rm BKT}^{(0)}\right)\simeq0.991g_*\ ,
\end{equation}
placing this reference point very close to the zero-temperature critical coupling \eqref{g-crit}. At $g=g_{\rm{BKT}}^{(0)}$, the canonical and microcanonical bootstrap bounds shown in Figure \ref{fig:tc-BKT} are, after extrapolation to large bootstrap level, approximately consistent with $T_c=T_{\rm BKT}^{(0)}$. This supports the low-temperature critical curve. Under the weak-winding assumptions of \eqref{TMQM_BKT<T_BKT}, the physical BKT coupling $g_{\rm BKT}$ is expected to lie closer to $g_*$ than the reference coupling $g_{\rm BKT}^{(0)}$. 
\subsection{Non-singlets and vortices}\label{sec:vortice}
We now turn to the microscopic MQM interpretation of the winding degrees of freedom underlying the continuum BKT description of Appendix \ref{app:gravity-description}. Worldsheet vortices have already played a central role in the analysis of the compact $c=1$ matrix model \cite{Gross:1990ub, Gross:1990md}. In the study of one-dimensional string theory on a circle, the vortex sector of the discretized theory was analyzed, and vortex proliferation was argued to lead to a Berezinskii-Kosterlitz-Thouless transition to a $c=0$ phase. In the continuous-circle matrix quantum mechanics considered in the same work, however, the exact free-energy calculation was restricted to the singlet sector, with the non-singlet sectors omitted. 

A direct relation between these two structures was proposed in the subsequent work \cite{Gross:1990md}. The degrees of freedom carried by the non-singlet sectors of the compact $c=1$ matrix model were identified with worldsheet vortices, establishing a direct connection between winding defects on the worldsheet and non-singlet representations in MQM. This provided the starting point for interpreting the BKT regime in terms of the non-singlet sector of the matrix model.

This correspondence was developed further in a more explicit representation-theoretic and diagrammatic form \cite{Boulatov:1991fp,Boulatov:1991xz}. In the discretized-worldsheet expansion of matrix-model Feynman diagrams, nontrivial color monodromy around the Euclidean-time circle produces faces whose boundaries wind around the compact target-space direction. For the adjoint representation, the relevant graphs contain two such winding faces, naturally interpreted as a vortex-antivortex pair.

In the continuum description reviewed in Appendix \ref{app:gravity-description}, the unit-winding sine-Liouville interaction is proportional to $\zeta e^{(2-R/\sqrt{\alpha'})\varphi} \cos(R\widetilde X/\alpha')$, combining operators of opposite winding numbers. Winding neutrality implies that its first nonvanishing contribution to the closed-worldsheet partition function is of order $\zeta^2$. At the next level, two box-antibox pairs decompose into the four biadjoint irreducible representations \eqref{biadj-irreps}. On the worldsheet side, this level is naturally associated, schematically, with configurations containing two $w=+1$ and two $w=-1$ insertions. These can be organized into two vortex-antivortex pairs or, through the fusion of equal-sign windings, into a pair of defects with $w=+2$ and $w=-2$. Thus the biadjoint basis and the winding basis need not coincide one-to-one. Clarifying this relation, and the gravity interpretation of the representation-dependent biadjoint splittings \eqref{biadjoint-finiteN-expansion}, would be an interesting extension of the correspondence. More general non-singlet representations generate correspondingly more complicated configurations of winding defects.

The adjoint spectrum suggests a refinement of the vortex-antivortex picture. Slightly away from criticality, the Regge tower extends up to $n\sim n_{\max}$ \eqref{regge-phi4}, with a crossover scale of $\Delta_{\max}\equiv\Delta_{n_{\max}}$. In the Euclidean description, the low-lying Regge states may be interpreted as bound or internal excitations of a vortex-antivortex pair, with $\Delta_{\max}$ playing the role of a dissociation threshold and an energy $\Delta_{\max}/2$ associated with each separated vortex. This interpretation is tentative and deserves further investigation.

After continuing to Lorentzian signature, the same crossover separates short folded strings from the long-string regime \cite{Klebanov:2026vmk}. Interestingly, at coupling $g^{(0)}_{\rm BKT}$, the quartic model gives $n_{\max}\simeq1.1$, suggesting that only the lowest adjoint state remains in the Regge regime. A similar pattern appears in the cubic model near its reference BKT temperature. Although $n_{\max}$ characterizes a smooth crossover, these estimates suggest that the Regge window has shrunk to roughly the lowest adjoint level in both models. Whether this observation is related to the BKT transition remains an interesting open question.

The correspondence between non-singlet representations and worldsheet vortices suggests that the microscopic parameters $(g,T)$ of ungauged MQM determine a particular family of theories within the larger space of continuum couplings. While the temperature fixes the microscopic compactification radius $R=(2\pi T)^{-1}$, the relation between the remaining continuum couplings and the microscopic theory is only partially understood. Denoting the appropriate renormalized Liouville cosmological constant by $\mu_{\rm grav}$, the expected dictionary takes the schematic form
\begin{equation}
    \mu_{\rm grav}=\mu_{\rm grav}(g,T)\ ,\qquad \zeta=\zeta(g,T)\ ,\qquad \ldots\ .
\end{equation}
An important part of this dictionary can already be identified from our results. In the low-temperature $c=1$ regime, the Liouville cosmological constant is identified, up to normalization, with the thermally corrected Fermi-level separation $\mu_T$ defined in \eqref{thermal-mu}, extending the standard zero-temperature correspondence. In the high-temperature $c=0$ regime, the corresponding scaling variable is the distance to criticality in the dimensionally reduced matrix model, proportional to $g-g_c(T)$ at fixed temperature, as follows from the $5/2$ singularity in \eqref{highT-c0-singularity}. Determining the microscopic origin of the continuum winding couplings remains an important step toward completing the dictionary between ungauged MQM and continuum gravity. The results obtained here may provide a starting point, at least for the unit-winding coupling $\zeta$.
\section{Thermal bootstrap for MQM}\label{sec:bootstrap}
Matrix bootstrap is a powerful numerical method to compute correlators in the large $N$ matrix models by constructing a hierarchy of semidefinite-programming problems \cite{Lin:2020mme, Han:2020bkb,Lin:2025srf,Cho:2024kxn}. It allows non-perturbative calculations that apply directly in the large $N$ limit by imposing the large $N$ factorization. In this section, we apply the matrix bootstrap in the thermal state to study the thermodynamics of the ungauged one-matrix quantum mechanics. 
We show that bootstrap can bound (and in most cases appears to converge towards) the limiting temperature $T_c(g)$. 
We also demonstrate the power of precision bootstrap by presenting the bootstrap results for several observables of interest to high truncation level, showing excellent agreement with high/low-temperature expansions in the respective regimes. We use the open-sourced packages \texttt{SDPA-DD} and \texttt{SDPA-GMP} to solve the semi-definite programming problems. We also use \texttt{SDPA-DD-OMP}, an improvement over \texttt{SDPA-DD}, to accelerate the computations.\footnote{\texttt{SDPA-DD-OMP} is available at \textcolor{blue}{https://github.com/Canonical111/sdpa-dd-omp.}}
\subsection{Canonical ensemble bootstrap}

Here we review the essentials of the thermal bootstrap specialized for the canonical ensemble. For more details, we refer the readers to \cite{Cho:2024kxn}. To specify a canonical ensemble, one straightforward way is to specify the temperature by imposing the KMS condition on the Euclidean two-point correlators. Bootstrap methods based on this are discussed in \cite{Cho:2025vws}.

An equivalent formulation of the KMS condition is the energy-entropy balance (EEB) inequality \cite{Araki:1977px}:
\begin{equation}
    \beta \frac{\ev{O^\dagger [H,O]}_\beta}{\ev{O^\dagger O}_\beta} \ge  \log \frac{\ev{O^\dagger O}_\beta}{\ev{O O^\dagger}_\beta}\ .
\end{equation}
This was first discussed in the bootstrap context of \cite{Fawzi:2023fpg}, where the authors reformulated this inequality in a manner suitable for semi-definite programming. It was then applied to the matrix quantum mechanics context in \cite{Cho:2024kxn}, see also \cite{Adams:2025nww}. (A short derivation of the EEB inequality was given in \cite{Cho:2025vws}.)

We have also made the following improvements\footnote{We thank Minjae Cho, Barak Gabai, and Zechuan Zheng for discussions and related collaboration.} of the previous methods. 
\begin{enumerate}
    \item Used the discrete symmetries of MQM to reduce the number of unknowns and to block diagonalize the positivity matrices. This includes the $\mathbb{Z}_2$ symmetry $X \to -X$ and the time-reversal symmetry.
    
    \item We have organized the constraints by level $\Lambda$ where $\Lambda(X) = 1$ and $\Lambda(P) = 2$. This choice is motivated by demanding that acting on a level $\Lambda(\mathcal{O})$ operator $\mathcal{O}$ with the Hamiltonian produces an operator of level $\Lambda([H,\mathcal{O}]) \le \Lambda(\mathcal{O}) + 1$.  \cite{Lin:2024vvg, Lin:2025srf}
  
    \item We have included multi-adjoint positivity (see Section \ref{sec:multi-adjoint-positivity}). We consider a general word made up of products of $X_{ij}$ and $P_{ij}$ and then decompose that word into irreps of SU($N$); this can be simplified using Schur-Weyl duality. This is explained in more detail in the next subsection. These multi-trace constraints add new conditions for non-linear relaxation.
\end{enumerate}
\subsubsection{Multi-adjoint positivity}\label{sec:multi-adjoint-positivity}
Consider a word $\mathbf{W}$ consisting of $k$ letters. Each letter $W_i$ could be any operator that transforms in the adjoint of SU($N$) %
\begin{equation}
   \mathbf{W}=W_1\otimes W_2\otimes\ldots\otimes W_k\ .
\end{equation}
For example $(W_1)=X^3P-\mathbb 1 \, \tr (X^3 P)$. There is a natural inner product on this space
\begin{equation}
    \langle\mathbf V|\mathbf W\rangle =\langle{V_1, V_2, \ldots, V_k}|{W_1, \ldots, W_k}\rangle=\tr(V_1^\dagger W_1)\tr(V_2^\dagger W_2)\ldots \tr(V_k^\dagger W_k)\ . 
\end{equation}
Now we can consider two permutation group elements $\sigma, \tau \in S_k$ which act on a tensor
\begin{equation}
    (\sigma, \tau) T^{i_1 i_2 \ldots i_k}_{j_1 j_2 \ldots j_k} =T^{i_{\sigma(1)} i_{\sigma(2)} \ldots i_{\sigma(k)}}_{j_{\tau(1)} j_{\tau(2)} \ldots j_{\tau(k)}}\ .
\end{equation}
So $\sigma$ acts on the fundamental indices whereas $\tau$ acts on the anti-fundamental indices. Then we have
\begin{equation}
    \resizebox{\textwidth}{!}{$
    \begin{aligned}
        \bra{V_1, V_2, \ldots, V_k} U({\sigma',\tau'} )^\dagger U({\sigma,\tau}) \ket{W_1, \ldots, W_k} =\bra{V_1, V_2, \ldots, V_k} U({(\sigma') \inv \cdot\sigma,(\tau')\inv\cdot\tau})\ket{W_1, \ldots, W_k}\ .
    \end{aligned}
    $}
\end{equation}

Now let us define $\eta=(\sigma')\inv\cdot\sigma$ and $\bar\eta=(\tau')\inv\cdot\tau$. This inner product depends only on $\eta$ and $\bar\eta$, so we denote it by the shorthand $\bra{\mathbf V} U(\eta, \bar\eta) \ket{\mathbf W}$.

Now let us consider the large-$N$ limit. The overlap $\langle\mathbf V|U(\eta,\bar\eta)|\mathbf W\rangle$ contains $k$ fundamental and $k$ antifundamental index lines.  %
Let us follow one fundamental line; the permutation sends $a \to \eta(a)$. Returning along the corresponding antifundamental line applies \(\bar\eta^{-1}\).  So $a$ is sent to
\begin{equation}
    a\longmapsto \bar\eta^{-1}\eta(a)\ .
\end{equation}
Consequently, the closed color loops are in one-to-one correspondence with the cycles of the permutation group element $\pi=\bar\eta^{-1}\eta.$ Let $c(\pi)$ denote the number of cycles of $\pi$. Each closed color loop contributes one factor of $N$, while the normalized inner product contains an overall factor $N^{-k}$. Therefore
\begin{equation}
    \langle\mathbf V|U(\eta,\bar\eta)|\mathbf W\rangle\propto  N^{c(\pi)-k}\ ,
\end{equation}
The number of cycles determines the power of $N$, whereas the labeled wiring determines the noncommutative order inside each trace.

Since a permutation of $k$ objects has at most $k$ cycles, the leading order answer occurs when every cycle has length one:
\begin{equation}\label{eta-equals-etabar}
    c(\bar\eta^{-1}\eta)=k \quad \Longleftrightarrow\quad \eta=\bar\eta\ .
\end{equation}
It is important that \eqref{eta-equals-etabar} does not require either permutation to be the identity.  %
\begin{equation}
    \langle\mathbf V|U(\rho,\rho)|\mathbf W\rangle=\prod_{a=1}^k\tr\left(V_a^\dagger W_{\rho^{-1}(a)}\right)\ .
\end{equation}
For example, for $k=2$ and choosing $\rho = (12)$, 
\begin{equation}
    \langle\mathbf V|U(\rho,\rho)|\mathbf W\rangle =\tr(V_1^\dagger W_2)\,\tr(V_2^\dagger W_1)\ .
\end{equation}
Now let us consider the vectors
\begin{equation}\label{character-vector}
    \ket{\mathbf V}_R=\sum_{\sigma\in S_k}\chi_R(\sigma)\,U(\sigma,\sigma)\ket{\mathbf V}\ .
\end{equation}
We form their overlap:
\begin{equation}
    _{R'}\langle{\mathbf V}|{\mathbf W}\rangle_R=\sum_{\sigma,\tau\in S_k}\overline\chi_{R'}(\sigma)\chi_R(\tau)\,\bra{\mathbf V}U(\sigma^{-1}\tau,\sigma^{-1}\tau)\ket{\mathbf W}\ .
\end{equation}
Set $\rho=\sigma^{-1}\tau.$ Using character orthogonality,
\begin{equation}
    \sum_{\sigma\in S_k}\overline\chi_{R'}(\sigma)\chi_{R}(\sigma\rho)=\frac{k!}{d_R}\chi_R(\rho) \delta_{R,R'}\ ,
\end{equation}
where $d_R$ is the dimension of $R$. Therefore
\begin{equation}
    _R\langle{\mathbf V}|{\mathbf W}\rangle_R=\frac{k!}{d_R}\sum_{\rho\in S_k}\chi_R(\rho)\, \bra{\mathbf V}U(\rho,\rho)\ket{\mathbf W}=\frac{k!}{d_R}\sum_{\rho\in S_k}\chi_R(\rho)\,\prod_{a=1}^k\tr\left(V_a^\dagger W_{\rho^{-1}(a)}\right)\ .
\end{equation}
The resulting matrix is positive semidefinite because it is, by construction, the Gram matrix obtained by taking overlaps of the vectors in \eqref{character-vector}. We illustrate these formulas for small values of $k$ in Appendix \ref{app:bootstrap-details}.

\subsubsection{Non-linear relaxation \label{sec:bootstrap:nlr}}

Recall that when we bootstrap matrix quantum mechanics in the Hamiltonian formalism, cyclicity of the trace generates relations between single traces and multi-traces \cite{Han:2020bkb}. Multi-traces then factorize into products of single traces due to large $N$ factorization. This means that imposing such relations generically leads to polynomial constraints in the single traces. These constraints make the semi-definite-programming problem non-linear and non-convex. 

In principle, one can scan over all non-linear variables, but this is prohibitively costly at higher levels. A solution to this problem was proposed in \cite{Kazakov:2021lel}, which is to perform non-linear relaxation on double traces. Here we systematize all such constraints on multi-traces; together with our multi-adjoint positivity constraints, one can view this as a Lasserre hierarchy \cite{Lasserre:2001mom} for multi-traces.

In practice, we must perform non-linear relaxation on multi-traces. This means we define a function $\mathcal{L}$ which takes a multi-trace and assigns it a new variable, e.g., $\tr A_1\tr A_2\ldots\tr A_k\rightarrow \mathcal{L}(A_1|A_2|\ldots|A_k)$. This function satisfies various properties:
\begin{equation}\label{Lunordered}
\begin{aligned}
    \mathcal{L}(A_1|A_2|\ldots|A_k) &=\mathcal{L}(A_{\sigma(1)}|A_{\sigma(2)}|\ldots|A_{\sigma(k)})\ ,\quad\sigma\in S_k \ , \\
    \mathcal{L}(A_1|A_2|\ldots|A_k)^* &=\mathcal{L}(r(A_1)|r(A_2)| \ldots|r(A_k))\ ,\\
    \mathcal{L}(a_{1}a_2|A_2|\ldots|A_k) &=\mathcal{L}(a_2a_{1}|A_2| \ldots| A_k)-\mathcal{L}([a_2,a_{1}]|A_2|\ldots|A_k)\ , \\
    \mathcal{L}(1|A_2|\ldots| A_k) &=\mathcal{L}(A_2|\ldots| A_k)\ , \\
    \mathcal{L}(1) &=1\ .
\end{aligned}
\end{equation}
Notice that \eqref{Lunordered} neglects commutators of multi-traces, which is only true at infinite $N$ in quantum mechanics, since the traces themselves are in general non-commuting quantum operators. Here $r(A)$ denotes reversal of the operator ordering in a word $A$, corresponding to Hermitian conjugation for words constructed from $X$ and $P$.

We replace each whole
multi-trace product by a variable $\mathcal L$, with the conventions of
\eqref{Lunordered}. In the relaxed problem, $\mathcal L(A|B)$ need not
equal $\mathcal L(A)\mathcal L(B)$. The number of variables is therefore increased, while the constraints become linear equalities and positive semidefinite matrix conditions. We give some examples of this in Appendix \ref{app:bootstrap-small-k}. Besides reducing the size of the islands somewhat, systematically implementing these constraints appears to also have the benefit of significantly reducing numerical instability. Nevertheless, for some values of $g$ and at higher levels it was necessary to use a high precision solver like \texttt{SDPA-GMP-OMP} or to use facial reduction to identify and remove directions that caused the numerical instability.

\subsubsection{Thermal energy}
\begin{figure}[h!]
    \centering
\includegraphics[width=1\linewidth]{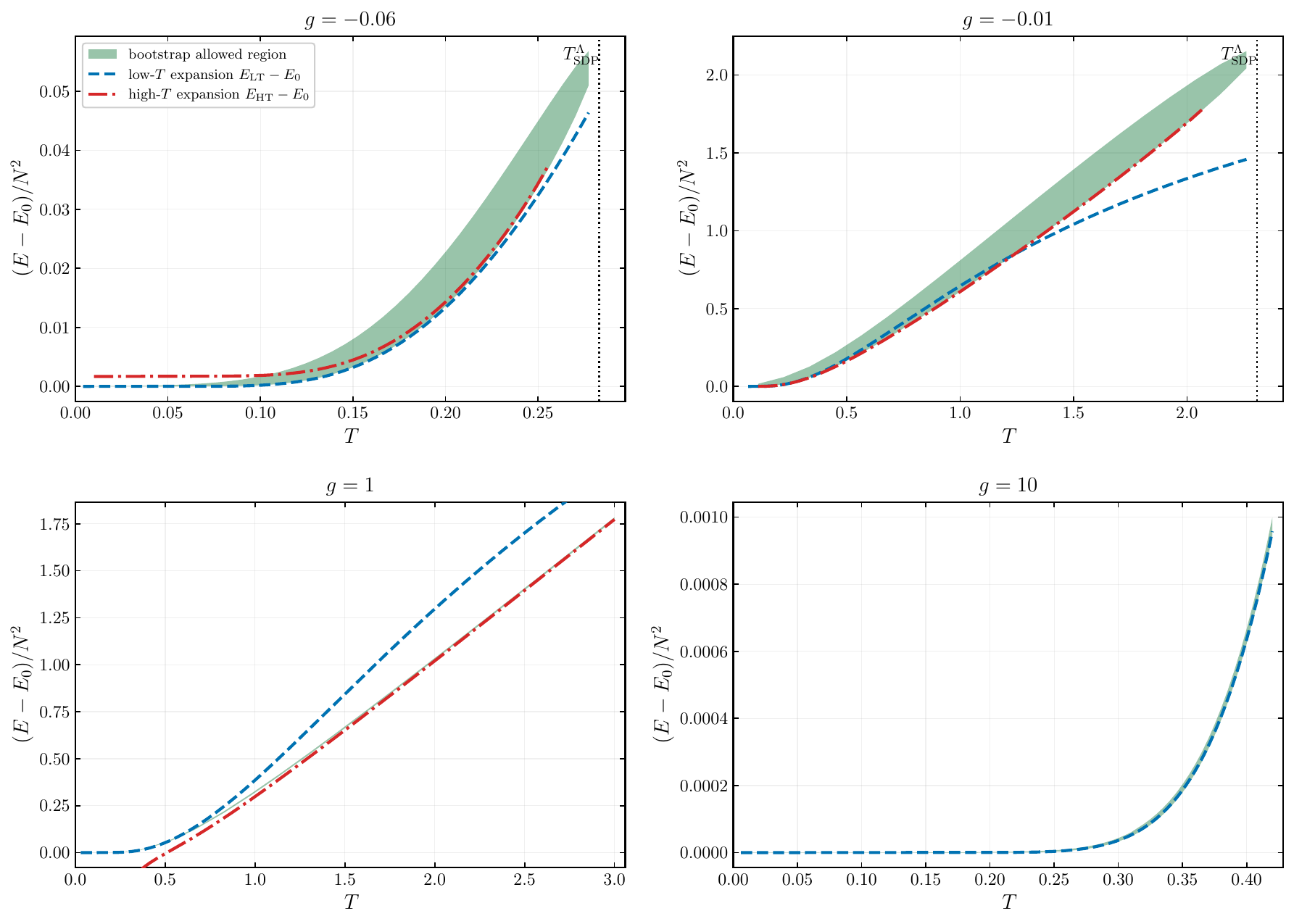}
    \caption{Comparison of the bootstrap results of energy density (subtracted the ground state energy \eqref{MQM-planar-gs-quartic}) with the low-temperature expansion \eqref{lowT-energy} and the high-temperature approximation obtained from \eqref{highT-full-free-energy}, for $g=\{-0.06,-0.01,1,10\}$. For negative $g$, the vertical dashed lines indicate the finite-level bootstrap upper bounds $T_{\rm SDP}^{\Lambda}$. For $g=10$, only the low-temperature approximation is shown, since the high-temperature expansion is not reliable at such large coupling.}
    \label{fig:energy-compare}
\end{figure}

As a first demonstration, we compute the thermal energy (to planar order) as a function of temperature. Figure \ref{fig:energy-compare} shows the $\Lambda=14$ bootstrap results for planar energy density at $g=\{-0.06,-0.01,1,10\}$, compared with the low- and high-temperature approximation. To demonstrate the precision, we included the precise numbers in Table \ref{tab:energy-gm006} for $g=-0.06$.

At low temperature, we include the contribution from the first two adjoint gaps in Equation \eqref{lowT-energy}. At high temperature, the energy is obtained by differentiating the dimensionally reduced free energy \eqref{highT-full-free-energy} with respect to the temperature as in \eqref{thermal-observables-from-F}. In particular, this includes the contribution of the hard modes perturbatively. 

For $g>0$, the quartic MQM has no thermal phase transition, and it is stable for any temperature. Bootstrap constraints on thermodynamic quantities, such as energy, have been studied in \cite{Cho:2024kxn, Adams:2025nww}. For $g < 0$, there is a phase transition. The results were shown to be in good agreement with the low-temperature expansion.  We also included the bootstrap results for quadratic Casimir in \ref{app:casimir}.

\subsubsection{Extraction of \texorpdfstring{$T_c(g)$}{}}
\label{sec:tc-extraction}
The EEB condition specifies the criteria for the ensemble to be in thermal equilibrium at some temperature $T$. Failure to satisfy the condition indicates that such thermal ensemble does not exist. Therefore, bootstrap provides a method to determine this limiting temperature by tracking when the bounds cease to exist.

Given a bootstrap level $\Lambda$, we use $T_\textrm{SDP}^\Lambda(g)$ to denote the point beyond which the bounds become unattainable at the coupling $g$. This represents an upper bound on the true $T_c$. We expect to have $T_\textrm{SDP}^\Lambda(g) \rightarrow T_c(g)$ from above as $\Lambda \rightarrow \infty$. We impose the following general ansatz to model the convergence as $\Lambda \rightarrow \infty$: 
\begin{align}\label{extrapolation-fittings}
    T_\textrm{SDP}^\Lambda(g)&=  T_\textrm{SDP}^\infty(g) + A e^{-\Lambda/\Lambda_0} + B e^{-\Lambda/\Lambda_1}\ , \nonumber \\
    T_\textrm{SDP}^\Lambda(g) &=  T_\textrm{SDP}^\infty(g)+ \left(\frac{\Lambda}{\Lambda_0}\right)^{-p} + \left(\frac{\Lambda}{\Lambda_1}\right)^{-q}\ ,\qquad p,q>0\ .
\end{align}

For the canonical bootstrap, we impose only one exponential or one power term in the ansatz above due to the limited data points.  Table \ref{tab:tc-data} shows the $T_\textrm{SDP}^\Lambda(g)$ and the comparison with the perturbative high-temperature results and the low-temperature predictions. We use the canonical bootstrap data for $\Lambda=\{8,10,12,14,16\}$ for the fitting, and the error bars represent the discrepancy in the two fitting ansatz. 
We see that the EEB bootstrap gives critical temperature consistent with the high-temperature prediction for all $|g| \lesssim 0.06$ at the few-percent level. At $g \rightarrow g_*$, the EEB results do not converge well to the low-temperature prediction at $\Lambda=16$, even after extrapolating to large $\Lambda$. It seems possible that the systematic uncertainty in the large $\Lambda$ extrapolation is larger than what we quantified; obtaining more precise bootstrap estimates in the regime $g_*\leq g\lesssim-0.06$ remains a goal for future work.

In Section \ref{sec:alternative-extraction-tc}, we discuss an alternative method to estimate the critical temperature based on the microcanonical bootstrap method, which allows us to achieve a higher level and obtain results closer to the low-temperature prediction after extrapolation.

\subsection{Microcanonical ensemble bootstrap}
In the microcanonical ensemble, we fix the energy of the system by specifying 
\begin{equation}\label{ho}
    \ev{H \O}=E\ev{\O}.
\end{equation}
Here, $E$ is a chosen number that we choose to fix, and equation \eqref{ho} essentially becomes an additional relation on the moments. More precisely, we assume that $\frac{1}{N^4} \ev{(H-E)^2}$ is down by powers of $1/N$ in the large $N$ limit, so that we may use large $N$ factorization. 
Practically, this means that we impose
\begin{equation}\label{cutoff-latent-energy}
    \mathcal L(\tfrac12 P^2+\tfrac12 X^2+gX^4
               \mid Q_1\mid\ldots\mid Q_n)
    =\mathcal E\,\mathcal L(Q_1\mid\ldots\mid Q_n),
    \qquad \mathcal E=\frac{E}{N^2},
\end{equation}
where the $Q_i$'s represent general multi-traces.

Similarly, we can also consider fixing the value of quadratic Casimir
\begin{equation}\label{c2o}
    \ev{C_2 \O}=C_2\ev{\O}.
\end{equation}
These equations \eqref{ho} and \eqref{c2o} introduce non-linearity in the bootstrap. For practical purposes, we consider only imposing \eqref{c2o} for $\mathcal{O}=1$. This means that we perform a 1-dimensional scan over $\mathcal{E}$, while minimizing/maximizing $\ev{C_2}$ using SDP. Note that the Casimir is a level 6 operator, whereas the Hamiltonian is level 4, so it is more advantageous to scan over $\mathcal{E}$ rather than $C_2$. (At higher levels, one could also impose the analog of \eqref{cutoff-latent-energy} for the Casimir; however, scanning over two variables at higher levels is rather expensive.)  

Then, we impose the standard Gram matrix positivity 
\begin{equation}
    \M_{ij}=\ev{\O^\dagger_i \O_j} \succeq 0\ .
\end{equation}
The result of the microcanonical bootstrap is shown in Figure \ref{fig:microcanonical_multi_lvl}, which shows the allowed range of $\ev{C_2}$ when fixing the energy at various couplings. Note that the maximum energy achievable with SU($N$) singlets, $E_\textrm{top}$, can be computed analytically as in \eqref{Etop-quartic}. 
\subsubsection{Alternative extraction of \texorpdfstring{$T_c$}{}} \label{sec:alternative-extraction-tc}

\begin{figure}[p]
    \centering
    \includegraphics[width=0.95\linewidth]{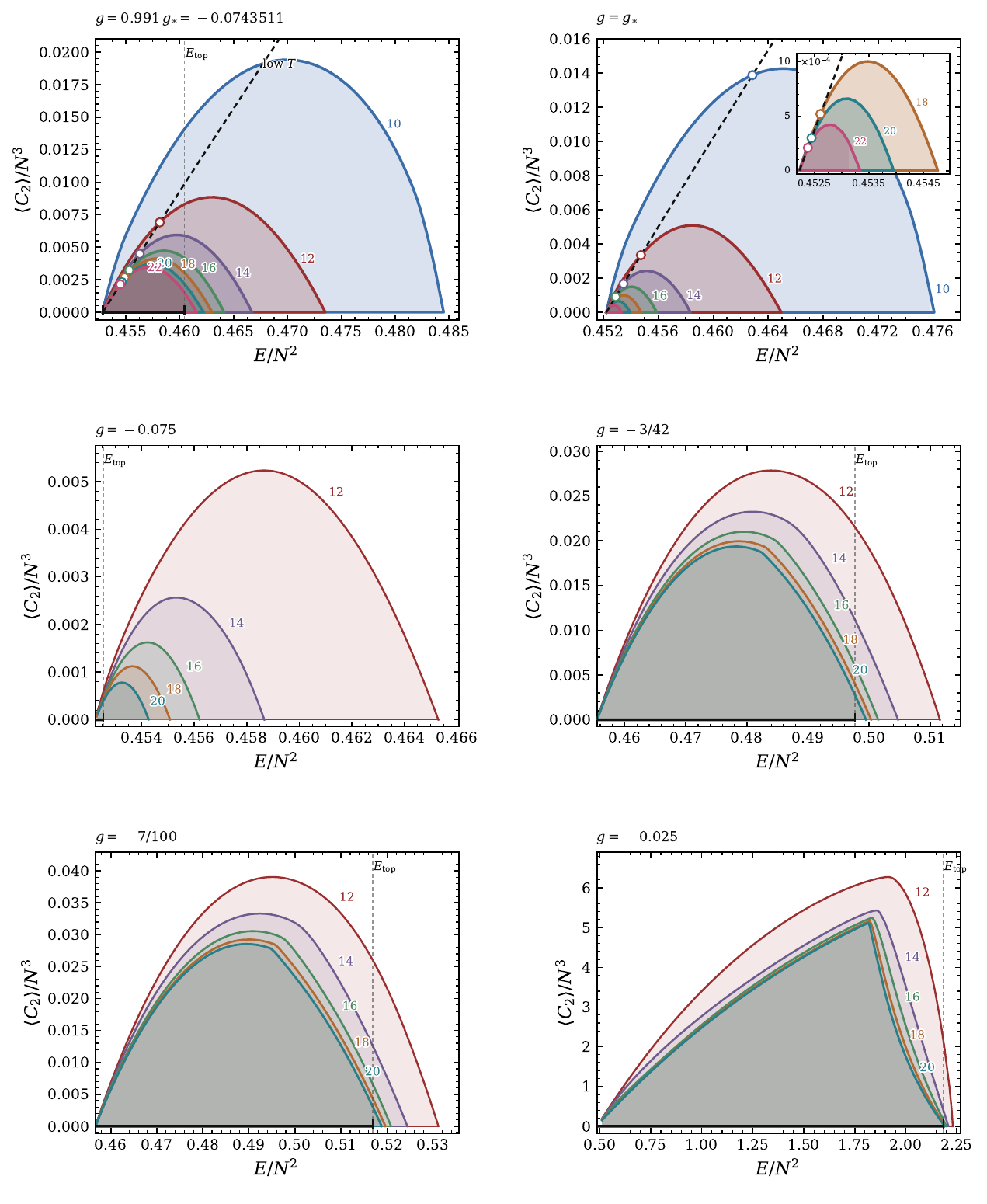}
    \caption{Microcanonical bootstrap islands at six couplings, with axes
    $E/N^2$ and $\langle C_2\rangle/N^3$. The labels on the curves indicate the bootstrap level. The top row shows $g=0.991g_*$ and $g=g_*$; the inset resolves the highest three levels at $g=g_*$. Black dashed curves in the top row are the low-temperature approximation \eqref{lowT-E-vs-C2}; their intersections with the island boundaries give upper estimates of $T_c$. Vertical gray dashed lines indicate $\mathcal{E}_{\rm top}$, the maximum energy in the singlet sector \eqref{Etop-quartic}. The remaining panels show $g=-3/40$, $-3/42$, $-7/100$, and $-1/40$. Curiously, the boundary of the island develops an increasingly pronounced kink as the coupling moves toward zero. Understanding the physical interpretation of the kink is an interesting future direction.}
    \label{fig:microcanonical_multi_lvl}
    \label{fig:island-gm3d40}
    \label{fig:island-gm3d42}
    \label{fig:island-gm7d100}
    \label{fig:island-gm025}
\end{figure}

\begin{figure}[h!]
    \centering
\includegraphics[width=0.49\columnwidth]
{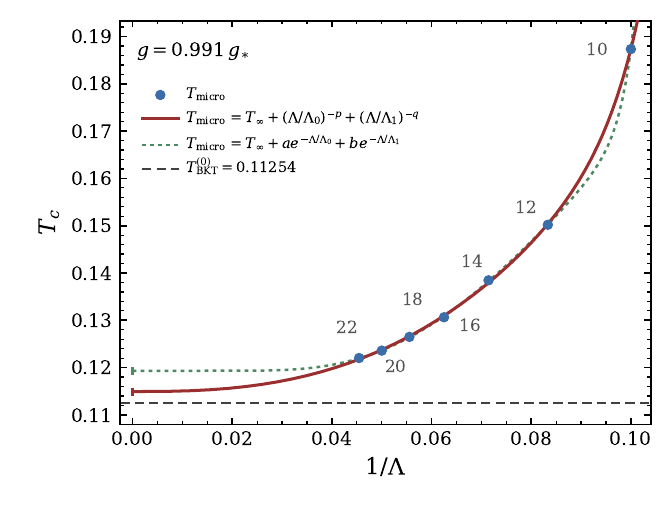}
    \includegraphics[width=0.49\columnwidth]{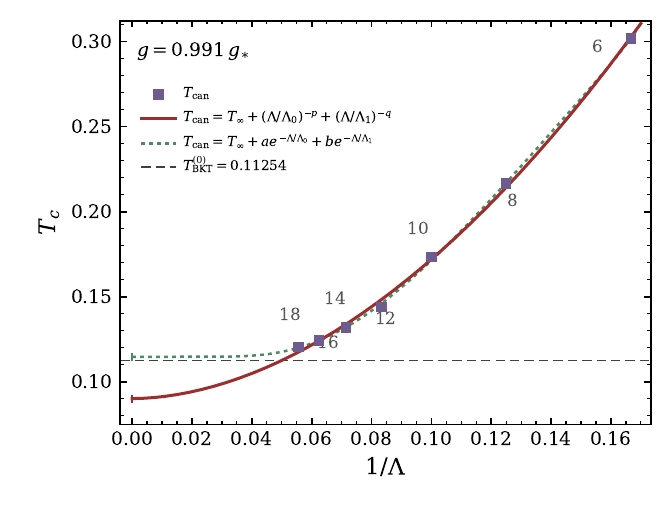}
    \caption{Critical temperature estimates at $g= g_{\rm BKT}^{(0)} \simeq 0.991g_*$ for different $\Lambda$ in the microcanonical (left) or canonical (right) bootstrap. After a large $\Lambda$ extrapolation with the ansatz in \eqref{extrapolation-fittings}, the results appear to be approximately consistent with a $T_c\approx T^{(0)}_{\rm BKT}\approx 0.11254$, where $T^{(0)}_{\rm BKT}$ is the continuum value (BKT point is discussed in Section \ref{sec:BKT-point}).}
    \label{fig:tc-BKT}
\end{figure}
\begin{figure}[h!]
    \centering
    \includegraphics[width=0.6\columnwidth]{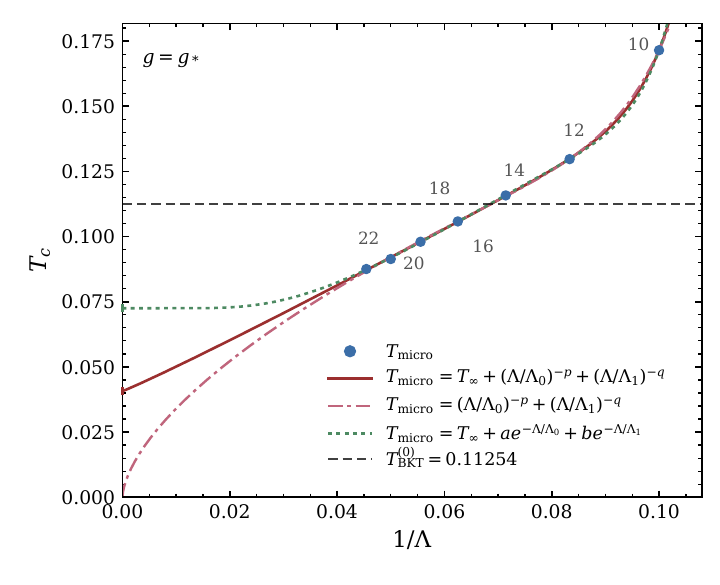}
    \caption{
    Critical-temperature estimates at $g=g_*$ as a function of
    $1/\Lambda$. The points are obtained by intersecting the microcanonical bootstrap islands with the low-temperature expansion. We show two-power (solid red) and two-exponential (dotted green) fits with free intercepts, together with a two-power fit constrained to $T_c(g_*)=0$ (dashed red). The horizontal dashed line marks the Gaussian BKT temperature. The dependence on the fitting ansatz illustrates the uncertainty in extrapolating to $\Lambda\to\infty$.}
    \label{fig:gc-tc-extrapolation}
\end{figure}
\begin{table}[h!]
    \centering
    \begin{tabular}{cccc}
    \toprule
    $\lceil\Lambda\rceil_2$
        & microcanonical
        & $T_{\mathrm{SDP}}$ (odd)
        & $T_{\mathrm{SDP}}$ (even) \\
    \midrule
        10 & 0.171501  & 0.182949 & 0.162695 \\
        12 & 0.129719  & 0.153010 & 0.128063 \\
        14 & 0.115798  & 0.125013 & 0.113879 \\
        16 & 0.105822 & 0.110884 & 0.102322 \\
        18 & 0.098019  & --       & --       \\
        20 & 0.091394 & --&-- \\
        22 & 0.087559& --&--\\
    \midrule
        $\infty$ (power)
        & 0.04876
        & 0.08856
        & 0.07697\\
        $\infty$ (exponential)
        &0.07249&0.10616 & 0.09360\\

    \bottomrule
    \end{tabular}
    \vspace{0.5ex}
    \caption{Finite-level estimates of the critical temperature at $g=g_*$ from the canonical and microcanonical bootstrap. The second column gives microcanonical estimates obtained by intersecting the bootstrap islands with the low-temperature expansion, while the last two columns give canonical-bootstrap upper bounds on $T_c$ at odd and even truncation levels. The notation $\lceil\Lambda\rceil_2$ denotes rounding $\Lambda$ up to the smallest even integer. The extrapolations to $\Lambda\to\infty$ use the ansatz in \eqref{extrapolation-fittings}, retaining a single power-law or exponential correction for the canonical bootstrap and two correction terms for the microcanonical bootstrap. The canonical fits also include data at $\Lambda=7$ and $\Lambda=8$ for odd and even levels, respectively.}
    \label{tab:tc-versus-level} 
\end{table}

We now discuss another bootstrap method to extract the critical temperature based on the microcanonical bootstrap. At $g$ slightly above $g_c$, we can estimate $T_c$ using the low-temperature expansion. Namely, we consider the curve 
\begin{equation}\label{lowT-E-vs-C2}
    \mathcal{E}-\mathcal{E}_0=\Delta_1 \mathcal{C}_2+(\Delta_2-\Delta_1)(\mathcal{C}_2)^{\Delta_2/\Delta_1}+\ldots\ ,
\end{equation}
where $\mathcal{E}\equiv E/N^2$ and $\mathcal{C}_2\equiv\braket{C_2}_\beta/(N(N^2-1))$. This curve is obtained by starting with the low-temperature expansion \eqref{lowT-energy} and \eqref{Casimir-lowT}, keeping only the first two terms, and then solving for $\mathcal{E}$ in terms of $\mathcal C_2$. In other words, we obtain a curve in the plane of $\mathcal{E}$ and $\mathcal C_2$, parametrized by $T$.

We then identify the intersection of this curve and the boundary of the allowed region from the microcanonical bootstrap and denote $T_{\rm micro}$ to be the corresponding temperature.  This gives an alternative estimate for $T_c$, motivated by the agreement between the low-temperature expansion and the thermal bootstrap in their common regime, illustrated in Figure \ref{fig:energy-compare} and Table \ref{tab:energy-gm006}. Note that this is an estimation of $T_c$ instead of a rigorous bound as in Section \ref{sec:tc-extraction}. Whether the physical critical endpoint lies on the boundary in the $\Lambda\to\infty$ limit remains an open question. The main advantage of this method is that it allows us to reach higher bootstrap levels with comparable computational resources. 

Figure \ref{fig:tc-BKT} compares the canonical and microcanonical estimates at $g=g^{(0)}_{\rm BKT}\simeq 0.991g_*$ and shows the extrapolated $T_c$ is consistent with $T^{(0)}_{\rm BKT}$ in both methods. Figure \ref{fig:gc-tc-extrapolation} shows the finite-level analysis at $g=g_*$. Besides the two fitting functions in \eqref{extrapolation-fittings}, we also present the two-power ansatz without the intercept, corresponding to $T_c(g_*)=0$. We show that this ansatz is also in agreement with the finite-level bootstrap data, demonstrating that our numerical results are still fully consistent with the analytical prediction of Section \ref{sec:backreaction} that $T_c(g_*)=0$. This also illustrates the large systematic uncertainty when extrapolating to $\Lambda\rightarrow \infty$. We present more details of canonical and microcanonical bootstrap in Table \ref{tab:tc-versus-level}. 
However, we note that neither the canonical nor the microcanonical method conclusively establishes $T_c(g_*)=0$, which remains an interesting target for future bootstrap studies.

The shape of the microcanonical bootstrap island further supports the analytical prediction of $T_c(g_*)=0$. Suppose instead that $T_c > 0$. Then energies larger than the ground state energy $\mathcal{E}_0$ must be allowed. On the other hand, from Section \ref{sec:singlet-sector}, we have $\mathcal{E}_0 = \mathcal{E}_{\rm top}$ at $g_*$, so at $\mathcal C_2=0$ the island must have zero width. Hence, the only way to achieve $\mathcal E>\mathcal E_0$ is to have $\mathcal{C}_2>0$. So in this scenario, the allowed region in the plane of $\mathcal{E}$ and $\mathcal{C}_2$ should exhibit an ``overhang''. However, as we report in Figure \ref{fig:microcanonical_multi_lvl}, we found no such evidence of an overhang, even up to high levels; we interpret this as evidence that $T_c(g_*)=0$.

\section{Monte Carlo simulation of the MQM-inspired Villain model}\label{app:MQMinspired-Villain}
In this section, we study the BKT transition on a regular lattice using Monte Carlo simulations. For the standard link factor, such simulations can be found, for example, in \cite{Otsuka:2023far, Okabe:2025rps}, and for the original Villain model in \cite{Janke:1993va}. In both models, one finds that the lattice transition temperature lies below its continuum, zero-fugacity estimate, $T^{\rm lat}_{\rm BKT}<T_{\rm BKT}^{(0)}$. This shift is due to short-distance effects, in particular the nonzero vortex fugacity, and follows from RG analysis \cite{Kosterlitz:1974nba, Amit:1979ab}. Our goal is to repeat these calculations for the version of Villain model where the link factor is taken to be the Green function \eqref{MQMvillain} of matrix quantum mechanics (to our knowledge, this has not been done before, although similar models were considered before \cite{Jaster:1997grain}). 
By varying $\beta$ we will find a BKT transition on a square lattice, which is qualitatively similar to those for more conventional link factors. Thus, as expected \cite{Gross:1990ub, Gross:1990md}, there is nothing pathological about the MQM link factor \eqref{MQMvillain}. Our results are consistent with the fact that,  
in the $\beta\rightarrow \infty$ limit and on a regular lattice, this model was recently proven to flow to the 2D continuum Gaussian field \cite{Armstrong:2023arxiv}.   
\subsection{Lattice action}
The local Boltzmann weight is given by the Euclidean time propagator appearing on each edge of the graphs in matrix quantum mechanics, \eqref{MQMvillain}; see e.g. \cite{Klebanov:1991qa} for a review.  %
The target space Euclidean time has a periodicity given by the inverse temperature $\tau\sim\tau+\beta$.
On a periodic $L_x\times L_t$ square lattice we simulate 
\begin{equation}
    Z(\beta)=\int\prod_s\frac{d\tau_s}{\beta}\prod_{\langle ss'\rangle}G_\beta(\tau_s-\tau_{s'})\ ,
\end{equation}
In MQM, one would also sum over the geometry of the lattice, but for this simulation we {\it fix} the lattice geometry to be a torus. We refer to this fixed-lattice theory as the MQM-inspired Villain model \cite{Villain:1974ir}, since it retains the compact $U(1)$ structure of the Villain model but uses the MQM link kernel instead of the standard Gaussian weight.

\begin{figure}[t]
    \centering
    \includegraphics[width=\linewidth]
    {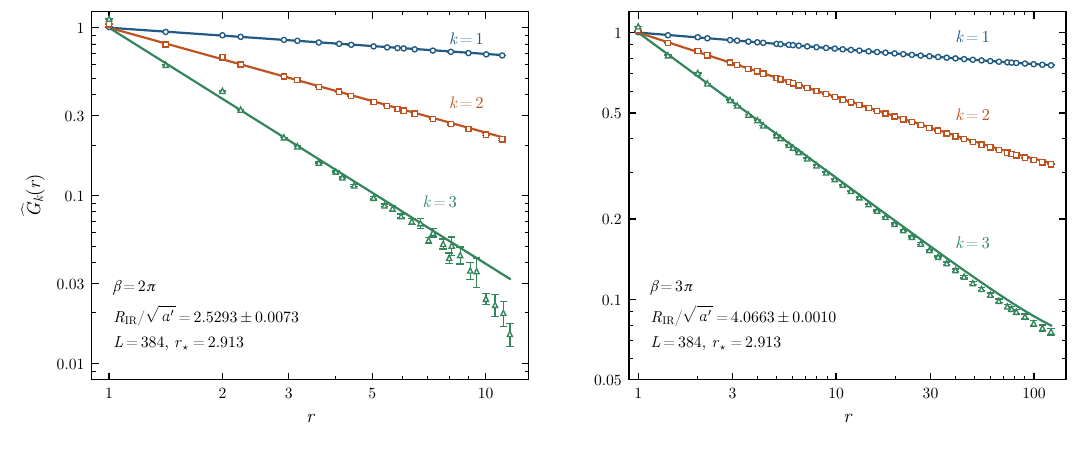}
    \caption{Primary 2-pt functions at two values of $\beta$: (a) $\beta=2\pi$ and (b) $\beta=3\pi$. We have normalized the 2-pt function so that it agrees with the CFT normalization at $r_\star\simeq3$. The CFT curves use values of $R_{\rm IR}/\sqrt{\alpha'}$ inferred independently from the charge probability distribution.}
    \label{fig:primary2pt}
\end{figure}

In the above partition function, one thinks of the periodic variable $\tau$ as living on the vertices of the lattice and being integrated over in the path integral. To simulate the system, it will be more convenient to change variables and work with the Matsubara frequencies $\omega_j=2\pi j/\beta$:
\begin{equation}
    \widetilde{G}_\beta(j)=\frac{1/\beta}{1+(2\pi j/\beta)^2}\ . %
\end{equation}
After integrating out the compact fields, we obtain an exactly conserved integer-current model with $j_\ell\in\mathbb Z$ living on the links:
\begin{equation}
    Z(\beta)=\sum_{\{j_\ell\in\mathbb Z\}}\prod_{s\, \in\,  \text{vertices}}\delta_{(\nabla\cdot j)_s,0}\prod_{\ell\,\in\,\text{edges}} \widetilde{G}_\beta(j)=\frac{1}{\beta^{|E|}}\sum_{\{j_\ell\in\mathbb Z\}}\prod_s\delta_{(\nabla\cdot j)_s,0}\,e^{-S_{\mathrm{cur}}[j]}\ ,
\end{equation}
here $|E|=2L_xL_t$ is the number of edges of the periodic square lattice and
\begin{equation}
    S_{\mathrm{cur}}[j]=\sum_{\ell\,\in\,\text{edges}}\log\left[1+\left(\frac{2\pi j_\ell}{\beta}\right)^2\right]\ .
\end{equation}
This has the advantage that the Monte Carlo only involves discrete variables. Explicitly, if $j_{s,\mu}$ is oriented from $s$ to $s+\mu$, then the discrete conservation law is
\begin{equation}
    (\nabla\cdot j)_s\equiv\sum_{\mu=x,t}\bigl(j_{s,\mu}-j_{s-\mu,\mu}\bigr)=0\ . %
\end{equation}
We sample the resulting theory using the Metropolis updates. %
To test whether the resulting theory flows to the $c=1$ boson CFT, we compute the lattice two-point function:
\begin{equation}
    G_k(\mathbf d)\equiv\left\langle e^{2\pi i k\tau_0/\beta} e^{-2\pi i k\tau_{\mathbf d}/\beta}\right\rangle=\frac{1}{Z(\beta)}{\displaystyle\sum_{\{j_\ell\in\mathbb Z\}}\prod_\ell \widetilde{G}_\beta(j)\prod_s\delta_{(\nabla\cdot j)_s,\,k(\delta_{s,\mathbf d}-\delta_{s,0})}}\ .
\end{equation}
We define the discrete distance of $\mathbf{d}$ as
\begin{equation}
    d=\sqrt{\min(d_x,L_x-d_x)^2+\min(d_t,L_t-d_t)^2}\ .
\end{equation}
We then average over $\mathbf{d}$ which have distances in some bin.\footnote{More precisely, for logarithmically spaced bin boundaries $a_n$, let $\mathcal A_n=\{\mathbf d:a_n\leq\rho(\mathbf d)<a_{n+1}\}$ and $N_n=|\mathcal A_n|$. Each displacement receives equal weight:
\begin{equation}
    G_k(r_n)=\frac{1}{N_n}\sum_{\mathbf d\in\mathcal A_n}G_k(\mathbf d)\ ,\qquad r_n=\exp\left[\frac{1}{N_n}\sum_{\mathbf d\in\mathcal A_n}\log\rho(\mathbf d)\right]\ .
\end{equation}}
We compare these momentum correlators with those in the compact-boson CFT
with infrared radius $R_\text{IR}$:
\begin{equation}
    S_{\rm CFT}=\frac{1}{4\pi\alpha'}\int\d^2x\,(\partial X)^2\ , \qquad X \sim X+ 2\pi R_\text{IR}\ .
\end{equation}    
For operators of momentum $k$, the scaling dimensions and two-point function are
\begin{equation}
    {\Delta_k} =\frac{\alpha' k^2}{2R^2_\text{IR}}\ ,
    \qquad G_k(r)\propto r^{-2\Delta_k}\quad (1\ll r\ll L)\ .
\end{equation}
The comparison against the continuum prediction is shown in Figure \ref{fig:primary2pt}. (At finite $r/L$, we could use the compact-boson torus correlator, averaged over the same displacement bins as the data, although this makes little difference in our plots.) 

\begin{figure}[t]
    \centering
    \includegraphics[width=0.6\linewidth]{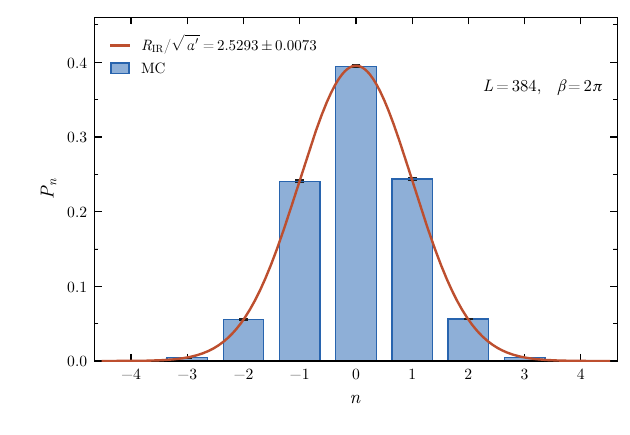}
    \caption{A histogram of the charge distribution as measured from the lattice data. The best fit value from the $Q_x$ charge alone is $R_{\rm IR}/\sqrt{\alpha'} = 2.529 \pm 0.007$, whereas combining the $Q_x$ and $Q_y$ data gives $R_{\rm IR}/\sqrt{\alpha'} = 2.514\pm 0.001$. 
    }
    \label{fig:histogram}
\end{figure}

\begin{figure}[t]
    \centering
    \includegraphics[width=0.6\linewidth]{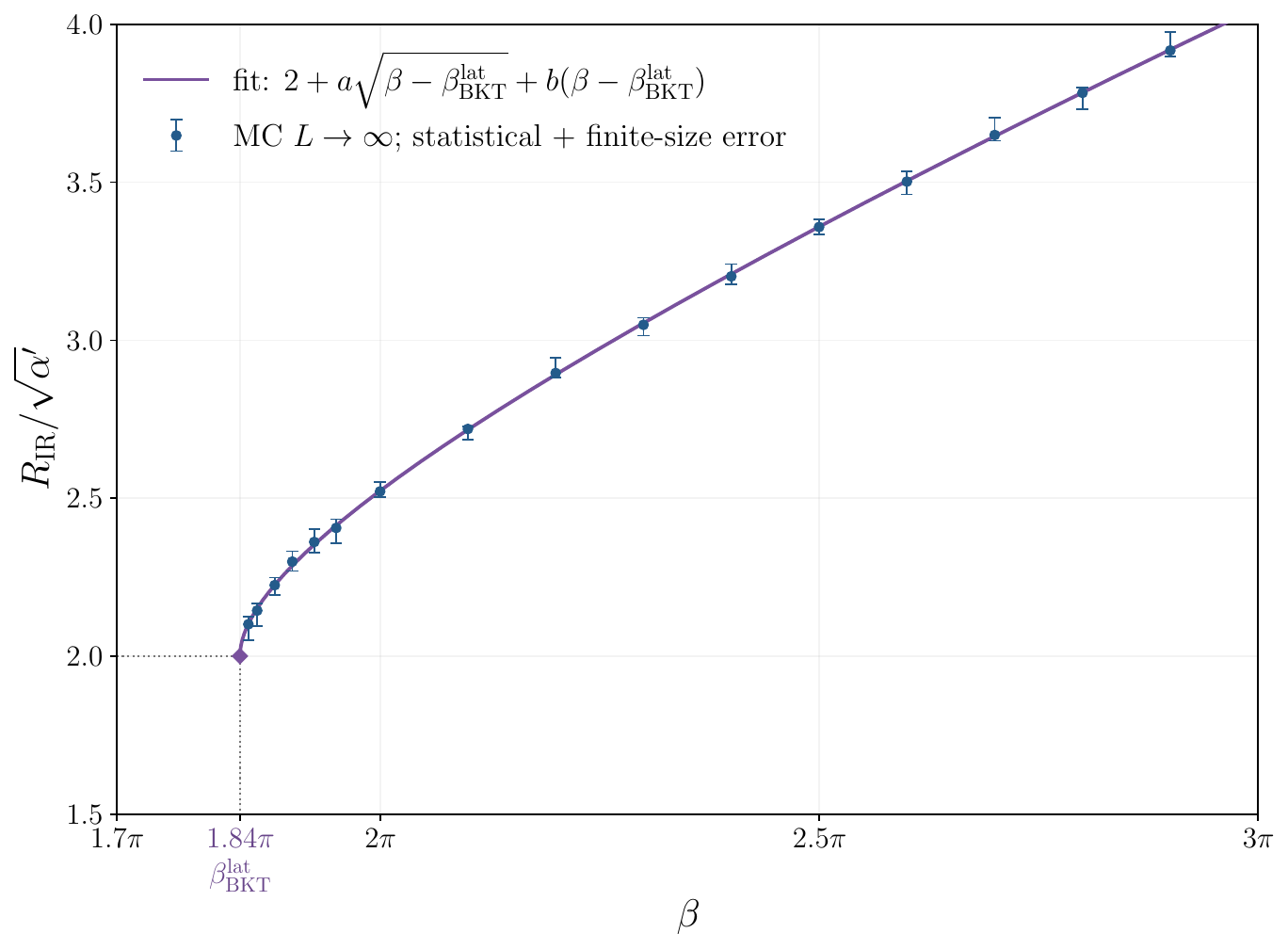}
    \caption{Infrared radius $R_{\rm IR}/\sqrt{\alpha'}$ versus the microscopic parameter $\beta$ (the circumference in the lattice action). Blue points show Monte Carlo estimates extrapolated to infinite lattice size;  error bars include both one standard deviation of statistical uncertainty and finite size systematics. Purple curve shows a fit with ansatz \eqref{RGansatz},  
    which gives $\beta^{\rm lat}_{\rm BKT} \approx 1.84\pi$.}
    \label{fig:radius-vs-beta-villain}
\end{figure}

In principle, measuring the two-point functions of the primaries $V_k$ allows us to infer the infrared boson radius $R_{\rm IR}$ and check its consistency across different values of $k$. Here we instead extract $R_{\rm IR}$ from fluctuations of the conserved integer currents in the lattice representation. We define
\begin{equation}
    Q_x=\sum_x j_{(x,y_0),y}\ ,\qquad Q_y=\sum_y j_{(x_0,y),x}\ .
\end{equation}
Current conservation makes these charges independent of the positions $y_0$ and $x_0$ of the respective cuts. Both charges take integer values since each $j$ is an integer. Now in the continuum theory, define $\theta=X/R_{\rm IR}$. Taking $y$ as Euclidean time, $Q_x$ is the momentum conjugate to the spatial zero mode $\theta_0$ on a circle of circumference $L_x$. Its rotor Hamiltonian is
\begin{equation}
    H_x^{\rm rotor}=-\frac{\pi\alpha'}{L_x R_{\rm IR}^2}\frac{\partial^2}{\partial\theta_0^2}\ , \qquad Q_x=-i\frac{\partial}{\partial\theta_0}\ .
\end{equation}
Evolution around the Euclidean-time circle of length $L_y$ therefore gives
\begin{equation}
    P(Q_x=n) \propto \operatorname{Tr}\left(\Pi_n e^{-L_yH_x}\right)=\exp\!\left[-\pi\alpha'\frac{L_y}{L_x}\frac{n^2}{R_{\rm IR}^2}\right].
\end{equation}
We compare this prediction to a histogram from the Monte Carlo simulation in Figure \ref{fig:histogram}. For $\beta=2\pi$ we find reasonably good agreement with the continuum compact scalar field with $R_{\rm IR}/\sqrt{\alpha'}=2.5293\pm0.0073$. Similarly, for $\beta = 3\pi$ we find $R_{\rm IR}/\sqrt{\alpha'}=4.0663\pm0.0010$ \footnote{The reported errors are statistical errors; there are also unquantified systematic errors from finite $L$/lattice effects.}. Both of these values of $\beta$ are clearly on the $c=1$ side of the BKT transition. These values indicate that $R_{\rm IR}(\beta)/ \sqrt{\alpha'}$ is not linear in $\beta$, as shown in Figure \ref{fig:radius-vs-beta-villain}. We then estimate the BKT transition temperature by extrapolating our MC results to the BKT transition point $R_{\rm IR}/\sqrt{\alpha'}=2$, where the vortex operator becomes marginal. We use the ansatz 
\begin{equation}\label{RGansatz}
    \frac{R_{\rm IR}}{\sqrt{\alpha'}}=2+a\sqrt{\beta-\beta^{\rm lat}_{\rm BKT}}+b(\beta-\beta^{\rm lat}_{\rm BKT})\ ,
\end{equation}
which is motivated by the structure of RG flow near the BKT transition \cite{Kosterlitz:1974nba, Amit:1979ab} while also incorporating the expected linear relation between $R_{\rm IR}$ and $\beta$ when both become large. The resulting estimate of the BKT transition point is $\beta^{\rm lat}_{\rm BKT}\approx 1.84\pi\approx 5.78$.


We also performed a preliminary Monte Carlo simulation of this model on a quartic random lattice with fixed area, for either spherical or toroidal topology, which should be equivalent to the sum over the MQM Feynman diagrams with fixed number of vertices. This exploration suggested that there is less renormalization of the radius in the random lattice case than for the fixed lattice. It is an interesting open problem to perform a high-precision Monte Carlo simulation that computes 
the BKT transition temperature on a random lattice with the MQM link factor \eqref{MQMvillain} (for earlier work with a different link factor, see \cite{Catterall:1993vs}). This would help to determine the value of $T_{\rm BKT}$ in the phase diagram in Figure \ref{fig:phase-diagram}.
\section{Discussion}\label{sec:discussion}
The main result of this work is the phase diagram of the ungauged MQM, which is shown schematically in Figure
\ref{fig:phase-diagram}. The first thermal bootstrap study of the critical stability curve $T_c(g)$ was carried out in \cite{Cho:2024kxn}, and we use a combination of methods to determine its structure with greater precision. Its low- and high-temperature branches exhibit the characteristic singularities of $c=1$ and $c=0$ gravity, respectively. The two branches are separated by a BKT transition at a special point $\left(g_{\rm BKT},T_{\rm BKT}\right)$ on the critical curve.\footnote{A qualitatively similar phase diagram appears in MQM with a double-trace deformation \cite{Sugino:1994zr, Gubser:1994yb}. These works studied a zero-temperature transition driven by varying the couplings. Their critical curve separates the $c=1$ and branched polymer phases, with a special point where the Liouville potential was argued to be $e^{2\phi}$ instead of $\phi e^{2\phi}$ \cite{Gubser:1994yb,Klebanov:1994pv}.} Combining analytic approximations with canonical and microcanonical matrix bootstrap, we obtain a quantitative description of this curve over a broad range of temperatures. Non-singlet states play a central role in its temperature dependence: at low temperatures they backreact on the singlet Fermi sea, while at high temperatures their effects are incorporated into the effective one-matrix integral obtained by dimensional reduction. Moreover, subleading $1/N$ corrections to multi-adjoint energies are crucial for the correct understanding of the planar thermodynamics. Our analysis may be extended to other metastable matrix potentials with similar physics to the quartic potential \eqref{MQM-Hamiltonian-def}, such as the cubic potential, as well as to matrix quantum mechanics with other global symmetry groups, including SO$(N)$ and USp$(2N)$. However, the critical curve and the BKT temperature are not universal and depend on the form of potential.

Our continuum RG analysis suggests that $T_{\rm BKT}$ lies not far from the Gaussian value $T_{\rm BKT}^{(0)}=1/(4\pi\sqrt{\alpha'})$, where the unit winding operator becomes marginal. The BKT transition in the $c=1$ matrix model thus has some similarity with the Hagedorn/deconfinement transition in critical string theory, which also occurs near the temperature where the unit winding operators become relevant \cite{Kogan:1987jd, Sathiapalan:1986db, Atick:1988si}. This analogy does not seem to be perfect, since in MQM the free energy is of the order $N^2$ on both sides of the phase transition, but we feel that it deserves further investigation.

Another problem that deserves further investigation is a possible relation between thermal MQM and the two-dimensional black hole discovered in \cite{Witten:1991yr}. Motivated by the FZZ duality between the $SL(2,\mathbb R)_k/U(1)$ with $k=9/4$ cigar CFT and sine-Liouville theory \cite{FZZ,Hikida:2008pe}, the authors of \cite{Kazakov:2000pm} proposed that MQM with a strong winding condensate describes the Euclidean cigar background at $R_{\rm BH}=3\sqrt{\alpha'}/2$, corresponding to $T_{\rm BH}=\frac{4}{3} T_{\rm BKT}^{(0)}$. However, the recent analysis of the sine-Liouville phase diagram \cite{Kostov:2025awi,Junca:2026uqv} suggests that this identification needs to be reconsidered. In particular, these works reveal a massless flow between theories with different compactification radii, calling into question the fixed-radius assumption underlying the original black-hole proposal of \cite{Kazakov:2000pm}. Our phase diagram of ungauged MQM (Figure \ref{fig:phase-diagram}) contains no separate black-hole critical phase: at $T_{\rm BH}$, the real thermal critical curve is expected to belong to the $c=0$ universality class. Understanding whether the black-hole background can be reached via a different winding deformation of MQM remains an interesting open problem.

An especially interesting extension and application of the methods developed here is to the symmetric double-well potential $V(X)=-\frac{1}{2}X^2+gX^4, \, g>0$, whose two-sided Fermi sea describes two-dimensional type 0B string theory with $\hat c=1$ \cite{Douglas:2003up,Takayanagi:2003sm}. We expect an analogous finite-temperature critical curve $T_c(g)$ (or $g_c(T)$) separating the two-cut and one-cut phases, with a BKT transition between the $\hat c=1$ and $\hat c=0$ supergravity regimes along this curve. Unlike the metastable quartic model studied here, the double-well theory is well defined at any temperature, so the critical curve would correspond to the merging of two eigenvalue cuts rather than a boundary of stability. The work of Matytsin and Zaugg on the bosonic infinite matrix chain \cite{Matytsin:1996tx,Matytsin:1997uj} provides a precedent for such a transition, while $\hat c=0$ supergravity also arises in the double-scaling limit of the Gross--Witten--Wadia matrix model \cite{Klebanov:2003wg}.

Another natural extension is to introduce a chemical potential for the quadratic Casimir of SU($N$), which commutes with the Hamiltonian. In forthcoming work \cite{KlebanovInPrep}, we will study the deformation $H\to H+\nu C_2/N$, extending the two-dimensional phase diagram considered here to a critical surface in $(g,T,\nu)$.

\subsection{Some open questions}
We conclude with several questions that arise directly from our calculations and can be addressed by further numerical and analytic studies. 
\begin{enumerate}

\item \emph{What is the microscopic dictionary between non-singlet representations and worldsheet vortices?}
In ungauged MQM, the contributions of all non-singlet representations are fixed by the microscopic theory, whereas the continuum description involves winding couplings whose values remain unknown. Can these couplings and the renormalization of the compactification radius be derived directly from MQM, allowing a quantitative prediction of the physical BKT temperature? More generally, how are different multi-adjoint representations encoded in configurations of interacting worldsheet vortices, and is the adjoint Regge crossover observed in \cite{Klebanov:2026vmk} related to vortex-antivortex dissociation?

\item \emph{Can the BKT transition and its universal critical behavior be established directly in MQM?}
The low- and high-temperature analyses give the characteristic $c=1$ and $c=0$ singularities of the planar free energy, respectively, but do not determine the crossover between them or the location of the BKT point. Can analytic methods or matrix bootstrap distinguish these two universality classes and identify the transition, for example through the critical scaling of matrix moments or susceptibilities? In particular, does the limiting free energy along the critical curve exhibit a BKT-type essential singularity, schematically $\delta\mathcal F(T)\sim t^p\exp\left(-\frac{b}{\sqrt{t}}\right),\, t=\frac{T-T_{\mathrm{BKT}}}{T_{\mathrm{BKT}}}\to 0^+$, or is this behavior modified by coupling to two-dimensional gravity? Can one construct a double-scaling limit near the BKT point that captures the universal crossover, including possible logarithmic corrections from the marginal winding interaction?

\item \emph{Can one find a bootstrap method to get a lower bound on $T_c(g)$?} The bootstrap methods we discussed in this paper only give an upper bound on the critical temperature. A related question is if one can rule {\it in} energies $E/N^2$ in the microcanonical ensemble, as opposed to just ruling {\it out} energies.

\item \emph{What is the allowed set of microcanonical energies and representations of SU$(N)$?} For the case of SU$(N)$ singlets, we gave a precise characterization of this set in Section \ref{sec:singlet-sector}. %
Can one characterize this shape for more general irreps, especially large irreps that have ``smooth'' Young diagrams?
A simpler version of this request is to understand analytically the allowed region in $(\mathcal E,\mathcal{C}_2)$ that we obtained by numerical bootstrap in Figure \ref{fig:microcanonical_multi_lvl}.

\end{enumerate}

The new results reported in this paper strengthen our conviction that the ungauged MQM is a good benchmark model, which is relatively simple yet highly non-trivial. We hope that our study provides motivation for using similar methods to investigate more complicated large $N$ models, such as the ungauged BFSS/BMN matrix models \cite{Maldacena:2018vsr, Berkowitz:2018qhn}. 
\section*{Acknowledgements}
We thank Alexander Artemev, Panos Betzios, Minjae Cho, Barak Gabai, David Gross, Andrei Katsevich, David Kutasov, Juan Maldacena, Emil Martinec, Steve Shenker, Xi Yin, and Zechuan Zheng for useful discussions. IRK and PM are supported in part by the National Science Foundation grant PHY-2609862, Simons Foundation grant 917464 and Simons Foundation International grant SFI-MPS-QCD-00014867-01 (Simons Collaboration on Confinement and QCD Strings). IRK's work was performed in part at the Aspen Center for Physics, which is supported by the National Science Foundation grant PHY-2210452. JY is partially supported by the National Science Foundation under award No. 2016245 and 2440805 and by the DOE Office of Science under grant agreement DE-SC0025934.
\section*{AI statement}
Generative AI models were used to assist with the Monte Carlo codes, producing numerical figures, certain perturbative calculations, and manuscript revisions. 
\appendix
\section{Singlet particle-hole spectrum and quartic asymptotics}\label{app:singlet-WKB}

The exact free-fermion reduction provides a useful microscopic description of the singlet excitations. In this appendix we make the particle-hole structure explicit and derive the first $1/N$ splitting of the planar spectrum. We then collect the weak-coupling and near-critical expansions of the planar ground-state energy and Fermi-surface frequency used in the main text.
\subsection{Particle-hole spectrum near the Fermi surface}

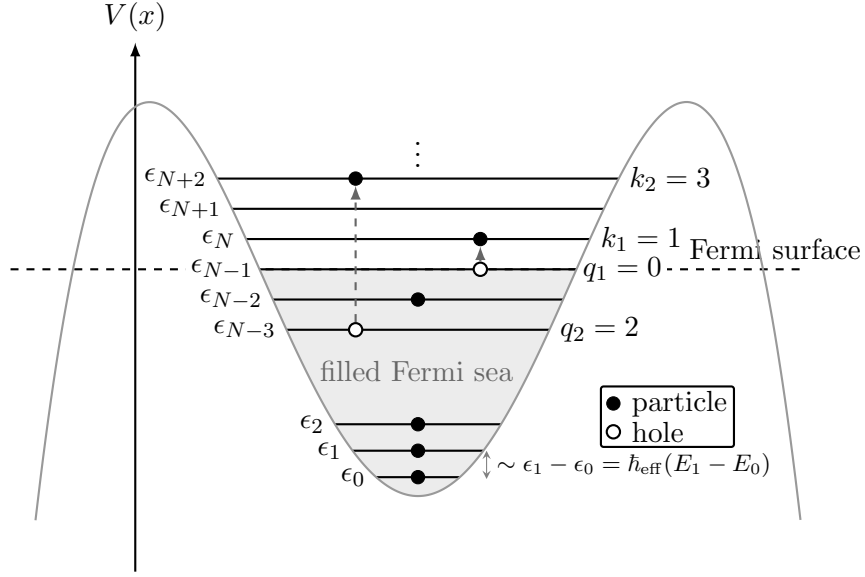
\begin{figure}[t]
    \centering
    \begin{tikzpicture}[
    xscale=0.55, yscale=0.5,
    level/.style={thick, black},
    potential/.style={thick, gray!80},
    particle/.style={circle, draw=black, fill=black, inner sep=0pt, minimum size=5pt},
    hole/.style={circle, draw=black, fill=white, thick, inner sep=0pt, minimum size=5pt},
    declare function={
        V(\x) = -6 + 0.5*\x*\x - 0.006*\x*\x*\x*\x;
        tp(\E) = sqrt( (0.5 - sqrt(0.25 - 0.024*(\E + 6))) / 0.012 );
    }
    ]
    
        \fill[gray!15] plot[domain=-3.812:3.812, smooth, samples=100] (\x, {V(\x)}) 
                       -- (-3.812, 0) -- cycle;
        
        \draw[-latex, thick] (-6.8, -8.0) -- (-6.8, 6.0) node[above] {$V(x)$};
        
        \draw[dashed, thick] (-9.8, 0) -- (9.2, 0) node[pos=1.09, above left] {Fermi surface};
        \node[gray!80!black, anchor=west, align=left] at (-2.6, -2.7) {filled Fermi sea};

        \draw[level] ({-tp(2.4)}, 2.4) node[left] {$\epsilon_{N+2}$} -- ({tp(2.4)}, 2.4) node[right] {$k_2=3$};
        \draw[level] ({-tp(1.6)}, 1.6) node[left] {$\epsilon_{N+1}$} -- ({tp(1.6)}, 1.6) node[right] {};
        \draw[level] ({-tp(0.8)}, 0.8) node[left] {$\epsilon_{N}$}   -- ({tp(0.8)}, 0.8) node[right] {$k_1=1$};
        
        \draw[level] ({-tp(0)}, 0) node[left, fill=white, inner sep=2pt] {$\epsilon_{N-1}$} -- ({tp(0)}, 0) node[right, fill=white, inner sep=2pt] {$q_1=0$};
        \draw[level] ({-tp(-0.8)}, -0.8) node[left] {$\epsilon_{N-2}$} -- ({tp(-0.8)}, -0.8) node[right] {};
        \draw[level] ({-tp(-1.6)}, -1.6) node[left] {$\epsilon_{N-3}$} -- ({tp(-1.6)}, -1.6) node[right] {$q_2=2$};
        
        \node at (0, 3.25) {$\vdots$}; 

        \draw[<->, >=stealth, gray, thin] (1.65, -5.55) -- (1.65, -4.8) node[midway, right, black] {\footnotesize $\sim \epsilon_1-\epsilon_0=\hbar_{\rm eff}(E_1-E_0)$};
        
        \draw[level] ({-tp(-4.1)}, -4.1) node[left] {$\epsilon_2$} -- ({tp(-4.1)}, -4.1);
        \draw[level] ({-tp(-4.8)}, -4.8) node[left] {$\epsilon_1$} -- ({tp(-4.8)}, -4.8);
        \draw[level] ({-tp(-5.5)}, -5.5) node[left] {$\epsilon_0$} -- ({tp(-5.5)}, -5.5);
        
        \draw[potential] plot[domain=-9.2:9.2, smooth, samples=150] (\x, {V(\x)});
        
        \node[particle] at (0, -5.5) {};
        \node[particle] at (0, -4.8) {};
        \node[particle] at (0, -4.1) {};
        \node[particle] at (0, -0.8) {};
        
        \node[hole] (h1) at (1.5, 0.) {};
        \node[particle] (p1) at (1.5, 0.8) {};
        \draw[-latex, dashed, thick, gray!80!black] (h1.north) -- (p1.south);
        
        \node[hole] (h2) at (-1.5, -1.6) {};
        \node[particle] (p2) at (-1.5, 2.4) {};
        \draw[-latex, dashed, thick, gray!80!black] (h2.north) -- (p2.south);
        
        \draw[thick, fill=white, rounded corners=1pt] (4.4, -4.7) rectangle (7.6, -3.1);
        \node[particle] at (4.7, -3.55) {}; \node[right] at (4.9, -3.55) {particle};
        \node[hole] at (4.7, -4.3) {}; \node[right] at (4.9, -4.3) {hole};
    \end{tikzpicture}
    \caption{Particle-hole excitations of the filled Fermi sea in the MQM potential $V(x)$. Particles (solid) and holes (open) are indexed by their respective distances $k_i$ and $q_i$ from the Fermi surface, with dashed arrows indicating the pair creation corresponding to the singlet excitation energy.}
    \label{fig:Fermi-see}
\end{figure}

The free-fermion description gives a simple microscopic organization of the singlet excitations \cite{Brezin:1977sv,Marchesini:1979yq}. Denote the physical one-particle energies by $E_n=N\epsilon_n$. A finite-energy excitation of the filled Fermi sea is obtained by removing $r$ fermions from occupied levels and placing them above the Fermi surface. We label such a state by
\begin{equation}
    0\leq q_1<\ldots<q_r\ , \qquad 1\leq k_1<\ldots<k_r\ ,
\end{equation}
where the holes lie at levels $N-q_i-1$ and the particles at levels $N+k_i-1$, as illustrated in Figure \ref{fig:Fermi-see}. Its exact gap above the ground state is
\begin{equation}\label{singlet-spectrum}
    \Delta_{\mathrm{sing}}(\{k_i\},\{q_i\})=\sum_{i=1}^{r}\left(E_{N+k_i-1}-E_{N-q_i-1}\right)=N\sum_{i=1}^{r}\left(\epsilon_{N+k_i-1}-\epsilon_{N-q_i-1}\right)\ .
\end{equation}
For example, the lowest one is $\omega_N\equiv E_{N}-E_{N-1}$.

For fixed $k_i,q_i$ at large $N$, only the local one-particle spectrum near the Fermi surface is probed. It is convenient to introduce $\xi_n=(n+\frac{1}{2})/N.$ At leading semiclassical order, the scaled energy $e^{(0)}(\xi)$ is determined by the Bohr-Sommerfeld condition
\begin{equation}\label{singlet-Bohr-Sommerfeld}
    \mathcal I\left(e^{(0)}(\xi)\right)=\xi\ ,
\end{equation}
where $\mathcal I(e)$ is the action variable defined in \eqref{singlet-action}. Differentiating \eqref{singlet-Bohr-Sommerfeld} and using \eqref{frequency-def} gives
\begin{equation}
    \frac{\d e^{(0)}}{\d\xi}=\omega(e)\ ,\qquad\frac{\d^2e^{(0)}}{\d\xi^2}=\omega(e)\,\partial_e\omega(e)\ ,
\end{equation}
with $e=e^{(0)}(\xi)$. Since $\xi_{N+k-1}=1+(k-\frac{1}{2})/N$ and $\xi_{N-q-1}=1-(q+\frac{1}{2})/N$, expanding \eqref{singlet-spectrum} around the Fermi surface gives
\begin{equation}\label{singlet-spectrum-WKB}
    \begin{aligned}
        \Delta_{\mathrm{sing}}(\{k_i\},\{q_i\})=\omega_F\sum_{i=1}^{r}(k_i+q_i)+\frac{1}{2N}\left[\omega\partial_e\omega\right]_{\mathrm F}\sum_{i=1}^{r}(k_i+q_i)(k_i-q_i-1)+\mathcal O\left(N^{-2}\right)\ .
    \end{aligned}
\end{equation}
Importantly, the order-$1/N$ term follows entirely from the curvature of the leading semiclassical spectrum; genuine higher-order WKB corrections to the individual fermion levels first enter the gaps at order $1/N^2$.

The first term of \eqref{singlet-spectrum-WKB} reproduces the planar tower \eqref{singlet-planar-spectrum}. The particle-hole labels have a standard combinatorial interpretation: $(k_i-1,q_i)$ are the Frobenius coordinates of a partition of $s$. Hence, for fixed $s$ and sufficiently large $N$, the planar level $s\,\omega_F$ has multiplicity $p(s)$, the number of partitions of $s$. This is precisely the counting obtained from the linearized collective-field description, where the singlet excitations are organized as bosonic normal modes \cite{Jevicki:1979mb,Das:1990kaa}.

The second term of \eqref{singlet-spectrum-WKB} shows that this planar degeneracy is generically lifted at finite $N$. For the quartic potential, the scaling properties of the Fermi sea imply
\begin{equation}\label{quartic-frequency-derivative}
    \omega\partial_e\omega\big|_{\mathrm F}=g\,\omega'(g)\ ,
\end{equation}
and therefore
\begin{equation}\label{singlet-spectrum-quartic}
    \Delta_{\mathrm{sing}}=\omega(g)\sum_i(k_i+q_i)+\frac{g\omega'(g)}{2N}\sum_i(k_i+q_i)(k_i-q_i-1)+\mathcal O\left(N^{-2}\right)\ .
\end{equation}
For example,
\begin{equation}\label{lowest-singlet-splittings}
    \begin{aligned}
        \Delta_{\rm sing}(\{1\},\{0\})&=\omega(g)+\mathcal O\left(N^{-2}\right)\ ,
        \\
        \Delta_{\rm sing}(\{2\},\{0\})&=2\omega(g)+\frac{g\omega'(g)}{N}+\mathcal O\left(N^{-2}\right)\ ,
        \\
        \Delta_{\rm sing}(\{1\},\{1\})&=2\omega(g)-\frac{g\omega'(g)}{N}+\mathcal O\left(N^{-2}\right)\ .
    \end{aligned}
\end{equation}
Thus the first level is non-degenerate and has no order-$1/N$ correction, whereas the two states at planar energy $2\omega$ split symmetrically. At the next level the three configurations $(3,0)$, $(2,1)$, and $(1,2)$ acquire respectively the shifts $+3g\omega'/N$, $0$, and $-3g\omega'/N$. The first state involving two particle-hole pairs appears at $s=4$, in agreement with $p(4)=5$. These examples also illustrate how the order-$1/N$ splittings cancel within a planar-degenerate multiplet in the thermal trace, consistently with the even-power large-$N$ (topological) expansion of the singlet free energy in \eqref{MQM-gs}.

The expansion \eqref{singlet-spectrum-WKB} is controlled for fixed particle-hole labels on any regular Fermi surface. Its approach to the critical endpoint is non-uniform, however, because $\omega_F$ vanishes while its derivatives with respect to the Fermi energy become singular. The corresponding near-critical scaling is discussed in the following subsection.

\subsection{Weak-coupling and critical expansions}
\paragraph{Weak coupling.} The exact quartic expressions \eqref{frequency-analyt} and \eqref{MQM-planar-gs-quartic}, together with the parametric relation \eqref{g-muF-as-modulus-k}, determine the planar ground-state energy and the singlet excitation scale throughout the one-cut branch. At weak coupling, $k\to0$, one may expand $g(k)$ around the origin, invert this relation perturbatively to obtain $k=k(g)$, and substitute the resulting series into $\mathcal E_0(k)$ and $\omega(k)$. This gives
\begin{equation}\label{planar-ground-state-small-g}
    \mathcal E_0(g)=\frac{1}{2}+\frac{g}{2}-\frac{17}{16}g^2+\frac{75}{16}g^3-\frac{3563}{128}g^4+\frac{12507}{64}g^5-\frac{3132399}{2048}g^6+\frac{26469553}{2048}g^7+\mathcal O\left(g^8\right)\ ,
\end{equation}
and
\begin{equation}\label{singlet-frequency-expansion}
    \omega(g)=1+3g-\frac{51}{4}g^2+\frac{375}{4}g^3-\frac{53445}{64}g^4+\frac{262647}{32}g^5-\frac{21926793}{256}g^6+\frac{238225977}{256}g^7+\mathcal O\left(g^8\right)\ .
\end{equation}
The Gaussian values $\mathcal E_0(0)=1/2$ and $\omega(0)=1$ are recovered smoothly.

The coefficients of the planar weak-coupling expansion may also be generated directly from the exact parametric solution, without explicitly constructing the inverse function $k(g)$ order by order. Writing
\begin{equation}
    z=k^2\ ,\qquad g(z)=-\frac{z}{4}A(z)\ ,\qquad A(z)=(1+z)^{-3/2}\, {}_2F_1\left(-\frac{1}{2},\frac{1}{2};2;z\right)\ ,
\end{equation}
we have $A(0)=1$ and hence $g'(0)=-1/4\neq0$, so that $z(g)$ is locally well defined around $g=0$. For
\begin{equation}
    \mathcal E_0(z(g))=\sum_{n\geq0}c_{0,n}g^n\ ,
\end{equation}
the Lagrange-B\"urmann inversion formula gives
\begin{equation}\label{E0-Lagrange-inversion}
    c_{0,n}=\frac{(-4)^n}{n}\left[\frac{\mathcal E_0'(z)}{A^n(z)}\right]_{z^{n-1}}=\frac{(-4)^n}{n}\oint_{\mathcal C_0}\frac{\d z}{2\pi i}\,\frac{\mathcal E_0'(z)}{z^nA^n(z)}\ , \qquad n\geq1\ .
\end{equation}
Here, $[\ldots]_{z^{n-1}}$ denotes the coefficient of $z^{n-1}$. Thus any fixed weak-coupling coefficient $c_{0,n}$ can be obtained algebraically from the expansion around $z=0$, without explicitly inverting $g(z)$. Equation \eqref{E0-Lagrange-inversion} is also useful for studying the large-order behavior of these coefficients, which is controlled by the critical singularity of the parametric solution discussed below.

The expansions \eqref{planar-ground-state-small-g} and \eqref{singlet-frequency-expansion} admit several complementary perturbative checks. Since the singlet sector is described by free fermions, the physical one-particle energies $E_n=N\epsilon_n$ are those of the quartic oscillator with coupling $g/N$ and have the expansion
\begin{equation}\label{En-perturbative}
    E_n=n+\frac{1}{2}+\sum_{r\geq1}\left(\frac{g}{N}\right)^r P_r(n)\ ,
\end{equation}
where $P_r(n)$ is a polynomial of degree $r+1$. For example,
\begin{equation}
    \begin{aligned}
        P_1(n)&=\frac{3}{4}(2n^2+2n+1)\ ,\\
        P_2(n)&=-\frac{1}{8}(34n^3+51n^2+59n+21)\ ,\\
        P_3(n)&=\frac{3}{16}(125n^4+250n^3+472n^2+347n+111)\ .
    \end{aligned}
\end{equation}
Summing \eqref{En-perturbative} over the occupied levels $n=0,\ldots,N-1$ and retaining the leading $N^2$ contribution reproduces \eqref{planar-ground-state-small-g} through order $g^3$ (and also gives the subleading large-$N$ corrections order by order in $g$). The same formulas give
\begin{equation}
    \omega_N=E_N-E_{N-1}=1+3g-\frac{51}{4}g^2+\frac{375}{4}g^3+\mathcal O\left(g^4,N^{-2}\right)\ ,
\end{equation}
and hence reproduce the first terms of the planar frequency \eqref{singlet-frequency-expansion}. Higher-order polynomials $P_r(n)$ can be generated efficiently using the Bender--Wu recursion and its Mathematica implementation \cite{Sulejmanpasic:2016fwr}, providing a practical high-order check of the exact parametric expansion.

A separate check comes from the planar vacuum-diagram calculation of Section \ref{sec:small-g}: its zero-temperature limit \eqref{E0-small-g-from-thermal} reproduces the ground-state expansion \eqref{planar-ground-state-small-g} through the third order computed there. The matrix-oscillator formalism of Appendix \ref{app:operator-formalism} provides an additional direct check of the first perturbative coefficients.

\paragraph{Critical coupling.} We parametrize the approach to the critical point by $g=g_*(1-\delta)$, with $\delta\to0^+$ and $g_*$ given in \eqref{g-crit}. It is also useful to introduce the energy difference between the potential maximum and the Fermi level, $\mu\equiv V_{\max}-\mu_F$, where $V_{\max}=-1/(16g)$, and the logarithm $L_\mu\equiv\log(12\sqrt{2}\pi/\mu)$. Expanding the exact elliptic solution \eqref{g-muF-as-modulus-k} gives
\begin{equation}\label{delta-mu-critical}
    \delta=\frac{\mu}{\sqrt{2}\pi}\left(L_\mu+1\right)-\frac{\mu^2}{24\pi^2}\left(12L_\mu^2+6L_\mu+17\right)+\mathcal O\left(\mu^3L_\mu^3\right)\ .
\end{equation}
In particular, $\delta\sim\mu L_\mu/(\sqrt{2}\pi)$, or equivalently
$\mu\sim\sqrt{2}\pi\delta/\log(1/\delta)$ up to logarithmic corrections. This is the familiar logarithmic scaling violation of the $c=1$ matrix model \cite{Gross:1990ub}.

The Fermi-surface frequency vanishes logarithmically,
\begin{equation}\label{omega-critical-mu}
    \omega_F=\frac{\sqrt{2}\pi}{L_\mu}-\frac{2\mu}{L_\mu}+\frac{7\mu}{3L_\mu^2}+\mathcal O\left(\mu^2\right)\ .
\end{equation}
For the quartic potential both endpoints become critical and $V''_{\mathrm{max}}=-2$, so that $n_{\rm end}=2$ and $\alpha'=1/2$ in \eqref{frequency-critical}; in our conventions the corresponding scale is $\Lambda=12\sqrt{2}\pi$. Thus the classical period diverges while the characteristic spacing of the singlet particle-hole spectrum collapses as $1/L_\mu$.

The planar ground-state energy instead remains finite. Its near-critical expansion is
\begin{equation}\label{E0-critical-mu}
    \mathcal E_0=\underbrace{\frac{57\pi}{280\sqrt{2}}}_{\mathcal E_0(g_*)}+\frac{9\mu}{560}\left(L_\mu+1\right)+\frac{\sqrt{2}\mu^2}{4480\pi}\left(-192L_\mu^2+790L_\mu+353\right)+\mathcal O\left(\mu^3L_\mu^2\right)\ .
\end{equation}
Eliminating $\mu$ through \eqref{delta-mu-critical} and expanding for $L_\mu\gg1$ gives
\begin{equation}\label{E0-critical-nonanalytic}
    \mathcal E_0=\mathcal E_0(g_*)+\frac{9\pi}{280\sqrt{2}}\delta-\frac{39\pi}{280\sqrt{2}}\delta^2+\frac{\pi}{\sqrt{2}}\frac{\delta^2}{L_\mu}-\frac{3\pi}{2\sqrt{2}}\frac{\delta^2}{L_\mu^2}+\mathcal O\left(\frac{\delta^2}{L_\mu^3}\right)\ .
\end{equation}
The first two terms in $\delta$ are regular at $g=g_{*}$, whereas the leading nonanalytic contribution is
\begin{equation}\label{E0-leading-singular}
    \mathcal E_{0,\rm{singular}}\sim\frac{\pi}{\sqrt{2}}\frac{\delta^2}{L_\mu}\ .
\end{equation}
The singular term \eqref{E0-leading-singular} controls the large-order behavior of the weak-coupling coefficients generated by \eqref{E0-Lagrange-inversion}. Retaining the normalization and logarithmic scale fixed by the exact quartic solution, one finds
\begin{equation}\label{E0-large-order}
    c_{0,n}\sim-\frac{\pi\sqrt{2}}{g_*^n}\,\frac{1}{n^3\log^2(12en)}\left[1+\mathcal O\left(\frac{\log\log n}{\log n}\right)\right]\ .
\end{equation}
The characteristic $n^{-3}(\log n)^{-2}$ behavior is the standard planar $c=1$ asymptotics \cite{Gross:1990ub}. Here the exact quartic solution also fixes the explicit prefactor and the potential-dependent scale $12e$ entering the logarithm. The convergence to \eqref{E0-large-order} is rather slow, as expected from the logarithmically suppressed corrections.

Thus the same critical region controls two complementary properties of the singlet sector: $\mathcal E_0$ remains finite but develops the nonanalyticity responsible for the large-order weak-coupling expansion, while $\omega_F$ vanishes logarithmically, producing an increasingly dense spectrum of low-energy singlet excitations.
\section{Matrix oscillator and operator perturbation theory}\label{app:operator-formalism}
This appendix collects the oscillator description of the free matrix model and the weak-coupling perturbation theory used in the main text. In oscillator language, singlet excitations are naturally represented by closed chains of creation operators, whereas non-singlet excitations are represented by open chains carrying uncontracted color indices. The same language provides a direct operator derivation of the first perturbative corrections to the singlet, adjoint, and biadjoint gaps.
\subsection{Canonical oscillator variables and closed/open chains}
For the operator calculation it is convenient to pass to canonically normalized matrices
\begin{equation}
    Q=\sqrt{N}\,X\ , \qquad \Pi=\sqrt{N}\,P\ , \qquad [\Pi_{ij},Q_{kl}]=-i\delta_{il}\delta_{jk}
\end{equation}
so that the quartic Hamiltonian becomes
\begin{equation}\label{canonical-quartic-H}
    H=\frac{1}{2}\Tr\left(\Pi^2+Q^2\right)+\frac{g}{N}\Tr Q^4\ .
\end{equation}
We introduce
\begin{equation}\label{creatop}
    Q_{ij}=\frac{1}{\sqrt{2}}\left(a_{ij}+a^\dagger_{ji}\right)\ ,\qquad \Pi_{ij}=-\frac{i}{\sqrt{2}}\left(a_{ij}-a^\dagger_{ji}\right)\ , \qquad [a_{ij},a^\dagger_{kl}]=\delta_{ik}\delta_{jl}\ .
\end{equation}
The free Hamiltonian is therefore
\begin{equation}\label{H0-oscillator}
    H_0=\frac{N^2}{2}+a^\dagger_{ij}a_{ij}\ ,
\end{equation}
and the Fock vacuum $\ket{0}$ is a $U(N)$ singlet.

The color structure of the Fock space is particularly transparent in this basis. Closed chains,
\begin{equation}\label{closed-chain}
    \mathcal O^{\rm{closed}}_n=\Tr\left[(a^\dagger)^n\right]\ ,
\end{equation}
and products of such operators create singlet states. Open chains,
\begin{equation}\label{open-chain}
    \mathcal O^{\rm{open}}_n(M)=M_{ji}(a^\dagger{}^n)_{ij}\ , \qquad \Tr M=0\ ,
\end{equation}
transform in the adjoint. More generally, several open chains can be projected with Young symmetrizers on their upper and lower endpoints; after removing all upper-lower traces, this produces the higher neutral representations discussed in Section \ref{sec:Non-singlets} and Appendix \ref{app:representation-characters}. This is the oscillator counterpart of the ``closed-string/open-string'' organization of the singlet and non-singlet sectors.

At the first oscillator level one has
\begin{equation}\label{level-one-states}
    \ket{S_1}=\frac{1}{\sqrt N}\Tr a^\dagger\ket{0}\ , \qquad \ket{A_1;ij}=\left(a^\dagger_{ij}-\frac{\delta_{ij}}{N}\Tr a^\dagger\right)\ket{0}\ ,
\end{equation}
which realize $N\otimes\overline N=\mathbb 1\oplus\mathrm{Adj}$. A convenient component of the adjoint is $a^\dagger_{12}\ket{0}$. 

At the next level, the natural planar oscillator representative of the second adjoint mode is the length-two open chain
\begin{equation}\label{A2-state}
    \ket{A_2;12}
    =
    \frac{1}{\sqrt N}
    a^\dagger_{1j}a^\dagger_{j2}\ket{0}\ .
\end{equation}
At finite $N$ this state is degenerate at $g=0$ with $a^\dagger_{12}\Tr a^\dagger\ket{0}$, whose planar interpretation is the lowest adjoint excitation dressed by a singlet excitation. More generally, the planar oscillator representative of the $n$-th adjoint mode is the normalized open chain
\begin{equation}\label{An-state}
    \ket{A_n;12}=N^{-\frac{n-1}{2}}a^\dagger_{1i_1}a^\dagger_{i_1i_2}\ldots a^\dagger_{i_{n-1}2}\ket{0}\ .
\end{equation}
At planar order, multiplying an open-chain adjoint state by one or more closed-chain singlet excitations gives the corresponding singlet-dressed adjoint states. Consequently, the adjoint can occur with nontrivial multiplicity at a fixed oscillator level; already at level two there are two copies, represented by $(a^{\dagger\,2})_{ij}\ket{0}$ and $a^\dagger_{ij}\Tr a^\dagger\ket{0}$. A systematic construction of these multiplicity spaces in the free matrix oscillator is given in \cite{Hatsuda:2006xr}.
\subsection{Quartic interaction in oscillator form}

With the contraction $\contraction{}{Q_{ij}}{}{Q_{kl}} Q_{ij}Q_{kl}=\frac{1}{2}\delta_{il}\delta_{jk},$ Wick's theorem gives the normal-ordering identity
\begin{equation}\label{normal-order-Q4}
    \Tr Q^4=:\Tr Q^4:+2N:\Tr Q^2:+:(\Tr Q)^2:+\frac{N^3}{2}+\frac{N}{4}\ .
\end{equation}
For $N=1$ this reduces to the standard one-dimensional identity $Q^4=:Q^4:+:3Q^2:+3/4$, providing a simple check of all coefficients.

Using \eqref{normal-order-Q4}, the Hamiltonian \eqref{canonical-quartic-H} can be written as \cite{Veneziano:2006dj,DePietri:2006hi}
\begin{equation}\label{Hamiltonian-quartic-operators}
    H=\frac{N^2(1+g)}{2}+\frac{g}{4}+(1+2g)a^\dagger_{ij}a_{ij}+gV\ ,\qquad V=V_{-4}+V_{-2}+V_0+V_2+V_4\ ,
\end{equation}
where
\begin{equation}\label{Vk}
    \begin{aligned}
        V_0
        &=
        \frac{1}{N}\left(a^\dagger_{ii}a_{jj}+a^\dagger_{ij}a^\dagger_{jk}a_{il}a_{lk}+\frac{1}{2}a^\dagger_{ji}a^\dagger_{lk}a_{jk}a_{li}\right)\ ,
        \\
        V_2
        &=(V_{-2})^\dagger=a^\dagger_{ij}a^\dagger_{ji}+\frac{1}{2N}a^\dagger_{ii}a^\dagger_{jj}+\frac{1}{N}a^\dagger_{ij}a^\dagger_{jk}a^\dagger_{kl}a_{il}\ ,
        \\
        V_4
        &=(V_{-4})^\dagger=\frac{1}{4N}a^\dagger_{ij}a^\dagger_{jk}a^\dagger_{kl}a^\dagger_{li}\ .
    \end{aligned}
\end{equation}
The subscript on $V_k$ denotes the change in oscillator number. Thus $V_0$ acts within a fixed free level, whereas $V_{\pm2}$ and $V_{\pm4}$ connect states whose occupation numbers differ by two and four, respectively. Accordingly, $V_0$ controls the first-order splittings, while the number-changing terms enter through virtual intermediate states at second order.

The constant term in \eqref{Hamiltonian-quartic-operators} gives
\begin{equation}
    \frac{E_{\rm gs}(g,N)}{N^2}=\frac{1}{2}+\frac{g}{2}+\frac{g}{4N^2}+\mathcal O\left(g^2\right)\ ,
\end{equation}
which is in agreement with the weak-coupling ground-state energy \eqref{ground-state-quartic}.
\subsection{Lowest singlet and adjoint gap}
The distinction between a closed and an open chain is already visible at first order. On the one-particle singlet state in \eqref{level-one-states}, the first term in $V_0$ gives one additional unit,
\begin{equation}
    \frac{1}{N}\Tr a^\dagger\Tr a\,\ket{S_1}=\ket{S_1}\ ,
\end{equation}
whereas it annihilates the off-diagonal adjoint component $a^\dagger_{12}\ket{0}$. Hence
\begin{equation}\label{first-sing-adj-gaps}
    \omega=1+3g+\mathcal O\left(g^2\right)\ ,\qquad\Delta_1=1+2g+\mathcal O\left(g^2\right)\ .
\end{equation}
The first formula agrees with the weak-coupling singlet frequency \eqref{singlet-frequency-expansion}, while the second agrees with the lowest MO eigenvalue \eqref{adjoint-lowest-planar-gaps}.

The same counting extends to a long open chain. The number operator contributes $2ng$, while the planar term $N^{-1}a^\dagger_{ij}a^\dagger_{jk}a_{il}a_{lk}$ acts on each of the $n-1$ neighboring pairs of links. This reproduces the first two terms of the perturbative MO spectrum, $\Delta_n=n+(3n-1)g+\mathcal O\left(g^2\right)$. For a closed planar chain there is no boundary and the corresponding nearest-neighbor counting is cyclic, giving $n+3ng+\mathcal O\left(g^2\right)=n\omega(g)+\mathcal O\left(g^2\right)$.

For the finite-$N$ analysis, $\Delta_1$ is exceptional: the adjoint occurs only once at oscillator level one, so ordinary non-degenerate perturbation theory applies. Higher oscillator levels contain several copies of the adjoint, which can mix at finite $N$. We therefore restrict the explicit nonplanar calculation below to $\Delta_1$.

As a nontrivial finite-$N$ check, consider the one-link adjoint. Since $V_2$ and $V_4$ change the unperturbed oscillator energy by $2$ and $4$, respectively, the second-order correction $\sim g^2$ to the gap is the excited-state shift minus the vacuum shift,
\begin{equation}\label{Delta1-second-order-general}
    [\Delta_1]_{g^2}=-\frac{1}{2}\left\|V_2a^\dagger_{12}\ket{0}\right\|^2-\frac{1}{4}\left\|V_4a^\dagger_{12}\ket{0}\right\|^2-\left(-\frac{1}{2}\left\|V_2\ket{0}\right\|^2-\frac{1}{4}\left\|V_4\ket{0}\right\|^2\right)\ .
\end{equation}
Direct contractions give
\begin{equation}\label{Delta1-norms}
    \begin{aligned}
        \left\|V_2a^\dagger_{12}\ket{0}\right\|^2
        &=2N^2+\frac{31}{2}+\frac{9}{N^2}\ ,
        &
        \left\|V_4a^\dagger_{12}\ket{0}\right\|^2
        &=\frac{N^2}{4}+\frac{9}{4}+\frac{3}{N^2}\ ,
        \\
        \left\|V_2\ket{0}\right\|^2
        &=2N^2+\frac{5}{2}\ ,
        &
        \left\|V_4\ket{0}\right\|^2
        &=\frac{N^2}{4}+\frac{5}{4}\ .
    \end{aligned}
\end{equation}
Substitution into \eqref{Delta1-second-order-general} yields
\begin{equation}\label{delta1-quartic-perturb-2}
    \Delta_1=1+2g-\left(\frac{27}{4}+\frac{21}{4N^2}\right)g^2+\mathcal O\left(g^3\right)\ .
\end{equation}
The planar term $-27g^2/4$ agrees with the MO calculation \eqref{adjoint-lowest-planar-gaps}, while the operator derivation additionally fixes the leading $1/N^2$ correction, $-21g^2/(4N^2)$, which is the first term in \eqref{genus-corrections-Delta1}.

The operator formalism can be extended systematically to higher orders in $g$, yielding the weak-coupling expansion of the singlet frequency $\omega$ (which is actually known analytically \eqref{singlet-frequency-expansion}) and of the planar adjoint gaps $\Delta_n$ given in \eqref{adjoint-lowest-planar-gaps}. Keeping the $N$ in contractions for the non-degenerate lowest adjoint state $\Delta_1$ additionally determines the nonplanar corrections to it given in \eqref{genus-corrections-Delta1}.
\subsection{Lowest biadjoint states and the two-impurity splitting}
The first genuinely two-impurity states in the bosonic Fock space occur at oscillator level two. For distinct color labels a convenient pair of normalized states is
\begin{equation}\label{Bpm-states}
    \ket{B_\pm}=\frac{1}{\sqrt{2}}\left(a^\dagger_{12}a^\dagger_{34}\pm a^\dagger_{14}a^\dagger_{32}\right)\ket{0}\ .
\end{equation}
They are components of the self-conjugate representations $R_{++}$ and $R_{--}$ \eqref{biadj-irreps}. For distinct color labels these components are automatically traceless, and since $R_{++}$ and $R_{--}$ first appear at oscillator level two with multiplicity one, the states \eqref{Bpm-states} can be treated by ordinary non-degenerate perturbation theory.

Acting with $V_0$ gives the first-order splitting
\begin{equation}
    \Delta_\pm=2(1+2g)\pm\frac{g}{N}+\mathcal O\left(g^2\right)\ .
\end{equation}
At second order the same norm method as above gives
\begin{equation}\label{Bpm-second-order-direct}
    [\Delta_\pm]_{g^2}=-\frac{27}{2}\mp\frac{6}{N}-\frac{63}{4N^2}\ .
\end{equation}
Combining this with \eqref{delta1-quartic-perturb-2},
\begin{equation}\label{Bpm-final}
    \Delta_\pm=2\Delta_1\pm\frac{1}{N}\left(g-6g^2\right)-\frac{21}{4N^2}g^2+\mathcal O\left(g^3\right)\ .
\end{equation}
This reproduces the general biadjoint structure \eqref{lowest-biadjoint-allorders}--\eqref{lowest-biadjoint-interactions} and determines directly the first weak-coupling coefficients
\begin{equation}
    A(g)=g-6g^2+\mathcal O\left(g^3\right)\ ,\qquad B(g)=-\frac{21}{4}g^2+\mathcal O\left(g^3\right)\ .
\end{equation}
Here the $1/N^2$ term in \eqref{Bpm-final} is the residual two-impurity interaction after the adjoint finite-$N$ corrections already contained in $2\Delta_1$ have been subtracted. The result for $A=h_{11;11}$ agrees with the perturbative MO expansion \eqref{h-perturbative-1}.

Two comments are in order. First, the mixed biadjoint representations $(2,1^2)$ and $(1^2,2)$ do not occur at oscillator level two, because the bosonic creators span a symmetric tensor power; they first appear at level three.

Second, the same construction extends systematically to the lightest states with planar energy $L\Delta_1$: one projects $L$ one-link adjoints onto the desired irreducible representation and applies perturbation theory in the resulting representation block. States involving higher MO modes are more involved, because at a given free oscillator level they coexist with states obtained by dressing lower non-singlet modes with closed-chain singlet excitations, and these states can mix at finite $N$. Their analysis therefore requires the complete representation block together with a proper identification of the corresponding planar MO branches. The oscillator formalism is thus especially direct for the lowest $L\Delta_1$ states that enter the low-temperature expansion.
\section{Representation sectors and free-oscillator characters}\label{app:representation-characters}
In this appendix we collect the representation-theoretic results relevant for the decomposition of the free matrix oscillator into its singlet and non-singlet sectors. For each representation $R$, we denote by $\chi_R(U)$ its $U(N)$ character, by $d_R=\chi_R(\mathbb{1})$ its dimension, and by $P^{(N)}_R(q)=Z_R(q)/Z_{\mathrm{sing}}(q)$ the corresponding reduced thermal character, together with its stable large-$N$ limit $P_R(q)=\lim_{N\to\infty}P_R^{(N)}(q)$.
\subsection{Twisted partition function and character projection}
After subtracting the oscillator zero-point energy, let
\begin{equation}
    \mathcal N=a^\dagger_{ij}a_{ij}\ , \qquad q=e^{-\beta}\ .
\end{equation}
Under conjugation, $a^\dagger$ transforms as $a^\dagger\mapsto Ua^\dagger U^\dagger$. The twisted free partition function is therefore
\begin{equation}\label{twisted-Z-product}
    Z(U,q)=\Tr_{\mathcal H}\left(q^{\mathcal N}U\right)=\prod_{i,j=1}^{N}\frac{1}{1-qz_i z_j^{-1}}\ ,
\end{equation}
where $z_i=e^{i\theta_i}$ are the eigenvalues of $U$.  Equivalently,
\begin{equation}\label{log-partition-fn}
    \log Z(U,q)=\sum_{n=1}^{\infty}\frac{q^n}{n}t_n\bar t_n\ ,\qquad t_n=\Tr U^n\ , \qquad \bar t_n=\Tr U^{-n}\ .
\end{equation}
Expanding in irreducible characters,
\begin{equation}\label{character-expansion}
    Z(U,q)=\sum_{R\in\mathcal R_0} \chi_R(U)Z_R(q)\ ,\qquad Z_R(q)=\int\mathrm dU\,\chi_R(U)^*Z(U,q)\ .
\end{equation}
The sector multiplicity is not included in $Z_R$: the full untwisted contribution of $R$ is $d_R Z_R$.

The exact singlet result is
\begin{equation}\label{Zsing-exact}
    Z_{\mathrm{sing}}(q)=\prod_{m=1}^{N}\frac{1}{1-q^m}\ ,
\end{equation}
which is the excitation partition function of the $N$ free fermions in the singlet sector \cite{Brezin:1977sv}. The exact adjoint factor is \cite{Boulatov:1991fp,Hatsuda:2006xr}
\begin{equation}\label{Padj-finiteN}
    P_{\mathrm{Adj}}^{(N)}(q)=\frac{Z_{\mathrm{Adj}}}{Z_{\mathrm{sing}}}=\frac{q-q^N}{1-q}=\sum_{k=1}^{N-1}q^k\ .
\end{equation}
Thus the finite-$N$ adjoint tower terminates at $N-1$; its large-$N$ limit is $q/(1-q)$.

The free Fock space at oscillator level $n$ is
\begin{equation}\label{fock-level-sym}
    \mathcal H_n\cong\mathrm{Sym}^n\left(\mathrm{End}(\mathbb C^N)\right)\cong\mathrm{Sym}^n\left(\mathbb 1\oplus\mathrm{Adj}\right)\ ,\qquad\dim\mathcal H_n=\binom{N^2+n-1}{n}\ .
\end{equation}
For example,
\begin{equation}\label{H2-decomp}
    \mathcal H_2=2\,\mathbb 1\oplus2\,\mathrm{Adj}\oplus(2,2)\oplus(1^2,1^2)\ .
\end{equation}
This explains directly why the two self-conjugate biadjoints begin at $q^2$, whereas the mixed biadjoints are absent at that level and begin only at $q^3$, as follows from the characters below.
\subsection{Neutral representations from pairs of Young diagrams}
We now collect the explicit representation-theoretic formulas used in the main text. As discussed in Section \ref{sec:Non-singlets}, in the stable range the neutral irreducible representations are labeled by pairs of Young diagrams. In this appendix we denote them by
\begin{equation*}
    R=(\lambda,\mu)\ ,\qquad |\lambda|=|\mu|=L\ ,\qquad R\in\mathcal R_0\ ,
\end{equation*}
where $\lambda$ and $\mu$ correspond respectively to the diagrams $Y$ and $\overline Y$ used in Section \ref{sec:Non-singlets}. The first diagram specifies the symmetry of the upper indices and the second that of the lower indices, with all upper--lower traces removed.

Writing
\begin{equation}
    \lambda=(\lambda_1,\ldots,\lambda_r)\ ,\qquad \mu=(\mu_1,\ldots,\mu_s)\ ,\qquad r+s\leq N\ ,
\end{equation}
the corresponding $U(N)$ highest weight is
\begin{equation}
    \bm{h}=\left(\lambda_1,\ldots,\lambda_r,\underbrace{0,\ldots,0}_{N-r-s},-\mu_s,\ldots,-\mu_1\right)\ .
\end{equation}
This gives a direct starting point for the dimensions and characters needed below.

The Weyl formulas are
\begin{equation}\label{weyl-dim-char}
    \begin{aligned}
        d_{(\lambda,\mu)}=\prod_{1\leq i<j\leq N}\frac{h_i-h_j+j-i}{j-i}\ ,\qquad \chi_{(\lambda,\mu)}(U)=\frac{\det\limits_{1\leq i,j\leq N}\left(z_j^{h_i+N-i}\right)}{\det\limits_{1\leq i,j\leq N}\left(z_j^{N-i}\right)}\ .
    \end{aligned}
\end{equation}

For low levels it is more convenient to use Schur functions.  We write
\begin{equation}
    s_\lambda=s_\lambda(U)\ ,\qquad \bar s_\lambda=s_\lambda(U^\dagger)\ .
\end{equation}
The composite character can be written in terms of skew Schur functions using the universal-character formula \cite{Koike:1989}
\begin{equation}\label{composite-schur}
    \chi_{(\lambda,\mu)}(U)=\sum_{\tau}(-1)^{|\tau|}s_{\lambda/\tau}(U)s_{\mu/\tau^{\mathrm T}}(U^\dagger)\ .
\end{equation}
Here $\tau$ runs over Young diagrams such that $\tau\subseteq\lambda$ and $\tau^{\mathrm T}\subseteq\mu$, $|\tau|$ denotes the number of boxes in $\tau$, and $\tau^{\mathrm T}$ is the transposed diagram.

For the low-rank representations, the Weyl character formula reduces to the familiar Schur characters associated with the corresponding Young diagrams. Expressed in terms of the power sums $t_n=\Tr U^n=\sum_i z^n_i$, they are
\begin{equation}\label{small-schurs}
    \resizebox{\textwidth}{!}{$
    \begin{aligned}
        s_1&=\sum_i z_i=t_1\ ,
        &
        s_2&=\sum_{i\leq j}z_iz_j=\frac{t_1^2+t_2}{2}\ ,
        &
        s_{1^2}&=\sum_{i<j}z_iz_j=\frac{t_1^2-t_2}{2}\ ,
        \\
        s_3&=\sum_{i\leq j\leq k}z_iz_jz_k=\frac{t_1^3+3t_1t_2+2t_3}{6}\ ,
        &
        s_{21}&=\sum_{i\ne j}z_i^2z_j+2\sum_{i<j<k}z_iz_jz_k=\frac{t_1^3-t_3}{3}\ ,
        &
        s_{1^3}&=\sum_{i<j<k}z_iz_jz_k=\frac{t_1^3-3t_1t_2+2t_3}{6}\ .
    \end{aligned}
    $}
\end{equation}

The same Young diagrams also organize the tensor-index symmetries. At fixed tensor rank $L$, a partition $\lambda\vdash L$ labels an irreducible symmetry type under permutations of the $L$ indices. The dimension $f^\lambda$ of the corresponding irreducible representation of the permutation group $S_L$ is given by the hook-length formula
\begin{equation}\label{hook-specht-correct}
    h_{ij}=\lambda_i-j+\lambda_j^{\prime}-i+1\ ,\qquad f^\lambda=\frac{L!}{\displaystyle\prod_{(i,j)\in\lambda}h_{ij}}\ .
\end{equation}

The same Young diagram $\lambda$ labels the associated irreducible polynomial representation $S_\lambda(\mathbb C^N)$ of $U(N)$, whose dimension is
\begin{equation}\label{hook-content}
    \dim S_\lambda\left(\mathbb C^N\right)=\prod_{(i,j)\in\lambda}\frac{N+j-i}{h_{ij}}\ .
\end{equation}
The identity
\begin{equation}
    \sum_{\lambda\vdash L}(f^\lambda)^2=L!\ ,
\end{equation}
expresses the decomposition of the regular representation of $S_L$ into its irreducible components.
\subsection{Representations through the triadjoint level}
Table \ref{tab:irreps} lists the exact dimensions and group characters of the neutral representations through $L=3$, together with their stable large-$N$ thermal characters.
\begin{table}[t]
    \centering
    \small
    \begin{tabular}{@{} c c l l l @{}} %
        \toprule
        $L$ & $R=(\lambda,\mu)$ & $d_R$ & $\chi_R(U)$ & $P_R(q)$ at stable large $N$ \\
        \midrule
        $0$ & $(\varnothing,\varnothing)$ & $1$ & $1$ & $1$ \\
        \midrule
        $1$ & $(1,1)$ & $N^2-1$ & $s_1\bar s_1-1$ & $\frac{q}{1-q}$ \\
        \midrule
        $2$ & $(2,2)$ & $\frac{1}{4}N^2(N+3)(N-1)$ & $s_2\bar s_2-s_1\bar s_1$ & $\frac{q^2}{(1-q)(1-q^2)}$ \\
        $2$ & $(2,1^2)$ & $\frac{1}{4}(N^2-1)(N^2-4)$ & $s_2\bar s_{1^2}-s_1\bar s_1+1$ & $\frac{q^3}{(1-q)(1-q^2)}$ \\
        $2$ & $(1^2,2)$ & $\frac{1}{4}(N^2-1)(N^2-4)$ & $s_{1^2}\bar s_2-s_1\bar s_1+1$ & $\frac{q^3}{(1-q)(1-q^2)}$ \\
        $2$ & $(1^2,1^2)$ & $\frac{1}{4}N^2(N+1)(N-3)$ & $s_{1^2}\bar s_{1^2}-s_1\bar s_1$ & $\frac{q^2}{(1-q)(1-q^2)}$\\
        \midrule
        $3$ & $(3,3)$ & $\frac{1}{36}N^2(N^2-1)(N+5)(N+1)$ & $s_3\bar s_3-s_2\bar s_2$ & $\frac{q^3}{(1-q)(1-q^2)(1-q^3)}$ \\
        $3$ & $(3,21)$  & $\frac{1}{18}N^2(N^2-4)(N+4)(N-1)$ & $s_3\bar s_{21}-s_2\bar s_1^{\,2}+s_1\bar s_1$ & $\frac{q^4}{(1-q)^2(1-q^3)}$ \\
        $3$ & $(3,1^3)$ & $\frac{1}{36}(N^2-1)(N^2-4)(N^2-9)$ & $s_3\bar s_{1^3}-s_2\bar s_{1^2}+s_1\bar s_1-1$ & $\frac{q^6}{(1-q)(1-q^2)(1-q^3)}$ \\
        $3$ & $(21,3)$ & $\frac{1}{18}N^2(N^2-4)(N+4)(N-1)$ & $s_{21}\bar s_3-s_1^{\,2}\bar s_2+s_1\bar s_1$ & $\frac{q^4}{(1-q)^2(1-q^3)}$\\
        $3$ & $(21,21)$ & $\frac{1}{9}(N^2-1)^2(N^2-9)$ & $s_{21}\bar s_{21}-s_1^{\,2}\bar s_1^{\,2}+2s_1\bar s_1-1$ & $\frac{q^3(1+q^2)}{(1-q)^2(1-q^3)}$ \\
        $3$ & $(21,1^3)$ & $\frac{1}{18}N^2(N^2-4)(N+1)(N-4)$ & $s_{21}\bar s_{1^3}-s_1^{\,2}\bar s_{1^2}+s_1\bar s_1$ & $\frac{q^4}{(1-q)^2(1-q^3)}$\\
        $3$ & $(1^3,3)$ & $\frac{1}{36}(N^2-1)(N^2-4)(N^2-9)$ & $s_{1^3}\bar s_3-s_{1^2}\bar s_2+s_1\bar s_1-1$ & $\frac{q^6}{(1-q)(1-q^2)(1-q^3)}$ \\
        $3$ & $(1^3,21)$ & $\frac{1}{18}N^2(N^2-4)(N+1)(N-4)$ & $s_{1^3}\bar s_{21}-s_{1^2}\bar s_1^{\,2}+s_1\bar s_1$ & $\frac{q^4}{(1-q)^2(1-q^3)}$\\
        $3$ & $(1^3,1^3)$ & $\frac{1}{36}N^2(N^2-1)(N-1)(N-5)$ & $s_{1^3}\bar s_{1^3}-s_{1^2}\bar s_{1^2}$ & $\frac{q^3}{(1-q)(1-q^2)(1-q^3)}$\\
        \bottomrule
    \end{tabular}
    \vspace{0.5em}
    \caption{Properties of neutral $U(N)$ representations up to $L=3$. The first two columns specify the level $L$ and representation $R=(\lambda,\mu)$. The remaining columns provide the exact dimension $d_R$, exact character $\chi_R(U)$, and reduced thermal character $P_R(q)$ in the stable large-$N$ limit. All expressions are simultaneously valid for $N\geq 6$, with conjugate pairs sharing identical dimensions and thermal characters.}
    \label{tab:irreps}
\end{table}
\subsection{Stable large-\texorpdfstring{$N$}{} thermal characters}
At large $N$, the Haar measure and the twisted oscillator partition function combine into a quadratic action for the Fourier modes of the eigenvalue density \cite{Sundborg:1999ue,Aharony:2003sx}. Using
\begin{equation}
    [\d U]\propto\exp\left[-\sum_{n=1}^{\infty}\frac{1}{n}\left|\Tr U^n\right|^2\right]\ , \qquad Z(U,q)=\exp\left[\sum_{n=1}^{\infty}\frac{q^n}{n}\left|\Tr U^n\right|^2\right]\ ,
\end{equation}
one finds
\begin{equation}
    [\d U]\,Z(U,q)\propto\exp\left[-\sum_{n=1}^{\infty}\frac{1-q^n}{n}\left|\Tr U^n\right|^2\right]\ .
\end{equation}
Defining
\begin{equation}
    \rho_n=\frac{1}{N}\Tr U^n\ , \qquad t_n=N\rho_n\ ,
\end{equation}
the effective action becomes
\begin{equation}\label{rho-Gaussian}
    S_{\mathrm{eff}}=N^2\sum_{n=1}^{\infty}\frac{1-q^n}{n}|\rho_n|^2\ .
\end{equation}
For $0<q<1$, the quadratic coefficients are positive, and the saddle is the uniform eigenvalue distribution, $\rho_n=0$ for $n\geq1$, with Gaussian fluctuations around it.

Consequently the Fourier modes $t_n$ are independent complex Gaussian variables, with moments
\begin{equation}\label{t-Gaussian-moments}
    \braket{t_n\bar t_m}_q=\delta_{nm}\frac{n}{1-q^n}\ , \qquad \braket{t_n^r\bar t_n^s}_q=\delta_{rs}\,r!\left(\frac{n}{1-q^n}\right)^r\ .
\end{equation}

The Gaussian normalization itself gives
\begin{equation}
    Z_{\mathrm{sing}}(q)\xrightarrow[N\to\infty]{}\prod_{n=1}^{\infty}\frac{1}{1-q^n}\ .
\end{equation}

Combining it with the character projection \eqref{character-expansion}, one obtains
\begin{equation}\label{PR-Gaussian}
    P^{(N)}_R(q)\equiv\frac{Z_R(q)}{Z_{\mathrm{sing}}(q)}\xrightarrow[N\to\infty]{}P_R(q)=\braket{\chi_R(U)^*}_q\ .
\end{equation}
For example, using the character of the $(2,2)$ representation together with the Gaussian moments \eqref{t-Gaussian-moments},
\begin{equation}
    \resizebox{\textwidth}{!}{$
    \begin{aligned}
        P_{(2,2)}(q)=\frac{1}{4} \braket{(t_1^2+t_2)(\bar{t}_1^2+\bar{t}_2)}_q-\braket{|t_1|^2}_q=\frac{1}{4}\left(\left\langle|t_1|^4\right\rangle_q+\left\langle|t_2|^2\right\rangle_q\right)-\left\langle|t_1|^2\right\rangle_q=\frac{q^2}{(1-q)(1-q^2)}\ .
    \end{aligned}
    $}
\end{equation}
The remaining results follow from the same calculation and are collected in Table \ref{tab:irreps}.

Summing over all neutral irreducible representations at fixed tensor rank $L$ gives
\begin{equation}
    \sum_{\lambda,\mu\vdash L}d_{(\lambda,\mu)}P_{(\lambda,\mu)}(q)=\frac{N^{2L}}{L!}\left(\frac{q}{1-q}\right)^L\left[1+\mathcal O\left(N^{-2}\right)\right]\ .
\end{equation}
More generally, the fixed-$L$ sum contains only powers $N^{2L-2h}$, reflecting the usual topological $1/N^2$ expansion of matrix models at the level of the representation-resolved partition function. The cases $L=2$ and $L=3$ are realized explicitly by the biadjoint and triadjoint sectors in Table \ref{tab:irreps}.
\paragraph{Finite-$N$ thermal factors.}
For some of the low-rank sectors, the exact finite-$N$ thermal factors can be expressed directly through principal specializations of Schur functions. The relevant identity is
\begin{equation}
    s_\lambda(q,q^2,\ldots,q^{N-1})=q^{|\lambda|+n(\lambda)}\prod_{(i,j)\in\lambda}\frac{1-q^{N-1+j-i}}{1-q^{h_{ij}}}\ , \qquad n(\lambda)=\sum_{r\geq1}(r-1)\lambda_r\ .
\end{equation}
We record here the corresponding results at the triadjoint level, in addition to the already presented results for the adjoint \eqref{adj-dim-and-character} and biadjoint \eqref{biadj-characters-1}-\eqref{biadj-characters-2} ones. For $N\geq6$,  the reduced thermal characters of the three self-conjugate representations are
\begin{equation}\label{triadjoint-finiteN-characters-1}
\begin{aligned}
    P^{(N)}_{(3,3)}(q)
    &=\frac{q^3(1-q^{N-1})(1-q^N)(1-q^{N+1})}{(1-q)(1-q^2)(1-q^3)}\ ,
    \\
    P^{(N)}_{(21,21)}(q)
    &=\frac{q^3(1-q^{N-3})(1-q^{N-1})\left(1+q^2-(1+q)q^N\right)}{(1-q)^2(1-q^3)}\ ,
    \\
    P^{(N)}_{(1^3,1^3)}(q)
    &=\frac{q^3(1-q^{N-5})(1-q^{N-1})(1-q^N)}{(1-q)(1-q^2)(1-q^3)}\ .
\end{aligned}
\end{equation}
The remaining six sectors give
\begin{equation}\label{triadjoint-finiteN-characters-2}
\begin{aligned}
    P^{(N)}_{(3,21)}(q)=P^{(N)}_{(21,3)}(q)
    &=\frac{q^4(1-q^{N-2})(1-q^{N-1})(1-q^N)}{(1-q)^2(1-q^3)}\ ,
    \\
    P^{(N)}_{(21,1^3)}(q)=P^{(N)}_{(1^3,21)}(q)
    &=\frac{q^4(1-q^{N-4})(1-q^{N-2})(1-q^N)}{(1-q)^2(1-q^3)}\ ,
    \\
    P^{(N)}_{(3,1^3)}(q)=P^{(N)}_{(1^3,3)}(q)
    &=\frac{q^6(1-q^{N-3})(1-q^{N-2})(1-q^{N-1})}{(1-q)(1-q^2)(1-q^3)}\ .
\end{aligned}
\end{equation}
Their stable large-$N$ limits reproduce the triadjoint entries of Table \ref{tab:irreps}. In particular, the three self-conjugate sectors $(3,3)$, $(21,21)$, and $(1^3,1^3)$ begin at order $q^3$, the conjugate pairs $(3,21)$ and $(21,1^3)$ begin at order $q^4$, while $(3,1^3)$ and its conjugate first appear at order $q^6$.
\section{Thermodynamics of a toy model %
}\label{app:non-interacting-ungauged-states}
The free oscillator in Section \ref{sec:MHO-thermo} illustrates that the thermal trace is controlled both by the non-singlet energies and by the representation-theoretic multiplicities. The latter are fixed entirely by the $U(N)$ symmetry, whereas the excitation energies depend on the dynamics. In the interacting theory, the planar adjoint gaps are determined by the MO equation \eqref{MO-eq}, while the non-singlet spectrum is not yet under complete analytic control: this applies both to $1/N$ corrections of the adjoint levels themselves and, more importantly here, to the residual interactions inside general multi-adjoint ``molecules". This motivates us to consider a simpler model in which the representation-theoretic organization of the ungauged MQM is retained, but these molecular interactions are switched off. The only assumption is that the energies of multi-adjoint molecules are additive in its ``atomic" constituents $\Delta_n$.\footnote{For the considered toy model, the origin and detailed form of the atomic spectrum is, in fact, immaterial. One may regard $\Delta_n$ simply as an arbitrary discrete set of atomic energies, not necessarily identified with the MO adjoint gaps. The number of levels can be finite with allowed degeneracies.}

It is convenient to collect the atomic spectrum into the thermal sum
\begin{equation}\label{additive-z1}
    z_1(\beta)\equiv\sum_n e^{-\beta\Delta_n}\ .
\end{equation}
Here, each atomic state is included separately, so possible degeneracies are automatically counted in $z_1$. The partition function of the resulting non-interacting molecular model can then be calculated in closed form. In the stable large-$N$ limit, we find an exact expression
\begin{equation}\label{additive-partition-function}
    \boxed{\frac{1}{N^2}\log Z_{\mathrm{add}}(\beta)+\beta\mathcal E_0=\sum_{d=1}^{\infty}\frac{1}{d}\,\frac{z_1(d\beta)}{1+z_1(d\beta)}}\ .
\end{equation}
For the MQM application, $\mathcal E_0$ is the planar ground-state energy density \eqref{MQM-planar-gs-0}; more generally it is simply an additive vacuum-energy contribution. In the associated unitary matrix model, the absence of a Hagedorn instability follows from $z_1(\beta)/(1+z_1(\beta))<1$ whenever $z_1(\beta)$ is finite.

The derivation closely parallels the free-oscillator construction of Appendix \ref{app:representation-characters}. Instead of a single matrix oscillator of unit energy, consider first a set of non-interacting bosonic matrix ``letters" with energies $\epsilon_r$, each transforming by conjugation under $U(N)$. Their twisted Fock partition function is a generalization of \eqref{twisted-Z-product}
\begin{equation}\label{additive-generalized-twisted-product}
    Z_f(U,\beta)=e^{-\beta N^2\mathcal{E}_0}\prod_{r}\prod_{i,j=1}^{N}\frac{1}{1-e^{-\beta\epsilon_r}z_i z_j^{-1}}\ ,
\end{equation}
where we explicitly restored the ground state.

Taking the logarithm exactly as in \eqref{log-partition-fn} gives
\begin{equation}\label{additive-generalized-twisted}
    \log Z_f(U,\beta)+\beta N^2\mathcal{E}_0=\sum_{d=1}^{\infty}\frac{f(d\beta)}{d}\, t_d\bar t_d\ ,\qquad  f(\beta)\equiv\sum_r e^{-\beta\epsilon_r}\ .
\end{equation}
Thus $q^d$ in the free oscillator is replaced by $f(d\beta)$ simply because a cycle of length $d$ carries the Boltzmann weight $e^{-d\beta\epsilon_r}$. The same single-particle cycle structure appears in the unitary-matrix description of free large-$N$ gauge theories \cite{Sundborg:1999ue,Aharony:2003sx}.

For a generic atomic spectrum, the function $f(\beta)$ need not admit a realization in terms of physical letters with positive integer multiplicities. In that case it should be understood as a formal or ``virtual" alphabet. The character construction only requires the quantities $f(d\beta)$ appearing in \eqref{additive-generalized-twisted}.

To fix the letter weight $f$, we project \eqref{additive-generalized-twisted} onto the adjoint sector. As reviewed in Appendix \ref{app:representation-characters}, in the stable large-$N$ limit the Haar measure makes the modes $t_d=\Tr U^d$ independent Gaussian variables. With the insertion \eqref{additive-generalized-twisted}, their covariance becomes \eqref{t-Gaussian-moments}
\begin{equation}
    \braket{t_d\bar t_d}_f=\frac{\int \d U\,t_d\bar t_d\,Z_f(U,\beta)}{\int \d U\,Z_f(U,\beta)}=\frac{d}{1-f(d\beta)}\ .
\end{equation}
In particular, using $\chi_{\rm adj}(U)=t_1\bar t_1-1,$ the reduced adjoint partition function is
\begin{equation}\label{additive-adjoint-character}
    P_{\rm adj}^{(f)}(\beta)\equiv\frac{Z_{\rm adj}^{(f)}(\beta)}{Z_{\rm sing}^{(f)}(\beta)}=\frac{f(\beta)}{1-f(\beta)}\ .
\end{equation}
The atomic spectrum of the additive model is prescribed by \eqref{additive-z1}, this fixes the effective one-letter weight uniquely,
\begin{equation}\label{additive-f}
    P_{\rm adj}^{(f)}(\beta)=z_1(\beta) \qquad\Longrightarrow\qquad f(\beta)=\frac{z_1(\beta)}{1+z_1(\beta)}\ .
\end{equation}

Finally, setting $U=\mathbb 1$ in \eqref{additive-generalized-twisted}, so that $t_d=\bar t_d=N$, and using \eqref{additive-f} immediately reproduces \eqref{additive-partition-function}, completing the proof.

The partition function in \eqref{additive-partition-function} has a natural plethystic form,
\begin{equation}\label{additive-plethystic}
    Z_{\rm add}(\beta)=e^{-\beta N^2\mathcal E_0}\left(\mathrm{PE}[f]\right)^{N^2},\qquad\mathrm{PE}[f]\equiv\exp\left[\sum_{d=1}^{\infty}\frac{f(d\beta)}{d}\right].
\end{equation}
Related Fock-space and cycle constructions appear in the second-quantized description of symmetric products \cite{Dijkgraaf:1996xw}. Similar plethystic structures also appear in representation-theoretic generating functions \cite{Getzler1998}, in the organization of knot invariants and topological-string amplitudes \cite{Labastida:2001ts}, and in the counting of single- and multi-trace operators in large-$N$ gauge theories \cite{Benvenuti:2006qr}. A closely related dual single-letter transformation appears in giant-graviton expansions \cite{Murthy:2022ien,Budzik:2026vqe}. In Murthy's convention, $\widehat f=f/(1-f)$, whose inverse reproduces \eqref{additive-f}, while the Koszul-dual convention differs by a minus sign.

The virtual-letter interpretation introduced above becomes explicit upon formally expanding
\begin{equation}
    f=z_1-z_1^2+z_1^3-\ldots\ .
\end{equation}
The alternating terms implement the inclusion--exclusion required to reconstruct the prescribed atomic spectrum after the free Fock completion; they do not represent physical states with negative multiplicity. In particular,
\begin{equation}\label{additive-product}
    \mathrm{PE}[f]=\prod_{k=1}^{\infty}\prod_{n_1,\ldots,n_k}\left[1-e^{-\beta(\Delta_{n_1}+\ldots+\Delta_{n_k})}\right]^{(-1)^k}\ .
\end{equation}
A simple check is provided by an equidistant atomic spectrum, $\Delta_n=\omega n,\,n=1,2,\ldots .$ In this case
\begin{equation}
    z_1(\beta)=\frac{e^{-\beta\omega}}{1-e^{-\beta\omega}}\ , \qquad f(\beta)=e^{-\beta\omega}\ ,\qquad \frac{1}{N^2}\log Z_{\rm add}+\beta\mathcal E_0=-\log\left(1-e^{-\beta\omega}\right)\ .
\end{equation}
Thus an equidistant tower of atomic gaps is mapped exactly to the planar thermodynamics of a matrix oscillator of frequency $\omega$. In particular, at the free MQM point $g=0$, where $\Delta_n=n$, this reproduces \eqref{free-energy-oscillator}.

The thermal energy of the model is
\begin{equation}\label{additive-energy}
    \mathcal E_{\rm add}=\mathcal E_0+\sum_{d=1}^{\infty}\frac{1}{[1+z_1(d\beta)]^2}\sum_n\Delta_n e^{-d\beta\Delta_n}\ .
\end{equation}
At low temperature,
\begin{equation}\label{additive-low-temperature-logZ}
    \frac{1}{N^2}\log Z_{\rm add}+\beta\mathcal E_0=e^{-\beta\Delta_1}+e^{-\beta\Delta_2}-\frac12e^{-2\beta\Delta_1}+\mathcal O\left(e^{-\beta\Delta_3},e^{-\beta(\Delta_1+\Delta_2)},e^{-3\beta\Delta_1}\right).
\end{equation}
For the quartic MQM the coefficient of $e^{-2\beta\Delta_1}$ in the full theory, is instead (see \eqref{lowT})
\begin{equation}\label{additive-full-comparison}
    -\frac12-\beta A-\frac{\beta}{2}B+\frac{\beta^2A^2}{4}+\ldots\ .
\end{equation}
The additional terms originate from the residual interaction of the lowest two-adjoint molecule. Since $A=h_{11;11}=g+\mathcal O\left(g^2\right)$, the first difference from the additive model near the free point appears at order $\beta ge^{-2\beta}$. For fixed atomic gaps, \eqref{additive-partition-function} is convergent for every finite $\beta>0$, so any finite-temperature instability of the full MQM must involve physics absent from this fixed-spectrum description. Near $g_*$ the background dependence of the adjoint gaps already provides the leading backreaction mechanism from Section \ref{sec:backreaction}, while away from the extreme low-temperature regime the omitted molecular interactions are expected to become increasingly important for the full planar thermodynamics.

It is also useful to examine the opposite high-temperature limit $\beta\to0$. We now specialize to an infinite tower of positive atomic energies with asymptotic density  
\begin{equation}
    \rho_{\rm at}(\Delta)\sim C\Delta^{p-1}\ , \qquad p>0\ , \qquad \Delta\to\infty\ .
\end{equation}
Then
\begin{equation}
    z_1(\beta)\sim C\,\Gamma(p)\beta^{-p}\ ,\qquad f(\beta)=1-\frac{\beta^p}{C\,\Gamma(p)}+\ldots\ , \qquad \beta\to 0\ .
\end{equation}
The WKB spectrum $\Delta_n\sim\omega n$ corresponds to $p=1$ and $C=1/\omega$, while the Regge spectrum $\Delta_n\sim\sqrt{n/\alpha'}$ corresponds to $p=2$ and $C=2\alpha'$.

The high-temperature behavior can be extracted from a Mellin representation
\begin{equation}
    \sum_{d=1}^{\infty}\frac{f(d\beta)}{d}=\frac{1}{2\pi i}\int_{c-i\infty}^{c+i\infty}\limits\d s\, \mathcal M_f(s)\,\zeta(s+1)\,\beta^{-s}\ ,\qquad \mathcal M_f(s)\equiv \int_0^\infty\limits \d u\,u^{s-1}f(u)\ ,\qquad c>0\ .
\end{equation}
Near $s=0$ it is convenient to separate
\begin{equation}
    \mathcal M_f(s)=\Gamma(s)+\int_0^\infty\limits\d u\,u^{s-1}\bigl[f(u)-e^{-u}\bigr]\ .
\end{equation}
The double pole at $s=0$ produces the universal $\log(1/\beta)$ term and the spectrum-dependent constant $\gamma_\Delta$, while the small-$u$ behavior $f(u)=1-[C\,\Gamma(p)]^{-1}u^p+\ldots$ gives a pole at $s=-p$. Consequently,
\begin{equation}\label{additive-high-temperature}
    \frac{1}{N^2}\log Z_{\rm add}+\beta\mathcal E_0=\log\frac{1}{\beta}+\gamma_\Delta -\frac{\zeta(1-p)}{C\,\Gamma(p)}\,\beta^p+\ldots\  ,\qquad \gamma_\Delta\equiv\int_0^\infty\limits\frac{\d u}{u}\left[\frac{z_1(u)}{1+z_1(u)}-e^{-u}\right]\ .
\end{equation}
The logarithm is universal for such an infinite tower, whereas $\gamma_\Delta$ depends on its complete spectrum. The first power correction is fixed by the asymptotic density. For the WKB tower it is $\omega\beta/2$, while for the Regge tower it is $\beta^2/(24\alpha')$.

Finally, the toy model clarifies a limitation of the thermal counting used in \cite{Cho:2024kxn,Adams:2025nww}. Treating the adjoint excitations as independent matrix oscillators gives
\begin{equation}
    \frac{1}{N^2}\log Z_{\rm Fock}+\beta\mathcal E_0=-\sum_{n=1}^{\infty}\log{\left(1-e^{-\beta\Delta_n}\right)}=\sum_{d=1}^{\infty}\frac{z_1(d\beta)}{d}\ ,
\end{equation}
instead of \eqref{additive-partition-function}. The difference is already present at $g=0$: the former does not reproduce the exact matrix-oscillator partition function \eqref{partition-func-MHO}, with the mismatch beginning at the two-adjoint level. The failure lies in the unrestricted adjoint Fock-space counting. A complete thermal description must instead respect the $U(N)$ representation structure and include the omitted multi-adjoint interactions.
\section{Backreaction of the eigenvalue density}\label{app:backreaction}
In Section \ref{sec:backreaction}, the leading low-temperature backreaction of the non-singlet sector was obtained by varying the collective-field free energy. Here we derive the same effective potential directly from the response of the zero-temperature singlet density to a general deformation of the one-particle potential. This provides an independent check of the construction and relates it directly to the low-temperature one-point functions of Section \ref{sec:low-T-moments}.

Consider ungauged MQM with a general potential $V(x)$. The thermal partition function is
\begin{equation*}
    Z_T[V]=\int_{X(\beta)=X(0)}\mathcal D X\, \exp\left[-N\int_0^\beta\limits\d\tau\, \Tr\left(\frac{1}{2}\dot X^2+V(X)\right)\right]\ .
\end{equation*}

At low temperature, its planar free energy has the spectral expansion
\begin{equation*}
    \mathcal{F}_T[V]\equiv-\frac{T}{N^2}\log Z_T[V]=\mathcal{E}_0[V]-T\sum_ne^{-\beta\Delta_n[V]}+\ldots\ ,
\end{equation*}
where $\mathcal{E}_0[V]$ is the planar singlet ground-state energy density and $\Delta_n[V]$ are the adjoint gaps in the background defined by $V$.

Differentiating with respect to the potential gives the thermal eigenvalue density,
\begin{equation}
    \rho_T[V](x)=\braket{\tr\delta(x-X)}_\beta=\frac{\delta\mathcal{F}_T[V]}{\delta V}\qquad \text{with}\qquad \rho_0[V](x)=\frac{\delta\mathcal E_0[V]}{\delta V(x)}\ ,
\end{equation}
and therefore
\begin{equation}\label{thermal-density-correction}
    \rho_T[V](x)-\rho_0[V](x)=\sum_ne^{-\beta\Delta_n[V]}\frac{\delta\Delta_n[V]}{\delta V(x)}+\ldots\ .
\end{equation}

At planar order, the dependence of the adjoint gaps on the potential enters through the equilibrium density $\rho_0$. Indeed, the MO equation takes the form
\begin{equation*}
    \Delta_n\Phi_n(x)=\int\d z\,\rho_0(z)\frac{\Phi_n(x)-\Phi_n(z)}{(x-z)^2}\ .
\end{equation*}
Defining
\begin{equation}\label{app-backreaction-definitions}
    y_n\equiv e^{-\beta\Delta_n}\ , \qquad Q_n(z)\equiv\frac{\delta\Delta_n}{\delta\rho_0(z)}\ , \qquad \chi_0(x,z)\equiv\frac{\delta\rho_0(x)}{\delta V(z)}=\frac{\delta^2\mathcal E_0}{\delta V(x)\delta V(z)}\ ,
\end{equation}
the chain rule gives
\begin{equation}\label{susceptibility}
    \frac{\delta\Delta_n[V]}{\delta V(x)}=\int\d z\, Q_n(z)\chi_0(z,x)\ .
\end{equation}
By its definition as a second functional derivative of $\mathcal E_0$, the susceptibility is symmetric, $\chi_0(z,x)=\chi_0(x,z)$. Hence,
\begin{equation}
    \rho_T[V](x)-\rho_0[V](x)=\int\d z\,\chi_0(x,z)\,\delta V(z)+\ldots\ ,\qquad \delta V(z)=\sum_ny_nQ_n(z)\ .
\end{equation}

This is precisely the linear response to the potential deformation $V\to V_{\rm eff}=V+\delta V$. Therefore, to leading order in the low-$T$ expansion, the thermal eigenvalue density coincides with the singlet ground-state density in the effective potential $V_{\rm eff}$,
\begin{equation}
    \rho_T(x)=\rho_0[V_{\rm eff}](x)+\dots\ .
\end{equation}
Here the effective potential is
\begin{equation}\label{drhoW}
    V_{\rm{eff}}(x,T)=V(x)+\sum_ne^{-\beta\Delta_n}\frac{\delta\Delta_n}{\delta\rho_0(x)}=V(x)+\frac{\delta}{\delta\rho(x)}\left[-T\sum_{n}e^{-\Delta_n[\rho]/T}\right]\Biggl|_{\rho=\rho_0}.
\end{equation}
This reproduces the effective potential obtained from the collective-field free energy \eqref{backreaction-free-energy}, whose saddle-point equation \eqref{backreaction-saddle} determines the equilibrium eigenvalue density. Thus the linear-response and collective-field approaches give equivalent descriptions of the leading low-temperature backreaction.

The same linear-response argument provides a useful check on the single-trace moments. Introducing sources through
\begin{equation}
    V(x)\longrightarrow V(x)+\sum_{k\geq1}J_{2k}x^{2k}\ ,
\end{equation}
we have
\begin{equation}
    \left\langle\tr X^{2k}\right\rangle_\beta=\frac{\partial\mathcal{F}_T}{\partial J_{2k}}=\int\d x\,x^{2k}\rho_T(x)\ .
\end{equation}
Using \eqref{thermal-density-correction}, the leading thermal correction becomes
\begin{equation*}
    \left\langle\tr X^{2k}\right\rangle_\beta-\left\langle\tr X^{2k}\right\rangle_{T=0}=\sum_n\frac{\partial\Delta_n}{\partial J_{2k}}e^{-\beta\Delta_n}+\ldots\ ,
\end{equation*}
in agreement with the spectral low-temperature expansion of the one-point functions derived in Section \ref{sec:low-T-moments}.
\section{High-temperature matching}\label{app:highT}
This appendix gives the hard-mode calculation underlying Section \ref{sec:highTexpansion}. We integrate out the nonzero Matsubara modes through second order in $\lambda=g/\beta$, keeping their $\beta$ dependence exact. This determines the quadratic static coupling through two loops, the leading corrections to the quartic single- and double-trace interactions, and the zero-mode-independent contribution to the free energy. The resulting reduced theory can then be compared directly with the fixed-temperature small-$g$ expansion of Section \ref{sec:small-g} before taking the high-temperature limit. The final subsection summarizes the symbolic-algebra calculation used to extend the weak-coupling expansion of $T_c(g)$ to higher orders.
\subsection{Matsubara sums}
The hard-mode diagrams appearing in the dimensional reduction can be evaluated either directly as sums over nonzero Matsubara frequencies or, equivalently, in Euclidean time using the thermal propagator with its zero-mode contribution removed. We collect here the basic sums and propagator integrals that will be used below.

The free thermal propagator \eqref{MQMvillain} may be written as 
\begin{equation*}
    G_\beta(\tau)=\frac{1}{\beta}\sum_{n\in\mathbb Z}\frac{e^{-i\omega_n\tau}}{\omega_n^2+1}=\frac{1}{2}\frac{\cosh\left(\beta/2-\tau\right)}{\sinh(\beta /2)}\ , \qquad \omega_n=\frac{2\pi n}{\beta}\ , \qquad 0\leq\tau<\beta\ .
\end{equation*}
Since the static mode is retained explicitly in the reduced theory, hard-mode diagrams involve the zero-mode-subtracted propagator $G_\beta^{\mathrm h}(\tau)=G_\beta(\tau)-1/\beta$. Introducing the dimensionless Euclidean time
$t=\tau/\beta$ and
\begin{equation}
    D_n(\beta)\equiv(2\pi n)^2+\beta^2\ ,
\end{equation}
it is convenient to remove an overall factor of $\beta$ and define the dimensionless kernel
\begin{equation}\label{hard-time-propagator}
    \mathcal G_{\mathrm h}(t;\beta)\equiv\frac{1}{\beta}G_\beta^{\mathrm h}(\beta t)=\sum_{n\neq0}\frac{e^{-2\pi int}}{D_n(\beta)}=\frac{1}{2\beta}\frac{\cosh\left[\beta\left(\frac{1}{2}-t\right)\right]}{\sinh(\beta/2)}-\frac{1}{\beta^2}\ ,\qquad 0\leq t<1\ .
\end{equation}

For diagrams involving a single independent Matsubara frequency, it is useful to introduce
\begin{equation}\label{Jk-functions}
    J_k(\beta)\equiv\sum_{n\neq0}\frac{1}{D_n(\beta)^k}\ ,\qquad k=1,2,\ldots\ .
\end{equation}
The first member of this family is simply the coincident hard propagator,
\begin{equation}
    J_1(\beta)=\mathcal{G}_{\mathrm h}(0;\beta)=\frac{1}{2\beta}\coth\frac{\beta}{2}-\frac{1}{\beta^2}\ .
\end{equation}
All higher $J_k$ follow by differentiation,
\begin{equation}\label{Jk-recursion}
    J_k(\beta)=\frac{(-1)^{k-1}}{(k-1)!}\left(\frac{\partial}{\partial\beta^2}\right)^{k-1}J_1(\beta)\ .
\end{equation}
Equivalently, their high-temperature expansion can be written in the general form
\begin{equation}\label{Jk-zeta-expansion}
    J_k(\beta)=2\sum_{\ell=0}^{\infty}(-1)^\ell\binom{k+\ell-1}{\ell}\frac{\zeta(2k+2\ell)}{(2\pi)^{2k+2\ell}}\beta^{2\ell}\ ,\qquad |\beta|<2\pi\ .
\end{equation}
In particular,
\begin{equation}
    J_2(\beta)=\int_0^1\limits\mathrm dt\,\mathcal{G}_{\mathrm h}^2(t;\beta)=\frac{1}{4\beta^4}\left[\beta\coth\frac{\beta}{2}+\frac{\beta^2}{2\sinh^2(\beta/2)}-4\right]\ ,
\end{equation}
and the first two sums have the expansions
\begin{equation}\label{Jk-highT}
    \begin{aligned}
        J_1(\beta)
        &=\frac{1}{12}-\frac{\beta^2}{720}+\frac{\beta^4}{30240}-\frac{\beta^6}{1209600}+\mathcal O\left(\beta^8\right)\ ,
        \\
        J_2(\beta)
        &=\frac{1}{720}-\frac{\beta^2}{15120}+\frac{\beta^4}{403200}-\frac{\beta^6}{11975040}+\mathcal O\left(\beta^8\right)\ .
    \end{aligned}
\end{equation}

Diagrams with several independent Matsubara frequencies involve genuine convolutions rather than the single-frequency sums $J_k$. The first one appears in the melon diagram. Defining
\begin{equation}\label{K3-function}
    K_3(\beta)\equiv\sum_{\substack{n,p\neq0\\n+p\neq0}}\frac{1}{D_n(\beta)D_p(\beta)D_{n+p}(\beta)}=\int_0^1\limits\mathrm dt\,\mathcal{G}_{\mathrm h}^3(t;\beta)\ ,
\end{equation}
where the second equality follows from Fourier orthogonality, the time integral gives
\begin{equation}
    K_3(\beta)=\frac{1}{24\beta^6}\left[48-18\beta\coth\frac{\beta}{2}-\frac{3\beta^2}{\sinh^2(\beta/2)}+2\beta^2\right]\ .
\end{equation}
Its high-temperature expansion is
\begin{equation}\label{K3-highT}
    K_3(\beta)=\frac{1}{30240}-\frac{\beta^2}{604800}+\frac{\beta^4}{15966720}-\frac{691\beta^6}{326918592000}+\mathcal O\left(\beta^8\right)\ .
\end{equation}

The connected vacuum graph in which all four hard propagators join the two quartic vertices similarly involves the four-propagator convolution
\begin{equation}\label{K4-function}
    K_4(\beta)\equiv\sum_{\substack{n_1,n_2,n_3\neq0\\n_1+n_2+n_3\neq0}}\frac{1}{D_{n_1}(\beta)D_{n_2}(\beta)D_{n_3}(\beta)D_{n_1+n_2+n_3}(\beta)}=\int_0^1\limits\mathrm dt\,\mathcal{G}_{\mathrm h}^4(t;\beta)\ .
\end{equation}
Evaluating the integral gives
\begin{equation}
\resizebox{\textwidth}{!}{$
    \begin{aligned}
        K_4(\beta)=\frac{1}{768\beta^8}\Bigg[\frac{18\beta^4}{\sinh^4(\beta/2)}+24\beta(\beta^2+48)\coth\frac{\beta}{2}+\frac{12\beta^2}{\sinh^2(\beta/2)}\left(5\beta\coth\frac{\beta}{2}-16\right)-256(\beta^2+9)\Bigg]\ ,
    \end{aligned}$}
\end{equation}
with the expansion
\begin{equation}\label{K4-highT}
    K_4(\beta)=\frac{1}{241920}-\frac{\beta^2}{2851200}+\frac{8797\beta^4}{435891456000}-\frac{631\beta^6}{653837184000}+\mathcal O\left(\beta^8\right)\ .
\end{equation}

Unlike the single-frequency sums $J_k$, the quantities $K_3$ and $K_4$ are genuine multi-frequency convolutions; in particular,
$K_3\neq J_3$ and $K_4\neq J_4$.
\subsection{Matching of the static action}
Recall that on the symmetric saddle the reduced action was written in \eqref{reduced-soft-action} as
\begin{equation*}
    \frac{S_{\mathrm{red}}}{N^2}=\frac{r}{2}\,c_2+u\,c_4+v\,c_2^2 +\ldots\ ,\qquad c_n\equiv\tr X_0^n\ ,\qquad c_{2n+1}=0\ .
\end{equation*}
Here we evaluate all contributions through two interaction vertices, i.e. through $\mathcal O\left(\lambda^2\right)$, retaining their full dependence on $\beta$. This contains somewhat more information than is required for \eqref{highT-matched-couplings}: the determination of $T_c(g)$ through order $|g|^3$ only requires the reduced action through $\mathcal O\left(\beta^4\right)$. The higher powers of $\beta$ generated by the two-vertex expressions below are therefore only partial contributions to the effective action at those orders.

The interaction between the static and hard modes is conveniently organized according to the number of external $X_0$ insertions,
\begin{equation}
    S_{\mathrm{int}}=S_{\mathrm{int}}^{(0)}+S_{\mathrm{int}}^{(1)}+S_{\mathrm{int}}^{(2)}\ ,
\end{equation}
where
\begin{equation}\label{Effective-interactions}
    \begin{aligned}
        S_{\mathrm{int}}^{(0)}
        &=N^2\lambda\int_0^1\limits\mathrm{d}t\,\tr Y^4\ ,
        \\
        S_{\mathrm{int}}^{(1)}
        &=4N^2\lambda\int_0^1\limits\mathrm{d}t\,\tr(X_0Y^3)\ ,
        \\
        S_{\mathrm{int}}^{(2)}
        &=N^2\lambda\int_0^1\limits\mathrm{d}t\,\tr\left(4X_0^2Y^2+2X_0YX_0Y\right)\ .
    \end{aligned}
\end{equation}
The term containing three static fields and one hard field vanishes because $Y$ has no zero-frequency component. The reduced action follows from \eqref{highT-effective-action-definition}
\begin{equation}\label{highT-cumulant-expansion}
    S_{\mathrm{eff}}-S_0=\left\langle S_{\mathrm{int}}\right\rangle_0-\frac{1}{2}\left\langle \left(S_{\mathrm{int}}\right)^2\right\rangle_{0,c}+\frac{1}{6}\left\langle \left(S_{\mathrm{int}}\right)^3\right\rangle_{0,c}+\ldots\ .
\end{equation}

Collecting the terms with two external static fields in the cumulant expansion gives
\begin{equation}\label{quadratic-cumulants}
    (S_{\mathrm{eff}}-S_0)\Big|_{X_0^2}=\left\langle S_{\mathrm{int}}^{(2)}\right\rangle_0-\left\langle S_{\mathrm{int}}^{(0)}S_{\mathrm{int}}^{(2)}\right\rangle_{0,c}-\frac{1}{2}\left\langle\left(S_{\mathrm{int}}^{(1)}\right)^2\right\rangle_{0,c}+\mathcal O\left(\lambda^3\right).
\end{equation}

\begin{figure}[t]
    \centering
    \begin{tikzpicture}[
        hard/.style={thick},
        zero/.style={thick,dashed},
        vertex/.style={circle,fill=black,inner sep=1.7pt},
        scale=0.75 %
    ]
        \begin{scope}[shift={(0,0)}]
            \node[vertex] (v) at (0,0) {};
            \draw[zero] (-1.4,0) -- (v);
            \draw[zero] (v) -- (1.4,0);
            \draw[hard] (0,0.62) circle (0.62);
            \node at (0,-1.05) {\small one-loop snail};
            \node at (0,-1.65) {\scriptsize $\langle S_{\rm int}^{(2)}\rangle_0$};
        \end{scope}

        \begin{scope}[shift={(4.2,0)}]
            \node[vertex] (v1) at (0,0) {};
            \node[vertex] (v2) at (0,1.1) {};
            \draw[zero] (-1.4,0) -- (v1);
            \draw[zero] (v1) -- (1.4,0);
            \draw[hard] (0,0.55) circle (0.55);
            \draw[hard] (0,1.65) circle (0.55);
            \node at (0,-1.05) {\small two-loop double-snail};
            \node at (0,-1.65) {\scriptsize $-\langle S_{\rm int}^{(0)}S_{\rm int}^{(2)}\rangle_{0,c}$};
        \end{scope}

        \begin{scope}[shift={(8.8,0)}]
            \node[vertex] (v1) at (-0.65,0) {};
            \node[vertex] (v2) at (0.65,0) {};
            \draw[zero] (-1.95,0) -- (v1);
            \draw[zero] (v2) -- (1.95,0);
            \draw[hard] (0,0) circle (0.65);
            \draw[hard] (v1) -- (v2);
            \node at (0,-1.05) {\small two-loop melon};
            \node at (0,-1.65) {\scriptsize $-\frac12\langle(S_{\rm int}^{(1)})^2\rangle_{0,c}$};
        \end{scope}

        \begin{scope}[shift={(13.4,0)}]
            \node[vertex] (v1) at (-0.65,0) {};
            \node[vertex] (v2) at (0.65,0) {};
            \draw[zero] (-1.85,0.60) -- (v1);
            \draw[zero] (-1.85,-0.60) -- (v1);
            \draw[zero] (v2) -- (1.85,0.60);
            \draw[zero] (v2) -- (1.85,-0.60);
            \draw[hard] (0,0) circle (0.65);
            \node at (0,-1.05) {\small four-point bubble};
            \node at (0,-1.65) {\scriptsize $-\frac12\langle(S_{\rm int}^{(2)})^2\rangle_{0,c}$};
        \end{scope}
    \end{tikzpicture}
    \caption{Hard-mode topologies used in the matching. Dashed lines denote external zero modes and solid lines denote nonzero Matsubara propagators. The quadratic coupling is computed through two loops. At the same order in $\lambda$, the four-point bubble renormalizes the quartic coupling and generates a double-trace interaction. Double-line color indices are suppressed.}
    \label{fig:highT-diagrams}
\end{figure}
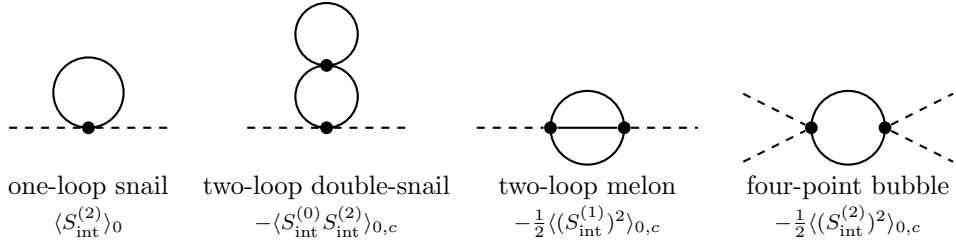

The diagrams contributing through two interaction vertices are shown in Figure \ref{fig:highT-diagrams}. The quadratic coupling receives a one-loop contribution from the snail diagram,
\begin{equation}\label{one-loop-quadratic}
    \frac{1}{N^2}\left\langle S_{\mathrm{int}}^{(2)}\right\rangle_0=4\lambda\beta^2J_1(\beta)c_2\ .
\end{equation}
At second order, the double-snail contributes
\begin{equation}\label{double-snail-result}
    -\frac{1}{N^2}\left\langle S_{\mathrm{int}}^{(0)}S_{\mathrm{int}}^{(2)}\right\rangle_{0,c}=-32\lambda^2\beta^6J_1(\beta)J_2(\beta)c_2\ ,
\end{equation}
while the melon gives
\begin{equation}\label{melon-result}
    -\frac{1}{2N^2}\left\langle\left(S_{\mathrm{int}}^{(1)}\right)^2\right\rangle_{0,c}=-8\lambda^2\beta^6 K_3(\beta)c_2\ .
\end{equation}

Combining the one- and two-vertex contributions gives the quadratic coupling
\begin{equation}\label{r-exact-highT}
    r=1+8\lambda\beta^2J_1(\beta)-16\lambda^2\beta^6\left[4J_1(\beta)J_2(\beta)+K_3(\beta)\right]+\mathcal O\left(\lambda^3\right)\ .
\end{equation}
The $\mathcal O(\lambda^2)$ contribution to $r$ starts at $\mathcal O(\beta^6)$ and hence does not enter \eqref{highT-matched-couplings}.

The first hard-mode correction to the quartic zero-mode action comes from the second cumulant,
\begin{equation}\label{quartic-cumulant}
   (S_{\mathrm{eff}}-S_0)\Big|_{X_0^4}=-\frac{1}{2}\left\langle\left(S_{\mathrm{int}}^{(2)}\right)^2\right\rangle_{0,c}+\mathcal O\left(\lambda^3\right)\ .
\end{equation}

Evaluating the planar contractions gives
\begin{equation}\label{quartic-full-matrix}
    -\frac{1}{2N^2}\left\langle\left(S_{\mathrm{int}}^{(2)}\right)^2\right\rangle_{0,c}=-\lambda^2\beta^4J_2(\beta)\left[8c_4+12(c_2)^2\right]\ .
\end{equation}
And therefore
\begin{equation}\label{quartic-matched-exact}
    u=\lambda-8\lambda^2\beta^4J_2(\beta)+\mathcal O\left(\lambda^3\right)\ , \qquad v=-12\lambda^2\beta^4J_2(\beta)+\mathcal O\left(\lambda^3\right)\ .
\end{equation}
Expanding \eqref{quartic-matched-exact} through $\mathcal O(\beta^4)$ gives the quartic and double-trace couplings quoted in \eqref{highT-matched-couplings}. The generated $v\,c_2^2$ term is a genuine planar double trace: its contraction has two color boundaries but still contributes at order $N^2$.

Although the two-vertex formulas above contain higher powers of $\beta$, they do not constitute the complete reduced theory beyond $\mathcal O(\beta^4)$. Starting at $\mathcal O(\beta^6)$, higher cumulants generate additional static operators, including sextic single- and multi-trace interactions. For the calculation of $T_c(g)$ through order $|g|^3$, only the terms through $\mathcal O(\beta^4)$ are required.
\subsection{Hard-mode contribution to the free energy}\label{app:free-energy-highT}
The zero-mode-independent dimensionless hard-mode contribution to $\beta \mathcal{F}$ is defined by
\begin{equation}
    \mathcal F_{\mathrm h}\equiv-\frac{1}{N^2}\log Z_{\mathrm h}\ ,\qquad Z_{\mathrm h}=\int\mathcal D Y\,e^{-S_{\mathrm h}-S_{\mathrm{int}}^{(0)}}\ .
\end{equation}
The Gaussian determinant is
\begin{equation}\label{F-hard-gaussian}
    \mathcal F_{\mathrm h}^{(0)}=\log
    \frac{\sinh(\beta/2)}{\beta/2}\ .
\end{equation}
The first interaction correction corresponds to the figure-eight topology of Figure \ref{fig:vacuum-diagrams-g2},
\begin{equation}
    \frac{1}{N^2}\left\langle S_{\mathrm{int}}^{(0)}\right\rangle_0=2\lambda\beta^4J_1^2(\beta)\ ,
\end{equation}
while at second order,  the double-snail and melon topologies of Figure \ref{fig:vacuum-diagrams-g2} give
\begin{equation}
    -\frac{1}{2N^2}\left\langle\left(S_{\mathrm{int}}^{(0)}\right)^2\right\rangle_{0,c}=-\lambda^2\beta^8\left[16J_1^2(\beta)J_2(\beta)+2K_4(\beta)\right]\ .
\end{equation}
Thus
\begin{equation}\label{F-hard-final}
    \boxed{\mathcal F_{\mathrm h}(\beta,\lambda)=\log\frac{\sinh(\beta/2)}{\beta/2}+2\lambda\beta^4J_1^2(\beta)-\lambda^2\beta^8\left[16J_1^2(\beta)J_2(\beta)+2K_4(\beta)\right]+\mathcal O\left(\lambda^3\right)}\ .
\end{equation}

The exact $\beta$ dependence also provides a nontrivial check of the reduced-theory construction. Combining the hard-mode contribution with the static effective action and re-expanding in the original coupling $g=\lambda\beta$ must reproduce the fixed-temperature perturbative free energy \eqref{small-g-planar-free-energy}. Free energy \eqref{highT-full-free-energy} takes the form
\begin{equation}
    \beta\mathcal F=\beta\mathcal F_0+\lambda A_1(\beta)+\lambda^2A_2(\beta)+\mathcal O\left(\lambda^3\right)\ .
\end{equation}
For this perturbative check the matrix-model free energy is expanded about the unit Gaussian saddle, $r=1$, $u=v=0$, for which $c_2^{(0)}=1$ and $c_4^{(0)}=2$; these should not be confused with the critical moments $c_{n,c}$ used in the endpoint analysis.

At first order, the static and hard-mode contributions combine to
\begin{equation}
    A_1(\beta)=2+4\beta^2J_1(\beta)+2\beta^4J_1^2(\beta)=\frac{\beta^2(1+q)^2}{2(1-q)^2}=\beta^2\mathcal F_{1,0}(q)\ ,
\end{equation}
in agreement with \eqref{small-g-F-functions-1}. At second order (the common argument $\beta$ of $J_k$ and $K_k$ is suppressed),
\begin{equation}
\begin{aligned}
    A_2(\beta)
    ={}&-18-32\beta^2J_1-16\beta^4J_1^2-28\beta^4J_2-32\beta^6J_1J_2-8\beta^6K_3-16\beta^8J_1^2J_2-2\beta^8K_4=
    \\
    ={}&\frac{\beta^3}{16(1-q)^4}\left[17q^4+40q^3-40q-17-\beta\left(32q^3+76q^2+32q\right)\right]=
    \beta^3\left[\mathcal F_{2,0}(q)+\beta\mathcal F_{2,1}(q)\right]\ .
\end{aligned}
\end{equation}
This reproduces \eqref{small-g-F-functions-2} and shows that, through the order checked explicitly, the reduced-theory calculation agrees with the ordinary fixed-temperature perturbative free energy through $\mathcal O(g^2)$ without expanding in $\beta$. The high-temperature expansion used for the critical curve \eqref{Tc-high-T} is obtained only afterwards by expanding these expressions at small $\beta$.

\subsection{Higher-order expansion of the critical curve}\label{app:highT-higher-orders}
The explicit calculation above determines $T_c(g)$ through order $|g|^3$. Starting at order $|g|^5$, additional single- and multi-trace operators enter the reduced theory and must be included consistently. We extend the calculation using a symbolic-algebra implementation of the high-temperature expansion, which constructs the reduced zero-mode action to the required order and solves the symmetric one-cut criticality conditions recursively. This gives the critical-temperature series through order $|g|^{21}$ ($12$ terms),
\begin{equation}\label{highT-Tc-long-series}
    \begin{aligned}
        T_c(g)={}&\frac{m^4}{48|g|}-\frac{4|g|}{3m^2}-\frac{704|g|^3}{15m^8}-\frac{90112|g|^5}{21m^{14}}-\frac{3407872|g|^7}{7m^{20}}-\frac{357526667264|g|^9}{5775m^{26}}
        \\
        &-\frac{1475609700073472|g|^{11}}{175175m^{32}}-\frac{1050506042338705408|g|^{13}}{875875m^{38}}-\frac{5628684268881112465408|g|^{15}}{31906875m^{44}}
        \\
        &-\frac{46459760684030570333732864|g|^{17}}{1747370625m^{50}}-\frac{2224295841082284611299258138624|g|^{19}}{544597178125m^{56}}
        \\
        &-\frac{23937305390569799744098474714464256|g|^{21}}{37577205290625m^{62}}+\mathcal O\left(\frac{|g|^{23}}{m^{68}}\right).
    \end{aligned}
\end{equation}
The first three terms coincide with the explicit calculation in \eqref{Tc-high-T}, while the remaining coefficients require the higher-order multi-trace construction. All the remaining terms are expected to be negative. %

The successive coefficients indicate a characteristic coupling scale of order $|g|/m^3\simeq0.075$--$0.08$. With 12 terms, however, this should be regarded as indicative rather than as a determination of the radius of convergence. In particular, since $g_{\mathrm{BKT}}$ and $g_*$ are numerically very close, the present series does not allow us to decide which of these scales, if either, ultimately controls its continuation.
\section{Expectation values of simple observables}\label{app:simple-observables}
This appendix collects explicit results and consistency checks for several thermal observables used in the main text. We first derive the energy and simple single-trace moments from the thermal free energy, and then turn to the quadratic Casimir, for which we give a more detailed analysis in the weak-coupling, low-temperature, and high-temperature regimes. All expectation values in this Appendix are taken in the ordinary thermal ensemble
\begin{equation}
    \braket{\mathcal O}_{\beta}=\frac{1}{Z(\beta)}\Tr\left(\mathcal O e^{-\beta H}\right)\ , \qquad Z(\beta)=\Tr e^{-\beta H}\ .
\end{equation}
\subsection{Thermal energy}\label{app:thermal-energy}
The planar energy density is obtained from $\mathcal E(\beta,g)=\partial_\beta\left[\beta\mathcal F(\beta,g)\right]$ \eqref{thermal-observables-from-F}. Using $q=e^{-\beta}$ and the weak-coupling free energy \eqref{small-g-planar-free-energy}, the result through $O(g^3)$ can be organized as
\begin{equation}\label{energy-finite-temp-perturbation}
    \mathcal E(\beta,g)=\frac{1+q}{2(1-q)}+\sum_{r=1}^{3}g^r\sum_{s=0}^{r}\beta^s\,\mathcal E_{r,s}(q)+\mathcal O\left(g^4\right)\ .
\end{equation}
At first order,
\begin{equation}\label{small-g-E-functions-1}
    \mathcal E_{1,0}(q)=\frac{(1+q)^2}{2(1-q)^2}\ ,\qquad\mathcal E_{1,1}(q)=-\frac{2q(1+q)}{(1-q)^3}\ .
\end{equation}
At second order,
\begin{equation}\label{small-g-E-functions-2}
    \resizebox{\textwidth}{!}{$
    \begin{aligned}
        \mathcal E_{2,0}(q)=-\frac{(1+q)(17q^2+40q+17)}{16(1-q)^3}\ , \quad \mathcal E_{2,1}(q)=\frac{q(11q^2+19q+11)}{4(1-q)^4}\ , \quad \mathcal E_{2,2}(q)=\frac{q(1+q)(4q^2+27q+4)}{2(1-q)^5}\ .
    \end{aligned}
    $}
\end{equation}
At third order,
\begin{equation}\label{small-g-E-functions-3}
    \resizebox{\textwidth}{!}{$
    \begin{aligned}
        &\mathcal E_{3,0}(q)=\frac{(1+q)^2(75q^2+224q+75)}{16(1-q)^4}\ , &&\mathcal E_{3,1}(q)=-\frac{q(1+q)(121q^2+234q+121)}{8(1-q)^5}\ ,
        \\
        &\mathcal E_{3,2}(q)=-\frac{q(19q^4+210q^3+415q^2+210q+19)}{2(1-q)^6}\ , &&\mathcal E_{3,3}(q)=-\frac{q(1+q)(4q^4+115q^3+365q^2+115q+4)}{3(1-q)^7}\ .
    \end{aligned}
    $}
\end{equation}
The thermal energy exhibits the same organization as free energy \eqref{small-g-F-general-pattern}. One finds that at order $g^r$ the explicit dependence on $\beta$, apart from that contained in $q=e^{-\beta}$, is polynomial of degree $r$
\begin{equation}\label{small-g-E-general-pattern}
    \mathcal E_{r,s}(q)=\frac{e_{r,s}(q)}{(1-q)^{r+s+1}}\ ,\qquad q^{r+s+1}e_{r,s}(q^{-1})=e_{r,s}(q)\ ,\qquad 0\leq s\leq r\ .
\end{equation}
Thus the same reciprocity holds,
\begin{equation}\label{small-g-E-reciprocity}
    \mathcal E_{r,s}(q^{-1})=(-1)^{r+s+1}\mathcal E_{r,s}(q)\ .
\end{equation}
As for the free energy, the numerator is reciprocal with respect to the formal degree $r+s+1$; for $s=0$ this is its ordinary degree, while for $s\geq1$ its ordinary degree is $r+s$. Hence any all-orders reciprocity of the free energy is inherited automatically by the thermal energy. The $q^0$ part of \eqref{energy-finite-temp-perturbation} contains the interacting ground-state energy and reproduces \eqref{E0-small-g-from-thermal}, while its low-temperature expansion agrees with \eqref{lowT-energy} upon inserting the corresponding weak-coupling spectral data.

The high-temperature limits follow directly from the exact planar quartic zero-mode matrix model. Restoring the mass and defining $\lambda=gT/m^4$, the weak-coupling expansion of the BIPZ free energy \eqref{quartic-MM-free-energy} gives, for $|\lambda|\ll1$,
\begin{equation}\label{E-highT-small-lambda}
    \mathcal E=T\left[1-2\lambda+36\lambda^2-864\lambda^3+\mathcal O\left(\lambda^4\right)\right]+\mathcal O\left(\frac{m^2}{T}\right)\ ,
\end{equation}
in agreement with the high-temperature limit of the perturbative expansion \eqref{energy-finite-temp-perturbation}. For $g>0$, the opposite limit $\lambda\gg1$ yields
\begin{equation}\label{E-highT-large-lambda}
    \mathcal E=\frac{3}{4}T+\frac{m^2}{6\sqrt{3g}}\,T^{1/2}-\frac{m^4}{48g}+\mathcal O\left(T^{-1/2}\right)\ .
\end{equation}
Nonzero-Matsubara corrections first enter at the orders indicated in the respective remainders.
\subsection{Single-trace moments}\label{app:thermal-moments}
The single-trace moments entering the low-temperature formulas of Section \ref{sec:low-T-moments} can be evaluated explicitly on the zero-temperature quartic background. For $g<0$, the planar density \eqref{density-MQM} gives
\begin{equation}\label{planar-moments}
    \begin{aligned}
        \left\langle\mathrm{tr}\,X^{2n}\right\rangle_{\mathrm{T}=0}
        &=
        \int_{-a}^{a}\limits\d x\, x^{2n}\rho(x)=\left(\frac{k^2}{-2g(k)(1+k^2)}\right)^{n+1}\frac{B\left(n+\frac{1}{2},\frac{3}{2}\right)}{\pi\sqrt{1+k^2}}{}_2F_1\left(-\frac{1}{2},n+\frac{1}{2};n+2;k^2\right)\ .
    \end{aligned}
\end{equation}
Here $k$ is the elliptic modulus defined in \eqref{elliptic-modulus}, $B(a,b)=\Gamma(a)\Gamma(b)/\Gamma(a+b)$ is the beta function, and ${}_2F_1$ denotes the hypergeometric function. In particular, the cases $n=1,2$ provide the zero-temperature expectation values entering
\eqref{trX2-low-T} and \eqref{trX4-low-T}.

At finite temperature, the moments can instead be obtained directly from the perturbative free energy. For the quartic moment, the Hellmann--Feynman relation \eqref{thermal-observables-from-F} gives
$\langle\tr X^4\rangle_\beta=\partial_g\mathcal F(\beta,g)$. Using
\eqref{small-g-planar-free-energy}, we find
\begin{equation}\label{small-g-X4}
    \braket{\tr X^4}_\beta=\mathcal F_{1,0}+2g\left(\mathcal F_{2,0}+\beta\mathcal F_{2,1}\right)+3g^2\left(\mathcal F_{3,0}+\beta\mathcal F_{3,1}+\beta^2\mathcal F_{3,2}\right)+\mathcal O\left(g^3\right)\ ,
\end{equation}
where the functions $\mathcal F_{r,s}(q)$ are given in
\eqref{small-g-F-functions-1}-\eqref{small-g-F-functions-3}.

The quadratic moment then follows directly from the thermal virial relation \eqref{thermal-virial-relation}. In the $m=1$ normalization, $\braket{\tr X^2}_\beta=\mathcal E(\beta,g)-3g\braket{\tr X^4}_\beta.$ Equations \eqref{energy-finite-temp-perturbation} and \eqref{small-g-X4},  thus determine $\mathcal E$ through $\mathcal{O}\left(g^3\right)$ and $\braket{\tr X^4}$ through $\mathcal{O}(g^2)$, and hence $\braket{\tr X^2}$ through $\mathcal{O}\left(g^3\right)$.

At sufficiently low temperature, the spectral expansion of
Section \ref{sec:lowTexpansion} provides a complementary description.
In particular, the leading thermal corrections to these moments are
controlled by the response of the low-lying adjoint gaps to the
corresponding sources, as derived in Section \ref{sec:low-T-moments}.
The agreement of the two descriptions in their overlap gives an
independent consistency check of the thermal free-energy calculation.

At high temperature, the single-trace moments can be obtained from the dimensionally reduced free energy of Section \ref{sec:highTexpansion}. Keeping the leading correction generated by the nonzero Matsubara modes, we find
\begin{equation}\label{highT-X2-X4}
\begin{aligned}
    \frac{m^2}{T}\braket{\tr X^2}_\beta
    &=\frac{4}{3}\frac{2s+1}{(s+1)^2}+\frac{m^2}{T^2}\frac{s+5}{36(s+1)}+\mathcal O\left(\frac{m^4}{T^4}\right)\ ,
    \\
    \frac{m^4}{T^2}\braket{\tr X^4}_\beta
    &=4\,\frac{3s+1}{(s+1)^3}+\frac{m^2}{T^2}\frac{4(s+2)}{9(s+1)^2}+\mathcal O\left(\frac{m^4}{T^4}\right)\ .
\end{aligned}
\end{equation}
Here we restored the oscillator mass and denote $s\equiv\sqrt{1+48\lambda}$. The first terms in \eqref{highT-X2-X4} are the strict dimensional-reduction results obtained from the planar BIPZ saddle \cite{Brezin:1977sv}. The terms of order $m^2/T^2$ are the leading corrections from the nonzero Matsubara modes. They include the Gaussian hard-mode contribution and the one-loop renormalization of the quadratic coupling in the reduced matrix model. Higher corrections follow from the $\mathcal O\left(\beta^4\right)$ matching of Section \ref{sec:highTexpansion} and Appendix \ref{app:highT}.
\subsection{Quadratic Casimir}\label{app:casimir}
We use throughout the normalization of Section \ref{sec:Non-singlets}, in particular $C_2(\mathrm{Adj})=N$, and define the normalized thermal observable $\mathcal C_2\equiv\frac{\braket{C_2}_\beta}{N(N^2-1)}$. The purpose of this subsection is to give several explicit calculations of $\mathcal C_2$ and to show how the weak-coupling, low-temperature, and dimensionally reduced descriptions overlap.
\subsubsection{Thermal expectation value of the Casimir at \texorpdfstring{$g=0$}{g=0}}\label{App:thermal-Casimir}
We compute the thermal expectation value of the quadratic Casimir $C_2$ in the ungauged theory at $g=0$. The calculation follows directly from the character expansion \eqref{character-expansion}. Since the group Laplacian acts on an irreducible character as $\Delta_G\chi_R(U)=-C_2(R)\chi_R(U)$, the Casimir insertion is generated by $\Delta_G$. At the identity,
\begin{equation}
    \braket{C_2}_\beta=-\frac{\Delta_G Z(U,q)\big|_{U=\mathbb 1}}{Z(\mathbb 1,q)}\ .
\end{equation}

One can verify the Laplacian relation by Taylor series expanding the characters near the identity. We write $U=e^{i \epsilon^A T_A}$ with $\Tr_{\rm fund}(T_A T_B)=\frac{1}{2}\delta_{AB}$. Expanding the character to second order (the linear term vanishes since $\Tr_RT_A=0$),
\begin{equation}
    \chi_R(e^{i \epsilon^A T_A})=d_R-\frac{1}{2}\epsilon^A\epsilon^B\,\Tr_R(T_A T_B)+\ldots=d_R\left(1-\frac{C_2(R)}{2(N^2-1)}|\epsilon|^2+\ldots\right),
\end{equation}
where we used $\Tr_R(T_A T_B)=T_2(R)\,\delta_{AB}$ with $T_2(R)=C_2(R)d_R/(N^2{-}1)$. Summing $\partial_A^2$ over all $N^2{-}1$ generators confirms $\Delta_G\,\chi_R\big|_{U=\mathbb{1}}=-C_2(R)\,d_R$.
Since $\nabla\log Z\big|_{U=\mathbb{1}}=0$, this gives 
\begin{equation}\label{C2-from-laplacian}
    \braket{C_2}_\beta=-\Delta_G\log Z(U,q)\Big|_{U=\mathbb 1}\ .
\end{equation}
\paragraph{Evaluating the Laplacian.}
Using \eqref{log-partition-fn} with $t_n=\Tr U^n$, we need $\Delta_G |t_n|^2\big|_{U=\mathbb{1}}$. At $U=\mathbb{1}$:
\begin{equation}
    \partial_A\,\Tr U^n\big|_{U=\mathbb{1}} = i n\,\Tr T_A=0\ ,\qquad \partial_A^2\,\Tr U^n\big|_{U=\mathbb{1}}=-n^2\,\Tr T_A^2=-\frac{n^2}{2}\ .
\end{equation}
By the product rule (the cross term $2\,\partial_A t_n\,\partial_A \bar{t}_n$ vanishes),
\begin{equation}
    \partial_A^2\,|t_n|^2\big|_{U=\mathbb{1}} = N\left(-\frac{n^2}{2}\right)+N\left(-\frac{n^2}{2}\right)=-n^2 N\ .
\end{equation}
Summing over the $N^2-1$ generators of $SU(N)$,
\begin{equation}
    \Delta_G\,|t_n|^2\big|_{U=\mathbb{1}}=-n^2 N(N^2-1)\ .
\end{equation}
Note that the $U(\mathbb{1})$ direction does not contribute, since $|t_n|^2$ is invariant under $U\to e^{\i\alpha}U$. Therefore,
\begin{equation}
    \Delta_G\,\log Z\Big|_{U=\mathbb{1}}=\sum_{n=1}^{\infty}\frac{q^n}{n}\cdot\bigl(-n^2 N(N^2-1)\bigr)=-N(N^2-1)\sum_{n=1}^{\infty} n\,q^n=-N(N^2-1)\frac{q}{(1-q)^2}\ .
\end{equation}
The result is
\begin{equation}\label{C2-thermal}
    \braket{C_2}_{\beta}\Big|_{g=0}=N(N^2-1)\frac{q}{(1-q)^2} = \frac{N(N^2-1)}{4\sinh^2(\beta/2)}\ .
\end{equation}
The same result can be organized by irreducible representation rather than by oscillator level. Using the free theory partition function $Z_0$ \eqref{partition-func-MHO} and stable large-$N$ thermal characters $P_R(q)$ of Appendix \ref{app:representation-characters}, the first terms are
\begin{equation}\label{C2-sector-series}
    \frac{\braket{C_2}_\beta}{N}=\frac{Z_{\rm sing}}{Z_{0}}\left[(N^2-1)\frac{q}{1-q}+(N^4-N^2)\frac{q^2}{(1-q)(1-q^2)}+\mathcal{O}\left(q^3\right)\right]\ .
\end{equation}
The first term is the adjoint contribution, while the second comes from the two self-conjugate biadjoint representations $R_{++}$ and $R_{--}$. As a cross-check, using\footnote{For a neutral representation $R$ \eqref{general-irrep}, with $|Y|=|\overline{Y}|\equiv L$, the quadratic Casimir is
\begin{equation}\label{irreps-casimirs}
    C_2(R)=LN+\frac{1}{2}\left(\kappa_Y+\kappa_{\overline{Y}}\right)\ ,\qquad\kappa_Y=\sum_{i\geq1}\lambda_i(\lambda_i-2i+1)=2\sum_{(i,j)\in Y}(j-i) \ .
\end{equation}
Here $\lambda_i$ is the length of the $i$-th row of $Y$, while $(i,j)$ labels the box in row $i$ and column $j$.}
\begin{equation}\label{biadj-casimirs}
    C_2(R_{++})=2(N+1)\ , \qquad C_2(R_{--})=2(N-1)\ ,
\end{equation}
together with the dimensions \eqref{biadj-self-conj-dims}, their combined Casimir weight is in agreement with \eqref{C2-sector-series}
\begin{equation}
    \frac{1}{N}\big(C_2(R_{++})d_{++}+C_2(R_{--})d_{--}\big)=N^4-N^2\ .
\end{equation}

\paragraph{Thermal energy and $\mathcal{C}_2$ vs.\ $\mathcal{E}$.}
In the free ungauged theory,
\begin{equation}\label{E-thermal}
    E-E_0=N^2\frac{q}{1-q}=\frac{N^2}{e^\beta-1}\ .
\end{equation}
Writing the energy density as $\mathcal E\equiv E/N^2$, with the free ground-state value $\mathcal E_0=1/2$, we can eliminate $q$ between
\eqref{C2-thermal} and \eqref{E-thermal}. This gives an exact result
\begin{equation}\label{C2-vs-E}
    \mathcal C_2=\left(\mathcal E-\mathcal E_0\right)\left(\mathcal E-\mathcal E_0+1\right)=\mathcal E^2-\frac{1}{4}\ .
\end{equation}
\subsubsection{Casimir at weak coupling}\label{app:small-g-casimir}

We now develop a weak-coupling expansion of the thermal Casimir at arbitrary temperature. Since $C_2$ commutes with the Hamiltonian, its dependence on the quartic coupling satisfies
\begin{equation}
    \partial_g\braket{C_2}_\beta=-\beta\braket{C_2\,\partial_g H}_{c}\ .
\end{equation}
where the subscript $c$ denotes the connected thermal correlator. Evaluating the resulting connected thermal correlators by Wick contractions gives, through third order in $g$,
\begin{equation}\label{C2-small-g}
    \mathcal{C}_2=\frac{q}{(1-q)^2}+\sum_{r=1}^{3}(-g)^r\sum_{s=1}^{r}\beta^s\, \mathcal P_{r,s}(q,N)+\mathcal O\left(g^4\right)\ ,
\end{equation}
where first order contribution does not depend on $N$
\begin{equation}\label{C2-P-functions-1}
    \mathcal P_{1,1}(q,N)=\frac{2q(1+q)^2}{(1-q)^4}\ ,
\end{equation}
and at second order
\begin{equation}\label{C2-P-functions-2}
    \begin{aligned}
        \mathcal P_{2,1}(q,N)
        &=\frac{3q(1+q)}{4(1-q)^5}\left[9q^2+20q+9+\frac{7q^2+20q+7}{N^2}\right]\ ,
        \\
        \mathcal P_{2,2}(q,N)
        &=\frac{q}{(1-q)^6}\left[2q^4+21q^3+41q^2+21q+2+\frac{9q(q^2+3q+1)}{N^2}\right]\ .
    \end{aligned}
\end{equation}
At third order,
\begin{equation}\label{C2-P-functions-3}
    \resizebox{\textwidth}{!}{$
    \begin{aligned}
        \mathcal P_{3,1}(q,N)
        &=\frac{q(1+q)^2}{8(1-q)^6}\left[337q^2+904q+337+\frac{789q^2+3096q+789}{N^2}\right]\ ,
        \\
        \mathcal P_{3,2}(q,N)
        &=\frac{3q(1+q)}{4(1-q)^7}\left[18q^4+239q^3+522q^2+239q+18+\frac{14q^4+415q^3+1354q^2+415q+14}{N^2}\right]\ ,
        \\
        \mathcal P_{3,3}(q,N)
        &=
        \frac{q}{6(1-q)^8}\left[8q^6+281q^5+1654q^4+2792q^3+1654q^2+281q+8+\frac{273q^5+2646q^4+5160q^3+2646q^2+273q}{N^2}\right]\ .
    \end{aligned}
    $}
\end{equation}

A further structural property is the reappearance of the $q\leftrightarrow q^{-1}$ reciprocity encountered earlier in the free and thermal energies, \eqref{small-g-F-general-pattern} and
\eqref{small-g-E-general-pattern}. Namely,
\begin{equation}\label{C2-general-pattern}
    \mathcal P_{r,s}(q,N)=\frac{p_{r,s}(q,N)}{(1-q)^{r+s+2}}\ , \qquad q^{r+s+2}p_{r,s}(q^{-1},N)=p_{r,s}(q,N)\ ,
\end{equation}
one equivalently has
\begin{equation}\label{C2-reciprocity}
    \mathcal P_{r,s}(q^{-1},N)=(-1)^{r+s+2}\mathcal P_{r,s}(q,N)\ .
\end{equation}
Remarkably, the planar and $1/N^2$ parts in \eqref{C2-P-functions-1}--\eqref{C2-P-functions-3} satisfy this relation separately. Together with the corresponding structure of the free energy, this suggests that the reciprocity persists genus by genus in the full topological expansion. The thermal-energy relation then follows automatically from $\mathcal E=\partial_\beta(\beta\mathcal F)$.

The expansion \eqref{C2-small-g} is exact in $\beta$ through the displayed order in $g$. Only the planar and $1/N^2$ terms appear through the order $\mathcal{O}\left(g^3\right)$, contributions of order $1/N^4$ can first appear at $\mathcal{O}\left(g^4\right)$. Its low- and high-temperature limits agree with the independent spectral and dimensionally reduced calculations, respectively, as we show in the following subsections.
\subsubsection{Casimir at low temperature}
We now apply the low-temperature spectral expansion of Section \ref{sec:low-T-planar-thermodynamics} to the quadratic Casimir. Since $C_2$ acts as a constant $C_2(R)$ in each irreducible sector,
\begin{equation}
    \braket{C_2}_{\beta}=\frac{\displaystyle\sum_{R\in R_0}C_2(R)d_R Z_R(\beta)}{\displaystyle\sum_{R\in R_0}d_RZ_R(\beta)}\ .
\end{equation}

Keeping the same adjoint and biadjoint states as in \eqref{lowT}, and using $C_2(\mathrm{Adj})=N$ together with \eqref{biadj-casimirs}, gives
\begin{equation}\label{Casimir-lowT}
    \mathcal C_2(\beta)\equiv\frac{\braket{C_2}_\beta}{N(N^2-1)}=e^{-\beta\Delta_1}+e^{-\beta\Delta_2}+\left[1-3\beta A-\beta B+\frac{\beta^2A^2}{2}\right]e^{-2\beta\Delta_1}+\ldots\ .
\end{equation}
Here $A$ and $B$ are the residual biadjoint interaction coefficients defined in \eqref{lowest-biadjoint-allorders}--\eqref{lowest-biadjoint-interactions}, whose contributions to planar thermodynamics were discussed in \ref{sec:low-T-planar-thermodynamics}.

A useful check is provided by the fixed-temperature weak-coupling result \eqref{C2-small-g}. At the first thermal winding, only the lowest adjoint state contributes, and before taking the planar limit
\begin{equation}\label{C2-lowT-q-check}
    \left[\mathcal{C}_2\right]_{q}=e^{-\beta(\Delta_1-1)}\ .
\end{equation}
Substituting the finite-$N$ weak-coupling expansion of $\Delta_1$ \eqref{adj-gaps}-\eqref{genus-corrections-Delta1} reproduces the complete coefficient of $q$ in \eqref{C2-small-g} through $\mathcal{O}\left(g^3\right)$, including its $1/N^2$ part.

The second winding gives a more nontrivial planar check, since both the second adjoint mode and the lowest biadjoint states contribute:
\begin{equation}\label{C2-lowT-q2-check}
    \begin{aligned}
        \left[\mathcal C_2\right]_{q^2}={}&e^{-\beta(\Delta_2^{(0)}-2)}+\left(1-3\beta A-\beta B+\frac{\beta^2A^2}{2}\right)e^{-2\beta(\Delta_1^{(0)}-1)}=
        \\
        ={}&2-12\beta g+\left(\frac{111}{2}\beta+33\beta^2\right)g^2-\left(450\beta+\frac{1149}{4}\beta^2+\frac{115}{2}\beta^3\right)g^3+\mathcal{O}\left(g^4\right)\ .
    \end{aligned}
\end{equation}
This agrees exactly with the planar $q^2$ coefficient obtained by expanding \eqref{C2-small-g}. Thus the low-temperature representation expansion and the fixed-temperature weak-coupling calculation give the same result in their common regime of validity.

We do not extend \eqref{Casimir-lowT} systematically to order $1/N^2$. At this order, higher adjoint levels contain several states in the same representation, which mix at finite $N$. For example, the second adjoint mode mixes with the lowest adjoint dressed by a singlet excitation. Therefore the $1/N^2$ correction cannot be obtained from the individual gaps $\Delta_1$, $\Delta_2$, and $\Delta_\pm$ alone; one must first diagonalize the corresponding mixing problem.
\subsubsection{Casimir at high temperature}\label{sec:casimir_highT}
We now consider the high-temperature limit. The dimensional reduction developed in Section \ref{sec:highTexpansion} applies directly to the Casimir. Using the planar one-cut solution \eqref{highT-density} with the leading reduced couplings $r_{\rm eff}=1$ and $u=\lambda$, we obtain
\begin{equation}\label{C2-highT-zero-mode}
    \mathcal C_2=\left(\frac{T}{m}\right)^2 c_2^\star+\mathcal O\left(T^0\right)=\frac{4}{3}\left(\frac{T}{m}\right)^2\frac{2\sqrt{1+48\lambda}+1}{\left(\sqrt{1+48\lambda}+1\right)^2}+\mathcal O\left(T^0\right)\ , \qquad \lambda=\frac{gT}{m^4}\ .
\end{equation}
Here $c_2^\star$ is the planar zero-mode saddle moment introduced near \eqref{double-trace-mass}. Expanding \eqref{C2-highT-zero-mode} for small $\lambda$ gives
\begin{equation}\label{C2-zero-mode-small-lambda}
    \frac{m^2}{T^2}\mathcal C_2=1-8\lambda+144\lambda^2-3456\lambda^3
    +96768\lambda^4+\mathcal O\left(\lambda^5,\frac{m^2}{T^2}\right)\ .
\end{equation}

This planar result can be compared directly with the fixed-temperature weak-coupling expansion \eqref{C2-small-g}, which was obtained at finite $N$. Taking its high-temperature limit at fixed $\lambda$ gives
\begin{equation}\label{C2-small-g-highT}
    \frac{m^2}{T^2}\mathcal C_2=1-8\lambda+\left(144+\frac{96}{N^2}\right)\lambda^2-\left(3456+\frac{7488}{N^2}\right)\lambda^3+\mathcal O\left(\lambda^4,\frac{m^2}{T^2}\right)\ .
\end{equation}
The planar part of \eqref{C2-small-g-highT} agrees term by term with \eqref{C2-zero-mode-small-lambda}. The terms proportional to $1/N^2$ are the first nonplanar corrections and are not captured by the planar zero-mode density used in \eqref{C2-highT-zero-mode}; reproducing them directly within the dimensionally reduced description would require the genus-one correction to the zero-dimensional matrix integral. Through the orders displayed, the finite-$N$ correction terminates at order $1/N^2$; contributions of order $1/N^4$ can first appear at order $\lambda^4$, consistently with the finite-$N$ weak-coupling expansion \eqref{C2-small-g}.

For $g<0$, the leading static saddle terminates at $\lambda_c=-1/48$, corresponding to $T_c^{(0)}=m^4/(48|g|)$. Near this endpoint, the leading high-temperature contribution can be written as
\begin{equation}
    \mathcal C_2=\frac{4}{3}\left(\frac{T_c^{(0)}}{m}\right)^2\left[1-3\left(1-\frac{T}{T_c^{(0)}}\right)+2\left(1-\frac{T}{T_c^{(0)}}\right)^{3/2}\right]+\mathcal O\left(T^0\right)\ .
\end{equation}
Thus the leading Casimir remains finite at the endpoint, while its first nonanalytic contribution is proportional to $\left(1-T/T_c^{(0)}\right)^{3/2}$. The nonzero-Matsubara corrections shift the endpoint away from $T_c^{(0)}$, as discussed in Section \ref{sec:highTexpansion}.

It is also useful to characterize the high-temperature relation between the Casimir and the energy. For $|\lambda|\ll1$, combining \eqref{C2-zero-mode-small-lambda} with the energy expansion \eqref{E-highT-small-lambda}, and using the fixed-temperature weak-coupling result \eqref{C2-small-g} to retain the first quantum correction, gives
\begin{equation}
    \mathcal C_2=\frac{\mathcal E^2}{m^2}-\frac{1}{4}-\frac{4g}{m^6}\mathcal E^3+\mathcal O\left(\frac{g^2\mathcal E^4}{m^{10}}\right)\ , \qquad\frac{|g|T}{m^4}\ll1\ .
\end{equation}
At $g=0$, this reduces to the exact relation \eqref{C2-vs-E}. In fact, the terms displayed through first order in $g$ follow directly from the fixed-temperature weak-coupling expansion and are exact in $\beta$; the high-temperature expansion only organizes the terms omitted at $\mathcal O\left(g^2\right)$ and beyond.

For $g>0$ and $\lambda\gg1$, the Casimir itself behaves as
\begin{equation}\label{C2-highT-large-lambda}
    \mathcal C_2=\frac{2}{3\sqrt{3g}}\,T^{3/2}-\frac{m^2}{12g}\,T+\frac{m^4}{48\sqrt{3}\,g^{3/2}}\,T^{1/2}+\mathcal O\left(T^0\right)\ .
\end{equation}
Combining this with \eqref{E-highT-large-lambda} gives
\begin{equation}
    \mathcal C_2=\frac{16}{27\sqrt{g}}\,\mathcal E^{3/2}-\frac{17m^2}{81g}\, \mathcal E+\frac{103m^4}{1944g^{3/2}}\, \mathcal E^{1/2}+\mathcal O\left(\frac{m^6}{g^2}\right)\ ,\qquad \frac{gT}{m^4}\gg1\ .
\end{equation}

Thus the weak-$\lambda$ and strong-$\lambda$ (quartic-dominated) regimes exhibit the distinct asymptotic relations $\mathcal C_2\sim\mathcal E^2/m^2$ and $\mathcal C_2\sim\mathcal E^{3/2}/\sqrt g$, respectively.

\begin{figure}[t]
    \centering
\includegraphics[width=1.\linewidth]{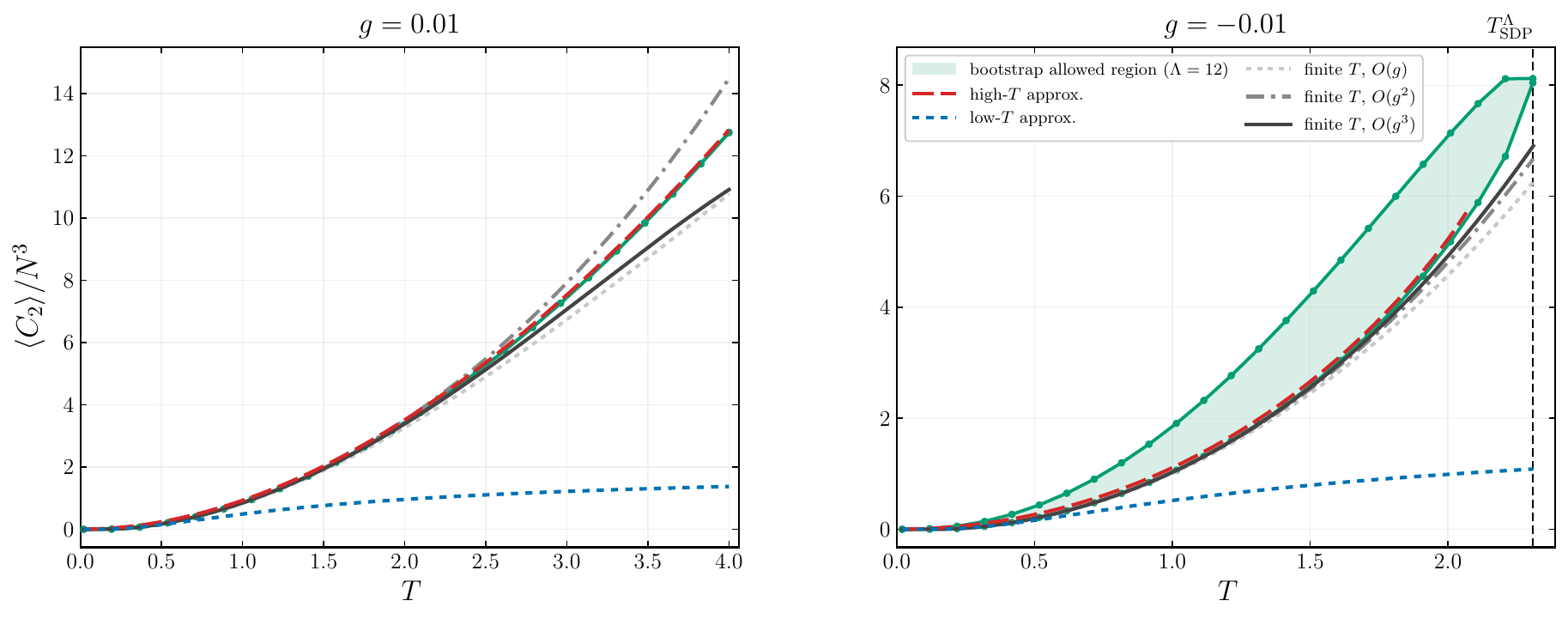}
    \caption{Casimir as a function of temperature at $g=\pm 0.01$. The green region shows the bootstrap allowed range at $\Lambda=12$ . The red curve is computed by reducing to the matrix integral as in \eqref{C2-highT-zero-mode}. The blue curve is obtained from the low-temperature expansion, including the two lowest adjoint states in \eqref{Casimir-lowT}. The gray to black curves represent the perturbative calculation at different orders of $g$ in \eqref{C2-small-g}.} 
    \label{fig:Casimir}
\end{figure}
\section{Review of the continuum gravity description}\label{app:gravity-description}
For completeness, we review here the continuum worldsheet description underlying the BKT interpretation used in Section \ref{sec:BKT}. The low-temperature branch is described by the compact $c=1$ string, namely a compact boson coupled to Liouville gravity \cite{Gross:1990ub,Moore:1992ga,Hsu:1992cm}. As the temperature is increased, winding modes become increasingly important and drive the BKT transition. On the strong-vortex side, the compact matter sector becomes massive and decouples from the long-distance worldsheet dynamics, leaving $c=0$ pure gravity.
\paragraph{c=1.} The compact worldsheet scalar satisfies
\begin{equation}
    X\sim X+2\pi R\ ,\qquad R=\frac{\beta}{2\pi}\ .
\end{equation}
The corresponding conformal-gauge action is
\begin{equation}\label{c1-gravity-action}
    \begin{aligned}
        S_{c=1}={}&\frac{1}{4\pi\alpha'}\int\d^2z\sqrt{\hat g}\,\left(\hat\nabla X\right)^2+\frac{1}{4\pi}\int\d^2z\sqrt{\hat g}\,\left[\left(\hat\nabla\phi\right)^2+2\hat{\mathcal R}\phi+4\pi\mu_1 e^{2\phi}\right]+S_{\rm{gh}}\ .
    \end{aligned}
\end{equation}
Here $\hat{g}_{ab}$ is background metric with curvature $\hat{\mathcal R}$, $\phi$ is the dimensionless Liouville field, $S_{\rm gh}$ is the action of ghosts, and $\mu_1$ is the renormalized bulk cosmological constant. For $c=1$ matter one has $Q=2$ and hence $c_L=1+6Q^2=25$, so that the total worldsheet central charge vanishes, $c+c_L+c_{\mathrm{gh}}=1+25-26=0$ \cite{David:1988hj,Distler:1988jt}. More precisely, the theory is obtained as the renormalized $b\to1$ limit of ordinary Liouville theory with $Q=b+b^{-1}$ and interaction $\mu_L e^{2b\phi}$. Since the exact Liouville amplitudes depend on the combination $\pi\mu_L\gamma(b^2)$, with $\gamma(x)=\Gamma(x)/\Gamma(1-x)$, the renormalized $c=1$ cosmological constant is defined, up to normalization, by $\mu_1=\lim_{b\to1}\pi\mu_L\gamma(b^2)$. Keeping $\mu_1$ finite and nonzero requires $\mu_L$ to diverge as $b\to1$ \cite{Dorn:1994xn,Zamolodchikov:1995aa,Balthazar:2017mxh}.\footnote{At $b=1$, the renormalized cosmological perturbation is also commonly represented by the puncture operator proportional to $\phi e^{2\phi}$, obtained from the $b\to1$ resonance limit of $e^{2b\phi}$ \cite{Polchinski:1990mf, Seiberg:1990eb}. This operator can be defined as $\frac{\partial}{\partial b} e^{2b \phi}$ evaluated at $b=1$ \cite{Duplantier:2012saa, Chatterjee:2024phq}.}

The Liouville zero mode fixes the dependence of the sphere free energy on the cosmological constant. For generic $b$, the shift
\begin{equation}
    \phi\longrightarrow\phi-\frac{1}{2b}\log\mu
\end{equation}
removes $\mu$ from the Liouville interaction and gives the scaling of the genus-zero (planar) free energy $F_0\sim\mu^{Q/b}$ \cite{Distler:1988jt}. More explicitly, integration over the Liouville zero mode produces
\begin{equation}
    F_0(\mu)\propto\Gamma\left(-\frac{Q}{b}\right)\mu^{Q/b}
\end{equation}
up to $\mu$-independent factors \cite{Goulian:1990qr}. At $c=1$, one has $b\to1$ and $Q/b\to2$, so that the Gamma function develops a pole. Indeed, writing $Q/b=2+\epsilon$, one has $\Gamma(-2-\epsilon)\mu^{2+\epsilon}\sim-\frac{\mu^2}{2\epsilon}-\frac{1}{2}\mu^2\log\mu+\ldots$; after the subtraction of the divergent analytic $\mu^2$ contribution, the universal nonanalytic part is
\begin{equation}
    F^{c=1}_{0}\sim\mu_1^2\log\mu_1\ .
\end{equation}
This logarithmic scaling violation is reproduced by the original $c=1$ double-scaling analyses \cite{Brezin:1989ss,Gross:1990ay,Ginsparg:1990as} and, for the compact theory at finite radius, in \cite{Gross:1990ub}.
\paragraph{Sine-Liouville perturbation and the BKT transition.} The winding operators are
\begin{equation}\label{winding-operators}
    \mathcal V_{\pm w}=\exp\left[\pm i\frac{wR}{\alpha'}\widetilde X\right]\ , \qquad \widetilde X=X_L-X_R\ , \qquad w=1,2,\ldots\ ,
\end{equation}
with (anti)holomorphic conformal weights
\begin{equation}
    h_{X,w}=\bar h_{X,w}=\frac{w^2R^2}{4\alpha'}\ .
\end{equation}
For closed-worldsheet vacuum amplitudes, integration over the zero mode of the dual scalar imposes winding neutrality, $\sum_i w_i=0$. For a charge-conjugation invariant deformation, the operators with opposite winding numbers have equal couplings and can be combined into a cosine. Their Liouville dressing is fixed by the David-Distler-Kawai (DDK) condition that the integrated operator has conformal weights $(1,1)$ \cite{David:1988hj,Distler:1988jt},
\begin{equation}
    e^{\left(2-wR/\sqrt{\alpha'}\right)\phi}\cos\left(\frac{wR}{\alpha'}\widetilde X\right)\ .
\end{equation}

As the temperature increases, the $w=1$ winding sector is the first to become marginal without Liouville dressing. Retaining this pair gives a Liouville-dressed sine-Gordon interaction for the dual scalar $\widetilde X$, called the unit-winding sine-Liouville perturbation of compact $c=1$ gravity \cite{Gross:1990ub,Moore:1992ga,Hsu:1992cm}
\begin{equation}
    S=S_{\rm{SL}}+\ldots\ , \qquad S_{\rm{SL}}=S_{c=1}+\zeta\int\d^2z\sqrt{\hat g}\, e^{\left(2-R/\sqrt{\alpha'}\right)\phi}\cos\left(\frac{R}{\alpha'}\widetilde X\right)\ .
\end{equation}
The ellipsis denotes additional interactions that may arise in the continuum limit of the MQM random-graph expansion. Even if only the unit-winding interaction is retained, its second-order contribution generates winding-neutral operators. In particular, near the Gaussian BKT radius $R=2\sqrt{\alpha'}$, the operator product of opposite windings induces a Liouville-dependent counterterm proportional to $\zeta^2\phi(\hat\nabla X)^2$ \cite{Schmidhuber:1993qe}. This term is required to maintain conformal invariance with respect to the fiducial worldsheet metric. Under a shift of the Liouville field corresponding to a change of physical scale, it produces a renormalization of the compact-boson kinetic term.  We parametrize the resulting running by
\begin{equation}
    S_X(\ell)=\frac{z_X(\ell)}{4\pi\alpha'}\int \d^2z\,\sqrt{\hat g}\,(\hat\nabla X)^2\ , \qquad R_{\rm eff}(\ell)=R\sqrt{z_X(\ell)}\ .
\end{equation}
Here $\ell=\log(L/a)$ is the worldsheet RG time, with $a$ the microscopic cutoff and $L$ the coarse-graining scale. Increasing $\ell$ corresponds to flowing toward the infrared, while the microscopic MQM temperature $T=1/\beta$ remains fixed.

Near the Gaussian BKT point, the leading gravitationally dressed RG equations, describing the flow with respect to the physical worldsheet scale rather than the fiducial metric, take the form \cite{Schmidhuber:1993qe}
\begin{equation}\label{BKT-RG}
\begin{aligned}
    \frac{\d\zeta}{\d\ell}
    &=\left(2-\frac{R_{\rm eff}}{\sqrt{\alpha'}}\right)\zeta+\ldots\ ,\\
    \frac{\d R_{\rm eff}}{\d\ell}
    &=-\pi^2\sqrt{\alpha'}\,\zeta^2+\ldots\ .
\end{aligned}
\end{equation}
The second equation describes the renormalization of the compact-boson kinetic term induced by winding fluctuations. To determine the critical trajectory, define
\begin{equation}
    \epsilon(\ell)\equiv\frac{R_{\rm eff}(\ell)}{\sqrt{2\alpha'}}-\sqrt{2}\ .
\end{equation}
To leading BKT order, \eqref{BKT-RG} gives
\begin{equation}
    \frac{\d\epsilon}{\d\ell}=-\frac{\pi^2}{\sqrt{2}}\zeta^2\ , \qquad \frac{\d\zeta}{\d\ell}=-\sqrt{2}\epsilon\zeta\ , \qquad \epsilon^2-\frac{\pi^2}{2}\zeta^2=\mathrm{constant}\ .
\end{equation}
In this case, “constant” means that it is fixed along the selected RG trajectory. For positive real $\zeta$, the critical separatrix approaching the marginal Gaussian fixed point satisfies $\epsilon=\pi\zeta/\sqrt{2}$. Along this trajectory,
\begin{equation}
    \zeta(\ell)=\frac{\zeta_{\rm UV}}{1+\pi\zeta_{\rm UV}\ell}\ , \qquad R_{\rm eff}(\ell)=2\sqrt{\alpha'}\left(1+\frac{\pi}{2}\zeta(\ell)+\ldots\right)\ .
\end{equation}
Thus $\zeta_{\rm IR}=0$ and $h_{X,1}^{\rm IR}=1$, with the winding coupling vanishing logarithmically slowly.

At the continuum matching scale, the separatrix condition becomes
\begin{equation}
    R\sqrt{z_X^{\rm UV}}=2\sqrt{\alpha'}\left(1+\frac{\pi}{2}\zeta_{\rm UV}+o(\zeta_{\rm UV})\right)\ ,
\end{equation}
which gives \eqref{BKT-shift} upon using $R=1/\left(2\pi T_{\rm BKT}\right)$. Determining the initial continuum couplings $\zeta_{\rm UV}$ and $z_X^{\rm UV}$ from the microscopic ungauged MQM remains an open problem.
\paragraph{c=0.} On the strong-vortex side of the separatrix, the compact matter field becomes massive and decouples from the long-distance worldsheet dynamics. The infrared theory is therefore pure two-dimensional gravity, i.e. the $c=0$ universality class \cite{Hsu:1992cm}. In conformal gauge it is described by
\begin{equation}
    S_{c=0}=\frac{1}{4\pi}\int \d^2z\sqrt{\hat g}\,\left[\left(\hat\nabla\phi_0\right)^2+Q_0\hat{\mathcal R}\phi_0+4\pi\mu_0 e^{2b_0\phi_0}\right]+S_{\mathrm{gh}}\ ,
\end{equation}
where
\begin{equation}
    b_0=\sqrt{\frac{2}{3}}\ ,\qquad Q_0=b_0+b_0^{-1}=\frac{5}{\sqrt6}\ .
\end{equation}
Indeed, $c_L=1+6Q_0^2=26$, so that the total worldsheet central charge again vanishes, $c_L+c_{\mathrm{gh}}=0$, in the absence of a gapless matter sector \cite{David:1988hj,Distler:1988jt}. Applying the same Liouville zero-mode analysis to the sphere free energy gives
\begin{equation}
    F^{c=0}_0(\mu_0)\propto\Gamma\left(-\frac{Q_0}{b_0}\right)\mu_0^{Q_0/b_0}=\Gamma\left(-\frac{5}{2}\right)\mu_0^{5/2}\ .
\end{equation}
In contrast to the $c=1$ case, $\Gamma(-5/2)$ is finite, so there is no resonance and hence no additional logarithmic scaling violation. Equivalently, the pure-gravity string susceptibility exponent is $\gamma_{\mathrm{str}}=-1/2$, so that $F_0\sim\mu_0^{2-\gamma_{\mathrm{str}}}$ \cite{Knizhnik:1988ak}. The same $5/2$ singularity arises in the double-scaling limit of the one-matrix model describing $c=0$ gravity \cite{Brezin:1990rb,Douglas:1989ve,Gross:1989vs}, and it is precisely the critical behavior found from the dimensionally reduced matrix integral in the high-temperature regime in Section \ref{sec:highTexpansion}.

The continuum description therefore provides the expected universality-class picture along the critical curve: the low-temperature branch is governed by compact $c=1$ matter coupled to Liouville gravity, the change of universality class is associated with the unit-winding BKT flow, and the strong-vortex branch flows to $c=0$ pure gravity.

\section{Bootstrap details}\label{app:bootstrap-details}
We give explicit examples of the multi-adjoint positivity constraints for
small $k$, then describe the choice of operators and the implementation of
nonlinear relaxation at finite bootstrap level.
The numerical tables comparing bootstrap results with the low- and
high-temperature approximations are collected in the final subsection.

\subsection{Small \texorpdfstring{$k$}{} examples}
\label{app:bootstrap-small-k}
We illustrate the general multi-adjoint positivity conditions described in Section \ref{sec:multi-adjoint-positivity} for small $k$. We also discuss the implementation of non-linear relaxation for these values of $k$.
For $k=0$, the vectors are singlet operators $\tr O_i$. Their planar Gram matrix is
\begin{equation}\label{singletPos}
    \mathcal M^s_{ij}=\evs{\tr O_i^\dagger}\evs{\tr O_j}\ ,\qquad \mathcal M^s\succeq0\ .
\end{equation}
Here large-$N$ factorization has been used to write the expectation value of the product as a product of expectation values. We include the identity among the singlet operators. Although this matrix is an outer product due to large $N$ factorization, its positivity is a nontrivial condition after nonlinear relaxation, as discussed below.

For $k=1$, the vectors are the traceless parts of the open words $O_i$, whose overlaps give the adjoint Gram matrix
\begin{equation}\label{adPos}
    \mathcal M^{\mathrm{ad}}_{ij}
    =\evs{\tr O_i^\dagger O_j}
     -\evs{\tr O_i^\dagger}\evs{\tr O_j}\ ,
    \qquad \mathcal M^{\mathrm{ad}}\succeq0\ .
\end{equation}
The subtraction separates adjoint positivity from positivity of the ordinary single-trace Gram matrix, which contains both singlet and adjoint contributions.

For $k=2$, the vectors are symmetric or antisymmetric tensor products of two traceless open words. Let $\mathbf V=(V_1,V_2)$ and $\mathbf W=(W_1,W_2)$ denote two such pairs. Up to a common positive normalization, the Gram-matrix entries are
\begin{equation}\label{app:bootstrap-biadjoint-entries}
    \mathcal M_\pm(\mathbf V,\mathbf W)=\evs{\tr V_1^\dagger W_1}\evs{\tr V_2^\dagger W_2}\pm\evs{\tr V_1^\dagger W_2}\evs{\tr V_2^\dagger W_1}\ .
\end{equation}
The plus sign gives the symmetric block and the minus sign gives the antisymmetric block. In particular, a diagonal entry obeys
\begin{equation}\label{Mpm}
    \evs{\tr V_1^\dagger V_1}\evs{\tr V_2^\dagger V_2}\pm\evs{\tr V_1^\dagger V_2}\evs{\tr V_2^\dagger V_1}\geq0\ .
\end{equation}
For any chosen collection of pairs, we form a Gram matrix from their symmetric or antisymmetric projections, retaining all mutual overlaps \eqref{app:bootstrap-biadjoint-entries}. Checking only the diagonal inequalities \eqref{Mpm} would discard some of these constraints. The same construction applies for larger $k$, with one Gram matrix for each irreducible representation of $S_k$.

The two Young diagrams used in Section \ref{sec:Non-singlets} describe the fundamental and antifundamental indices separately, while the planar Gram matrices are classified by their common permutation group $S_k$. Decomposing the tensor product of the two representations under this group gives the positivity blocks.  For $k=2$, $R_{++}$ and $R_{--}$ give the symmetric block, while $R_{+-}$ and $R_{-+}$ give the antisymmetric block. %

For $k \ge 3$, a single pair of Young diagrams can contribute to several positivity blocks. For example, suppose both diagrams have shape $(2,1)$. Each carries the two-dimensional irreducible representation $V_{(2,1)}$ of $S_3$. Under simultaneous permutations of their fundamental and anti-fundamental indices, the tensor product decomposes 
\begin{equation}
    V_{(2,1)}\otimes V_{(2,1)}
    =
    V_{(3)}\oplus V_{(1,1,1)}\oplus V_{(2,1)}\ .
\end{equation}
The three terms are respectively the symmetric, antisymmetric, and mixed-symmetry representations. Thus the pair $((2,1),(2,1))$ contributes to all three positivity blocks. 
At $k=2$, the tensor product associated with each fixed pair of Young diagrams is one-dimensional and therefore irreducible under the common $S_2$. Each pair consequently contributes to a single permutation-symmetry sector.  (Of course, each permutation-symmetry sector may contain multiple operators and could be further decomposed using other symmetries of the model, such as the $\mathbb{Z}_2$ symmetry.) %

As we described in Section \ref{sec:bootstrap:nlr}, we then deal with expectation values of multi-traces by performing non-linear relaxation. For example, the $k=0,1,2$ positivity conditions \eqref{singletPos}, \eqref{adPos} and \eqref{Mpm} give
\begin{align}
    \mathfrak{M}_{ij}^{k=0} &=\mathcal{L}(O_i^\dagger| O_j)\ , %
    \\
    \mathfrak{M}_{ij}^{k=1} &=\mathcal{L}(O_i^\dagger O_j) -\mathcal{L}(O_i^\dagger| O_j)\ , %
    \\
    \mathfrak{M}_{ij}^{k=2,\pm}
    &={}
    \mathcal{L}(O_{i,1}^\dagger O_{j,1}
                |O_{i,2}^\dagger O_{j,2})
    \pm\mathcal{L}(O_{i,1}^\dagger O_{j,2}
                  |O_{i,2}^\dagger O_{j,1})
    \nonumber\\
    &-\mathcal{L}(O_{i,1}^\dagger O_{j,1}
                  |O_{i,2}^\dagger|O_{j,2})
     -\mathcal{L}(O_{i,2}^\dagger O_{j,2}
                  |O_{i,1}^\dagger|O_{j,1})
    \nonumber\\
    &\mp\mathcal{L}(O_{i,1}^\dagger O_{j,2}
                   |O_{i,2}^\dagger|O_{j,1})
     \mp\mathcal{L}(O_{i,2}^\dagger O_{j,1}
                   |O_{i,1}^\dagger|O_{j,2})
    \nonumber\\
    &+(1\pm1)\mathcal{L}(O_{i,1}^\dagger|O_{i,2}^\dagger
                         |O_{j,1}|O_{j,2}),
\end{align}
The $k=0$ and $k=1$ constraints were used in the bootstrap of \cite{Lin:2025srf}; for $k=0$ it is important to include the identity $O_1 = 1$ as a possible singlet operator, see their equations (B6) and (B7). The $k\ge 2$ constraints are new and not implied by the lower constraints.

To illustrate the difference, let us consider applying it to our traceless words $V_1, V_2$. Adjoint positivity
\begin{equation}
    \mathfrak{M}_{k=1}  = 
    \begin{bmatrix}
        \mathcal{L} (V_1^\dagger V_1) & \mathcal{L} (V_1^\dagger V_2) \\
        \mathcal{L} (V_2^\dagger V_1)   & \mathcal{L} (V_2^\dagger V_2)
    \end{bmatrix}
    \succeq 0 \ .
\end{equation}
Positivity implies that the determinant of this matrix
\begin{equation} \label{eq:ad1}
    \mathcal{L}(V_1^\dagger V_1)\mathcal{L}(V_2^\dagger V_2) - \mathcal{L}(V_1^\dagger V_2)\mathcal{L}(V_2^\dagger V_1) \ge 0.
\end{equation}
On the other hand, the $k=2$ positivities give
\begin{align}
    \mathfrak{M}^{k=2,\pm} &=\mathcal{L}(V_1^\dagger V_1|V_2^\dagger V_2) \pm \mathcal{L}(V_1^\dagger V_2|V_2^\dagger V_1) \\
    \mathfrak{M}^{k=2,\pm} &\ge 0.
\end{align}
This is not equivalent to \eqref{eq:ad1}. %

\subsection{Choice of operators at finite level}
\label{app:irrep-first-cutoff}
For adjoint operators, we assign a level by weighting each $X$ by 1 and weighting $P$ by 2 \cite{Lin:2024vvg}. For multi-trace operators, one would ideally follow this convention as well. However, at high levels this would necessitate a significant amount of non-linear relaxation; for example at level 20 we could have an operator such as $(\tr X^2)^{10}$ and we would have to consider rather large irreps. This motivated the choice of a different weighting scheme for operators whose norms are multi-traces.

Let $\ell(S)$ denote the word weight, assigning weight one to $X$ and weight two to $P$. For a product of $t\geq1$ nontrivial normalized traces, define
\begin{equation}
    \Lambda(\tr S_1\cdots\tr S_t)=\sum_{\alpha=1}^{t}\ell(S_\alpha)+3(t-1).
\end{equation}
The choice of $3(t-1)$ was motivated by the following consideration: using the canonical commutation relations/cyclicity the number of traces increases by 1 while lowering $\Lambda(S) \to \Lambda(S)-3$ since an $X$ and a $P$ are removed. This operation preserves the overall $\Lambda$.

We select the vectors (of operators) entering the Gram matrices using the levels of their unprojected squared norms. Before subtracting traces, we consider operators  $(U_1\otimes U_2\otimes\cdots\otimes U_k)$, possibly multiplied by $s$ nontrivial normalized traces $\tr S_1 \, \tr S_2 \, \cdots \, \tr S_s$. 
The squared norm of such an operator contains $k+2s$ traces, we include it if
\begin{equation}\label{cutoff-seed-grade}
    2\sum_{a=1}^{k}\ell(U_a)+2\sum_{\alpha=1}^{s}\ell(S_\alpha)+3(k+2s-1)\leq\Lambda.
\end{equation}
This includes non-constant SU$(N)$ singlet operators by setting $k=0$. The identity operator is included separately.

We next split every open word into its traceless part and its normalized trace times the identity. Expanding the tensor product produces vectors with different numbers of adjoint factors.  In this work we only considered (ensembles of) states which are $\mathbb{Z}_2$ invariant. Hence for $\mathbb{Z}_2$ odd words we do not need to perform trace subtraction. This also implies that at planar order its singlet projection is negligible. We project the remaining adjoint tensors onto the $S_k$ representations and retain all linearly independent projected permutations. In a representation of dimension greater than one, a single ordering of a given set of words need not span all independent projected vectors.

We form each Gram matrix from vectors with the same number of adjoint factors, the same $S_k$ representation, and the same number of odd words, retaining all mutual overlaps even when the vectors originate from different initial tensors. Vectors with different numbers of odd words are orthogonal at planar order: each surviving pairing would otherwise contain an odd trace. Identity factors in the open tensor merely reproduce vectors with fewer open factors and need not be enumerated separately.

Through $\Lambda=20$, no tensor with more than four open words satisfies \eqref{cutoff-seed-grade}. At $\Lambda=22$, the only tensor with five open words is $X^{\otimes5}$ without additional traces. Its fully symmetric projection is the only nonzero one and is included together with all nonzero projections of the lower-order tensors.

There is an important subtlety that trace subtraction brings. To illustrate it let us consider $\Lambda = 10$. The vector $X\otimes X^2$ is included because its squared norm has level $6 + 3(2-1) = 9$. However, after trace subtraction, we can generate a triple trace $z$ which would be level 12: %
\begin{equation}\label{app:bootstrap-shared-variables}
    y=\mathcal L(X^2|X^4)\ ,\qquad
    z=\mathcal L(X^2|X^2|X^2)\ .
\end{equation}
At $\Lambda=10$ we keep such variables $z$ that are generated by such trace subtractions. We also derive constraints on these extra variables generated by trace subtractions by applying symmetry, conjugation, and microcanonical fixed-energy relations. Cyclicity/commutation relations and equations of motion are imposed for correlation functions with level strictly at most $\Lambda$. 

We also impose positivity relations for these extra variables generated by trace relations. For example at level 10, trace subtraction generates the adjoint operator $(\tr X^2)X$, with squared  norm $z$, and the symmetric and antisymmetric projections of $X\otimes(X^2-\mathbb 1\,\tr X^2)$, each with squared norm $(y-z)/2$ for normalized projectors. Positivity then gives
\begin{equation}\label{app:bootstrap-shared-positivity}
    z\geq0\ ,\qquad y-z\geq0.
\end{equation}

\subsection{Numerical tables}\label{app:bootstrap-tables}
\setlength{\parskip}{0.5ex}
\captionsetup{font={sf,small},skip=5pt}
In this subsection, we collect the numerical data from the canonical bootstrap underlying Figure \ref{fig:energy-compare}
and the critical-temperature analysis of Section \ref{sec:tc-extraction}.
The energy bounds are at $\Lambda=14$; the critical-temperature bounds
are located by bisection to precision $0.001$.

\medskip
\centering
\begingroup
\renewcommand{\arraystretch}{0.94}
\begin{tabular}{ccccc}
\toprule
$T$ & lower bound & upper bound & low-$T$ approx. & high-$T$ approx.\\
\midrule
0.020 & 0.464495 & 0.464550 & 0.464495 & 0.466181 \\
0.040 & 0.464495 & 0.464614 & 0.464495 & 0.466195 \\
0.060 & 0.464496 & 0.464832 & 0.464496 & 0.466209 \\
0.080 & 0.464518 & 0.465275 & 0.464519 & 0.466232 \\
0.100 & 0.464687 & 0.466195 & 0.464690 & 0.466353 \\
0.120 & 0.465281 & 0.467885 & \textcolor{blue}{0.465284} & 0.466826 \\
0.140 & 0.466644 & 0.470629 & 0.466642 & \textcolor{blue}{0.468005} \\
0.160 & 0.469091 & 0.474703 & 0.469066 & 0.470234 \\
0.180 & 0.472851 & 0.480254 & 0.472768 & 0.473775 \\
0.200 & 0.478081 & 0.487261 & 0.477865 & 0.478803 \\
0.220 & 0.484889 & 0.495551 & 0.484403 & 0.485446 \\
0.240 & 0.493400 & 0.504725 & 0.492372 & 0.493863 \\
0.260 & 0.503889 & 0.514141 & 0.501725 & -- \\
0.280 & 0.517833 & 0.521991 & 0.512392 & -- \\
\bottomrule
\end{tabular}
\endgroup
\captionof{table}{Planar energy $E/N^2$ at $g=-0.06$: canonical EEB bootstrap bounds and low- and high-temperature approximations. Blue entries mark the approximate crossover of their regimes of reliability. Dashes indicate that the high temperature theory is not reliable.}
\label{tab:energy-gm006}
\medskip
\medskip
\begingroup
\renewcommand{\arraystretch}{0.94}
\begin{tabular}{ccccc}
\toprule
$g$ & $T_{\mathrm{SDP}}^{\Lambda=16}$
    & $T_{\mathrm{SDP}}^{\Lambda\to\infty}$
    & high-$T$ approx. & low-$T$ approx.\\
\midrule
-0.0082500 & 2.623 & $2.583\pm0.027$ & 2.514226 & -- \\
-0.0155000 & 1.388 & $1.368\pm0.015$ & 1.323241 & -- \\
-0.0200000 & 1.069 & $1.051\pm0.012$ & 1.014610 & -- \\
-0.0227500 & 0.935 & $0.919\pm0.011$ & 0.884837 & -- \\
-0.0261111 & 0.806 & $0.791\pm0.011$ & 0.762166 & -- \\
-0.0300000 & 0.692 & $0.676\pm0.010$ & 0.653061 & -- \\
-0.0322222 & 0.640 & $0.623\pm0.010$ & 0.601850 & -- \\
-0.0383333 & 0.523 & $0.505\pm0.009$ & 0.489296 & -- \\
-0.0444444 & 0.435 & $0.418\pm0.009$ & 0.404402 & -- \\
-0.0505556 & 0.365 & $0.347\pm0.008$ & 0.336581 & -- \\
-0.0566667 & 0.304 & $0.288\pm0.008$ & 0.279459 & -- \\
-0.0627778 & 0.250 & $0.238\pm0.006$ & 0.228446 & -- \\
-0.0688889 & 0.196 & $0.188\pm0.004$ & 0.179004 & 0.154464 \\
-0.0720000 & 0.164 & $0.157\pm0.004$ & 0.152410 & 0.140011 \\
-0.0740000 & 0.134 & \textcolor{blue}{$0.127\pm0.004$} & 0.133970 & 0.119556 \\
-0.0750264 & 0.102 & \textcolor{blue}{$0.085\pm0.008$} & 0.123892 & 0 \\
\bottomrule
\end{tabular}
\endgroup
\captionof{table}{Canonical EEB bootstrap upper bounds on $T_c(g)$ at $\Lambda=16$, their extrapolated estimates, and high$-T$ and low$-T$ approximations. The high-temperature values use the eleven-term truncation of \eqref{highT-Tc-long-series}, through order $|g|^{19}$. Blue entries highlight the discrepancy with the low-temperature prediction near $g_*$. Dashes indicate that the low-temperature predictions outside are not reliable.}
\label{tab:tc-data}
\clearpage

\bibliographystyle{JHEP}
\bibliography{MyBib}
\end{document}